\documentstyle[epsf]{elsart}

\begin{document}

\newcommand{\Scat}{S}
\newcommand{\ScatOrb}{S^o}
\newcommand{\ScatOrbDag}{S^{o*}}
\newcommand{\ScatOrbDagQ}{S^{o\dagger}}
\newcommand{\Action}{{\cal S}_{\rm eff}}
\newcommand{\Lkern}{L}
\newcommand{\Nch}{N_{\rm ch}}
\newcommand{\NL}{N_1}
\newcommand{\NR}{N_2}
\newcommand{\VL}{V_1}
\newcommand{\VR}{V_2}
\newcommand{\Lambdamatrix}{\Lambda}
\newcommand{\Greenmatrix}{{\cal G}}
\newcommand{\Wmatrix}{W}

\newcommand{\Ec}{E_c}

\def\gtrsim{\:{{}^{{}_{\displaystyle >}}_{\displaystyle \sim}}\:}
\def\lesssim{\:{{}^{{}_{\displaystyle <}}_{\displaystyle \sim}}\:}          

\begin{frontmatter}
\title{Quantum Effects in Coulomb Blockade}
\author[Aleiner]{I.L. Aleiner,}
\author[Brouwer]{P.W. Brouwer,}
\author[Glazman]{L.I. Glazman}

\address[Aleiner]{Department of Physics, SUNY at Stony Brook,
  Stony Brook, NY 11794, USA}
\address[Brouwer]{Laboratory of Atomic and Solid State Physics,
Cornell University, Ithaca, NY 14853-2501, USA}
\address[Glazman]{Theoretical Physics
Institute, University of Minnesota, Minneapolis, MN 55455, USA}

\begin{abstract}
  We review the quantum interference effects in a system of interacting
  electrons confined to a quantum dot. The review starts with a
  description of an isolated quantum dot. We discuss the
  Random Matrix theory (RMT) of the one-electron states in the dot,
  present the universal form of the interaction Hamiltonian
  compatible with the RMT, and derive the leading corrections to the 
  universal interaction Hamiltonian. Next, we discuss a theoretical 
  description of a dot connected to leads via point contacts. Having 
  established the theoretical framework to describe such an open system,
  we discuss its transport and thermodynamic properties. We review the
  evolution of the transport properties with the increase of the
  contact conductances from small values to values $\sim
  e^2/\pi\hbar$. In the discussion of transport, the emphasis is put
  on mesoscopic fluctuations and the Kondo effect in the conductance.

\end{abstract}
\end{frontmatter}
\newpage
\tableofcontents
\newpage

\section{Introduction}
Conventionally, 
electric transport in bulk materials is characterized
by the conductivity $\sigma$. Then, the conductance $G$ of a
finite-size 
sample of dimensions $L_x \times L_y \times L_z$ can be found by
combining the
conductances of its smaller parts, $G= \sigma L_y L_z/L_x$. This
description, however, is applicable only at sufficiently high
temperatures,
at which the conductivity can be treated as a local quantity. It was
discovered about two decades ago~\cite{meso,meso1} that the quantum
corrections to the conductivity are non-local on the scale of the
(temperature dependent) dephasing length
$L_\varphi$, which is much larger than the elastic mean free
path. 
If the sample is small, or the temperature low, so that $L_{\varphi}$ 
exceeds the sample size,
the concept of conductivity loses its meaning. Moreover, the
conductance $G$ acquires significant sample-to-sample fluctuations and can no 
longer be treated as a self-averaging quantity. For a temperature 
$T$ below the Thouless energy $E_T = \hbar/\tau_D$,
these sample-to-sample fluctuations are of the 
order of the conductance quantum $e^2/2\pi\hbar$,
independent of sample size or mean free path ($\tau_D$ is the time 
it takes for an electron to diffuse through the system). This 
phenomenon is called
``Universal Conductance Fluctuations'' (UCF), and it was studied in
great detail both theoretically and experimentally,
see~\cite{AronovSharvin,mesoreview,mesoreview1,mesoreview2,mesoreview3} 
for a review. It is important to emphasize
that the lack of self-averaging of the conductance is a feature of
{\it mesoscopic} samples, {\it i.e.}, samples with linear dimensions
smaller than $L_\varphi$ but still much larger than the Fermi wave
length.

\begin{figure}
\epsfxsize=0.8\hsize
\centerline{\epsffile{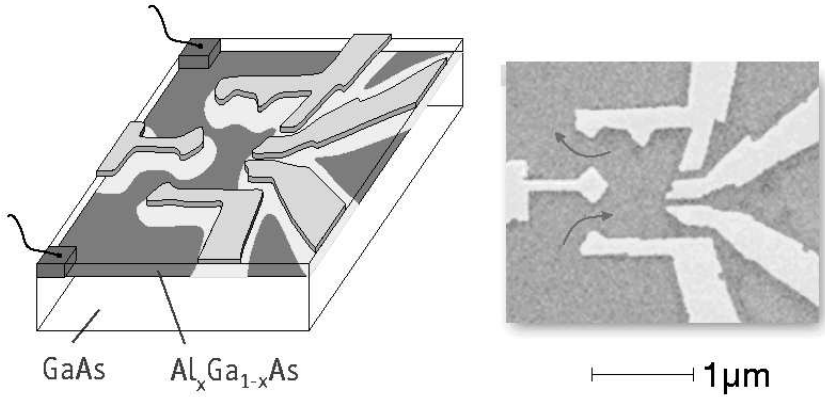}}
\caption{Micrograph of the quantum dot~\protect\cite{atom}. By
changing the voltages on the gates conductances of point contacts and
the shape of the dot can be changed.}
\label{Fig1}
\end{figure}

These ideas that defined the field of mesoscopic physics were
initially developed in the context of transport
through disordered metals.
A convenient way of creating more controllable mesoscopic samples was
developed with the use of semiconductor heterostructures, see
\cite{dots} for a review. As the electron motion
in these systems is quantized in the direction perpendicular to the plane of
the heterostructure, a two-dimensional electron gas is formed at the
interface between the layers.  With
the help of additional electrostatic confinement, usually in the form
of metal gates deposited on top of the hetrostructure, the 
two-dimensional electron gas (2DEG) can be tailored to form a
finite-size sample, referred to as a quantum dot.  
Its size, shape, and
connection with the rest of the 2DEG can be 
controlled
by means of a sophisticated system of electrostatic gates, see
Fig.~\ref{Fig1}.

Links connecting the quantum dot to the 2DEG (quantum point contacts)
are characterized by the number $\Nch$ of electron modes (or channels)
propagating through the contact at the Fermi level, and by the set of
transmission probabilities for these modes. With the increase of the
point contact cross-section, the number of modes, $\Nch$, is
increasing. If the boundaries of the point contact are smooth on the
length scale set by the electron wavelength, then the contact acts as
an electron wave guide~\cite{Glazman88}: all the propagating modes,
except the one which opened last, do not experience any appreciable
backscattering.\footnote{We refer the reader to Ref.~\cite{Kramer} for
a pedagogical presentation of the theory of adiabatic electron
transport} For the last mode, the transmission coefficient varies
from $0$ to $1$ in the crossover from the evanescent to propagating
behavior~\cite{mesoreview1}. In the specific case of a
gate-controlled contact, see Fig.~\ref{Fig1}, the distance between the
gate and the depleted edge of the 2DEG is typically large compared to
the electron wavelength in the electron gas~\cite{LGIL}. The curvature
radius of the edge of the 2DEG at the contact is of the order of
this distance, and therefore the smoothness condition is satisfied.

A statistical ensemble of quantum dots can be obtained by slightly
varying the shape of the quantum dot, the Fermi energy, or the
magnetic field, keeping the properties of its contacts to the
outside world fixed.
Statistical properties of the conductance for such an ensemble 
depend on the
number of modes, $\Nch$, in the junctions and their transparency. 
The ``usual'' UCF theory
adequately describes the conductance through a disordered or chaotic
dot in the limit of large number of modes, $\Nch\gg 1$. 
When all channels in the two point contacts are transparent, the universal
value of the root mean square (r.m.s.) conductance fluctuations 
$\langle \delta G^2 \rangle^{1/2} \simeq e^2/2\pi\hbar$
is much smaller than the average conductance
$\langle G\rangle \simeq \Nch e^2/2\pi\hbar$. The distribution of
conductances is essentially Gaussian, and can be found theoretically
by means of the
standard diagrammatic expansion for disordered systems~\cite{AGD}, or
with Random Matrix Theory (RMT), see Refs.~\cite{Beenakker97} and
\cite{Alhassid2000} for reviews. Upon closing the
point contacts ({\it i.e.}, decreasing $\Nch$), the average conductance
decreases whereas the r.m.s. fluctuations retains its universal 
value.  At
$\Nch\simeq 1$, the fluctuations and the average of the conductance are of
the same order, and the conductance distribution function is no 
longer Gaussian. Still, if one neglects the interaction between the
electrons, the statistics of the conductance can be readily
obtained within RMT \cite{Beenakker97,BarangerMello,Jalabert}.

The RMT of transport through quantum dots can be easily justified 
for non-inte\-racting electrons. However, real electrons are charged, 
and thus interact with each other. It is known from the theory
of bulk disordered systems, that the electron-electron interaction
results in corrections to the conductance~\cite{Altshuler85} of the
order of unit conductance quantum $e^2/2\pi\hbar$. For a quantum dot 
connected to a 2DEG by
single-mode junctions, this correction is not small anymore in
comparison to the average conductance $\langle G \rangle$.

We see that at $\langle G\rangle\simeq e^2/2 \pi\hbar$ and low
temperatures, both mesoscopic
fluctuations and interaction effects become strong. Quantum
interference and interaction effects
can not be separated, and there is no
obvious expansion parameter to use for building a theory
describing the statistics of the conductance in this
regime.
This paper reviews the current understanding of physical
phenomena in a wide class of mesoscopic systems -- quantum
dots -- under the conditions in which
effects of electron-electron interaction are strong.

A first step to approach this problem is to disregard the mesoscopic
fluctuations altogether. It means that one neglects the random spatial
structure of the wave functions in the dot and the randomness of the
energy spectrum; the discrete energy spectrum with spacing $\Delta$
between the one-electron levels in a closed dot is replaced by a
continuous spectrum with the corresponding macroscopic
density of states. For
realistic systems, this approach can be justified sometimes at not-so-low
temperatures. The remaining problem of accounting for the Coulomb
interaction is non-perturbative at small conductance 
$G \lesssim e^2/2\pi\hbar$, and
therefore still not trivial. It constitutes the essence of the
so-called Coulomb blockade phenomenon~\cite{AverinLikharev,dots}.

\begin{figure}
\epsfxsize=0.8\hsize
\centerline{\epsffile{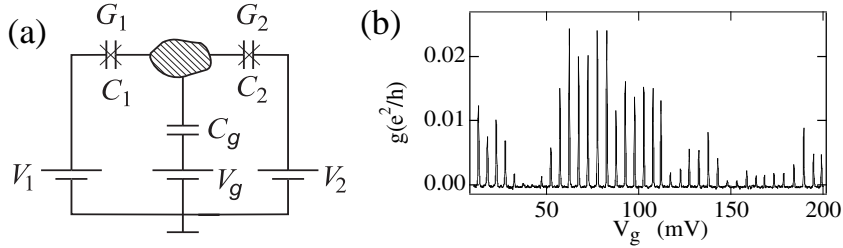}}
\caption{(a): Schematic view of the dot with the gate electrode; (b):
  conductance of the dot as a function of the gate voltage 
(from Ref.~\protect\cite{Folk96}).} 
\label{Fig2}
\end{figure}

The Coulomb blockade manifests itself most profoundly (see, e.g.,
Ref.~\cite{Kouwenhoven,dots}) in 
oscillations of the dot's conductance $G(V_g)$ with the
variation of the voltage $V_g$ on a gate, which is capacitively coupled
to the dot, see Fig.~\ref{Fig2}. The
resulting dependence $G(V_g)$ exhibits equidistant Coulomb blockade
peaks separated by deep minima (Coulomb blockade valleys).  The peaks
occur at the charge degeneracy points, {\it i.e.}, specific
values of $V_g$ at which changing the dot charge by a single quantum
$e$ does not cost any energy. The idea of the Coulomb blockade was 
suggested in the
early experimental paper~\cite{Zeller} in the late 1960's, though
the term ``Coulomb blockade'' was coined only two decades
later~\cite{slovo}. 
A quantitative theory in terms of rate equations describing the transport
through a blockaded quantum dot or metal grain at $G \ll
e^2/2\pi\hbar$, 
was formulated in Ref.~\cite{Shekhter,Kulik}, and was generalized to
systems with a controllable gate in Ref.~\cite{slovo}. This theory 
is commonly
referred to as the ``orthodox theory''. The main conclusion of the
orthodox theory is that the conductance through a blockaded grain at
low temperatures is exponentially suppressed. The first experiments on 
gated metallic Coulomb blockade systems were reported in Ref.~\cite{Fulton}.

At larger values of the dot's conductance (corresponding to larger
values of the conductances $G_1$ and $G_2$ of the point contacts
connecting the dot to the outside world), the orthodox theory and,
eventually, the phenomenon of Coulomb blockade break down because of
quantum fluctuations of the charge in the dot. The fluctuations
destroy the Coulomb blockade at the value $G\sim e^2/\pi\hbar$.
Quantum fluctuations affect both transport properties (such as linear
conductance and $I$-$V$ characteristics) and thermodynamics ({\it
e.g.}, differential capacitance). The $I$-$V$ characteristic in the
Coulomb blockade valleys at relatively small conductance was
considered in~\cite{AverinOdintsov} in second order perturbation
theory in $G$; later the corresponding non-linear contribution to the
current found in~\cite{AverinOdintsov} became known as ``inelastic
co-tunneling''~\cite{AverinNazarov90}.  Corrections induced by charge
fluctuations at $G\ll e^2/\pi\hbar$ were calculated
in~\cite{MatveevGlazman90} for the thermodynamic characteristics, and
in~\cite{Schoeller,Averin94,Furusaki95,Konig97} for the linear conductance. At
$G\ll e^2/\pi\hbar$ the problem of quantum charge fluctuations was
mapped~\cite{Matveev91} on the multi-channel Kondo
problem~\cite{LarkinMelnikov,NozieresBlandin}, which in some cases can
be solved exactly~\cite{Andrei84,Tsvelik84}.

Another limit which allows for a comprehensive solution, is the case
of a point contact with one almost open conducting channel linking 
the dot to the lead (conductance $G$ is close to the conductance
quantum)~\cite{Flensberg,Matveev95,Furusaki95}. In this limit, and still
neglecting mesoscopic fluctuations ($\Delta=0$), the Coulomb
blockade almost vanishes: the contribution
periodic in the gate voltage $V_g$ to the
differential capacitance, which is the prime signature of Coulomb
blockade, is small compared to the average
capacitance~\cite{Matveev95}. Moreover, it was
shown within this model that the Coulomb blockade disappears
completely in the case where the channel is fully transparent.

In all of the aforementioned works the effects of mesoscopic
fluctuations, associated with the finite size and
the discrete energy spectrum of the isolated dot (nonzero level
spacing $\Delta$),
were ignored. However, these fluctuations drastically
affect the results. In open dots, mesoscopic fluctuations 
reinstate periodic oscillations of the differential capacitance and
conductance as a function
of the gate voltage $V_{g}$, as a remnant the charge discreteness
in the dot, even in the limit of 
completely transparent channels in the point contacts. 
If the dot is almost closed, the
chaotic nature of the wave functions in the dot gives rise to 
mesoscopic fluctuations of observable quantities. Both
in the Coulomb blockade valleys and at the peaks, the conductance
turns out to have large fluctuations, and is very sensitive 
to a magnetic field.

Our main goal in this paper is to review the rich variety of effects
associated with the finite level spacing $\Delta$, 
the chaotic nature of the wave
functions, and the Coulomb interaction for both closed and open dots.
Here is a one-paragraph guide to the review.
Sections~\ref{sec:2},~\ref{sec:3}, and~\ref{sec:4} are devoted,
respectively, to the physics of an isolated quantum dot, transport
through a dot weakly connected to leads, and thermodynamic and
transport properties of a dot connected to lead(s) by one or two
almost
reflectionless quantum point contacts. We begin in Section~\ref{sec:2}
with the formulation of the model describing the statistical
properties of the system.  We clarify the conditions of the
equivalence of the usual diagrammatic technique and Random Matrix
representation, and establish the validity of the capacitive
interaction approximation. The central result of this part can be
found in Subsection~\ref{sec:hfinal}, where we present the universal
form of the interaction Hamiltonian describing the dot, and estimate
the accuracy of the universal description. In Section~\ref{sec:3} we
review the theory of mesoscopic fluctuations of the conductance in
peaks and valleys of the Coulomb blockade,\footnote{There is certain
overlap
with Ref.~\protect\cite{Alhassid2000}} and the theory of the Kondo effect
in quantum dots. Weak dot-lead coupling allows us to use the familiar
technique of the perturbation theory in tunneling strength. 
We believe it makes
Section~\ref{sec:3} easy to read. Nevertheless, a very pragmatic
reader who wants to grasp the main results first, without going
through the details, may start by reading the summary
in Subsection~\ref{sec:tdependence}.  Section~\ref{sec:4} deals with the
similar phenomena in quantum dots connected to leads by transparent or
almost transparent point contacts, where the interplay between quantum
charge fluctuations and mesoscopic physics becomes especially
interesting. Methods described and used in Section~\ref{sec:4} are
technically
more involved than in Section~\ref{sec:3}, and include the bosonization
technique along with the effective action formalism. To make the
navigation through the material easier, we decided to start this
Section with a presentation of the main results for the conductance 
through a
dot and for its thermodynamics, see Subsection~\ref{sec:overall}. Along
with providing the general picture for the conductance and differential
capacitance, this Subsection points to formulas valid in various
important limiting cases, which are derived later in
Subsections~\ref{sec:cap}--\ref{sec:g}.
Subsections~\ref{sec:finitesize}--\ref{sec:open} review the main
physical ideas built into the theory, while the details of the rigorous
theory are presented in Subsection~\ref{sec:rigorous}. Last but not
least, we illustrate the application of theoretical results by briefly
reviewing the existing experimental material, see
Subsections~\ref{sec:3.1}--\ref{sec:Kondo}, and~\ref{sec:dos}.

\section{The model}
\label{sec:2}

In Subsections~\ref{sec:2.1} and \ref{sec:mf} we discuss the
electronic properties of an isolated quantum dot. Under the 
assumption that electron motion inside the dot is chaotic, 
we establish a hierarchy of energy scales for the free-electron 
spectrum: Fermi energy, Thouless energy, and level spacing. 
The correlation between electronic eigenstates and
eigenvalues can be described by Random Matrix Theory (RMT) if the 
difference between the corresponding energy eigenvalues is smaller 
than the Thouless energy. In subsection \ref{sec:interaction} we 
proceed with the inclusion of effects of the interaction between 
the electrons. It is shown that the two-particle interaction matrix 
elements also  have hierarchical structure. The largest matrix elements 
correspond to charging of the quantum dot, as described by the 
charging energy in the
constant interaction model. Finally, open quantum dots are
discussed in Subsection~\ref{sec:2.2}, where
we introduce the Hamiltonian of the junctions connecting the dot with
the leads and relate the parameters of this Hamiltonian to the
characteristics of the quantum point contacts. 

\subsection{Non-interacting electrons in an isolated dot: Status of RMT}
\label{sec:2.1}
Let us start from a picture of non-interacting electrons, postponing
the introduction of electron-electron interactions to Sec.\
\ref{sec:interaction}.
The dot is described by the Hamiltonian
\begin{equation}
\hat{H}_{F}=\int d\vec{r}\left[\frac{1}{2m}\vec{\nabla}
\hat{\psi}^\dagger\vec{\nabla}\hat{\psi} +
U\left(\vec{r}\right)\hat{\psi}^\dagger\hat{\psi}\right],
\label{HF}
\label{eq:2.1}
\end{equation}
where the electron creation (annihilation) operators obey canonical
anticommutation relations 
\begin{equation}
  \left\{\hat{\psi}\left(\vec{r}_1\right),
\hat{\psi}\left(\vec{r}_2\right)\right\}=0,\ \
\left\{\hat{\psi}^\dagger\left(\vec{r}_1\right),
\hat{\psi}\left(\vec{r}_2\right)\right\}= \delta\left(\vec{r}_1
-\vec{r}_2 \right).
  \label{eq:anticomm}
\end{equation}
The potential $U(\vec{r})$ describes the
confinement of electrons to the dot, as well as the random potential
(if any) inside the dot.\footnote{We put $\hbar=1$ in all the
intermediate formulae.} We will neglect the spin-orbit
interaction; The one-particle Hamiltonian then is diagonal in spin
space, and for the time being we omit the spin indices.

The eigenfunctions of the Hamiltonian (\ref{eq:2.1}) are Slater 
determinants built from single-particle states with wavefunctions
defined by the Schroedinger equation
\begin{equation}
\left[-\frac{\vec{\nabla}^2}{2m}
 +
U\left(\vec{r}\right)\right]\phi_\alpha\left(\vec{r}\right)
=\varepsilon_\alpha \phi_\alpha\left(\vec{r}\right).
\label{eq:2.2} 
\end{equation}
In this orthonormal basis 
\begin{equation}
\hat{H}_{F}=\sum_\alpha
\varepsilon_\alpha\hat{\psi}^\dagger_\alpha\hat{\psi}_\alpha,
\label{eq:2.3} 
\end{equation}
where the fermionic operators $\hat\psi_{\alpha}$ are defined as
$\hat{\psi}_\alpha \equiv\int d\vec{r} \hat{\psi}\left(\vec{r}\right) 
\phi_\alpha^*\left(\vec{r}\right)$ and have the usual anticommutation
relations, cf.\ Eq.\ (\ref{eq:anticomm}).

Each particular eigenstate depends sensitively on the details of the random
potential $U(\vec{r})$, which is determined by the shape of the 
quantum dot. However, we will not be interested in the precise value
of observables that depend on the detailed realization of the potential
$U(\vec{r})$. Instead, our goal is a statistical description of the
various response functions of the system with respect to 
external parameters such as magnetic field, gate voltage, etc., and
of the correlations between the response functions at different 
values of those parameters. Hereto, the statistical properties of a 
response function are first related to the correlation function
of the eigenstates of the Hamiltonian (\ref{eq:2.2}). Then,
we can employ the known results for statistics of the eigenvalues
and eigenvectors in a disordered or a chaotic
system~\cite{AltshulerShklovskii,Berry,Efetovbook,MirlinReview}. 

For a disordered dot, the correlation functions are found by an 
average of the proper quantities over the realizations of the random
potential. Such averaging can be done by means of the standard
diagrammatic technique for the electron Green functions
\begin{equation}
{\cal G}^{R,A}(\varepsilon, \vec{r}_1, \vec{r}_2)=
\sum_\alpha\frac{\phi_\alpha^*(\vec{r}_2)\phi_\alpha(\vec{r}_1)}
{\varepsilon -\varepsilon_\alpha\pm i0},
\label{eq:2.4}
\label{GF}
\end{equation}
where plus and minus signs correspond to the retarded (R) and advanced
(A) Green functions respectively. 

The spectrum of one-electron energies is fully characterized
by the density of states
\begin{equation}
\nu (\varepsilon ) =
\sum_\alpha \delta \left(\varepsilon - \varepsilon_\alpha\right)
=\frac{1}{2\pi i} \int d\vec{r} \left[
{\cal G}^{A}(\varepsilon, \vec{r}, \vec{r}) -
{\cal G}^{R}(\varepsilon, \vec{r}, \vec{r})
\right],
\label{eq:2.5}
\end{equation}
where the last equality follows immediately from the definition
(\ref{eq:2.4}).
Even though the density of states is a strongly oscillating function of
energy, its average is smooth,
\begin{equation}
\langle\nu (\varepsilon )\rangle = \frac{1}{\Delta (\varepsilon)},
\label{eq:2.6}
\end{equation}
where ${\Delta (\varepsilon)}$ is the mean one-electron level spacing.
It varies on the characteristic scale of the order of the Fermi energy
$E_F$, measured, say, from the conduction band edge, which is the
largest energy scale in the problem. Since we are interested in
quantities associated with a much smaller energy scale, we can neglect the
energy dependence of the mean level spacing. The average in
Eq.~(\ref{eq:2.6}), denoted by brackets $\langle\dots\rangle$,
is performed over
the different realizations of the random potential for the case of a 
disordered dot, or over an energy strip of width $\gg \Delta$, but 
$\ll E_F$ for a clean ({\em i.e.}, ballistic) system.

\begin{figure}
\epsfxsize=0.8\hsize
\centerline{\epsffile{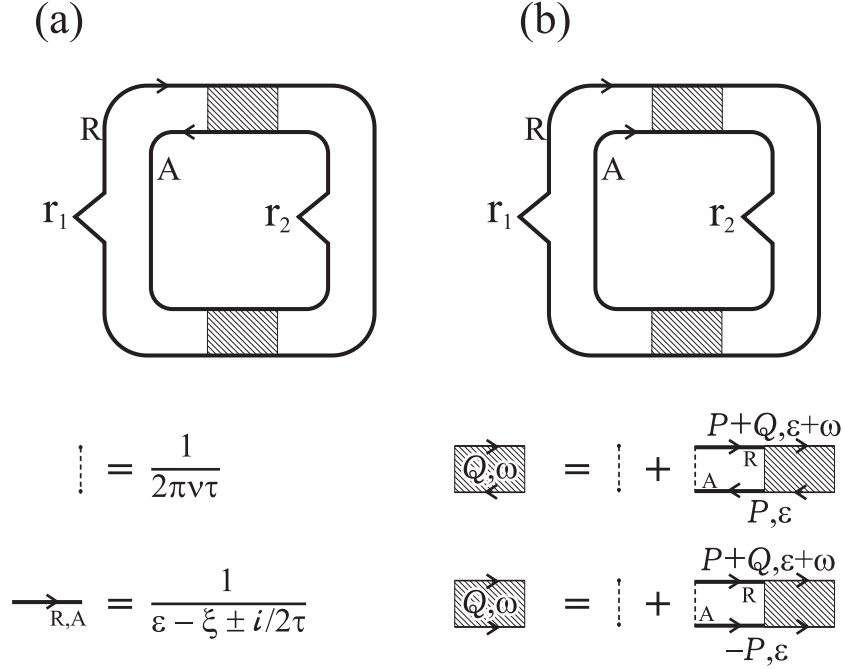}}
\caption{Diagrammatic expansion of the correlation of the density of
states for the disordered system~\protect\cite{AltshulerShklovskii}.
In the presence of magnetic field (unitary ensemble, $\beta =2$) 
the Cooperon diagram (b) is suppressed. In the figure, the symbols
$P$ and $Q$ refer to the momenta of the electrons; $P$ is the large
momentum, $|P-k_F | \ll k_F$, $Q$ is the small momentum
$|Q| \ll k_F$; $\xi = P^2/2m - \mu$, where $\mu$ is the chemical
potential; in two dimensions,
the mean free time $\tau$ is related to the diffusion constant $D$
through $D = v_F^2 \tau/2$, $v_F$ being the Fermi velocity.}
\label{Fig3}
\end{figure}

The average density of states does not carry any information about the
correlations between the energies of different eigenstates. Such
information is contained in the correlation functions for the electron
energy spectrum. Probably, the most important example is the two-point
correlation function ${\cal R}^{(2)}(\omega)$,
\begin{equation}
{\cal R}^{(2)}(\omega) = 
\Delta^2 \langle\nu (\varepsilon )
\nu (\varepsilon +\omega )\rangle - 1.
\label{eq:2.7}
\end{equation}
To calculate ${\cal R}^{(2)}(\omega)$, one substitutes the expression
(\ref{eq:2.5}) for the density of states in terms of
the Green functions $G^{R}$ and $G^{A}$ into Eq.~(\ref{eq:2.7}),
\begin{equation}
{\cal R}^{(2)}(\omega)=
\frac{\Delta^2}{2\pi^2}{\mathrm Re}\int d\vec{r}_1d\vec{r}_2
\langle
{\cal G}^{R}\left(\varepsilon +\omega, \vec{r}_1, \vec{r}_1\right)
{\cal G}^{A}\left(\varepsilon, \vec{r}_2, \vec{r}_2\right)
\rangle -1.
\label{eq:2.8}
\end{equation}
When the dot contains many electrons --- which is the case of
interest here --- the size of the dot and the region of integration 
in Eq.\ (\ref{eq:2.8}) greatly exceed the Fermi wavelength. In that
case, averaging of the products ${\cal
  G}^R{\cal G}^A$ using the diagrammatic technique yields two
contributions to the integrand in (\ref{eq:2.8}), proportional to 
the squares of diffuson and Cooperon propagators in coinciding
points, respectively, see Fig.~\ref{Fig3}.
The result of such a calculation~\cite{AltshulerShklovskii},
\begin{equation}
{\cal R}^{(2)}(\omega)=\frac{\Delta^2}{\beta \pi^2}
\mbox{Re}\,
\sum_{\gamma_n}
\frac{1}{\left(i\omega + \gamma_n\right)^2},
\label{eq:2.10a}
\end{equation}
is expressed in terms of the eigenvalues
$\gamma_n$ of the classical diffusion operator,
\begin{equation}
-D\vec{\nabla}^2 f_n(\vec{r}) =\gamma_n f_n(\vec{r}),
\label{eq:2.9}
\end{equation}
supplemented by von Neumann boundary conditions at the boundary of the
dot.  Here $D$ is the electron diffusion coefficient in the dot, and
the Dyson symmetry parameter $\beta=1$ ($2$) in the presence (absence)
of time reversal symmetry. Ensembles of random systems possessing and
not possessing this symmetry are called Orthogonal and Unitary
ensembles, respectively. Being derived by means of diagrammatic
perturbation theory, Eq.~(\ref{eq:2.10a}) and Eq.~(\ref{eq:2.10})
below are valid for energy differences $\omega \gg \Delta$ only. The
exact results, valid for all $\omega$, are also available, see, e.g.,
Refs.\ 
\cite{Efetovbook,MirlinReview,KravtsovMirlin,AndreevAltshuler,Benreview};
however, the approximation (\ref{eq:2.10a}) is sufficient for the
present discussion.  An expression similar to Eq.\ (\ref{eq:2.10a}) is
believed to be valid for a chaotic and ballistic quantum dot. The only
difference with the diffusive case is that instead of the eigenvalues
of the diffusion operator one has to use eigenvalues $\gamma_n$ of a
more general (Perron-Frobenius) operator\footnote{The 
corresponding non-linear {$\sigma$}-model was first 
suggested in Ref.~\protect\cite{Muzykantskii95}, however, 
this classical (Perron-Frobenius) operator was 
erroneously identified with the Liouville operator.} 
of the classical relaxational
dynamics~\cite{Agam96};\footnote{This 
identification does not properly handle the repetitions in
periodic orbits~\protect\cite{Bogomolny}, and therefore is applicable
only for the systems, where all the periodic orbits are unstable.}
in this case the eigenvalues $\gamma_n$ can be complex with
${\mathrm Re}\,\gamma_n>0$, see also Appendix~\ref{ap:0}.

The relation of ${\cal R}^{(2)}(\omega)$ to the dynamics of a particle
in a closed volume allows one to draw an important conclusion
regarding the universality of ${\cal R}^{(2)}(\omega)$. Note that
a spatially
uniform particle density satisfies the relaxational
dynamics or the diffusion equation. Because of particle number
conservation, such a solution is time independent. Therefore, the
lowest eigenvalue $\gamma_0$ in Eq.~(\ref{eq:2.9}) must be zero,
independent of the shape of the dot and the strength of the disorder,
\begin{equation}
{\cal R}^{(2)}(\omega)= -\frac{\Delta^2}{\beta\pi^2\omega^2}
+ 
\frac{\Delta^2}{\beta\pi^2}{\mathrm Re}
\sum_{\gamma_n \neq 0}
\frac{1}{\left(i\omega + \gamma_n\right)^2}\quad .
\label{eq:2.10}
\end{equation}
The first term in Eq.~(\ref{eq:2.10}) is universal. One can
immediately see that if $\hbar\omega$ is much smaller than the Thouless
energy $E_T \equiv \hbar{\mathrm Re}\,\gamma_1$,  the remaining terms in
Eq.~(\ref{eq:2.10}) are small compared to the first one and can be 
neglected.

The universal part of the correlation function
(\ref{eq:2.10}) can be reproduced within a model where the Hamiltonian
(\ref{HF}) is replaced by a hermitian matrix with random entries
\begin{equation}
\hat{H}_F = 
\sum_{\alpha,\gamma}
{\cal H}_{\alpha\gamma}\hat{\psi}^\dagger_\alpha\hat{\psi}_\gamma\quad .
\label{eq:2.11}
\end{equation}
The coefficients ${\cal H}_{\alpha\gamma}$ in Eq.~(\ref{eq:2.11}) 
form a real ($\beta =1$) or complex ($\beta=2$) random
Hermitian matrix ${{\cal H}}$ of size $M\times M$,  
belonging to the Gaussian ensemble
\begin{equation}
\langle{\cal H}_{\alpha\gamma}{\cal H}_{\alpha^\prime\gamma^\prime}\rangle
=\frac{M\Delta^2}{\pi^2}
\left[
\delta_{\alpha\gamma^\prime}\delta_{\alpha^\prime\gamma}
+ \left(2/\beta-1\right)
\delta_{\alpha\alpha^\prime}\delta_{\gamma\gamma^\prime}
\right]
, \quad
\beta=1,2. 
\label{eq:2.12}
\end{equation}
The Hamiltonian (\ref{eq:2.11}), with the distribution 
of matrix elements (\ref{eq:2.12})
reproduces the universal part of the spectral statistics of the 
microscopic Hamiltonian
(\ref{HF}) in the limit $M \to \infty$ of large matrix size. 
The condition $\Delta \ll E_T$, 
provides the significant region $\omega  \ll E_T$
where the non-universal part of the
two level correlator ${\cal R}_2(\omega)$ could be neglected. and
where  the replacement of the microscopic
Hamiltonian by the matrix model (\ref{eq:2.11}), is meaningful. 
This condition may be reformulated
as the requirement that the {\em dimensionless conductance} $g$ of
the dot be large, where
\begin{equation}
g\equiv \frac{\hbar\gamma_1}{\Delta} = \frac{E_T}{\Delta} \gg 1.
\label{eq:2.13}
\label{g}
\end{equation}
A large value of $g$ indicates that the dot can be treated as a
good conductor.

We now present explicit expressions for the
dimensionless conductance $g$ for a disc-shaped dot
for two simple models: (i) a dot of radius $R$ exceeding the electron
mean free path $l$, and (ii) a ballistic dot of radius $R$ with a
boundary which scatters electrons diffusively, see
Ref.~\cite{diffboundary,diffboundary1} and references therein.

For a diffusive dot, one needs to solve Eq.~(\ref{eq:2.9}) to obtain
$\gamma_1=Dx_1/R^2$, where $x_1\approx 2.40$ is the first
zero~\cite{Janke} of the Bessel function, $J_0(x_1)=0$. Using the
electron density of states in two-dimensional electron gas, one finds
$\Delta=2\hbar^2/mR^2$, and the conductance is
\[
g=\frac{x_1}{4}k_Fl=\frac{2\pi\hbar}{e^2}\frac{x_1}{4\pi R_\Box}.
\]
Here $k_F$ is the Fermi wave vector, and $R_\Box$ is the resistance
per square of the two-dimensional electron gas. Note that $g$ is
independent of the radius of the dot in the case of the diffusive
electron motion.

For the model (ii), the eigenvalue ${\mathrm Re}\,\gamma_1\approx
0.38v_F/R$ of the Liouville operator with the diffusive boundary
conditions was found in Ref.~\cite{diffboundary}, and one finds
\[
g=0.19k_FR.
\]

\subsection{Effect of a weak magnetic field on statistics
of one-electron states in the dot}
\label{sec:mf}

Physical properties of a mesoscopic system are manifestly random. The
statistics of the random behavior can be studied by a measurement of the
fluctuations caused by the application of an external magnetic field 
$B$. The role of such a field is primarily to alter the quantum
interference pattern. 
As a result, the
characteristic fields which  significantly affect the properties
of a mesoscopic system are rather weak in a classical sense:
 the cyclotron radius of an
electron remains much larger than the linear size of the dot, and the
effect of the magnetic fields on classical trajectories can be neglected.

A magnetic field can be included in the random matrix model of
Eq.\ (\ref{eq:2.11}). 
In order to include the magnetic field, we have to take the
 matrix ${\cal H}_{mn}$ in Eq.~(\ref{eq:2.11}) from a crossover
ensemble that interpolates between the orthogonal ($\beta =1$) and
unitary ($\beta=2$) ensembles
\cite{Mehtabook,PandeyMehta,MehtaPandey}:
\begin{equation}
{\cal H}_{\alpha\gamma} = \left(1+X^2\right)^{-1/2}
\left({\cal H}_{\alpha\gamma}^s + 
i X {\cal H}_{\alpha\gamma}^a\right).
\label{eq:2.98}
\end{equation}
Here ${\cal H}_{mn}^{s(a)}$ is the random realization of independent Gaussian
real symmetric (antisymmetric) $M\times M$ matrices,
\begin{equation}
\langle{\cal H}_{\alpha\gamma}^{a(s)}{\cal H}_{\alpha^\prime\gamma^\prime}^{a(s)}\rangle
=\frac{M\Delta^2}{\pi^2}
\left[
\delta_{\alpha\alpha^\prime}\delta_{\gamma\gamma^\prime}
\pm
\delta_{\alpha\gamma^\prime}\delta_{\alpha^\prime\gamma}
\right]
, 
\label{eq:2.99}
\end{equation}
where ``$+$'' (``$-$'') sign corresponds to the symmetric (antysimmetric)
part of the Hamiltonian.  In Eq.~(\ref{eq:2.98}), $X$ is a real parameter
proportional to the magnetic field (the precise relation between 
$X$ and $B$ is given below), so that the time reversal symmetry is 
preserved,
\begin{equation}
{\cal H}_{\alpha\gamma}(B)={\cal H}_{\gamma\alpha}(-B).
\label{eq:2.991}
\end{equation}
The prefactor $(1 + X^2)^{-1/2}$ in Eq.~(\ref{eq:2.98}) is chosen
in such a way that the mean level spacing $\Delta$ remains unaffected by
the magnetic field. 

The correlation function of elements of the Hamiltonian
(\ref{eq:2.98}) at different values of the magnetic field $B_{1,2}$ can be
conveniently written similarly to Eq.~(\ref{eq:2.12}) as
\begin{eqnarray} 
\label{eq:2.100}
\langle{\cal H}_{\alpha\gamma}(B_1){\cal H}_{\alpha' \gamma'}(B_2)\rangle
&=& \frac{M\Delta^2}{\pi^2}
\\ && \mbox{} \times
\left[\delta_{\alpha\gamma'}\delta_{\alpha'\gamma}
\left(1 - \frac{N_h^D}{4M}\right)
+\delta_{\alpha\gamma}\delta_{\alpha'\gamma'}
\left(1 - \frac{N_h^C}{4M}\right)
\right].\nonumber
\end{eqnarray}
The dimensionless quantities $N_h^{D,C}$, which characterize the effect 
of the magnetic field on the wave-functions and the spectrum of 
the closed dot, are related to the original parameters $X$ as 
\[
  N_h^{D} = 2 M \left( X_1 - X_2\right)^2,\ \
  N_h^{C} = 2 M \left( X_1 + X_2\right)^2,
\]
where the normalization is chosen in a way that the size of the
matrix $M$ does not enter into physical quantities in the physically
relevant limit $M \gg 1$. 

The relationship between the parameter $X$ and the real
magnetic field applied to the dot is best expressed with the help
of $N_h^{D}$ and $N_h^{C}$, 
\begin{equation}
N_h^{D} = \chi g
\left(\frac{\Phi_1 - \Phi_2}{\Phi_0}\right)^2,\quad
N_h^{C} = \chi g
\left(\frac{\Phi_1 + \Phi_2}{\Phi_0}\right)^2,
\label{eq:2.101}
\end{equation}
where $g\gg 1$ is the dimensionless conductance of the closed
dot, see Eq.\ (\ref{g}), 
$\Phi_{1(2)}$ is the magnetic flux through the dot at the 
magnetic field $B_{1(2)}$, $\Phi_0 = hc/e$ is the flux quantum,
and $\chi$ is a geometry dependent numerical factor of order unity. 
A derivation of Eq.~(\ref{eq:2.101}) is sketched at the end of
this subsection. We also refer the reader to Refs.\ 
\cite{FrahmPichard,Bohigas}. Here we give the values of the
constant $\chi$ for two specific cases: (i) $\chi=\pi/x_1\approx 1.31$ for a
disk-shaped diffusive dot, and (ii) $\chi\approx 2.78$ for a disc-shaped
ballistic dot with diffusive boundary scattering \cite{diffboundary1}.

The random matrix 
description (\ref{eq:2.98})---(\ref{eq:2.100}) is valid provided
$N_h^{D,C} \ll g$. At larger values of the magnetic field, a universal
description is no longer applicable. However, for most physical
quantities, the crossover between
the orthogonal and unitary ensembles takes place at $N_h^{D,C} 
\simeq 1$, {\em i.e.}, well within the regime where the RMT description 
is appropriate. 

A small magnetic field significantly affects the correlations of the
eigenvectors and eigenvalues of the random Hamiltonian
(\ref{eq:2.98}).
For the pure 
orthogonal and unitary ensembles ($\beta=1$ and $2$, respectively), 
the eigenvectors and eigenvalues are independent of each
other. Moreover, in the limit $M \to \infty$, 
the eigenvectors of different eigenstates are independently
distributed Gaussian variables,
\begin{equation}
\langle \psi_\alpha(i)\psi_\gamma^\ast(j) 
\rangle = \frac{\delta_{ij}\delta_{\alpha\gamma}}{M},
\quad
\langle \psi_\alpha(i)\psi_\gamma(j) 
\rangle
=\left(\frac{2}{\beta}-1\right)\frac{\delta_{ij}\delta_{\alpha\gamma}}{M}.
\label{eq:3.2.10}
\label{eq:2.102}
\end{equation}
In other words,
an eigenvector can be represented as 
\begin{equation}
\psi_\alpha (n) = \frac{1}{\sqrt{M}}
\frac{\left(u_n + it_\alpha v_n\right)}{\sqrt{1+t_\alpha^2}}
\label{eq:3.1.8}
\label{eq:2.103}
\end{equation}
where $M$ is the size of the matrix, $u, v$ are independent real
Gaussian variables, $\langle u_nu_m \rangle =\delta_{nm}$, $\langle
v_nv_m \rangle =\delta_{nm}$, $\langle u_nv_m \rangle =0$, and
$t_\alpha=0$ ($1$) for the orthogonal (unitary) ensemble 
\cite{PorterThomas,Mehtabook}.
Equation (\ref{eq:2.103}) is a consequence of the symmetry of the
distribution of the random matrices with respect to an arbitrary
rotation of the basis. At the crossover between those two ensembles,
this symmetry no longer exists, which leads to a departure of
the distribution of $\psi_{\alpha}(n)$ from a (real or complex)
Gaussian \cite{SommersIida94,FalkoEfetov94}, 
and to correlations of the values of the wave functions at
different sites \cite{FalkoEfetov96,FalkoEvetov96b}.  It was
noticed in Refs.~\cite{French88,Brouwer97} that the crossover
between the orthogonal and unitary ensembles can be described
by using the decomposition (\ref{eq:3.1.8})
where the parameter $t_\alpha$ is no longer fixed at the
extremal values $t=0$ or $t=1$, but a fluctuating
quantity $0 \le t \le 1$ with the distribution function 
\cite{SommersIida94,FalkoEfetov94,FalkoEfetov96,FalkoEvetov96b}
\begin{eqnarray}
W(t) &=&\left(\frac{\pi N_h^{C}}{8}\right)
\left(\frac{1-t^4}{t^3}\right)
\exp \left[-\left(\frac{\pi N_h^{C}}{16}\right)
\left(\frac{1-t^2}{t}\right)^2\right]\nonumber\\
&\times&
\left\{\phi\left(\frac{\pi N_h^{C}}{4}\right)
+
\left[
\left(\frac{1+t^2}{2t}\right)^2
-\frac{4}{\pi N_h^{C}}
\right]
\left[
1-\phi\left(\frac{\pi N_h^{C}}{4}\right)
\right]
\right\},
\nonumber\\
\phi(x)&=&\int_0^1 dy  \exp \left[-x \left(1-y^2\right)\right],
\label{eq:2.104}
\label{eq:3.1.14}
\end{eqnarray}
where $N_h^{C}$ is given by Eq.~(\ref{eq:2.101}) with $\Phi_1 = \Phi_2
\equiv\Phi$, being the magnetic flux through the dot.
We see, that the crossover between two ensembles occurs at
characteristic scale of the magnetic field $N_{h}^C \simeq 1$, or
using (\ref{eq:2.101}), at $\Phi = \Phi_{0}/(2\sqrt{\chi g})$.
Equation (\ref{eq:2.104}) completely describes the distribution of a
single eigenvector. In order to find the statistics of the quantities
contributed by several levels, it is necessary to find the joint
distribution of $t_\alpha$ of several levels, $W(t_1,t_2, \dots)$. The
exact form of such a distribution function is not known.

In the opposite limiting case when a physical result is contributed to
by a large number of levels, it is possible to use the 
conventional diagrammatic technique \cite{AGD}
to calculate different moments of the Green functions
\begin{equation}
{\cal G}^{R,A}_{ij}(\varepsilon)=
\sum_\alpha\frac{\psi_\alpha (i)\psi_\alpha^{\ast} (j)}
{\varepsilon -\varepsilon_\alpha\pm i0},
\label{eq:3.2.9}
\label{eq:2.105}
\end{equation}
see also Eq.~(\ref{GF}).
The diagrammatic calculation shown in Fig.~\ref{Fig10} yields 
\begin{eqnarray}
&& \left\langle
{\cal G}^{R}_{jk }(\varepsilon +\omega, B_1 )
{\cal G}^{A}_{lm }(\varepsilon,B_2)\right\rangle
=\frac{2\pi }{M^2 \Delta}\left(
\frac{\delta_{jm}\delta_{kl}}
{-i\omega + N_h^D\frac{\Delta}{2\pi} }
+\frac{\delta_{jl}\delta_{km}}
{-i\omega + N_h^C\frac{\Delta}{2\pi} }
\right)
,\nonumber \\
\label{eq:3.2.18}
\end{eqnarray}
where $N_h^{C}$ and $N_h^{D}$ are defined in Eq.~(\ref{eq:2.100}). 
The irreducible averages $\langle {\cal G}^{A}{\cal G}^{A}\rangle$ or
$\langle{\cal G}^{R}{\cal G}^{R}\rangle$ are smaller than
Eq.~(\ref{eq:3.2.18}) by factor of $\omega /(M\Delta)$ and can be
disregarded in the thermodynamic limit $M\to \infty$.  All the higher
moments can be expressed in terms of the second moments
(\ref{eq:3.2.18}) with the help of the Wick theorem.\footnote{It is
  noteworthy, that such a decomposition is legitimate only at energy
  scales greatly exceeding the level spacing $\Delta$. It is not valid
  for the calculation of the properties of a single wavefunction.}

Equation~(\ref{eq:3.2.18}) can serve as a starting point to relate 
the quantities $N_h^D$, $N_h^C$ to the real magnetic field, thus
providing a derivation of Eq.\ (\ref{eq:2.101}). To
establish such a relation, one needs to recalculate the average
(\ref{eq:3.2.18}) using the microscopic Hamiltonian (\ref{eq:2.1}).
The result is again given by Eq.\ (\ref{eq:2.10a}), where now the
magnetic field is introduced into the diffusion equation of Eq.\ 
(\ref{eq:2.9}) by the replacement of $\partial/\partial {\vec r}$
by the covariant derivative
$\partial/\partial{\vec r}+i(e/c\hbar){\vec A^\pm}({\vec r})$, and 
a modification of the boundary condition to ensure that the particle 
flux through the boundary remains zero. The vector-potentials 
${\vec A^\pm}({\vec r})$ are related to the magnetic fields 
${\vec B}_1$ and ${\vec B}_2$ by 
$\mbox{curl}\, {\vec A^\pm}={\vec B}_1\pm {\vec B}_2$. 
The corresponding lowest eigenvalues
$\gamma_0^-$ and $\gamma_0^+$ are identified with $(\Delta/2\pi)N_h^D$
and $(\Delta/2\pi)N_h^C$ respectively. A similar replacement can be also
made in the Liouville equation for the ballistic system.

{\begin{figure}[ht]
\epsfxsize=0.8\hsize
\centerline{\epsfbox{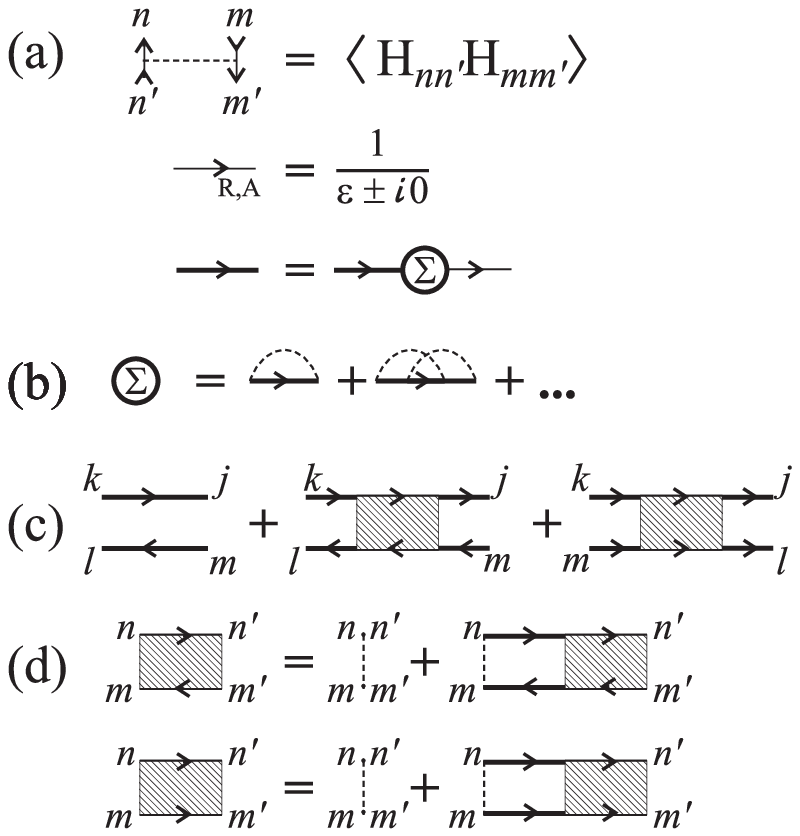}}
\caption{(a) analytic expressions for the  lines on the diagrams; 
(b) diagrams for self-energy $\hat{\Sigma}$. The second term in the
self-energy includes an intersection of the dashed lines ant it is
smaller than the first term by a factor $1/M$.
(c), (d) diagrammatic representation for irreducible averages
(\protect\ref{eq:3.2.18}). 
}
\vspace{0.5cm}
\label{Fig10}
\end{figure} 
}

We close this subsection with a more elaborate discussion 
of Eq.\ (\ref{eq:3.2.18}) and its physical interpretation. 
Hereto, we consider the case of
equal magnetic fields $B_1$ and $B_2$, {\em i.e.},\ $N_h^D = 0$. 
Then, one sees from Eq.\ (\ref{eq:3.2.18}) that a
weak magnetic field introduces new energy scale
\begin{equation}
\hbar/\tau_h^C=\Delta
N^C_h /2\pi
\label{eq:2.106}
\end{equation}
If this scale is smaller than the Thouless energy
$E_T$, the universal description holds, and we find that
the system under consideration is at the crossover between
the Orthogonal and Unitary ensembles. Phenomena associated with the
energy scale smaller than $\hbar/\tau_h^C$, are described effectively
by the unitary ensemble, whereas phenomena associated with larger
energy scales still can be approximately described by the orthogonal
ensemble. This conclusion allows for a simple semiclassical
interpretation of the relation (\ref{eq:2.101}) between $N_h^C$
and the real magnetic field $B$, as we now explain.

Consider a classical trajectory of an electron starting from
a point ${\bf r}$ and returning to the same point,
see Fig.~\ref{FigABflux}. The quantum mechanical amplitude for this
process contains an oscillating term $e^{i{\cal S}_{\rm cl}/\hbar}$, where
${\cal S}_{\rm cl}$ is the classical action along this trajectory. In a
weak magnetic field the classical trajectory does not change, while
the classical action acquires an Aharonov - Bohm phase:
\begin{equation}
{\cal S}_{\rm cl} \to {\cal S}_{\rm cl} + \frac{eB}{c}{\cal A}_{\rm cl},
\label{eq:2.107}
\end{equation}
where ${\cal A}_{\rm cl}$ is the directed area swept by the trajectory, see
Fig.~\ref{FigABflux}. If the time it takes for the electron to return
along the trajectory, $t$, would be of the order of the ergodic time
$\hbar/E_T$, the characteristic value of this area would be of the
order of the geometrical area of the dot $|{\cal A}_{\rm cl}| \simeq {\cal
  A}_{\rm dot}$
(the meaning of the Thouless energy $E_T$ is discussed in previous
subsections). If, however, $t \gg \hbar/E_T$, the electron trajectory
covers the dot $N_t \simeq tE_T/\hbar$ times. Each winding through the
dot adds a value of the order of ${\cal A}_{\rm dot}$ to ${\cal A}_{\rm cl}$; these
contributions, however, are of random signs, (an example is shown in
Fig.~\ref{FigABflux}). Therefore, the total area accumulated scales
with $\sqrt{N_t}$,
\begin{equation}
|{\cal A}_{\rm cl}|\simeq {\cal A}_{\rm dot}\sqrt{N_t} \simeq {\cal A}_{\rm dot}
\left(\frac{tE_T}{\hbar}
\right)^{1/2}.
\label{eq:2.108}
\end{equation} 
The orthogonal ensemble is different from the unitary ensemble by the 
fact that in the orthogonal ensemble the quantum mechanical amplitudes 
corresponding to a pair of time reversed trajectories have the same 
phase, while these phases are unrelated for the unitary ensemble. For
a small magnetic field, this means, that the trajectories which 
acquired an Aharonov-Bohm phase smaller than unity still effectively 
belong to the
orthogonal ensemble whereas, those which acquire a larger phase are
described by the unitary ensemble. To obtain the characteristic time
scale separating these two regimes, we require
\begin{equation}
\frac{eB |{\cal A}_{\rm cl}|}{c} \simeq \hbar.
\label{eq:2.109}
\end{equation}
Substituting estimate (\ref{eq:2.108}) into Eq.~(\ref{eq:2.109}), and 
using $\Phi=B{\cal A}_{\rm dot}$, one finds
\begin{equation}
\frac{1}{t} \simeq \frac{E_T}{\hbar}
\left(\frac{\Phi}{\Phi_0}\right)^2.
\label{eq:2.110}
\end{equation}
Up to a numerical coefficient, this estimate coincides with the
energy scale given by Eqs.~(\ref{eq:2.106}) and (\ref{eq:2.101}).  The
small energy scale physics is governed by long trajectories with a
typical accumulated flux larger than the flux quantum. For such
trajectories the interference between the time reversal paths is
already destroyed. At larger energy scales, the contributing
trajectories are shorter, and they do not accumulate enough magnetic
field flux to destroy the interference.

{\begin{figure}[ht]
\epsfxsize=0.6\hsize
\centerline{\epsfbox{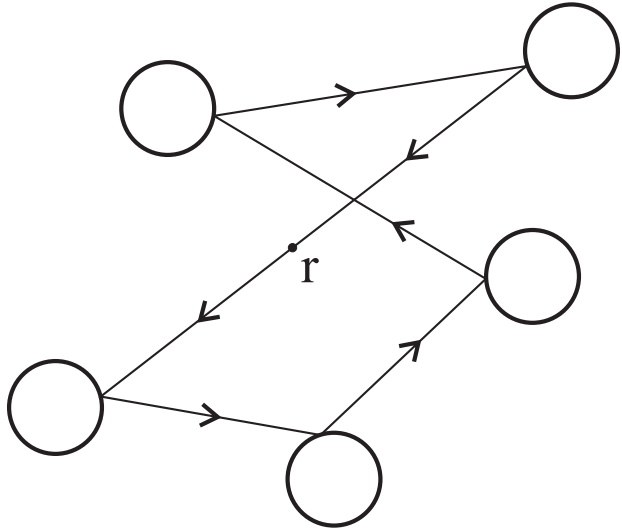}}
\caption{Example of the self-returning trajectories. 
}
\vspace{0.5cm}
\label{FigABflux}
\end{figure} 
}

\subsection{Interaction between electrons:  The universal description}
\label{sec:interaction}

Let us now turn to the discussion of interaction between electrons in
a quantum dot.  In the basis of eigenfunctions $\phi_\alpha$ of
the free-electron Hamiltonian, the two-particle interaction takes the
form:
\begin{equation}
\hat{H}_{\rm int}= \half\sum {\cal H}_{\alpha\beta\gamma\delta}
\psi^\dagger_{\alpha,\sigma_1} \psi^\dagger_{\beta,\sigma_2} 
\psi_{\gamma,\sigma_2}\psi_{\delta,\sigma_1}.
\label{eq:2.14}
\end{equation}
Hereinafter, we will write explicitly the spin indices $\sigma$
for  the fermionic operators. The generic matrix element of interaction
is:
\begin{equation}
{\cal H}_{\alpha\beta\gamma\delta}
=\int d\vec{r}_1 d\vec{r}_2 V(\vec{r}_1-\vec{r}_2)
\phi_{\alpha}(\vec{r}_1)
\phi_{\beta}(\vec{r}_2)
\phi_{\gamma}^*(\vec{r}_2)
\phi_{\delta}^*(\vec{r}_1).
\label{eq:2.15}
\end{equation}

Our goal is to show that the matrix elements of the interaction
Hamiltonian also have hierarchical
structure: Only a few of these elements are large and universal, whereas the
majority of them are proportional to the inverse dimensionless
conductance $1/g$, and thus 
small
~\cite{KamenevLevitov,Agam97,Blanter96,MirlinBlanter,AleinerGlazman98}.
As the result, Hamiltonian (\ref{eq:2.14}) will be separated into two
pieces:
\begin{equation}
\hat{H}_{\rm int}= \hat{H}_{\rm int}^{(0)} + \hat{H}_{\rm int}^{(1/g)}.
\label{eq:2.150}
\end{equation}
The first term here is universal, does not depend on the geometry of
the dot, and it does not fluctuate from sample to sample for samples
differing only by realizations of disorder.  The second term is small
as a power of $1/g$, and it fluctuates. This term only weakly affects the
low-energy ($E\lesssim E_T$) properties of the system.

The form of the universal term can be established using the
requirement of compatibility of this term with the Random Matrix model
(\ref{eq:2.11}). Since the Random Matrix distribution
is invariant with respect to an arbitrary rotation of the basis, the
operator $\hat{H}_{\rm int}^{(0)}$ may include only operators invariant
under such rotations. In the absence of the spin-orbit interaction,
there are three such operators:
\begin{equation}
 \hat{n} =\sum_{\alpha,\sigma} 
\hat{\psi}^\dagger_{\alpha,\sigma }
\hat{\psi}_{\alpha,\sigma };
\quad
\hat{\vec{S}} =\half\sum_\alpha
\hat{\psi}^\dagger_{\alpha,\sigma_1 }
\vec{\sigma}_{\sigma_1 \sigma_2}
\hat{\psi}_{\alpha,\sigma_2 };\quad
\hat{T} = \sum_\alpha
\hat{\psi}_{\alpha,\uparrow }
\hat{\psi}_{\alpha,\downarrow }.\label{eq:2.151}
\end{equation}
The operator $\hat{n}$ is the total number of particles, $\hat{\vec{S}}$
is the total spin of the dot, and operator $\hat{T}$ corresponds
to the interaction in the Cooper channel. (For the unitary ensemble
the operator $\hat{T}$ is not allowed.)

Gauge invariance requires that only the product $T^{\dagger} T$ of
the operators ${\hat T}$ and ${\hat T}^\dagger$ can enter into the 
Hamiltonian. At the same time, $SU(2)$
symmetry dictates that the Hamiltonian may depend only on
$({\hat{\vec S}})^2$, and not on separate components of the
spin vector. Taking into account that the initial interaction
Hamiltonian (\ref{eq:2.14}) is proportional to $\psi^4$, we find 
for its universal part:
\begin{equation}
\hat{H}_{\rm int}^{0}=\Ec \hat{n}^2 + J_S \left( \hat{\vec{S}}\right)^2 +
J_c\hat{T}^\dagger\hat{T}.
\label{eq:2.152}
\end{equation}
The three terms in the Hamiltonian (\ref{eq:2.152}) have a different
meaning.  The first two terms represent the dependence of the energy
of the system on the total number of electrons and total spin,
respectively.  Because both total charge and spin commute with the
free-electron Hamiltonian, these two terms do not have any dynamics for a
closed dot. We will see that the situation will change with the
opening of contacts to the leads. Finally, the third term corresponds
to the interaction in the Cooper channel, and it does not commute with
the free-electron Hamiltonian (\ref{eq:2.1}) or (\ref{eq:2.11}). This term is
renormalized if one considers contributions in higher order perturbation
theory in the interaction. For an attractive interaction, $J_c < 0$, the 
renormalization enhances
this interaction, eventually leading to the superconducting
instability. (We will not consider the case of an attractive
interaction here; for a recent review on the physics of small
superconducting grains, see Ref.~\cite{vonDelft}. Effects in larger
grains, $\Ec\gg\Delta$, are reviewed in Ref.~\cite{IntJ}.)
For the repulsive case, $J_c > 0$, this term renormalizes
logarithmically to zero. We will be dealing with the latter case throughout
the paper.

Constants in the Hamiltonian (\ref{eq:2.152}) are model-dependent.
In the remainder of this subsection we will show how to calculate them
for some particular interactions and discuss the structure of
the non-universal part ${\hat H}^{(1/g)}$.

At this point it is important to mention that universal Hamiltonian
(\ref{eq:2.152}) is defined within the Hilbert space of one-electron
states with energies of the order or less than Thouless energy.  The
matrix elements of this Hamiltonian are not just the matrix elements
of the interaction potential (\ref{eq:2.15}). This potential is
defined in a much wider energy strip which includes one-electron
states with energies $\simeq E_F$. Virtual transitions between the
low-energy ($\lesssim E_T$) sector and these high-energy states
renormalize the matrix elements of the universal Hamiltonian.  It
turns out~\cite{AGD} that the third term in the Hamiltonian
(\ref{eq:2.152}) corresponding to the Cooper channel of interaction,
can be strongly renormalized by such virtual transitions. 
In order to avoid this complication, we
will consider the case of the unitary ensemble in the discussion
below, where the ``bare'' matrix
elements corresponding to the Cooper channel are already suppressed by
a weak magnetic field (it is sufficient to thread a flux of the order
of unit quantum $\Phi_0$ through the cross-section of a dot), and
relegate the corresponding discussion of the interaction in the
Cooper channel to the Appendix~\ref{ap:1}. Omission of the Cooper
channel here is a
simplification that will not affect our principal conclusions, as we
will consider the case of repulsive interaction where the Cooper
channel, if it were included, is renormalized to zero anyway.

The statistics of the interaction
matrix elements can now be related to the properties
of the one-electron wave functions. The easiest way to study the statistics
of the wave-functions $\phi_\alpha (\vec{r})$ is to relate them to the Green
functions (\ref{GF})  and then use the diagrammatic technique for the
averaging of their products.  From Eq.~(\ref{GF}) we have
\begin{equation}
{\cal G}^A(\varepsilon; \vec{r}_1,\vec{r}_2 ) -
{\cal G}^R(\varepsilon; \vec{r}_1,\vec{r}_2 ) 
=2\pi i \sum_\alpha \phi_\alpha (\vec{r}_1)\phi_\alpha^*(\vec{r}_2)
\delta (\varepsilon - \varepsilon_\alpha).
\label{eq:2.17}
\end{equation}
At given energy $\varepsilon$, at most one eigenstate contributes to the
sum in Eq.~(\ref{eq:2.17}).  Furthermore, it is known that there is no
correlation between the statistics of levels and that of wave
functions in the lowest order in $1/g$, see, {\em e.g.},
Ref.~\cite{AltshulerShklovskii}, so we can neglect the level
correlations and average the $\delta$-function in Eq.~(\ref{eq:2.17})
independently.  As a result, we can estimate
\begin{equation}
\phi_\alpha (\vec{r}_1)\phi_\alpha(\vec{r}_2)^*
\theta(\delta/2-|\varepsilon_\alpha-\varepsilon|)
\approx 
\frac{1}{2\pi}\!\!\!
\int_{\varepsilon - \delta/2}^{\varepsilon + \delta/2}
d\varepsilon_1 \left[{\cal G}(\varepsilon_1; \vec{r}_1,\vec{r}_2 )
\right]_-,\,
\delta\ll\Delta,
\label{eq:2.18}
\end{equation}
where we introduced the notation
\begin{equation}
\left[ {\cal G} (\varepsilon; \vec{r}_1,\vec{r}_2 ) \right]_-=
-i\left[{\cal G}^A(\varepsilon; \vec{r}_1,\vec{r}_2 ) -
{\cal G}^R(\varepsilon;
\vec{r}_1,\vec{r}_2)\right].
\label{eq:2.180}
\end{equation}

Let us first calculate the average of the matrix element
(\ref{eq:2.15}). Because of the randomness of the wave functions, the
corresponding product in Eq.~(\ref{eq:2.15}) does not vanish only if
its indices are equal pairwise.  It is then readily expressed with the
help of Eq.~(\ref{eq:2.18}),
\begin{eqnarray}
&&\frac{\delta^2}{\Delta^2}\langle\phi_\alpha (\vec{r}_1)\phi_\beta (\vec{r}_2)
\phi_\gamma^* (\vec{r}_3)\phi_\delta^* (\vec{r}_4)\rangle
\nonumber\\
&&=\frac{1}{4\pi^2}
\int_{- \delta/2}^{\delta/2}d\varepsilon_1\int_{-
\delta/2}^{\delta/2}d\varepsilon_2
\left[\delta_{\alpha\delta}\delta_{\beta\gamma}
\langle
\left[ {\cal G}(\varepsilon_\alpha+\varepsilon_1; \vec{r}_1,\vec{r}_4 )\right]_-
\left[ {\cal G}(\varepsilon_\beta +\varepsilon_2;
\vec{r}_2,\vec{r}_3 )\right]_-
\rangle
\right. \nonumber \\
&&\left.+
\delta_{\alpha\gamma}\delta_{\beta\delta}
\langle
\left[ {\cal G}
(\varepsilon_\alpha+\varepsilon_1; \vec{r}_1,\vec{r}_3 )\right]_-
\left[ {\cal G}
(\varepsilon_\beta +\varepsilon_2;
\vec{r}_2,\vec{r}_4 )\right]_-
\rangle
\right]
\label{eq:2.19}
\end{eqnarray}
In deriving Eq.~(\ref{eq:2.19}) for $\alpha\neq\beta$ we have used the
relation
\begin{eqnarray*}
&&\langle\phi_\alpha (\vec{r}_1)\phi_\alpha(\vec{r}_2)^*
\phi_\beta (\vec{r}_1)\phi_\beta(\vec{r}_2)^*
\theta(\delta/2-|\varepsilon_\alpha-\varepsilon|)
\theta(\delta/2-|\varepsilon_\beta-\varepsilon|)\rangle\\
&&=\langle\phi_\alpha (\vec{r}_1)\phi_\alpha(\vec{r}_2)^*
\phi_\beta (\vec{r}_1)\phi_\beta(\vec{r}_2)^*
\rangle
\langle\theta(\delta/2-|\varepsilon_\alpha-\varepsilon|)
\theta(\delta/2-|\varepsilon_\beta-\varepsilon|)\rangle\\
&&=\frac{\delta^2}{\Delta^2}
\langle\phi_\alpha (\vec{r}_1)\phi_\alpha(\vec{r}_2)^*
\phi_\beta (\vec{r}_1)\phi_\beta(\vec{r}_2)^*\rangle.
\end{eqnarray*}
The result (\ref{eq:2.19}) can be justified also for $\alpha=\beta$
and $k_F|\vec{r}_1-\vec{r}_2| \gg 1$ (with $k_F$ being the Fermi
wavevector), see Ref.~\cite{MirlinBlanter}. The corrections to
Eq.~(\ref{eq:2.19}) are of the order or smaller than $1/g^2$, and will
be neglected since we are interested in the leading in $1/g$ terms.

In the leading approximation in $1/g$ needed to derive
$\hat{H}^{(0)}_{\rm int}$, the Green function entering into the products
in the above formula may be averaged independently.  Substituting
\begin{eqnarray}
  && \langle
\left[ {\cal G}
(\varepsilon_\alpha+\varepsilon_1; \vec{r}_1,\vec{r}_2 )
\right]_-
\rangle = \frac{2\pi}{\Delta {\cal V}_d} 
{\cal F}\left(k_F|\vec{r}_1-\vec{r}_2|\right); \label{eq:calF} \\
  && {\cal F}\left(k_F|\vec{r}|\right) = 
\langle e^{i\vec{k}\cdot\vec{r}}\rangle_{FS}, \nonumber
\end{eqnarray}
with $\langle \dots\rangle_{FS}$ denoting the average over the electron
momentum on the Fermi surface, into Eq.~(\ref{eq:2.19}), we find
\begin{equation}
{\left( {\cal V}_d\right)^2}
\langle\phi_\alpha (\vec{r}_1)\phi_\beta (\vec{r}_2)
\phi_\gamma^* (\vec{r}_3)\phi_\delta^* (\vec{r}_4)\rangle^{(0)}=
\label{eq:2.20} \\
\left[\delta_{\alpha\delta}\delta_{\beta\gamma}
{\cal F}_{14}
{\cal F}_{23}+
\delta_{\alpha\gamma}\delta_{\beta\delta}
{\cal F}_{13}
{\cal F}_{24}
\right],
\end{equation}
where we introduced the short-hand notation
\begin{equation}
{\cal F}_{ij}\equiv{\cal F}\left(k_F|\vec{r}_i-\vec{r}_j|\right),
\label{eq:2.200}
\end{equation}
and ${\cal V}_d$ is the volume of a $d$-dimensional grain.
Equation (\ref{eq:2.20}) does not depend on the disorder strength and 
corresponds to the universal limit. The coordinate dependence in
function ${\cal F}$  indicates simply that all the plane waves that 
are allowed by the conservation of energy are represented in the wave
function.

\subsubsection{Universal description for the case of short-range interaction}
\label{shortrange}

\addcontentsline{toc}{subsection}{
\protect\numberline{\ref{shortrange}}{\ 
Universal description for the case of short-range interaction}}

We start with the model case of a weak short range interaction
\begin{equation}
V(\vec{ r}) = \lambda \Delta {\cal V}_d
 \delta(\vec{ r}) 
\label{eq:2.16}
\end{equation} 
to illustrate the principle, and then discuss the realistic long range
Coulomb interaction. In Eq.~(\ref{eq:2.16}), ${\cal V}_d$ is the volume
of the dot in dimension $d=3$ or its area for $d=2$, and $\lambda \ll
1$ is the interaction constant. 

Averaging of the matrix elements (\ref{eq:2.15}) with the help of
Eq.~(\ref{eq:2.20}) allows us to derive the Hamiltonian
$\hat{H}^{(0)}_{\rm int}$ from Eq.~(\ref{eq:2.14}). Indeed, such averaging
yields
\begin{equation}
{\cal H}_{\alpha\beta\gamma\delta}^{(0)}= \lambda \Delta
\left(\delta_{\alpha\delta}\delta_{\beta\gamma}+
\delta_{\alpha\gamma}\delta_{\beta\delta}\right). 
\label{eq:2.24} 
\end{equation}
Now we substitute Eq.~(\ref{eq:2.24}) into Eq.~(\ref{eq:2.14}) and
rearrange the summation over spin indices with the help of identity
\[
2\delta_{\sigma_1\sigma_2}\delta_{\sigma_3\sigma_4}=
\delta_{\sigma_1\sigma_3}\delta_{\sigma_2\sigma_4}+
\vec{\sigma}_{\sigma_1\sigma_3}\vec{\sigma}_{\sigma_4\sigma_2},
\]
where $\vec{\sigma} =
\left(\hat{\sigma}^x,\hat{\sigma}^y,\hat{\sigma}^z\right)$ and
$\hat{\sigma}^k$ are the Pauli matrices in the spin space.  As a
result, the interaction Hamiltonian takes the universal form
(\ref{eq:2.151}), (\ref{eq:2.152}), with the coupling constants
\begin{equation}
\Ec = \quart \lambda \Delta; \quad J_S = -\lambda \Delta; \quad J_c =0.
\label{eq:2.240}
\label{eq:2.26}
\end{equation}
The vanishing of $J_c$ is a feature of the unitary ensemble, see also
Appendix~\ref{ap:1}. We see that for a weak {\it short-range} repulsive
interacton all the matrix elements in the universal part of the
Hamiltonian are smaller than the level spacing. Later we will see that
for the {\it Coulomb} interaction this is not the case, and the
constant $\Ec$ is large compared with $\Delta$.

We derived the universal Hamiltonian (\ref{eq:2.152}) in the leading
approximation, which corresponds to the limit $1/g\to 0$. In this
approximation, only the ``most diagonal'' matrix elements of the
interaction Hamiltonian are finite. In the first order in $1/g$, there
are two types of the corrections $H_{\alpha\beta\gamma\delta}^{(1/g)}$
to the universal Hamiltonian. First, the matrix elements of the
diagonal part of the Hamiltonian, Eq.~(\ref{eq:2.152}), acquire a
correction $\propto 1/g$. This correction exhibits mesoscopic
fluctuations and has a non-zero average value. Second, the
non-diagonal matrix elements of the interaction Hamiltonian for a
mesoscopic quantum dot become finite, but with zero average value.

\begin{figure}
\epsfxsize=0.6\hsize
\centerline{\epsffile{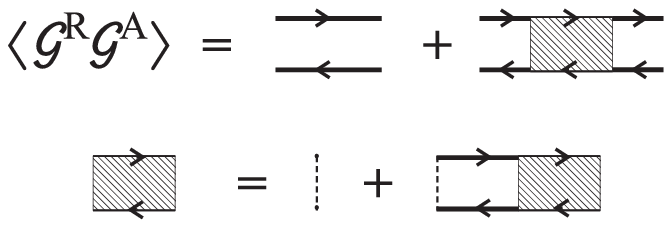}}
\caption{Diagrammatic expansion for the correlation of the Green functions
of the disordered system 
$\langle {\cal G^R}  {\cal G^A}\rangle$.}
\label{Fig4}
\end{figure}

We start with the evaluation of the average correction
$\langle H_{\alpha\beta\gamma\delta}^{(1/g)}\rangle$ to the matrix
elements $H_{\alpha\beta\gamma\delta}^{(0)}$ of the Hamiltonian
(\ref{eq:2.152}),
\begin{equation}
\langle H_{\alpha\beta\gamma\delta}^{(1/g)}\rangle=
\langle{H_{\alpha\beta\gamma\delta}}\rangle-
H_{\alpha\beta\gamma\delta}^{(0)}.
\label{eq:2.23}
\end{equation}
To this end, we have to find the $1/g$ contribution to the averages of
the products (\ref{eq:2.19}) of wave functions. This contribution can
be found from the corresponding irreducible product of the Green
functions. Using diagrams for the diffusive systems, see
Fig.~\ref{Fig4}, we find
\begin{eqnarray}
&&\langle
\left[ {\cal G}
(\varepsilon+\omega; \vec{r}_1,\vec{r}_2 )\right]_-
\left[ {\cal G}(\varepsilon;
\vec{r}_3,\vec{r}_4 )\right]_-
\rangle_{\rm ir} =
2 {\mathrm Re}\langle
{\cal G}^R(\varepsilon +\omega; \vec{r}_1,\vec{r}_2 )
{\cal G}^A(\varepsilon;
\vec{r}_3,\vec{r}_4 )
\rangle_{\mathrm ir}
\nonumber\\
&&= \frac{4\pi}{\Delta {\cal V}_d}
{\cal F}_{14}
{\cal F}_{23}
\sum_{\gamma_n > 0}
\frac{\gamma_n\  f_n(\vec{r}_1)  f_n(\vec{r}_2) }
{\left(\omega^2 + \gamma_n^2\right)}.
\label{eq:2.21}
\end{eqnarray}  
Here $\gamma_n$ and $f_n$ are defined in Eq.~(\ref{eq:2.9}),
and normalization condition 
\begin{equation}
\int  d\vec{r} f_n(\vec{r})  f_m(\vec{r}) = \delta_{nm},
\label{eq:2.210}
\end{equation}
is imposed.
Substituting Eq.~(\ref{eq:2.21}) into Eq.~(\ref{eq:2.19}), and using
the condition $\Delta \ll \gamma_n$ we will obtain the $1/g$
correction to the average product of the wave functions (\ref{eq:2.20}):
\begin{eqnarray}
&&\langle
\phi_\alpha (\vec{r}_1)\phi_\beta (\vec{r}_2)
\phi_\gamma^* (\vec{r}_3)\phi_\delta^* (\vec{r}_4)
\rangle^{(1/g)}=\delta_{\alpha\delta}\delta_{\beta\gamma}
{\cal F}_{13}
{\cal F}_{24}
\frac{\Delta}{\pi{\cal V}_d}
\sum_{\gamma_n > 0} 
\frac{\gamma_n\  f_n(\vec{r}_1)  f_n(\vec{r}_4) }
{\left(\varepsilon_\alpha-\varepsilon_\beta \right)^2 + \gamma_n^2}
\nonumber \\ && \mbox{}
+
\delta_{\alpha\gamma}\delta_{\beta\delta}
{\cal F}_{14}
{\cal F}_{23}
\frac{\Delta}{\pi{\cal V}_d}
\sum_{\gamma_n > 0} 
\frac{\gamma_n\  f_n(\vec{r}_1)  f_n(\vec{r}_3) }
{\left(\varepsilon_\alpha-\varepsilon_\beta \right)^2 + \gamma_n^2}
,
\label{eq:2.22}
\end{eqnarray}
where we used the short-hand notation (\ref{eq:2.200}).

We are interested in the interactions between the
electrons in states sufficiently close to the Fermi surface (within an
energy strip $\sim E_T$ around the Fermi level). It means that the
difference of one-electron eigenenergies entering in the denominators
of Eq.~(\ref{eq:2.22}) are much smaller than $\gamma_n$, and can be
neglected. Substituting Eq.~(\ref{eq:2.22}) into Eqs.~(\ref{eq:2.15})
and (\ref{eq:2.16}), and assuming $\left|\varepsilon_\alpha
-\varepsilon_\beta\right| \ll E_T$, we find
\begin{equation}
\langle H_{\alpha\beta\gamma\delta}^{(1/g)}\rangle = 
c_1\lambda \frac{\Delta}{g}
\left(\delta_{\alpha\delta}\delta_{\beta\gamma}+
\delta_{\alpha\gamma}\delta_{\beta\delta}\right);
\quad c_1= \frac{1}{\pi}\sum_{\gamma_n >0} \frac{\gamma_1}{\gamma_n}
\simeq 1,
\label{eq:2.27}
\end{equation}
where the dimensionless conductance of the dot $g$ is defined in
Eq.~(\ref{g}).  Thus, we have shown that in the metallic regime, $g\gg
1$, the non-universal corrections to the average interaction matrix
elements are small indeed.

\begin{figure}
\epsfxsize=0.8\hsize
\centerline{\epsffile{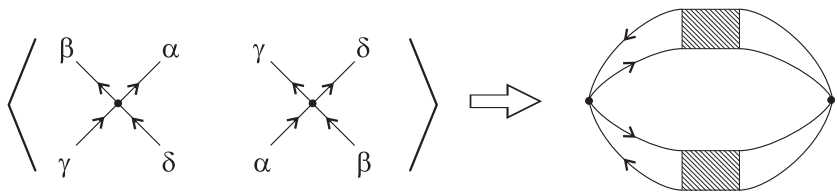}}
\caption{Diagrammatic expansion for the fluctuations of the matrix
element in the disordered systems.}
\label{Fig5}
\end{figure}

Now we turn to the mesoscopic fluctuations $\delta
H_{\alpha\beta\gamma\delta}$ of the matrix elements of the interaction
Hamiltonian. With the help of Eqs.~(\ref{eq:2.15}), (\ref{eq:2.16}),
(\ref{eq:2.18}), and of the diagrammatic representation for the disorder
average, Fig.~\ref{Fig5}, one obtains~
\cite{KamenevLevitov,Agam97,Blanter96,MirlinBlanter,AleinerGlazman98}:
\begin{equation}
\langle\left[\delta H_{\alpha\beta\gamma\delta}^{(1/g)}\right]^2\rangle = 
c_2 \lambda^2 \left(\frac{\Delta}{g}\right)^2,
\quad
c_2 =\frac{2}{\pi^2}
\sum_{\gamma_m\neq 0}
\left(\frac{\gamma_1}{\gamma_n}\right)^2
.
\label{eq:2.28}
\end{equation}
(This result is for the generic case where all indices $\alpha$,
$\beta$,
$\gamma$, and $\delta$ are different. If they are not all different, 
the fluctuations are bigger by a number of permutations
which that do not change the matrix element.
)
Equations (\ref{eq:2.27}) and (\ref{eq:2.28}) show that the typical
deviation of all the matrix elements from their universal values
(\ref{eq:2.24}) are indeed small as $1/g$.

All the above derivations of the $\propto 1/g$ corrections were
performed for diffusive systems. The final estimates in terms of
the conductance hold also for ballistic quantum dots with chaotic
classical dynamics of electrons, see Appendix~\ref{ap:0}.

\subsubsection{RMT for the intra-dot Coulomb interaction}
\label{RMTCoulomb}
\addcontentsline{toc}{subsection}{
\protect\numberline{\ref{RMTCoulomb}}{\ 
RMT for the intra-dot Coulomb interaction
}}

So far, we have shown that the matrix elements of the Hamiltonian of
weak, short-range interaction are small compared to
$\Delta$. Therefore any observable quantity involving electron states
within the energy strip $\sim E_T$ around the Fermi level, can be
calculated by the lowest-order in $\lambda$ perturbation theory. The
situation is different for the long-range Coulomb interaction
\begin{equation}
V(r) = \frac{e^2}{\kappa r},
\label{eq:2.29}
\end{equation}
where $\kappa$ is the dielectric constant of the medium containing the
electrons of the dot.\footnote{The constant $\kappa$ does not
account for the screening provided by the electrons of the dot. This
screening will be considered in detail below.}

For the simplest geometry, the corresponding matrix elements
$H_{\alpha\beta\gamma\delta}$ are inversely proportional to the {\it
linear} size of the dot, rather than to its volume (or area in
dimension $d=2$). If the linear size of the dot exceeds the screening
radius (or the effective Bohr radius $a_B=\kappa \hbar^2/e^2m^*$ in $d=2$), 
the matrix elements $H_{\alpha\beta\gamma\delta}$ exceed $\Delta$, and the
lowest-order perturbation theory in the interaction Hamiltonian fails.

In the theory of linear screening, it is well-known how to deal with
this difficulty. When calculating an observable quantity, one should
take into account the virtual transitions of low-energy electrons into
the high-energy states. If the gas parameter 
\begin{equation}
r_s\equiv\frac{e^2}{\kappa\hbar v_F}
\label{rs}
\end{equation}
(with $v_F$ being the electron Fermi
velocity) of the electron
system in the dot is small, then the random phase approximation (RPA)
allows one to adequately account for such virtual transitions. 
Accounting for these transitions yields,
in general, a retarded electron-electron
interaction~\cite{Altshuler85}. However, the characteristic
scale of the corresponding frequency dependence is of the order of
$\gamma_1$, see Eq.~(\ref{eq:2.9}).
That is why at the energy scale $|\varepsilon|\lesssim E_T$ we can
consider the interaction as instantantaneous one, and derive the
effective interaction Hamiltonian acting in this truncated space.

\begin{figure}
\epsfxsize=0.5\hsize
\centerline{\epsffile{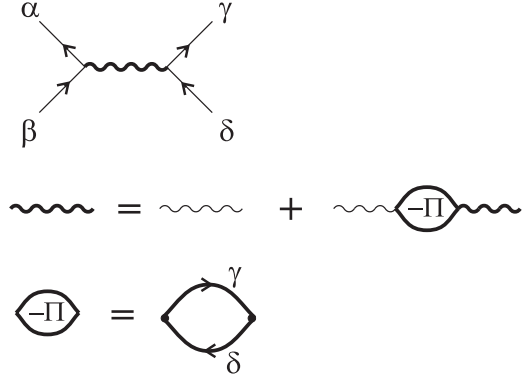}}
\caption{Matrix element in Random Phase approximation. Polarization
operator $\Pi$ involves all the transitions where at least one of the
states $\gamma,\ \delta $ is outside the energy strip of the
width of the Thouless energy around the Fermi level.}
\label{Fig6}
\end{figure}

A modified matrix element in RPA scheme is shown in Fig.~\ref{Fig6}. 
It involves
substitution of the bare potential (\ref{eq:2.29}) in
Eq.~(\ref{eq:2.15}) with the renormalized potential 
$V_{\rm sc}\left(\vec{r}_1,\vec{r}_2 \right)$. This potential is the
solution of the equation
\begin{eqnarray}
&& V_{\rm sc}\left(\vec{r}_1,\vec{r}_2 \right) =
V\left(\vec{r}_1-\vec{r}_2 \right)
- \int\!\!\int
d\vec{r}_3d\vec{r}_4
V\left(\vec{r}_1-\vec{r}_3 \right)
\Pi\left(\vec{r}_3,\vec{r}_4 \right)
V_{\rm sc}\left(\vec{r}_4,\vec{r}_2 \right),
\nonumber \\ && 
\label{eq:2.30}
\end{eqnarray}
where the integration over the
intermediate coordinates is performed within
the dot only. The polarization operator $\Pi$ should include only the states
where at least one electron or (hole) is outside the energy strip
of width $\varepsilon^* \lesssim E_T$ for which we derive the effective Hamiltonian,
\begin{equation}
\Pi\left(\vec{r}_1,\vec{r}_2 \right)=
2\sum_{
|\varepsilon_\beta -\varepsilon_\alpha| > \varepsilon^*
}
\frac{\phi_{\alpha}(\vec{r}_1)
\phi_{\beta}^*(\vec{r}_1)
\phi_{\beta}(\vec{r}_2)
\phi_{\alpha}^*(\vec{r}_2)}{|\varepsilon_\beta-\varepsilon_\alpha|}
\theta(-\varepsilon_\alpha\varepsilon_\beta).
\label{eq:2.31}
\end{equation}
Here all energies are measured from the Fermi level, and
$\theta(x)$ is the step function. The factor of two accounts for spin
degeneracy.  We can compare Eq.~(\ref{eq:2.31}) with the usual
frequency dependent operator $\Pi^R\left(\omega;\vec{r}_1,\vec{r}_2
\right)$ that involves all the electron states
\begin{equation}
\Pi^R\left(\omega;\vec{r}_1,\vec{r}_2 \right)=
2\sum_{
\varepsilon_\beta,\varepsilon_\alpha
}
\frac{\phi_{\alpha}(\vec{r}_1)
\phi_{\beta}^*(\vec{r}_1)
\phi_{\beta}(\vec{r}_2)
\phi_{\alpha}^*(\vec{r}_2)}{\varepsilon_\beta - \varepsilon_\alpha - \omega - i 0}
\theta(-\varepsilon_\alpha\varepsilon_\beta){\mathrm sgn}\varepsilon_\beta ,
\label{eq:2.32}
\end{equation}
and obtain
\begin{equation}
\Pi\left(\vec{r}_1,\vec{r}_2 \right)=\frac{1}{\pi}{\mathrm Im}
\int 
\frac{d\omega}{\omega}\Pi^R\left(\omega;\vec{r}_1,\vec{r}_2 \right)
\theta \left(|\omega| - \varepsilon^*\right)
.
\label{eq:2.33}
\end{equation}

On the next step, we  replace the polarization operator with its
average value~\cite{Altshuler85}
\begin{equation}
\langle\Pi^R\left(\omega;\vec{r}_1,\vec{r}_2 \right)\rangle=
\frac{2}{\Delta{\cal V}_d}
\sum_{\gamma_n > 0}
\frac{\gamma_n\  f_n(\vec{r}_1)  f_n(\vec{r}_2) }
{-i\omega + \gamma_n},
\label{eq:2.34}
\end{equation}
where $\gamma_n$ and $f_n$ for a diffusive system are defined in
Eq.~(\ref{eq:2.9}). The prefactor in Eq.~(\ref{eq:2.34}) is nothing but the
thermodynamic density of states per unit volume 
(area).\footnote{The mesoscopic fluctuations of the
polarization operator are related to the variation
of the density of states in the open sample of the size of
the order of $|r_1-r_2|$, corresponding discussion can
be found, e.g. in Ref.~\protect\cite{Narozhny}. The latter fluctuations
are small as $1/g$, and taking them into account here would be the
overstepping of accuracy of $1/g$ expansion for the matrix elements.}

Substituting Eq.~(\ref{eq:2.34}) into Eq.~(\ref{eq:2.33}) and taking
into account that $\varepsilon^* \lesssim\hbar\gamma_1$, we find
\begin{equation}
\Pi\left(\vec{r}_1,\vec{r}_2 \right)=
\frac{2}{\Delta{\cal V}_d}
\sum_{\gamma_n > 0}
f_n(\vec{r}_1) f_n(\vec{r}_2).
\label{eq:2.35}
\end{equation}
Now we can use the completeness of the solution set of the diffusion
equation,
\[
\sum_nf_n({\vec r}_1)f_n({\vec r}_2)=\delta ({\vec r}_1-{\vec r}_2),
\]
and the explicit form of the zero-mode solution
$f_0(\vec{r})=\theta_{\rm dot}(\vec{r})/\sqrt{{\cal V}_d}$ for this
equation, in order to present Eq.~(\ref{eq:2.35}) in the form:
\begin{equation}
\Pi\left(\vec{r}_1,\vec{r}_2 \right)=
\frac{2}{\Delta{\cal V}_d}\left[\delta\left(\vec{r}_1-\vec{r}_2
\right) - \frac{1}{{\cal V}_d}\right]\theta_{\rm dot}(\vec{r}_1)
\theta_{\rm dot}(\vec{r}_2).
\label{eq:2.36}
\end{equation}
Here $\theta_{\rm dot}(\vec{r}) =1$, if $\vec{r}$ belongs to the dot and
$\theta_{\rm dot}(\vec{r}) =0$ otherwise. The structure of
Eq.~(\ref{eq:2.36}) is easy to understand. The first term in brackets
characterizes local screening in the Thomas-Fermi approximation. The
last term in brackets subtracts the constant-potential contribution of
the zero  mode which cannot induce electron transitions between the
levels and therefore can not be screened.

Substituting the polarization operator (\ref{eq:2.36}) in
Eq.~(\ref{eq:2.30}), we find the equation for the self-consistent
potential in the Thomas-Fermi approximation:
\begin{eqnarray}
V_{\rm sc}\left(\vec{r}_1,\vec{r}_2 \right)&=&
V\left(\vec{r}_1-\vec{r}_2 \right)
\label{eq:2.37}\\
&-& \frac{2}{\Delta{\cal V}_d}\int
d\vec{r}_3
V\left(\vec{r}_1-\vec{r}_3 \right)
\left[V_{\rm sc}\left(\vec{r}_3,\vec{r}_2 \right)-
\frac{1}{{\cal V}_d}\int d\vec{r}_4
V_{\rm sc}\left(\vec{r}_4,\vec{r}_2 \right)\right].
\nonumber
\end{eqnarray}
The use of Thomas-Fermi approximation is justified if the linear size
of the dot exceeds the screening radius. The integrals here are taken
over the volume of the dot (or over the corresponding area in the $2d$ case). 
We can rewrite the integral equation (\ref{eq:2.37}) in the more
familiar differential form,
\begin{eqnarray}
-\nabla^2_{\vec{r}} V_{\rm sc}(\vec{r},\vec{r}_1)&=&\frac{4\pi e^2}{\kappa}
\left[\delta(\vec{r}-\vec{r}_1) 
-\frac{2}{\Delta{\cal V}_d^{\vphantom{2}}
  }\theta_{\rm dot}(\vec{r})V_{\rm sc}(\vec{r},\vec{r}_1)\right.
\label{eq:2.38}\\
&& \mbox{} \left.+ \frac{2}{\Delta{\cal V}^2_d}\theta_{\rm dot}(\vec{r})
\int d\vec{r}_2 V_{\rm sc}(\vec{r}_2,\vec{r}_1 )\theta_{\rm dot}(\vec{r}_2)\right],
\nonumber
\end{eqnarray}
supplemented with the requirement that the solution vanishes at
$r\to\infty$. In the case of a $2d$ dot, the $\nabla^2$ operator is
still acting in the three-dimensional space; the right-hand side of
Eq.~(\ref{eq:2.38}) must be multiplied by $\delta (z)$, where
coordinate $z$ is directed normal to the plane of the dot. The second
term in the right-hand side of Eq.~(\ref{eq:2.38}) represents the
familiar Thomas-Fermi screening. In the three-dimensional case, this
term is $\propto V_{\rm sc}({\vec r})/r_D^2$, with the screening radius $r_D$.

First, we find an approximate solution of Eq.~(\ref{eq:2.38}) inside
the dot. We consider only length scales larger than the screening radius.
This enables us to neglect the left-hand side of Eq.~(\ref{eq:2.38})
altogether. In addition, we will neglect the corrections of the order of
the level spacing, which are generated by the last term in the
right-hand side of Eq.~(\ref{eq:2.38}).
Then the solution takes the form:
\begin{equation}
V_{\rm sc}(\vec{r},\vec{r}_1)=\left[\frac{\Delta{\cal V}_d}{2}
\delta(\vec{r}-\vec{r}_1) + \bar{V}\right]
\theta_{\rm dot}(\vec{r})\theta_{\rm dot}(\vec{r}_1)
,
\label{eq:2.39}
\end{equation}
where the constant $\bar{V}$ will be determined later.

The easiest way to find $\bar{V}$ is to consider Eq.~(\ref{eq:2.38})
for $\vec r$ outside the dot (keeping $\vec r_1$ inside the dot). 
The right-hand side of Eq.~(\ref{eq:2.38}) then vanishes,
\begin{equation}
\vec{\nabla}^2_{\vec{r}}V_{\rm sc}(\vec{r},\vec{r}_1) =0.
\label{eq:2.390}
\end{equation}
The constant $\bar{V}$ defines for this Laplace equation the
boundary condition at the surface of the dot,
\begin{equation}
\label{eq:2.40}
V_{\rm sc}(\vec{r},\vec{r}_1)\big|_{\vec{r} \in S}=\bar{V}
\end{equation}
(the surface is defined unambiguously in the limit of zero screening
radius). After the solution $V_{\rm sc}(\vec{r},\vec{r}_1)$ of the Laplace equation
(\ref{eq:2.390}) is found, the constant $\bar{V}$ can be obtained by
integration of Eq.~(\ref{eq:2.38}) along the surface of the dot:
\begin{equation}
- \oint_{S}d\vec{S}\vec{\nabla}V_{\rm sc}(\vec{r},\vec{r_1}) =
\frac{4\pi e^2}{\kappa}.
\label{eq:2.41}
\end{equation}
For the two dimensional case, Eq.~(\ref{eq:2.41}) obviously reduces to
\begin{equation}
\int d\vec{r}
\stackrel{\leftrightarrow}{\partial}_z
V_{\rm sc}(\vec{r},\vec{r}_1)
 =\frac{4\pi e^2}{\kappa}, \quad 
\stackrel{\leftrightarrow}{\partial}_z \equiv
\left(\frac{\partial}
{\partial z}\big|_{z \to
-0} - 
\frac{\partial}
{\partial z}\big|_{z \to
+0}\right),
\label{eq:2.42}
\end{equation}
where the two-dimensional integration is performed over the area of
the dot ($z=0$).

Now the solution of Eqs.~(\ref{eq:2.390}) and (\ref{eq:2.40}) can be
formally found with the help of the Green function of the Laplace
equation
\begin{equation}
-\nabla^2_{\vec{r}_1} {\cal D}  (\vec{r}_1,\vec{r}_2)=
  \delta (\vec{r}_1 - \vec{r}_2);
\quad {\cal D}  (\vec{r}_1,\vec{r}_2)\big|_{\vec{r}_1 \in S} = 0
\label{eq:2.43}
\end{equation}
describing the electrostatic field outside the dot.  If the system
contains metallic gates, Eq.~(\ref{eq:2.43}) should be supplied with
additional Dirichlet boundary conditions on the surface $S_i$ of those
gates
\begin{equation}
{\cal D}  (\vec{r}_1,\vec{r}_2)\big|_{\vec{r}_1 \in S_i} = 0,
\label{eq:2.430}
\end{equation}
where index $i$ enumerates the corresponding gates.

One immediately finds from Eqs.~(\ref{eq:2.40}) and (\ref{eq:2.43})
\begin{equation}
V_{\rm sc}(\vec{r},\vec{r}_1)=
 \bar{V}\oint_{S}d\vec{S}_2\vec{\nabla}{\cal D}(\vec{r},\vec{r}_2)
\left[1-\theta_{\rm dot}(\vec{r})\right]
\label{eq:2.44}
\end{equation}
for the potential $V_{\rm sc}$ outside the dot. 
Substitution of Eq.~(\ref{eq:2.44}) into
Eq.~(\ref{eq:2.41}) yields
\begin{equation}
\bar{V} = \frac{e^2}{C}
\label{eq:2.45}
\end{equation}
for the constant $\bar{V}$. The geometrical capacitance of the dot is
given by formula valid for any shape of the dot
\begin{equation}
C = \frac{\kappa}{4\pi}
\left|\oint_{S}d\vec{S}_2\vec{\nabla_2}
\oint_{S}d\vec{S}_1\vec{\nabla_1}
{\cal D}(\vec{r}_1,\vec{r}_2)\right|.
\label{eq:2.46}
\end{equation}
In a two-dimensional system, Eq.~(\ref{eq:2.46}) is replaced with
\begin{equation}
C = \frac{\kappa}{4\pi}\left|
\int d\vec{r}_1\int d\vec{r}_2
\stackrel{\leftrightarrow}{\partial}_{z_1}
\stackrel{\leftrightarrow}{\partial}_{z_2}
{\cal D}(\vec{r}_1,\vec{r}_2)\right|,
\label{eq:2.47}
\end{equation}
where the two-dimensional integration is performed over the area of
the dot ($z=0$), and operator $\stackrel{\leftrightarrow}{\partial}_z$
is defined in Eq.~(\ref{eq:2.42}).  The concrete value of capacitance
$C$ is geometry-dependent. If the size of the dot is characterized by a single
parameter $L$, then $C$ is proportional to $L$ with some
coefficient, which depends on details of the geometry. In the practically
important case of a gated dot, the capacitance is $C\simeq L^2/4\pi d$,
where $L^2$ is the area of the dot, and $d\lesssim L$ is the distance from
the gate to the plane of the dot.

The charging energy, $e^2/2C$, is the
dominant energy scale for the matrix elements of the interaction
Hamiltonian, as we will now show. 
Corrections to this scale appear from terms of the
order of $\Delta$ in the effective interaction potential $V_{\rm sc}$.
In order to
find the matrix elements $V_{\alpha\beta\gamma\delta}$, we need to
know the screened potential $V_{\rm sc}(\vec{r}_1,\vec{r}_2)$ within the
dot. The constant part of this potential $\sim e^2/C$ contributes only
to the ``diagonal'' matrix elements (\ref{eq:2.15}), with $\alpha=\delta$;
$\beta=\gamma$. This part does not contribute to any other
matrix element, because of the orthogonality of the corresponding
one-electron wave functions. 

In order to find the non-diagonal matrix elements, we need to account
for smaller but coordinate-dependent terms in
$V_{\rm sc}(\vec{r}_1,\vec{r}_2)$. To accomplish that goal, we substitute
the potential (\ref{eq:2.39}), (\ref{eq:2.44}) into the left-hand side of
Eq.~(\ref{eq:2.38}) and treat it in the first order of perturbation
theory in $\Delta C/e^2\ll 1$. As the result, we obtain with the help
of Eq.~(\ref{eq:2.45})
\begin{equation}
V_{\rm sc}(\vec{r}_1,\vec{r}_2)= \frac{e^2}{C} +
\frac{{\cal V}_d \Delta}{2}\delta(\vec{r}_1-\vec{r}_2)
+ \tilde{V}(\vec{r}_1) \Delta + \tilde{V}(\vec{r_2})\Delta.
\label{eq:2.48}
\end{equation}
Here, the potential $\tilde{V}(\vec{r})$ appears due to the
finite size of the dot: a charge $e$ ``expelled'' from the point
$\vec{r}=\vec{r}_{1,2}$ in the process of screening, can not be pushed away
to infinity because of the dot's boundary. In the $3d$ case, this
charge forms a thin layer near the boundary of the dot; in the $2d$
case, it creates an inhomogeneous distribution over the whole area of
the dot. The explicit form of this potential [in units of level
spacing $\Delta$, cf.\ Eq.\ (\ref{eq:2.48})] is:
\begin{eqnarray}
\tilde{V}(\vec{r})&=&
\frac{{\cal V}_d}{8\pi C}
\oint_{S}
\delta(\vec{r}-\vec{r}_1)
d\vec{S}_1\vec{\nabla_1}
\oint_{S}d\vec{S}_2\vec{\nabla_2}
{\cal D}(\vec{r}_1,\vec{r}_2); \quad d=3;
\label{eq:2.49}
\\
\tilde{V}(\vec{r}) &=&
\frac{{\cal V}_d}{8\pi C}
\int d\vec{r}_1
\stackrel{\leftrightarrow}{\partial}_{z}
\stackrel{\leftrightarrow}{\partial}_{z_1}
{\cal D}(\vec{r},\vec{r}_1); \quad d=2,
\nonumber
\end{eqnarray}
where Green function ${\cal D}(\vec{r},\vec{r}_1)$ and the dot
capacitance $C$ are given by Eqs.~(\ref{eq:2.43}) and
(\ref{eq:2.45}), and $\stackrel{\leftrightarrow}{\partial}_z$ is
defined in Eq.~(\ref{eq:2.42}).  The importance of the additional
potential (\ref{eq:2.49}) for the structure of the matrix elements of
the interaction Hamiltonian was first noticed in Ref.\ 
\cite{Muzykantskii}.

The substitution of the first term of Eq.~(\ref{eq:2.48}) into
Eq.~(\ref{eq:2.15}) immediately yields the new value of the
interaction constant $\Ec$ in the $\hat n$-dependent part of the
universal Hamiltonian (\ref{eq:2.152}). This constant in fact is the
single-electron charging energy:
\begin{equation}
\Ec = \frac{e^2}{2C}.
\label{eq:2.50a}
\label{EC} 
\end{equation}
Note, that this value is much larger than the single electron level spacing: 
\[
\frac{\Ec}{\Delta} \simeq r_s \left(k_F L\right)^{d-1},\quad d=2,3,
\]
because of a large factor $k_F L \gg 1$. We intend to show now that
this large scale appears only in the charge part of the universal
Hamiltonian and it enters neither the spin channel nor the
non-universal part $\hat{H}_{\rm int}^{(1/g)}$. 

The constant  $J_S$ in Eq.~(\ref{eq:2.152})
originates from the two-body interaction term of the screened
potential. The use of the corresponding (second) term of
Eq.~(\ref{eq:2.48}) would yield $J_S= -\Delta/2$, which is an
overestimation of the spin-dependent interaction term. The correct
result
for $J_S$ in the random phase approximation reads:\footnote{
The additional smallness $r_s$ in the expression for $J_S$
arises because the main contribution to this constant comes
from the distances of the order of $1/k_F$;
at such distances the
approximation of the local part of the potential 
by  $\delta$ -function in Eq.~(\ref{eq:2.48}) is no longer valid
and one should replace:
\[
\frac{\Delta {\cal V}_d}{2} \delta(\vec{r})
\to \int \frac{d \vec{k}}{(2\pi)^d} \frac{V( k)}{1
+ 2V(k)/({\Delta {\cal V}_d})} e^{i\vec{k}\vec{r}}, 
\quad 
V(k) = \left\{
\matrix{\frac{4\pi e^2}{\kappa k^2}, & d=3; \cr
\frac{2\pi e^2}{\kappa k}, & d=2.
}\right.
\]
}
\begin{eqnarray}
&& - J_S = 
\int d \vec{r}{\cal F}
\left(
k_F|\vec{r}|
\right)^2
V_{\rm sc}(\vec{r})
=
\frac{\Delta }{\pi}\times
\left\{ \matrix{
r_s\ln\left(1+\frac{\pi }{2 r_s}\right); & \ \ d=3; \cr
\frac{r_s}{\sqrt{1-r_s^2}}\
\ln\left(
\frac{1+\sqrt{1-r_s^2}}
{1-\sqrt{1-r_s^2}}
\right); &\ \  d=2,
}\right. \nonumber \\
\label{eq:2.50}
\end{eqnarray}
where the gas parameter $r_s$ is given by Eq.~(\ref{rs}) and the
function ${\cal F}$ was introduced in Eq.\ (\ref{eq:calF}). It is
worthwhile to notice that at $r_s \gg 1$, the RPA result
(\ref{eq:2.50}) gives $J_S = -\Delta/2$, which forbids, in particular,
the Stoner instability. However, in the regime $r_s \gtrsim 1$, the RPA
scheme becomes non-reliable and the constant (\ref{eq:2.50}) should be
replaced with the corresponding Fermi-liquid constant. A universal
description still holds in this case provided that the Fermi liquid
description does not breakdown at length scales smaller than the
system size.

We now turn to the discussion of $1/g$ corrections. The first type of
such corrections results from the substitution of the third and
fourth terms of Eq.~(\ref{eq:2.48}) into Eq.~(\ref{eq:2.15}). One
thus finds \cite{Muzykantskii}:
\begin{equation}
\left[{\cal H}^{(1/g)}_1\right]_{\alpha\beta\gamma\delta}
= \delta_{\alpha\delta} {\cal X}_{\beta \gamma}^0+
\delta_{\beta\gamma} {\cal X}_{\alpha \delta}^0,
\quad {\cal X}_{\alpha \beta}^0 =
 \int d \vec{r} \tilde{V}(\vec{r})
\phi_{\alpha}(\vec{r}_2)
\phi_{\beta}^*(\vec{r}_2).
\label{eq:2.51}
\end{equation} 
The matrix elements in Eq.~(\ref{eq:2.51}) are random. Their
characteristic values can be easily found from Eqs.~(\ref{eq:2.48})
and (\ref{eq:2.22}), with the result
\begin{equation}
\langle\left[ {\cal X}^0_{\alpha \beta}\right]^2\rangle =
b_{00} \frac{\Delta^2}{g},
\label{eq:2.52}
\end{equation}
where $b_{00}$ is a geometry dependent numerical coefficient,
\begin{equation}
b_{00}=\frac{1}{\pi}\sum_{\gamma_n > 0} 
\frac{
\left|\int d \vec{r}\tilde{V}( \vec{r}) 
f_n( \vec{r})\right|^2\gamma_1 
}{\gamma_n},
\label{eq:2.53}
\end{equation}
and the potential $\tilde{V}$ is defined by Eq.~(\ref{eq:2.49}). The
coefficient $b_{00}$ is of the order of unity. According to
Eq.~(\ref{eq:2.51}), the matrix elements $\left[{\cal
    H}^{(1/g)}\right]_{\alpha\beta\gamma\delta}$ of the Hamiltonian
$\hat {H}^{1/g}$ with $\alpha=\delta$ or $\beta=\gamma$ are of the
order of $\Delta/\sqrt{g}$, contrary to the assumptions of
Ref.~\cite{Lewenkopf}. The rest of these matrix elements are smaller.
One can use the two-body part of the screened interaction potential
(\ref{eq:2.48}) to evaluate them~\cite{KamenevLevitov}. This part
coincides with the simple model (\ref{eq:2.16}), up to the constant
$\lambda$, which should be substituted by $1/2$. After the
substitution, one can use the results (\ref{eq:2.27}) and
(\ref{eq:2.28}).

In all the previous consideration, we assumed the potential on the
external gates to be fixed, see Eq.~(\ref{eq:2.430}). Such an
assumption was valid
because additional potentials created by those gates were implicitly
included into the confining potential $U(\vec{r})$ from
Eq.~(\ref{eq:2.1}). One, however, may be interested in a comparison
of the properties of the dot at two different sets of the gate
voltages. To address such a problem one has to include gates into the
electrostatic problem, {\em i.e.}, to find the correction
$\delta U(\vec{r})$ to the self-consistent confining potential due to
the variation of the gate potentials. This requires that the
Thomas - Fermi screening of the charge on the external gates by the
electrons in the dot has to
be considered. The resulting equation for this correction is
similar to Eq.~(\ref{eq:2.38}):
\begin{equation}
-\nabla^2\delta U(\vec{r})=
-\frac{2}{\Delta{\cal V}_d}\left[\theta_{\rm dot}(\vec{r})
\delta U(\vec{r})-
\frac{1}{{\cal V}_d}
\int d\vec{r}_1\theta_{\rm dot}(\vec{r}_1)\delta U(\vec{r}_1)\right],
\label{eq:2.530}
\end{equation}
supplemented with the requirement that the solution vanishes at
$r\to\infty$ and has the fixed value on the surface of each gate,
$S_i$, [compare with Eq.~(\ref{eq:2.430})]:
\begin{equation}
\delta U(\vec{r})\big|_{\vec{r} \in S_i} = - eV_g^{(i)},
\quad i=1,2,\dots,
\label{eq:2.531}
\end{equation}
where indices $i$ enumerate the gates, and $V_g^{(i)}$ is
the electrostatic potential on $i$-th gate. 

Solving Eqs.~(\ref{eq:2.530}) -- (\ref{eq:2.531}) involves the
same steps as in derivation of Eq.~(\ref{eq:2.48}) [the only
difference is that now there is no $\delta$-function in
Eq.~(\ref{eq:2.39}), the right hand side of Eq.~(\ref{eq:2.41})
vanishes, and the boundary conditions on the gate surfaces are given
by Eq.~(\ref{eq:2.531})]. The solution for the potential outside the
dot is obtained in terms of the Green function (\ref{eq:2.43}) --
(\ref{eq:2.430}), similarly to Eq.~(\ref{eq:2.44}):
\begin{equation}
\delta U (\vec{r})=
 \delta\bar{ U}\oint_{S}d\vec{S}_2\vec{\nabla}{\cal
 D}(\vec{r},\vec{r}_2) -
\sum_i
 eV_g^{(i)}\oint_{S_i}d\vec{S}_2\vec{\nabla}{\cal
 D}(\vec{r},\vec{r}_2),
\label{eq:2.532}
\end{equation}
where the zero-mode part of the potential of the dot
is given by
\begin{equation}
\label{eq:2.533}
 \delta\bar{ U} = -2\Ec{\cal N},
\quad {\cal N} = \sum_i {\cal N}_i,
\quad
e{\cal N}_i = C_i V_g^{(i)}.
\end{equation}
The geometrical capacitance of the dot $C$ is defined by
Eqs.~(\ref{eq:2.46}) and (\ref{eq:2.47}) and the geometrical
mutual capacitances between the dot and the gates are
\begin{equation}
C_i = \frac{\kappa}{4\pi}
\oint_{S_i}d\vec{S}_2\vec{\nabla_2}
\oint_{S}d\vec{S}_1\vec{\nabla_1}
{\cal D}(\vec{r}_1,\vec{r}_2),
\label{eq:2.534}
\end{equation}
for three-dimensional dots. To obtain the result for the
two-dimensional case, the surface integral over $S$ here
should be replaced as in Eq.~(\ref{eq:2.42}).
The resulting potential inside the dot is found similarly to
Eq.~(\ref{eq:2.48}) in the form
\begin{equation}
\delta U(\vec{r})= - 2 \Ec {\cal N} - \Delta
\left[
{\cal N}
\tilde{V}(\vec{r})
+\sum_i{\cal N}_i  \tilde{V}^{(i)}(\vec{r})
\right]
,
\label{eq:2.535}
\end{equation}
where $\tilde{V}(\vec{r})$ is defined in Eq.~(\ref{eq:2.49}) 
and potentials $\tilde{V}^{(i)}(\vec{r})$ are given by
\begin{eqnarray}
\tilde{V}^{(i)}(\vec{r})&=&
\frac{{\cal V}_d}{8\pi C_i}
\oint_{S}
\delta(\vec{r}-\vec{r}_1)
d\vec{S}_1\vec{\nabla_1}
\oint_{S_i}d\vec{S}_2\vec{\nabla_2}
{\cal D}(\vec{r}_1,\vec{r}_2); \quad d=3;
\label{eq:2.536}
\\
\tilde{V}^{(i)}(\vec{r}) &=&
\frac{{\cal V}_d}{8\pi C_i}
\int_{S_i} d\vec{r}_1
\stackrel{\leftrightarrow}{\partial}_{z}
\stackrel{\leftrightarrow}{\partial}_{z_1}
{\cal D}(\vec{r},\vec{r}_1); \quad d=2.
\nonumber
\end{eqnarray}
We see that the effect of the external gates has the same
hierarchical structure as the electronic interaction inside the dot: 
The largest scale $\Ec$ [the first term in
Eq.~(\ref{eq:2.535})] corresponds to a simple uniform shift of the
potential, while the much smaller energy scale $\Delta$ characterizes the
changes in the potential affecting the shape of the dot. Therefore,
the effect of the external gates can be separated into a universal part
$\delta U^{0}$ and a fluctuating non-universal part $\delta U^{1/g}$:
\begin{eqnarray}
&&\delta U^{0}_{\alpha\beta} = -2 \Ec{\cal N}\delta_{\alpha\beta};
\label{eq:2.537}\\
&& U^{(1/g)}_{\alpha\beta} = - {\cal N} {\cal X}^0_{\alpha \beta}
- \sum_j {\cal N}_j {\cal X}^j_{\alpha \beta}.
\label{eq:2.538}
\end{eqnarray}
In Eq.~(\ref{eq:2.538}), the ${\cal X}^j_{\alpha \beta}$ are random Gaussian
variables with second moments
\begin{equation}
\langle {\cal X}^i_{\alpha \beta}
{\cal X}^j_{\alpha \beta}
\rangle =
b_{ij} \frac{\Delta^2}{g},
\label{eq:2.539}
\end{equation}
where the geometry dependent coefficient  $b_{00}$ is given by
Eq.~(\ref{eq:2.53}) and all other coefficients are given by
\begin{eqnarray}
b_{ij}=\frac{1}{\pi}\sum_{\gamma_n > 0} 
\frac{\gamma_1}{\gamma_n} 
\int d \vec{r}\tilde{V}^{(i)}( \vec{r}) 
f_n( \vec{r})
\int d \vec{r}\tilde{V}^{(j)}( \vec{r}) 
f_n( \vec{r})
,
\label{eq:2.540}\\
b_{0j}=b_{j0}=\frac{1}{\pi}\sum_{\gamma_n > 0} 
\frac{\gamma_1}{\gamma_n} 
\int d \vec{r}\tilde{V}( \vec{r}) 
f_n( \vec{r})
\int d \vec{r}\tilde{V}^{(j)}( \vec{r}) 
f_n( \vec{r})
\nonumber
\end{eqnarray} 
for $i,j=1,2 \dots$. In general all these coefficients are of the
order of unity.

\subsubsection{Final form of the effective Hamiltonian}
\label{sec:hfinal}
\addcontentsline{toc}{subsection}{
\protect\numberline{\ref{sec:hfinal}}{\ Final form of the effective Hamiltonian}}

To summarize this subsection, we note that in the absence of
superconducting correlations, the universal part of the interaction
Hamiltonian, Eq.~(\ref{eq:2.152}), consists of two parts. The dominant
part depends on the dot's charge $\hat n$. The corresponding energy
scale $\Ec$, Eq.~(\ref{eq:2.50a}), is related to the geometric
capacitance of the dot $C$, Eq.~(\ref{eq:2.46}), and exceeds
parametrically the level spacing $\Delta$. [There are corrections
of order $\Delta$ to Eq.\ (\ref{eq:2.50a}) arising from the potentials
$\tilde V$ in Eq.\ (\ref{eq:2.48}) and from the local part of the screened
interaction.] The next part depends on
the total spin $\vec{S}$ of the dot. The corresponding energy scale
$J_S$ is smaller than $\Delta$. If the level spacing did not
fluctuate, then the smallness of $J_S$ would automatically mean that
the spin of the dot $S=0$ or $S=1/2$ depending on whether the number
of electrons in the dot is even or odd, respectively. Fluctuations of
the level spacing may lead to a violation of the strict periodicity in
this dependence \cite{Oreg,Baranger,Kurlyand}. 
However, the Stoner criterion~\cite{Ziman}
guarantees that the spin of the dot is not macroscopically large ({\it
{\em i.e.},}, does not scale with the volume of the dot). The strongest
non-universal correction to the Hamiltonian (\ref{eq:2.152}) comes from
the effect of the boundary of the dot on the interaction potential,
see Eq.~(\ref{eq:2.48}), and is of the order of $\Delta/\sqrt{g}$, where
$g$ is the dimensionless conductance of the dot (\ref{g}).  In
the presence of the gate, the universal part $\hat{H}^{(0)}$ of the
Hamiltonian (including the RMT representation (\ref{eq:2.11}) of its
free-electron part) and the two largest non-universal corrections
$\hat{H}^{1/g}$ have the following form:
\begin{eqnarray}
&&\hat{H}^{(0)}=
\sum_{\alpha,\gamma}
  {\cal H}_{\alpha\gamma}
  \psi^\dagger_{\alpha,\sigma} \psi_{\gamma,\sigma}
  + \Ec \left(\hat{n}-{\cal N}\right)^2
  + J_S \left(\hat{\vec{S}} \right)^2 - \Ec {\cal N}^2,
\label{eq:2.54a}\\
&&\hat{H}^{(1/g)}=
\sum_{\alpha,\beta}\psi^\dagger_{\alpha,\sigma}\psi_{\beta,\sigma}
\left[\left(\hat{n}-{\cal N}\right)
{\cal X}_{\alpha\beta}^0
-  \sum_j{\cal N}_j
{\cal X}_{\alpha\beta}^j
\right]
\nonumber\\
&&\quad\quad\quad\quad
+ \frac{1}{2}\sum_{\alpha\beta\gamma\delta}
 {\cal H}_{\alpha\beta\gamma\delta}^{(1/g)}
\psi^\dagger_{\alpha,\sigma_1} \psi^\dagger_{\beta,\sigma_2} 
\psi_{\gamma,\sigma_2}\psi_{\delta,\sigma_1}.
\label{eq:2.54}
\end{eqnarray}
The last term in Eq.~(\ref{eq:2.54a}) is a $c$-number and it can be
disregarded in all subsequent considerations.
The fluctuations of the matrix matrix elements ${\cal
  X}_{\alpha\beta}^{(j)}$ are given by Eq.~(\ref{eq:2.539}), the elements
${\cal H}_{\alpha\beta\gamma\delta}^{(1/g)}$ can be estimated from
Eqs.~(\ref{eq:2.27}) and (\ref{eq:2.28}) with $\lambda = 1/2$.
 Finally we should mention that for a
two-dimensional dot the relation  $\Ec\gg\Delta$ holds
only if the distance between the dot and gate exceeds the screening
radius in the dot.

\subsection{Inclusion of the  leads}
\label{sec:2.2}

So far we considered an isolated dot. We have established the validity
of the universal RMT description both for the one electron Hamiltonian
and for the interaction effects. Our goal now is to present a similar
description for the connection of the dot with the external leads.

We will assume that the leads contacting the dot are sufficiently
clean, and neglect the mesoscopic fluctuations of the local density of
states in the leads.  In other words, we neglect closed electron
trajectories in the leads that start at the
contacts.\footnote{Accounting for such trajectories would result in an
  additive contribution to the mesoscopic fluctuations of observable
  quantities; this contribution is negligibly small as long as the
  sheet conductance of the leads is large.} We also neglect the
electron-electron interaction in the leads.  In the case of
two-dimensional leads, this approximation is justified if the
dimensionless conductance of the leads greatly exceeds the number of
channels in the point contacts connecting leads to the quantum dot,
or, equivalently, when the resistance of the leads is much smaller
than the conductance of the point contacts.

We wish to keep the RMT description of the dot. On the other hand, we
saw that it is justified only for energy scales much smaller than
Thouless energy $E_T = g \Delta$, where $g$ is the dimensionless
conductance of the dot. The presence of point
contacts with a total number of propagating modes $\Nch$ broadens each
level by an amount of the order of $\Nch \Delta$. Hence, the 
use of the RMT description requires the condition
\begin{eqnarray}
\Nch \ll g,
\label{eq:2.55}
\end{eqnarray}
where the dimensionless conductance of the dot is defined by
Eq.~(\ref{g}).  Since $g \gg 1$, the condition (\ref{eq:2.55}) is
met even if more than one channel is open.

The conductance of a point contact connecting two clean conducting
continua can be related, by the Landauer formula
~\cite{FisherLee,Imry86,BuettikerIBM,StoneSzafer,BarangerStone},
to a scattering problem for electron
waves incident on the contact. These waves can be labeled by a
continuous wave number $k$ and by a set of discrete quantum numbers
$j$. The number $k$ accounts for the continuous energy spectrum of the
incoming waves, and $j$ defines their spatial structure in the
directions transverse to the direction of incidence. Although in
principle the number of discrete modes $j$ participating in scattering
is infinite, the number of relevant modes contributing to the
conductance is confined to the $\Nch$ modes that are propagating
through the contact; all other modes are evanescent and hardly contribute 
to the conductance. An adiabatic
point contact~\cite{mesoreview1,Glazman88,qpc} 
is an example adequately
described by a model having $\Nch$ modes in a lead.

Let us now turn to a quantitative description of the theory.
The total Hamiltonian of the system is given as a sum of three terms
\cite{LewenkopfWeidenmueller}, 
\begin{eqnarray}
\hat{H}_t = \hat{H} + \hat{H}_L  + \hat{H}_{LD},
\label{eq:2.56}
\end{eqnarray}
where the Hamiltonian of the closed dot, $\hat{H}$, will be taken in
the universal limit, see Eq.~(\ref{eq:2.54a}), $\hat{H}_L$ describes
the leads, and $\hat{H}_{LD}$ couples the leads and the dot.  A
separation into three terms, like Eq.~(\ref{eq:2.56}), is well known
in the context of the tunneling Hamiltonian formalism~\cite{Mahan},
and in nuclear physics \cite{MahauxWeidenmueller}. 
Its application to point
contacts coupling to quantum dots can be found, {\it e.g.}, in
Refs.~\cite{Beenakker97,LewenkopfWeidenmueller,IWZ,GMW}.  
The Hamiltonian of the leads reads
\begin{eqnarray}
  \hat{H}_L = v_F \sum_{j = 1}^{\Nch} \int \frac{dk}{2 \pi}
  k \hat \psi^\dagger_{j}(k) \hat \psi_{j}(k) ,
\label{eq:2.57}
\end{eqnarray}
where we have linearized the electron spectrum in the leads and
measure all the energies from the Fermi level. The vector $k\ll k_F$
is the deviation of longitudinal momentum in a propagating mode from
the Fermi wavevector $k_F$. 
[For the sake of simplicity, we assume the Fermi velocity to be
the same in all the modes; the general case can be reduced to
Eq.~(\ref{eq:2.57}) by a simple rescaling.] The Hamiltonian
(\ref{eq:2.57}) accounts for all the leads attached to the dot; the
lead index (in case the dot is coupled to more than one lead), 
the transverse mode index, and the spin index have
all been combined into the single index $j$, which is summed from $1$
to $\Nch$, $\Nch$ being the total number of propagating channels
in all the leads. (In this subsection, we reserve Greek and Latin
letters for labelling the fermionic states in the dot and in the leads
respectively.)

The Hamiltonian $\hat{H}_{LD}$ in Eq.~(\ref{eq:2.56}) describes the
coupling of the dot to the leads, 
\begin{eqnarray}
  \hat{H}_{LD} = \sum_{j = 1}^{\Nch} \sum_{\alpha = 1}^{M}
  \int \frac{dk}{2 \pi}
  \left[W_{\alpha j}\psi^\dagger_\alpha
  \psi_{j}(k) + {\rm h.c.}\right].\label{eq:2.58}
\end{eqnarray}
Here the coupling constants $W_{\alpha j}$ form a real $M\times
\Nch$ matrix $W$, and $M \to \infty$ is the size of the random
matrix describing the Hamiltonian of the dot. 
We emphasize that the matrix $W$ describes the point contacts,
not the dot. That is why this matrix is not random: all the
randomness is included in the matrix ${\cal H}$ from
Eq.~(\ref{eq:2.54a}). 

In the absence of electron-electron interactions in the dot, 
the Hamiltonian (\ref{eq:2.56}) of the combined system of dot 
and leads
can be easily diagonalized and the one-electron eigenstates can be
found. These eigenstates are best described by the 
$\Nch \times \Nch$ scattering matrix 
$S(\varepsilon)$, which relates the amplitudes $a^{\rm out}_j$
of out-going waves and $a^{\rm i}_j$ of in-going waves in the leads,
\begin{eqnarray}
a_i^{\rm out}= \sum_{j=1}^{\Nch}\Scat_{ij}(\varepsilon)a_j^{\rm in}.
  \label{eq:ainaout}
\end{eqnarray}
(For a precise definition of the amplitudes $a_j^{\rm in}$ and $a_j^{\rm
out}$ in terms of the lead states $\psi_j(k)$, and for a derivation
of the formulae presented below, we refer to appendix \ref{Ap:5}.)
The matrix $S$ is unitary, as required by particle conservation.
It can be expressed in terms of the matrices
$W$ and ${\cal H}$ that define the non-interacting part of the
Hamiltonian and a matrix $U$ that describes the boundary condition
at the lead-dot interface,\footnote{The states
$\psi_{j}(k)$ in Eq.\ (\ref{eq:2.57}) represent scattering states
in the leads, which are defined with the help of a suitably chosen
boundary condition at the lead-dot interface. These boundary
conditions may lead to a nontrivial scattering matrix $S_0
= U U^{\rm T}$ even in the absence of any lead-dot coupling $H_{LD}$,
see App.\ \ref{Ap:5} for details.}
\begin{eqnarray}
  \Scat(\varepsilon) =  U \left[1 - 
  2\pi i \nu W^{\dagger}
  \left({\varepsilon -{\cal H}
  +i\pi\nu WW^\dagger}\right)^{-1} W \right] U^{\rm T},
  \label{eq:2.60}
\end{eqnarray}
where $\nu =1/(2\pi v_F)$ is the one-dimensional density of states in
the leads, the $M\times M$ matrix ${\cal H}$ is formed by the elements
${\cal H}_{\alpha\gamma}$ of the Hamiltonian of the dot
(\ref{eq:2.54a}), and $U U^{\rm T}$ is the scattering matrix in
the absence of the coupling between the leads and the dot (i.e., when
$H_{LD} = 0$).
The scattering matrix $\Scat_{ij}(\varepsilon)$ describes scattering of 
electrons
from one lead to another, as well as backscattering of an electron
into the same lead. 
Equation (\ref{eq:2.60}) can be also rewritten in terms of the
(matrix) Green function of the closed dot
\begin{eqnarray}
  {\cal G}(\varepsilon) =
  (\varepsilon - {\cal H})^{-1},
\label{eq:2.600}
\end{eqnarray}
as
\begin{eqnarray}
\Scat(\varepsilon) = U \frac{
  1- i\pi \nu \Wmatrix^{\dagger}\Greenmatrix(\varepsilon) \Wmatrix}
  {1+i\pi \nu \Wmatrix^{\dagger}\Greenmatrix(\varepsilon) \Wmatrix}
  U^{\rm T}.
\label{eq:2.601}
\end{eqnarray}

The matrix $S$ describes both reflection from the
point contacts (i.e., scattering processes that involve the
backscattering from the contacts only, not the dot), and 
scattering that involves
(ergodic) exploration of the dot. Alternatively, 
the backreflection from the
point contacts into the leads can be described in terms of 
a reflection matrix $r_c$, which is related to the matrix $W$ as
\begin{equation}
  r_c = U
  {\pi^2 \nu - M \Delta W^{\dagger} W 
  \over
  \pi^2 \nu + M \Delta W^{\dagger} W} U^{\rm T}.
  \label{eq:rc}
\end{equation}
A contact is called ideal if $r_c = 0$, or, equivalently, 
$W^{\dagger} W = \pi^2 \nu/M \Delta$. For a single mode contact,
the transparency $|t_c|^2$ is given by $|t_c|^2 = 1 - |r_c|^2$.

For non-interacting electrons 
Eqs.~(\ref{eq:2.60}) -- (\ref{eq:rc}) essentially solve
the physical part of the problem since all the observable quantities
are expressed in terms of the scattering matrix. We now consider
three observables in more detail: the two-terminal conductance,
the tunneling density of states, and the ground state energy of
the dot in contact to the leads.

\begin{figure}
\epsfxsize=0.6\hsize
\hspace{0.2\hsize}
\epsffile{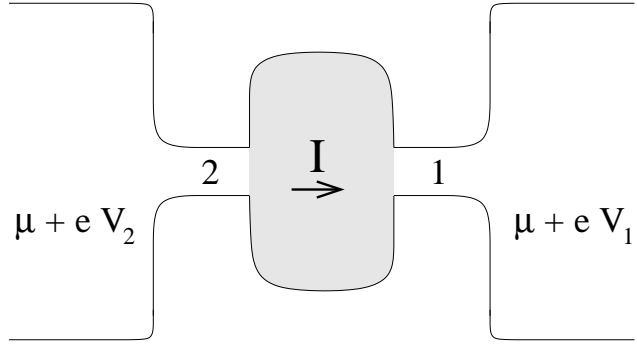}
\caption{Setup for a measurement of the two-terminal conductance.
The quantum dot (grey)
is connected to two leads, numbered $1$ and $2$,
that are connected to electron reservoirs. A current $I$ flows
through the dot as a function of the voltage difference $V_1 
- V_2$ between the reservoirs.}
\label{fig:Landauer}
\end{figure}

{\em Two-terminal conductance.}
The two-terminal conductance is defined
for a quantum dot that is connected to electron reservoirs
via two leads (numbered $1$ and $2$) 
with $\NL$ and $\NR$ channels each (where $\NL + \NR = \Nch$),
see Fig.\ \ref{fig:Landauer}. 
The electron reservoirs are held
at a constant chemical potential $\mu + e V_i$ ($i=1,2$). Labeling
the group of channels belonging to lead $1$ ($2$) 
with the index $1\leq j \leq \NL$ ($\NL+1\leq j \leq \Nch$),
the total current $I$ through, say, lead no.\ $1$ is given by
\begin{eqnarray*}
I = G \left(\VR - \VL\right),
\end{eqnarray*}
where the conductance $G$ is expressed 
in terms of the scattering matrix $S$ by the Landauer
formula \cite{Landauer,FisherLee,Imry86,BuettikerIBM}
\begin{eqnarray}
G = \frac{e^2}{2\pi\hbar}\int d \varepsilon 
\left(-\frac{\partial f_F}{\partial\varepsilon}\right)
\sum_{i=1}^{\NL}\sum_{j=\NL+1}^{\Nch}
\left|\Scat_{ij}(\varepsilon )\right|^2,
\label{eq:2.63}
\label{eq:2.64}
\end{eqnarray}
$f_F(\varepsilon) = 1/(1+e^{\varepsilon/T})$ being the Fermi distribution
function.  

It is convenient to use the unitarity of the scattering matrix and
rewrite Eq.~(\ref{eq:2.63}) as
\begin{eqnarray}
G = \frac{e^2}{2\pi\hbar}
\left[\frac{\NL\NR}{\Nch}- \int d \varepsilon 
\left(-\frac{\partial f_F}{\partial\varepsilon}\right)
{\rm Tr}\Lambdamatrix\Scat\Lambdamatrix\Scat^{\dagger}
\right],
\label{eq:2.63a}
\end{eqnarray}
where the $\Nch \times \Nch$ traceless matrix $\Lambda$ is defined as
\begin{eqnarray}
  \Lambda_{ij} = \delta_{ij} \times
  \left\{ \begin{array}{ll} - \displaystyle 
  {\NR\over \Nch \vphantom{M_M}},\quad &
  i=1,\dots,\NL, \\
  \hphantom{-} \displaystyle 
  {\NL^{\vphantom{M_M}}\over \Nch},\quad &
   i=\NL+1,\dots,\Nch . \end{array} \right.
\label{eq:2.63b}
\label{P}
\end{eqnarray}
The advantage of Eq.\ (\ref{eq:2.63a}) over the
more conventional form (\ref{eq:2.63})
of the Landauer formula is that it separates
the classical conductance $(2 e^2 / h) (N_1 N_2 / \Nch)$
and the quantum interference correction, the second term
on the r.h.s.\ of Eq.\ (\ref{eq:2.63a}). 

{\em Tunneling density of states.} An important particular 
case of Eq.~(\ref{eq:2.64}) is the strongly
asymmetric setup, where one of the contacts, say
the right one (no.\ $1$ in Fig.\ \ref{fig:Landauer}), has only 
one channel ($\NL = 1$), with a
very small transmission amplitude, $|t_c|^2 = 1 - |r_c|^2 \ll 1$. 
The corresponding
coupling matrix $W$ from Eq.~(\ref{eq:2.58}) then
acquires the form, see also Eq.\ (\ref{eq:rc})
\begin{eqnarray}
W_{\alpha j} = \delta_{\alpha j} \delta_{\alpha 1}\frac{t_c}{2}
\sqrt{\frac{\Delta M}{\pi^2\nu}} + \left(1-\delta_{1 j}\right)
W_{\alpha j}^{(2)}.
\label{eq:2.65}
\end{eqnarray}
Here the channel $j=1$ corresponds to the right contact, while
the channels $j=2,\ldots,\Nch$ correspond to the left contact,
and we have chosen the basis of the states inside the dot such
that the right contact is connected to ``site'' $\alpha=1$.
The matrix $\Wmatrix_{\alpha j}^{(2)}$ describes the
coupling to the left contact. Substituting Eq.~(\ref{eq:2.65})
into Eq.~(\ref{eq:2.60}) and using Eq.~(\ref{eq:2.65}) one obtains,
with the help of Eq.~(\ref{eq:2.64})
\begin{eqnarray}
G = G_1 \Delta M \int d\varepsilon \left(-\frac{\partial
f_F}{\partial\varepsilon}\right)
\nu_T(\varepsilon),
\label{eq:2.66}
\end{eqnarray}
where the tunneling conductance of the left contact is given 
by\footnote{The extra factor $2$ accounts for spin degeneracy. 
Strictly speaking, for electrons with spin $1/2$ one should set
$N_1 = 2$ and sum contributions to the tunneling density of states
for the two spin directions. In the presence of spin-rotation 
symmetry, however, the result is the same as that of Eqs.\
(\protect\ref{eq:2.68}) and (\ref{eq:2.69}) below.} 
\begin{eqnarray*}
G_1=\frac{e^2}{\pi\hbar}|t_c|^2.
\end{eqnarray*}

The quantity $\nu_T(\varepsilon)$ in Eq.~(\ref{eq:2.66}) 
is the tunneling density of states and is given by (in the next two
formulae we omit the superscript $(2)$ in $\hat{W}^{(2)}$):
\begin{eqnarray}
\nu_T(\varepsilon) &=& \nu
\left[\frac{1}
{\varepsilon -{\cal H}
+i\pi\nu WW^\dagger}
 WW^\dagger
\frac{1}
{\varepsilon -{\cal H}
-i\pi\nu WW^\dagger}
\right]_{11} \nonumber
\\ &=&
- \frac{1}{\pi}
  \mbox{Im}\,
\left[\frac{1}
{\varepsilon -{\cal H}
+i\pi\nu WW^\dagger}
\right]_{11}
\nonumber \\ &=&
 -\frac{1}{\pi}\mbox{Im} \left[\Greenmatrix_{o}^R(\varepsilon)
\right]_{11},
\label{eq:2.68}
\end{eqnarray}
a result which one can obtain also by treating the right contact in
the tunneling Hamiltonian approximation. Here, we introduced the 
Green functions for the dot connected to the leads, cf. Eq.~(\ref{eq:2.600})
\begin{equation}
 \Greenmatrix_o^{R(A)}=\left[\varepsilon - {\cal H} \pm i\pi\nu
 \Wmatrix\Wmatrix^\dagger
  \right]^{-1},
\label{eq:Gopen}
\end{equation}
where the ``$+$'' (``$-$'') sign corresponds to the retarded (advanced) Green
functions respectively.

An equivalent form of Eq.~(\ref{eq:2.68}) is obtained by
noticing from Eq.~(\ref{eq:2.600}) that $\partial{\cal
G}_{\alpha\beta}/ \partial{\cal H}_{11}={\cal G}_{\alpha 1}{\cal G}_{
1 \beta}$. One immediately finds from Eq.~(\ref{eq:2.68})
\begin{eqnarray*}
\nu_T(\varepsilon)=
\frac{1}{2\pi i}\frac{\partial}{\partial {\cal H}_{11}}
{\mathrm Tr }\ln \left(
 \frac{
1+ i\pi \nu \Wmatrix^{\dagger}\Greenmatrix(\varepsilon) \Wmatrix}
{1- i\pi \nu \Wmatrix^{\dagger}\Greenmatrix(\varepsilon) \Wmatrix}
\right),
\end{eqnarray*}
and with the help of Eq.~(\ref{eq:2.601}) and 
$\Scat\Scat^{\dagger}=1$, arrives at
\cite{BuettikerJCondensC}
\begin{eqnarray}
\nu_T(\varepsilon)=
-\frac{1}{2\pi i}
{\mathrm Tr } \Scat^\dagger
\frac{\partial {\Scat}}
{\partial {\cal H}_{11}}.
\label{eq:2.69}
\end{eqnarray}
This indicates that the tunneling density of states can be related to
the parametric derivative of the scattering matrix of the dot without 
the left (tunneling) contact.

{\em Free energy for non-interacting electrons.}
The thermodynamic potential $\Omega$ of the dot at chemical 
potential $\mu$ can be found in terms of the Green functions as
\begin{eqnarray}
  \Omega = -{T}
\int d\varepsilon \ln \left[ 1 + e^{-(\varepsilon - \mu)/T} \right]\,
\nu(\varepsilon),
  \label{eq:2.80}
\end{eqnarray}
where the total density of states ({\em i.e.}, the local density of states
integrated over the volume of the dot) $\nu(\varepsilon)$ is given
by [compare with Eq.\ (\ref{eq:2.68})]
\begin{eqnarray}
  \nu(\varepsilon) =  - {1 \over \pi}
\mbox{Im}\,
{\mathrm Tr}
\frac{1}
{\varepsilon -{\cal H}
+i\pi\nu WW^\dagger}
. \label{eq:2.70}
\end{eqnarray}
Similarly to the derivation of Eq.~(\ref{eq:2.69}), 
one can rewrite the latter formula in terms of the energy derivative
of the scattering matrix as \cite{BuettikerJCondensC}
\begin{eqnarray}
  \nu(\varepsilon) = {1 \over 2\pi i}
  {\mathrm Tr}\,
  \Scat^\dagger
  \frac{\partial {\Scat}}
  {\partial\varepsilon}
  \label{eq:OmegaDopen}
\end{eqnarray}
The average number of particles in the dot $\langle \hat{n} \rangle_q
= - \partial \Omega/\partial \mu$ 
(here $\langle \dots \rangle_q$ indicates the average over the quantum
state without disorder average)
can then be found as 
\begin{eqnarray}
\langle \hat{n} \rangle_q= 
\int d\varepsilon f_F(\varepsilon) \nu(\varepsilon),
\label{eq:2.71}
\end{eqnarray}
where $f_F(\varepsilon)$ is the Fermi function.

{\em Statistical properties of $S$.}  The calculation of the
statistical properties of the two-terminal conductance, the tunneling
density of states, and the ground state energy is now reduced to the
analysis of the properties of the scattering matrix for a chaotic
quantum dot, which is a doable, though not straightforward task.  The
available results are collected in an excellent review 
~\cite{Beenakker97} (see, in particular, Ch.\ 2 of that reference). 
Here, we mention a few results that are
needed in Sec.\ \ref{sec:5}.  For particles with spin, but in the
absence of spin-orbit scattering, the scattering matrix is block
diagonal, $\Scat = \mbox{diag}\,(\ScatOrb,\ScatOrb)$, where each block
$\ScatOrb$ is a matrix of size $\Nch^o = \Nch/2$, the total number of
orbital channels.  In the presence of spin-orbit scattering such a
block structure does not exist, and one commonly describes $S$ as a
$\Nch^{o}$-dimensional matrix of quaternions (denoted as $\ScatOrb$),
which are $2 \times 2$ matrices with special rules for complex
conjugation and transposition \cite{Mehtabook}.  For reflectionless
contacts all averages that involve $S$ (or $\Scat^{\dagger}$) only
vanish,
\begin{eqnarray}
   \langle \Scat_{i_1j_1}(\varepsilon_1) \Scat_{i_2j_2}(\varepsilon_2) \ldots
  \Scat_{i_nj_n}(\varepsilon_n) \rangle = 0,
  \ \ n = 1, 2, \ldots. \label{eq:60.7}
\end{eqnarray}
For nonideal contacts, the average of $S$ does not vanish, and is
given by the reflection matrix $r_c$ of the contact,
\begin{eqnarray}
  \langle \Scat_{i_1j_1}(\varepsilon_1) \Scat_{i_2j_2}(\varepsilon_2) \ldots
  \Scat_{i_nj_n}(\varepsilon_n) \rangle = \prod_{k=1}^{n} r_{c,i_kj_k},
  \ \ n = 1, 2, \ldots. \label{eq:60.7b}
\end{eqnarray}
Moments that involve both $S$ and $\Scat^{\dagger}$ have a
rather complicated dependence on energy (see, e.g., Ref.\
\cite{VWZ} for $\beta=1$). 
Here, we mention an approximate result for the second moment
$\langle{\ScatOrb_{i_1j_1}(\varepsilon_1) 
\ScatOrbDag_{i_2j_2}(\varepsilon_2)}\rangle$ in case $r_c=0$,
\begin{eqnarray}
  \langle {\ScatOrb_{i_1j_1}(\varepsilon_1)
  \ScatOrbDag_{i_2j_2}(\varepsilon_2)}\rangle &=&
  \frac{\delta_{i_1 i_2} \delta_{j_1 j_2} +
  (2/\beta - 1)
  \delta_{i_1 j_2} \delta_{j_1 i_2}}{\Nch^o + 2/\beta - 1
  + 2 \pi i (\varepsilon_2 - \varepsilon_1)/\Delta}, 
  \label{eq:avgSE}
\end{eqnarray}
where $\Delta$ is the mean level spacing of the 
dot.\footnote{For the symplectic ensemble,
for which $\beta=4$, $\ScatOrb_{ij}$ is a quaternion and 
the left hand side of Eq.\ (\ref{eq:avgSE}) 
has to be interpreted 
as the quaternion modulus of 
$\langle \ScatOrb_{i_1j_1}\ScatOrbDagQ_{i_2j_2}\rangle$, where
$\ScatOrbDagQ_{i_2j_2}$ is the hermitian conjugate of
$\ScatOrb_{i_2j_2}$. The same interpretation applies to Eqs.\
(\ref{eq:avgSEr}) and (\ref{eq:Savg}) below. 
In the r.h.s.\ of Eq.\ (\ref{eq:avgSEr}),
the complex conjugate $(r_c)^{*}_{i_2j_2}$ should be replaced by 
the hermitian conjugate $(r_c)^{\dagger}_{i_2j_2}$ for $\beta=4$.}
The approximation (\ref{eq:avgSE}) is valid for
$\varepsilon_1 - \varepsilon_2=0$ and for large
values of $|\varepsilon_2 - \varepsilon_1| \gg \Delta$.
It is also valid for arbitrary $|\varepsilon_1-\varepsilon_2|$ if $\Nch \gg 1$.
This is sufficient to estimate average and
fluctuation properties of the conductance for temperatures 
$T \gg \Nch \Delta$ (see Sec.\ \ref{sec:5}).

If $r_c$ is nonzero, one finds
\begin{eqnarray}
  \langle {\ScatOrb_{i_1j_1}(\varepsilon_1)
  \ScatOrbDag_{i_2j_2}(\varepsilon_2)}\rangle &=&
  (r_c)_{i_1j_1}^{\vphantom{\dagger}}
   (r_c)^{*}_{i_2j_2} \nonumber \\ && \mbox{} +
  {(1 - r_c^{\vphantom{\dagger}} r_c^{\dagger})_{i_1 i_2}
   (1 - r_c^{\dagger} r_c^{\vphantom{\dagger}})_{j_1 j_2}
  \over
  \Nch^o + 2/\beta - 1
  - \mbox{Tr}\, r_c r_c^{\dagger}
  + 2 \pi i (\varepsilon_2 - \varepsilon_1)/\Delta}
  \nonumber \\ && \mbox{} + 
  {(2/\beta-1) 
   (1 - r_c^{\vphantom{\dagger}} r_c^{\dagger})_{i_1 j_2}
   (1 - r_c^{\dagger} r_c^{\vphantom{\dagger}})_{j_1 i_2}
  \over
  \Nch^o + 2/\beta - 1
  - \mbox{Tr}\, r_c^{\vphantom{\dagger}} r_c^{\dagger}
  + 2 \pi i (\varepsilon_2 - \varepsilon_1)/\Delta}.
  \label{eq:avgSEr}
\end{eqnarray}
This approximate result is for large energy differences or
for the case $\Nch-\mbox{Tr}\, r_c^{\vphantom{\dagger}}
r_c^{\dagger}\gg 1$. 

In Sec.\ \ref{sec:5} we use the Fourier transform $\Scat(t)$ of the
scattering matrix, 
\begin{eqnarray}
  \Scat(\varepsilon) = \int_0^{\infty} dt \Scat(t) e^{i \varepsilon t},
  \label{eq:Stime}
\end{eqnarray}
for $\varepsilon$ in the upper half of the complex plane. The 
Fourier transform of the hermitian conjugate is defined with
negative times and for $\varepsilon$ in the lower half of the
complex plane,
\begin{eqnarray}
  \Scat^{\dagger}(\varepsilon) = \int_0^{\infty} dt \Scat^{\dagger}(-t)
  e^{-i \varepsilon t}.
\end{eqnarray}
Statistical averages for the scattering matrix in time-representation
can be obtained by Fourier transform of Eqs.\ (\ref{eq:60.7}) --
(\ref{eq:avgSE}), recalling that ensemble and energy averages are
equivalent. In particular, we find that the scattering from a nonideal
point contact with energy-independent $r_c\neq 0$ is instantaneous,
\begin{eqnarray}
  \Scat(t) = r_{c}\, \delta(t) + \mbox{fluctuating part},
  \label{eq:Sdirect26}
\end{eqnarray}
where the reflection matrix of the point contact $r_c$ is defined in
Eq.\ (\ref{eq:2.620}), and the ensemble average of the ``fluctuating
part'' (and of its integer powers) vanishes, see Eq.\ 
(\ref{eq:60.7b}).  For the correlator of $\Scat(t_1)$ and
$\Scat^{\dagger}(t_2)$ for $r_c = 0$, we find from Eq.\ (\ref{eq:avgSE})
\begin{eqnarray}
  \langle{\ScatOrb_{i_1j_1}(t_1)\ScatOrbDag_{i_2j_2}(t_2)}\rangle &=&
\frac{\Delta}{2\pi}
\left[\delta_{i_1 i_2} \delta_{j_1 j_2} +
 (2/\beta-1) \delta_{i_1 j_2} \delta_{j_1 i_2}
\right]
  \nonumber \\ && \mbox{} \times
  \delta(t_1 + t_2)e^{-(\Nch+2/\beta -1 ) t_1 \Delta/(2 \pi)},
  \label{eq:Savg}
\end{eqnarray}
where $\beta=1 (2,4)$ corresponds to the orthogonal
(unitary, symplectic) ensemble. 

{\em Statistical properties of $\Greenmatrix_o$.}  We will see in
Section~\ref{sec:dos}, that the tunneling conductance in the presence
of the interaction can not be expressed in terms of the parametric
derivative of $\Scat$, 
 as it was done in Eq.~(\ref{eq:2.69}).
The transport in this case will be related to the statistics of the
Green functions for the open dot (\ref{eq:Gopen}), and we give
those properties below. It is more convenient to write the results in
the time domain. For averages which include only retarded or only
advanced components one finds
\begin{eqnarray}
&& \langle\left[\Greenmatrix^R_o(t)\right]_{\alpha\gamma}\rangle=
-\langle\left[\Greenmatrix^A_o(t)\right]_{\alpha\gamma}\rangle=
-i\delta(t)\delta_{\alpha\gamma}\frac{\pi}{\Delta M},\nonumber \\
&& \langle \prod_j\Greenmatrix^R_o(t_j)\rangle
= \prod_j\langle\Greenmatrix^R_o(t_j)\rangle,
\label{eq:GopenRR}
\end{eqnarray}
which means that the attachment of the leads does not change the
average level
spacing in the dot.
For averages involving the retarded and advanced components,
we have for the case of reflectionless contacts
\begin{eqnarray}
  \langle\left[\Greenmatrix^R_o(t_1)\right]_{\alpha_1\gamma_1}
\left[\Greenmatrix^A_o(t_2)\right]_{\alpha_2\gamma_2}
\rangle
&=& \frac{2\pi}{M^2\Delta}
\left[\delta_{\alpha_1 \gamma_2} \delta_{\alpha_2 \gamma_1} +
 (2/\beta-1) \delta_{\alpha_1 \gamma_1} \delta_{\alpha_2\gamma_1}
\right]
 \nonumber \\ && \mbox{} \times
  \delta(t_1 + t_2)e^{-(\Nch+2/\beta -1 ) t_1 \Delta/(2 \pi)}.
\label{eq:GopenRA}
\end{eqnarray}
The derivation of Eq.~(\ref{eq:GopenRA}) may be performed similar to
that of Eq.~(\ref{eq:3.2.18}).

This concludes our brief review of the two-terminal conductance,
the tunneling density of states, and the ground state energy
for a quantum dot in the absence of electron-electron interactions.
The interaction between electrons leads to the Coulomb blockade, and
simple formulas as Eq.\ (\ref{eq:2.63}), (\ref{eq:2.69}), and 
(\ref{eq:OmegaDopen}) cease to be valid. With interactions, the answer 
very substantially depends on the conductance of the dot-lead
junctions. The consideration of this regime will be the subject of the
two remaining sections.

\section{Strongly blockaded quantum dots}
\label{sec:3}
In this Section, we discuss the regime of strong Coulomb blockade in
quantum dots. To allow for this regime, the conductances of the
contacts of the dot to the  leads $G_{1,2}$ must
be small in units of $e^2/\pi\hbar$. As it was already explained in
the Introduction, junctions to a quantum dot formed in a
two-dimensional electron gas of a semiconductor heterostructure are
well described by the adiabatic point contact model. Normally, a 
small conductance 
is realized only in single-mode junctions. Since there are two leads
connected to the dot, labeled $1$ and $2$,
we have total number of (orbital) channels 
$\Nch^o=2$,
and the transmission block of the S-matrix of the point contacts
has the form
\begin{equation}
t = \pmatrix{t_1 & 0 \cr  0 & t_2}.
\label{eq:3.1}
\end{equation}
The transmission amplitudes for the separate left and right point
contacts $t_{1,2}$ are related to the corresponding conductances by
Landauer formula (\ref{eq:2.63})
\begin{equation}
G_{1,2} = \frac{e^2}{\pi\hbar}\langle g_{1,2}\rangle, 
\quad \langle g_{1,2}\rangle =|t_{1,2}|^2,
\label{eq:3.2}
\end{equation}
where the extra factor of two comes from the spin degeneracy. We
recall that the coupling coefficients $t_{1,2}$
characterize only the point contact and are not random, see
discussion after Eq.~(\ref{eq:2.58}); they characterize
the average conductance of each point contact and not the 
total conductance of a given sample consisting of both the
two contacts and the quantum dot.

Since the conductances $G_{1}$ and $G_{2}$
are small, we can express the coupling
matrix $W$ relating the Hamiltonian of the closed quantum
dot to its scattering matrix in terms of $G_{1}$ and $G_{2}$,
see Eq.\ (\ref{eq:rc}),
\begin{equation}
\label{weak}
  W_{\alpha j} = \delta_{\alpha j} \delta_{j 1}
  \frac{1}{2 \pi} \sqrt{\Delta M \langle g_1 \rangle \over \nu}
  +
  \delta_{\alpha j} \delta_{j 2}
  \frac{1}{2 \pi} \sqrt{\Delta M \langle g_2 \rangle \over \nu}
  \label{eq:Wglgr}
\end{equation}
Because the coupling is weak, it is possible to construct a
perturbation theory of the conductance in terms of this coupling. This
perturbation theory differs substantially for the peaks and the
valleys of the Coulomb blockade. These two
cases will be considered separately.

\subsection{Mesoscopic fluctuations of Coulomb blockade peaks}
\label{sec:3.1}
In the weak tunneling regime, the charge on the dot is well defined
--- quantum fluctuations of the charge are small ---, and the particle
number $\hat{n}$ is quantized. In the course of an electron tunneling
on and off the dot, its charge varies by one. The activation
energy for the electron transport equals the difference between the
ground state energies of the Hamiltonian (\ref{eq:2.54a}) with two
subsequent values of $n$. A peak in the conductance as a function of
${\cal N}$ occurs at those values of ${\cal N}$ where the two ground
states are degenerate. If one neglects the level spacing $\Delta$, the
conductance peaks are equally spaced: ${\cal N}_{n+1}-{\cal
N}_{n}=1$. According to Eq.~(\ref{eq:2.54a}), the presence of a finite
(and fluctuating) level spacing yields a random contribution to the
spacings of the peaks. We will not discuss the mesoscopic fluctuations
of the peak spacings, where a significant disagreement between the
theory and experiments still exists (see Section V.E of the
review~\cite{Alhassid2000} for a discussion), and concentrate on the
amplitudes of the conductance peaks.

To describe the conductance peaks at low temperatures, $T\ll \Delta$,
it is sufficient to account only for the two charge states mentioned
above,
in which the dot carries $n$ and $n+1$ electrons, respectively,
and does not
have any particle-hole excitations. Even under these simplifying
circumstances the transport problem is not entirely trivial. Every
time the equilibrium number of electrons on the dot, $n$, is odd, its
state is spin-degenerate, which may result in the Kondo effect. The
transition between the spin-degenerate state and the singlet
state occurs just at the gate voltages ${\cal N}$ that correspond to
the points of charge degeneracy $n\leftrightarrow n+1$. At these
points, the quantum dot attached to the leads is in a state similar to
the mixed valence regime of an atom embedded in a metallic host
material. We postpone the discussion of these delicate many-body
phenomena till Section~\ref{sec:Kondo}, and confine ourselves here
to the results of the theory of rate equations. The applicability of
this theory sets a lower limit for the temperature in the 
discussion below: $T$ must exceed the
width $(g_1+g_2)\Delta$ of a discrete level broadened by the electron
escape from the dot. At such temperatures, the quantum coherence
between the electron states on the dot and within the leads is not
important. In this section we will also restrict ourselves to the
case that only a single discrete level in the dot contributes to
the current, {\em i.e.}, we restrict ourselves to the temperature 
interval
\begin{equation}
(g_1+g_2)\Delta \ll T\ll \Delta.
\label{eq:3.1.1}
\end{equation}
The case $\Delta \ll T \ll \Ec$ is briefly discussed in Subsection 
\ref{sec:tdependence}.

The upper bound in Eq.~(\ref{eq:3.1.1}) ensures that 
only the last occupied discrete
level in the dot contributes to the electron transport. 
In the absence of the interaction this level can be in four states:
we denote by $P_\sigma$ the probability 
of this level to be occupied  with one electron of given spin
projection $\sigma=\uparrow,\downarrow$,
by $P_0$ to be empty, and by $P_2$ to be occupied with two electrons.
Those probabilities are normalized as $P_0 + P_2 + P_\uparrow +
P_\downarrow =1$. If the level is adjusted in
resonance with the Fermi level in the leads,
then due to the electron transfer through the
dot, the state of the level changes between all of those states.

The picture is changed when the interaction between electrons is
included. As can be seen from Eq.~(\ref{eq:2.54}), only three 
states with the
number of electrons on the dot differing by 1, can be in resonance.
States with a different charge
always have an activation energy of the order of
the charging energy $\Ec$. As long as $\Ec\gg T$, these states do not
participate in the electronic transport through the dot. A
conductance peak is
characterized by the condition $|{\cal N} - {\cal N}^*| \lesssim T/\Ec \ll
1$. Depending on ${\cal N}^*$, one can distinguish two cases: \\
(i) ${\cal N}^* = 2j +1/2$, with $j$ being an integer, which
corresponds to states $|0\rangle$, $|\uparrow\rangle$, and
$|\downarrow\rangle$ in resonance, and $|2\rangle$ having an extra
energy $\Ec$;\\
(ii) ${\cal N}^* = 2j - 1/2$, which corresponds to states $|2\rangle$,
$|\uparrow\rangle$, and $|\downarrow\rangle$ in resonance and
$|0\rangle$ having an extra energy $\Ec$.  These limitations on the
Hilbert space of the level lead to the deviation from the simple
Breit-Wigner formula even in the temperature regime (\ref{eq:3.1.1}).

The form of the conductance peaks can be calculated using rate
equations.
We will present the master equation describing case (i).
Case (ii) is considered analogously and we will state the result after
Eq.~(\ref{eq:3.1.5}).
Considering the Hamiltonian (\ref{eq:2.58}) with the coupling
matrix $W$ given by Eq.\ (\ref{weak}), and in the Fermi Golden Rule
approximation, one finds the following rate 
equations~\cite{MatveevLG88,BeenakkerCB,Averin91}
\begin{eqnarray}
&&\frac{dP_\uparrow}{dt}=
\frac{\Delta}{2\pi\hbar}
\left\{g_1\left[f_1P_0-(1-f_1)P_\uparrow\right]
+g_2\left[f_2P_0-(1-f_2)P_\uparrow\right]\right\},
\label{eq:3.1.2}\\
&&\frac{dP_\downarrow}{dt}=
\frac{\Delta}{2\pi\hbar}
\left\{g_1\left[f_1P_0-(1-f_1)P_\downarrow\right]
+g_2\left[f_2P_0-(1-f_2)P_\downarrow\right]\right\},
\nonumber\\
&&\frac{dP_0}{dt}=
\frac{\Delta}{2\pi\hbar}
\left\{g_1\left[(1-f_1)(P_\uparrow+P_\downarrow)-2f_1P_0\right]
  \right. \nonumber \\ && 
\hphantom{\frac{dP_0}{dt}=\frac{\Delta}{2\pi\hbar}}
\left. \mbox{} 
+g_2\left[(1-f_2)(P_\uparrow+P_\downarrow)-2f_2P_0\right]\right\},
\nonumber
\end{eqnarray}
where $g_{1,2}$ are the exact conductances of the point contacts,
\begin{equation}
g_1 = \langle g_1 \rangle M 
|\psi( 1)|^2, 
\quad
g_2 = \langle g_2 \rangle M 
|\psi( 2)|^2,
\label{gf}
\end{equation}
and $\psi(n)$ are the RMT components of the $M$-component one-electron
wavefunction of the partially occupied level responsible for the
electron transport through the dot. It is readily seen that
$g_{1,2}$ experience strong mesoscopic fluctuations, which will be
important later.\footnote{The conductances $g_1$ ($g_2$) are
conductances for electron transport from reservoir $1$ ($2$)
through the point contact into 
a dot state with wavefunction $\psi$. They are fluctuating
quantities, because they depend on the local density of states
in the dot. In contrast, $\langle g_1 \rangle$
and $\langle g_2 \rangle$ in Eq.\ (\ref{eq:3.2}) are conductances
of the point contacts if connected to a (fictitious) electron
reservoir on either side. These
conductances are a property of the contact only, and do not
exhibit mesoscopic fluctuations.}
In Eq.\ (\ref{gf}), the normalization of the
wavefunction $\psi$ is such that $\langle |\psi(n)|^2 \rangle =
M^{-1}$.

Among the three equations (\ref{eq:3.1.2}) only two are independent,
and we have to supplement
those equations with the normalization condition
$P_0+P_\downarrow+P_\uparrow=1$ (recall that $P_2=0$, since double 
occupancy has an energy cost $\Ec \gg T$).
In the presence of a small bias $eV = \mu_1 - \mu_2$ applied
between the leads, the occupation factors $f_{1}$ and $f_{2}$ are
given by the Fermi distribution
functions of electrons taken at the 
energy $E \pm eV/2$, where $E= 2 \Ec({\cal N} - {\cal N}^* )$, which
is the energy it takes to put an electron into the unoccupied state with
the lowest energy in the dot. 

The current through, say, junction no.\ 1 is
\begin{equation}
I=\frac{e}{2\pi\hbar}\Delta g_1
\left[2f_1P_0-(1-f_1)(P_\uparrow+P_\downarrow)\right].
\label{eq:3.1.3}
\end{equation}

The stationary solutions for $P_{0}$, $P_{\uparrow}$ and
$P_{\downarrow}$ can be found from the rate
equations~(\ref{eq:3.1.2}). We substitute the result into
Eq.~(\ref{eq:3.1.3}), take the limit of low bias $V$, and
find that the conductance
of the system is given by
\begin{equation}
G({\cal N})=\frac{I}{V}{\Big |}_{V\to 0}=-\frac{e^2}{\pi\hbar}
\frac{g_1g_2}{g_1+g_2}\frac{\Delta}{T}
\frac{\partial f_F/\partial x}{1+f_F(x)}.
\label{eq:3.1.4}
\end{equation}
The dependence of the conductance on ${\cal N}$ comes through the
energy dependence of the Fermi distribution function $f_F(x)$:
\begin{equation}
f_F(x)=\frac{1}{e^x+1},\quad x=\frac{2 \Ec}{T}
\left({\cal N} -{\cal N}^*\right).
\label{eq:3.1.5}
\end{equation}
Note that the maximum of the conductance is slightly shifted (by $\sim
T/\Ec$) away from the point ${\cal N}={\cal N}^*$, and more
importantly, the conductance peak is not symmetric. This is the
consequence of correlations in transport of electrons with opposite
spins through a single discrete state on the dot. Because of the
repulsion, no more than one electron can reside in that state at any
time ($P_2=0$).  Case (ii) is also described by Eq.~(\ref{eq:3.1.4})
after the replacement $x \to -x$.

In the maximum, the function (\ref{eq:3.1.4}) takes the value 
of\footnote{The numerical prefactor in Eq.\ 
(\protect{\ref{eq:3.1.6}}) is different from that used in Refs.\
\cite{Efetovbook,FalkoEfetov96,Stone92,Alhassid98,Alhassid96,Alhassid962}.
The differences arises, because these references did not properly 
account for the 
two-fold spin degeneracy in the conditions of the Coulomb blockade.
However, this does not affect the results for the functional
form of the distribution functions for the fluctuations of the
conductance.  }
\begin{equation}
G_{\rm peak} =\left(3 - 2^{3/2}\right)\frac{e^2}{\pi\hbar} 
\left(\frac{g_1g_2}{g_1+g_2}\right)
\left(\frac{\Delta}{T}\right).
\label{eq:3.1.6}
\end{equation}
Equations (\ref{eq:3.1.4}) and (\ref{eq:3.1.6}) relate the conductance
through an interacting-electrons system to the single-particle
electron wave functions; the height of the conductance peaks is
expressed in terms of $g_{1,2}$. According to Eq.~(\ref{gf}), these
quantities experience mesoscopic fluctuations, which result in the
fluctuations of $G_{\rm peak}$. We discuss these fluctuations next.

Our goal is to compute the distribution function
\begin{equation}
W(G) = \langle 
\delta \left(G - G_{\rm peak} \right)
\rangle.
\label{eq:3.1.7}
\end{equation}
Most easily this task is accomplished for the pure orthogonal
$\beta=1$ or unitary $\beta=2$ ensemble~\cite{Stone92}.  In these
cases, the wave-functions are distributed according to
Eq.~(\ref{eq:2.102}).  Substituting Eq.~(\ref{eq:2.102}) into
Eq.~(\ref{gf}) and the result into Eq.~(\ref{eq:3.1.6}), we obtain
$G_{\rm peak}$ in terms of the average
conductances $\langle g_{1,2}\rangle$ and a single random variable
$\alpha$,
\begin{equation}
G_{\rm peak} =\alpha\left(3 - 2^{3/2}\right)
\frac{e^2}{\pi\hbar}\left(\frac{\Delta}{T}\right)
\frac{2
\langle g_1\rangle
\langle g_2 \rangle
}{\left(\langle g_1\rangle^{1/2}
+\langle g_2 \rangle^{1/2}
\right)^2
}
,
\label{eq:3.1.9}
\end{equation}
The distribution function for $\alpha$ is very sensitive to the
presence ($\beta=2$) or absence ($\beta=1$)
of a time-reversal symmetry breaking magnetic field, 
\begin{eqnarray}
W_{\beta=1}(\alpha) &=&\frac{e^{-\alpha}}
{\sqrt{ \pi \alpha}};
\label{eq:3.1.10}\\
W_{\beta=2}(\alpha) &=& \alpha \left(1 - a \right)^2
e^{- \alpha (1+a)} 
\left\{
K_0\left[\alpha (1- a)\right] +
\frac{1+a}{1-a} K_1\left[\alpha (1- a)\right]\right\}.
\nonumber
\end{eqnarray}
Here $K_0(x),\ K_1(x)$ are modified Bessel functions of the second 
kind~\cite{Ryzhik}, while the parameter $a$ characterizes the
asymmetry of the point contacts:
\begin{equation}
a= \left(\frac{
{\langle g_1 \rangle}^{1/2} - {\langle g_2 \rangle}^{1/2} 
}{{\langle g_1 \rangle}^{1/2} + {\langle g_2 \rangle}^{1/2} }\right)^2.
\label{eq:3.1.11}
\end{equation}
It is interesting that the distribution function in the orthogonal
case $\beta=1$ remains universal no matter how asymmetric contacts
are. In the unitary ensemble such a universality does not hold.
In the case of symmetric contacts, $\langle g_1\rangle=\langle
g_2\rangle$, this result was first obtained in \cite{Stone92}.
The general case was considered in \cite{Prigodin93}.

\begin{figure}
\epsfxsize=0.95\hsize
\centerline{\epsfbox{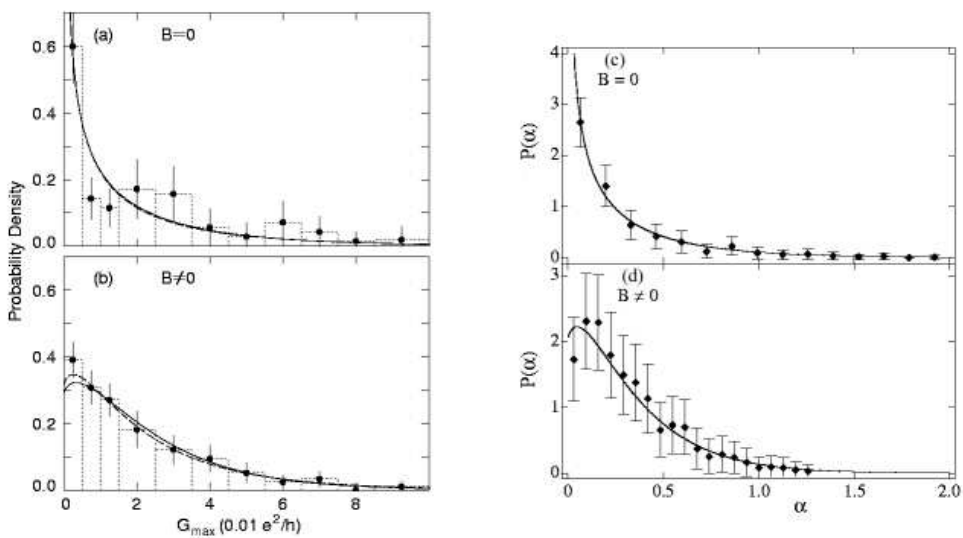}}
\caption{Histograms of conductance peak heights for orthogonal (a,c) and
unitary ensembles (b,d). Data (a,b) and (c,d) are taken from
\protect\cite{Chang96} and \protect\cite{Folk96} respectively. An
example of the raw data used for the graphs (c,d) is shown in
Fig.~\protect\ref{Fig2} (b).}
\label{Fig8}
\end{figure}

It is important to emphasize that the probability distributions
 (\ref{eq:3.1.10}) are strongly non-Gaussian and they are sensitive to
 the magnetic field.  These two features were checked experimentally
 in \cite{Chang96,Folk96}. The results were in a reasonable agreement
 with the theory,\footnote{The symmetry breaking occurs due to the
 orbital effect of the magnetic field, and one can neglect the role of
 Zeeman energy. Indeed, in experiments~\cite{Folk96} the
 crossover between the ensembles corresponds to a field $B\sim
 10$mT. Zeeman energy at such a field would be comparable to
 temperature only at $T\lesssim 2.5$mK, which is $10$ times lower than
 the base temperture in the measurements.} see
 Fig.~\ref{Fig8}.

It is also noteworthy that the average conductance depends on the
magnetic field:
\begin{eqnarray}
 && \langle G_{\rm peak}^{\beta}\rangle =\left(3 - 2^{3/2}\right)
\frac{e^2}{\pi\hbar}\left(\frac{\Delta}{T}\right)
\frac{
\langle g_1\rangle
\langle g_2 \rangle
}{\left(\langle g_1\rangle^{1/2}
+\langle g_2 \rangle^{1/2}
\right)^2
}F\left[2- \beta  + \left(\beta -1\right) a \right], \nonumber \\
\label{eq:3.1.12}
\end{eqnarray}
where the asymmetry parameter $a$ was defined in Eq.~(\ref{eq:3.1.11})
and the dimensionless function $F(x)$ is given by 
\begin{equation}
F(x)=\frac{1+x}{2x}
- \frac{(1-x)^2}{4 x^{3/2}}
{\mathrm arcosh}\left(\frac{1+x}{1-x}\right);
\quad F(1) = 1; \quad F(0) = \frac{4}{3}.
\label{eq:3.1.13}
\end{equation}
We see from Eqs.~(\ref{eq:3.1.12}) and Eq.~(\ref{eq:3.1.13}) that the
average conductance in the magnetic field (unitary case, $\beta=2$) is
{\em larger} than in the absence of the field (orthogonal case,
$\beta=1$). This phenomenon is analogous to the effect of negative
magnetoresistance due to the weak localization in bulk systems
\cite{magnetoresistance} with the same physics involved.  For the
average peak heights, the magnetoresistance is sensitive to the
asymmetry of the contacts (\ref{eq:3.1.11}) and in the limiting case
of a very asymmetric contacts, $a\to 1 $, the magnetoresistance
vanishes. 

So far, we have presented results for the peak height distributions for
ensembles of Coulomb-blockaded quantum dots with either fully
preserved or fully broken time-reversal symmetry. To investigate how
the crossover between those two ensembles happens for the temperature
regime (\ref{eq:3.1.1}), one has to use the
statistics of the wavefunctions in the crossover ensemble of random
Hamiltonians distributed according to Eq.~(\ref{eq:2.100}). This
statistics is described by Eqs.~(\ref{eq:2.103}) and
Eq.~(\ref{eq:2.104}).  To obtain the moments of the conductance at the
crossover regime, is a lengthy, albeit straightforward calculation
involving Eqs.~(\ref{gf}), (\ref{eq:3.1.6}), (\ref{eq:3.1.8}) and
(\ref{eq:3.1.14}).  We refer the readers to the original
papers~\cite{Alhassid98,Alhassid982} and review~\cite{Alhassid2000}
for the results.  (The peak height distribution function for the limit
$a \to 1$ corresponding to the case of strongly asymmetric contacts
was obtained earlier in Ref.~\cite{FalkoEfetov94}, see also
Ref.~\cite{Efetovbook}.

Besides the distribution functions of the peaks heights at a
fixed value of the magnetic field, another
quantity of interest is the correlation function of the peak
heights for different values of the magnetic field. It is 
defined as
\begin{eqnarray}
C(B_1,B_2) &=& \frac{\langle\delta G(B_1)\delta G(B_2) \rangle}
{\left(\langle\delta G(B_1)^2\rangle \langle \delta G(B_2)^2 \rangle
\right)^{1/2}}, 
\nonumber\\
 \delta G(B) &=& G_{\rm peak}(B) - \langle G_{\rm peak}(B).
\rangle 
\label{peaks}
\label{eq:3.1.15}
\end{eqnarray}
Physically, the correlator 
$C$ characterizes ``how fast'' the ensemble changes under
the effect of the magnetic field.

The peak-height correlation function (\ref{peaks}) 
was studied numerically in Refs.\
\cite{Alhassid96,Alhassid962}. For the unitary ensemble, $N_h^{C} \gg 1$,
the results were found to be well approximated as
\begin{equation}
C(B_1,B_2) \approx \frac{1}{\left(1 + 0.25 N_h^D\right)^2},
\label{eq:3.1.16}
\end{equation}
where the parameters $N_h^{D,C}$ are related to the difference and sum
of the magnetic fields by Eq.~(\ref{eq:2.101}).  Once again, the characteristic
magnetic field is determined by the condition $N_h \simeq 1$.

\subsection{Mesoscopic fluctuations of Coulomb blockade valleys}
\label{sec:3.2}
In this subsection, we study the mesoscopic fluctuations of the
conductance in the regime where the gate voltage ${\cal N}$ is tuned
away from the degeneracy point \cite{AleinerGlazman96}, $|{\cal N} -
{\cal N}^\ast| > T/\Ec$. Equations (\ref{eq:3.1.4}) and
(\ref{eq:3.1.5}) predict in this case the conductance to be
exponentially small $G({\cal N}) \propto \exp\left[-2\Ec|{\cal N} -
  {\cal N}^\ast|/T\right]$. However, this expectation is not correct.
The rate equations of Sec.~\ref{sec:3.1} take into account only
processes which are of the first order in the right or left
conductance. In these processes a charge is transferred from a lead
(initial state) to the dot (final state), or vise versa. In the
valleys such processes are exponentially suppressed because of the
high energy in the final state. On the other hand, higher order
processes allow for both the initial and final states to be in the
leads, so that the thermal exponent does not appear. Such
processes, commonly referred to as co-tunneling~\cite{AverinNazarov90},
are suppressed only algebraically, $G({\cal N}) \propto
1/\left[2\Ec|{\cal N} - {\cal N}^\ast|\right]$, and dominate the
transport already at moderate deviations from the conductance peaks.

Such second-order processes are shown in Fig.~\ref{Fig9}. The processes
(a),(b) and (c),(d) are electron-like and hole-like processes,
respectively. In an electron-like process, the intermediate state
corresponds to an extra electron placed on the dot. In this case the
electrostatic part of the energy of the virtual state is given by [see
Eq.~(\ref{eq:2.54a})],
\begin{equation}
E_e=2\Ec\left({\cal N}^\ast - {\cal N} \right),
\label{eq:3.2.1}
\end{equation}
where ${\cal N}^*$ is the half-integer corresponding to the Coulomb
Blockade peak. For a hole-like process, the intermediate state has
one less electron on the dot, so that the electrostatic part of
the energy of the virtual state is
\begin{equation}
E_h=2\Ec\left( {\cal N} - {\cal N}^\ast +1 \right).
\label{eq:3.2.2}
\end{equation}
In writing Eqs.~(\ref{eq:3.2.1}) and (\ref{eq:3.2.2}) we consider the
dimensionless gate voltages ${\cal N}$ in the interval ${\cal
  N}^*-1<{\cal N}<{\cal N}^*$.

{\begin{figure}
\epsfysize=0.5\hsize
\centerline{\epsffile{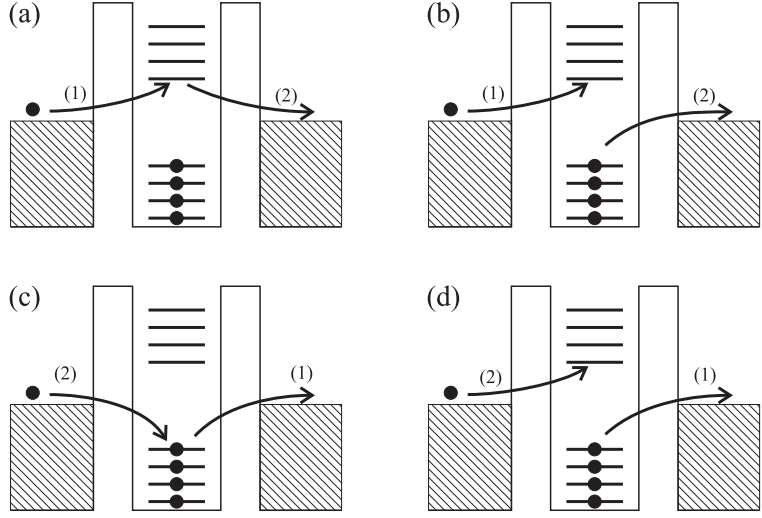}}
\hfill
\caption{Schematic picture of the second order processes contributing
to the conductance in the Coulomb blockade valleys. Processes (a), (b)
are electron-like and (c), (d) are hole-like. Elastic co-tunneling
\protect\cite{AverinNazarov90}
corresponds to the processes (a) and (c), and inelastic co-tunneling 
\protect\cite{AverinOdintsov,AverinNazarov90} is described by (b) and (d).}
\label{Fig9}
\end{figure}
}

Even though all the processes of Fig.~\ref{Fig9} are of the same order
in the tunneling constants, there is an important difference between
the transitions (a) and (c) on the one hand, and (b) and (d) on the
other hand: For the cases (a) and (c), the very
same electron (hole) that tunnels into the dot leaves the dot, whereas
in the cases (b) and (d), a different electron (hole) tunnels from 
the dot. The result is that in the cases (b) and (d), the final 
state differs from the initial state
by the excitation of an extra electron-hole pair (inelastic
co-tunneling). The phase volume for such a two-particle
excitation is proportional to $(\varepsilon_{\rm eh})^2$ (if
$\varepsilon_{\rm eh} \gg \Delta$), where
$\varepsilon_{\rm eh}$ is the energy of the electron-hole pair.
Since $\varepsilon_{\rm eh}$ can not exceed the temperature $T$, the 
inelastic
contribution vanishes at low temperature as $T^2$, even in the limit
$\Delta\to 0$.  The elastic contributions (a) and (c), in contrast,
are temperature independent. For that reason we here concentrate on 
the elastic co-tunneling and return to inelastic co-tunneling only 
at the end of the Section.

The tunneling conductance can be calculated according to the Golden
rule as
\begin{equation}
G = 2\frac{2\pi e^2}{\hbar}\nu^2 \left|A_e + A_h\right|^2,
\label{e-conductance}
\label{eq:3.2.3}
\end{equation}
where $\nu$ is the density of states per one spin 
in the leads and the extra factor of $2$ takes care of the spin 
degeneracy.
The amplitudes $A_e$ and $A_h$ correspond to the processes of
Fig.~\ref{Fig9}
(a) and (c), respectively. They can be calculated with the help
of the Hamiltonian (\ref{eq:2.58}), with Eq.\
(\ref{weak}) for the matrix $W$,
in second order perturbation
theory in the tunneling probabilities. The result is
\begin{eqnarray}
A_e &=& \sqrt{\langle g_1\rangle\langle g_2\rangle}
\left(\frac{\Delta M}{4\pi^2\nu}\right)
\sum_\alpha 
\frac{\psi_\alpha (1)\psi_\alpha^\ast (2)}{\varepsilon_\alpha + E_e}
\theta (\varepsilon_\alpha);
\nonumber\\
A_h&=&- \sqrt{\langle g_1\rangle\langle g_2\rangle}
\left(\frac{\Delta M}{4\pi^2\nu}\right)
\sum_\alpha 
\frac{\psi_\alpha (1)\psi_\alpha^\ast (2)}{-\varepsilon_\alpha + E_h}
\theta (-\varepsilon_\alpha),
\label{eq:3.2.4}
\end{eqnarray}
where $\varepsilon_\alpha$  and $\psi_\alpha(i)$
are the exact eigenenergies and eigenvectors of the noninteracting
Hamiltonian (\ref{eq:2.11}), and the argument $i=1,2$ of the
wave function $\psi$ labels the sites 
coupled to contacts $1$ and $2$, respectively, see 
Eqs.~(\ref{weak}).
The eigenenergies $\varepsilon_\alpha$ are measured from the upper filled
level in the dot, so that step function $\theta (\pm\varepsilon_\alpha)$
selects empty (occupied) orbital states for electron (hole) like
processes of Fig.~\ref{Fig9}a(c). The relative sign difference in the 
electron, $A_e$, and hole, $A_h$, amplitudes comes from the commutation
relation of the Fermion operators.

Substituting Eq.~(\ref{eq:3.2.4}) into Eq.~(\ref{eq:3.2.3}), we find
\begin{equation}
G = \frac{e^2}{4 \pi^3\hbar}
\langle g_2 \rangle
\langle g_1 \rangle
|F_e + F_h|^2,
\label{eq:3.2.6}
\end{equation}
where the dimensionless functions $F$ are given by
\begin{eqnarray}
  && F_e = M \Delta 
\sum_\alpha 
\frac{\psi_\alpha (1)\psi_\alpha^\ast (2)}{\varepsilon_\alpha + 
E_e
}\theta (\varepsilon_\alpha);\nonumber \\
  && F_h = M \Delta 
\sum_\alpha 
\frac{\psi_\alpha (1)\psi_\alpha^\ast (2)}{\varepsilon_\alpha 
 -
E_h
}\theta (-\varepsilon_\alpha)
.
\label{eq:3.2.7}
\end{eqnarray}
It is also possible, and more convenient for some applications, to
express $F$ in terms of the exact one electron 
Green functions of the dot  
\begin{eqnarray}
&& F_e=M \Delta\int_0^\infty \frac{d\varepsilon}{2\pi i}
 \frac{{\cal G}^{A}_{12}(\varepsilon) - 
{\cal G}^{R}_{12}(\varepsilon)}{\varepsilon +
E_e },
\nonumber \\
&& F_h=M \Delta\int^0_{-\infty} \frac{d\varepsilon}{2\pi i}
 \frac{{\cal G}^{A}_{12}(\varepsilon) - {\cal G}^{R}_{12}(\varepsilon)}{\varepsilon 
 - E_h },
\label{eq:3.2.8}
\end{eqnarray}
where the Green function is given by Eq.~(\ref{eq:3.2.9}).

Equations (\ref{eq:3.2.6}) -- (\ref{eq:3.2.8}) solve the part of the
problem depending on the interactions; the conductance is expressed in
terms of the single electron eigenfunctions and eigenvalues of the
isolated dot. What remains is to perform the statistical analysis
of the conductance in a fashion similar to what was done in
Sec.~\ref{sec:3.1} for the peak heights. Note, however, that there
is an important difference with the case of the peak height
statistics:
while the height of a conductance peak depended on 
the wavefunction of one level only, many levels with energies of the 
order of $E_{e(h)}$
contribute to the tunneling. The superposition of such a large number of
tunneling amplitudes significantly affects the conductance
fluctuations~\cite{AleinerGlazman96}. Despite these complications, 
expressions for the distribution function of the
conductance in the regime of elastic co-tunneling, and the 
correlation function of the conductance in a
magnetic field, can be found in closed form. We now describe
their calculation.

The cases of the pure orthogonal ($\beta=1$) or unitary ensembles
($\beta=2$) are the easiest to investigate. We use the fact that the
eigenvectors and eigenvalues of a random matrix in these cases are
independent of each other. Moreover, the eigenvectors of different
eigenstates are also Gaussian variables independent of each other, see
Eq.~(\ref{eq:3.2.10}).

Using Eq.~(\ref{eq:3.2.7}) and Eq.~(\ref{eq:3.2.10}), we immediately
find
\begin{equation}
\langle F_e\vphantom{o}
  F_e^\ast \rangle =\Delta^2
\sum_\alpha \theta (\varepsilon_\alpha)
{\left\langle
\frac{1}{\left(\varepsilon_\alpha + 
E_e\right)^2
}
\right\rangle}.
\label{eq:3.2.11}
\end{equation}
Since the averaged density of states does not depend on energy, and
since we consider $E_e,E_h\gg\Delta$, we can replace
the summation $\Delta\sum_\alpha$ by the integration 
$\int d\varepsilon_\alpha$ in
Eq.~(\ref{eq:3.2.11}). The resulting expressions are
\begin{eqnarray}
&& \langle F_eF_e^\ast \rangle =
\frac{\Delta}{E_e}, \quad
\langle F_eF_e\rangle=\left(\frac{2}{\beta}-1\right)\frac{\Delta}{E_e};
\nonumber\\
&& \langle F_hF_h^\ast \rangle =
\frac{\Delta}{E_h}, \quad
\langle F_hF_h\rangle=\left(\frac{2}{\beta}-1\right)\frac{\Delta}{E_h};
\nonumber\\
&& \langle F_eF_h^\ast \rangle
=\langle F_eF_h\rangle=0,
\label{eq:3.2.12}
\end{eqnarray}
where all the averages are found in the same fashion as
(\ref{eq:3.2.11}). 

Using Eq.~(\ref{eq:3.2.12}), we can find the average of the
conductance (\ref{eq:3.2.6}):
\begin{equation}
\langle G \rangle  = \frac{e^2}{4 \pi^3\hbar}
\langle g_2 \rangle
\langle g_1 \rangle
\left[\frac{\Delta}{E_e} + \frac{\Delta}{E_h}\right].
\label{eq:3.2.13}
\end{equation}
This expression was first obtained\footnote{Equation
(\protect\ref{eq:3.2.13}) is different from Eq.~(13) of
\protect\cite{AleinerGlazman96} by a factor of $2$, because of
algebraic error in this reference.} in \cite{AverinNazarov90} by a
different method of the ensemble averaging. (To calculate the
quantitites of the type of Eq.~(\ref{eq:3.2.12}), the authors of
Ref.~\cite{AverinNazarov90} used the technique developed in the
semiclassical theory of inhomogeneous
superconductors~\cite{deGennes,Shapoval}.) Notice that a magnetic
field has no effect on the average conductance; $\langle G\rangle$
remains the same for the unitary and orthogonal ensembles. We remind
the reader that Eq.~(\ref{eq:3.2.13}) is obtained within the RMT
theory, and therefore assumes $\Ec\ll E_T$.

However, as we show below, the fluctuations of the conductance are of
the order of the average conductance, so that the average 
(\ref{eq:3.2.13}) is not a good representative of the ensemble. 
In order to characterize the conductance
fluctuations, we need higher moments of the factors
$F_e$ and $F_h$ than those calculated in Eq.\ (\ref{eq:3.2.7}). 
Using Eq.~(\ref{eq:3.2.10}), we find
\begin{equation}
\langle (F_eF_e^\ast)^2 \rangle =
\left(4-\beta\right)
\Delta^4
\sum_{\alpha,\gamma} \theta (\varepsilon_\alpha)
\theta (\varepsilon_\gamma)
\left\langle
\frac{1}{\left(\varepsilon_\alpha + 
E_e\right)^2}
\frac{1}{\left(\varepsilon_\gamma + 
E_e\right)^2
}
\right\rangle.
\label{eq:3.2.14}
\end{equation}
Further simplification follows from the condition $E_e \gg \Delta$.  It
enables us to neglect the level repulsion and substitute
$\Delta^2\sum_{\alpha,\gamma} \to \int
d\varepsilon_\alpha d\varepsilon_\gamma$. One easily checks that the
corrections to this approximation are of the order of 
$( \Delta/E_e)^2\ln (E_e/\Delta) \ll
1$. As a result, we have
\begin{equation}
\langle (F_eF_e^\ast)^2 \rangle
= 2 \langle F_eF_e^\ast \rangle^2 +
\langle F_e^\ast F_e^\ast \rangle\langle F_eF_e \rangle,
\label{eq:3.2.15}
\end{equation}
where the irreducible averages are given in
Eq.~(\ref{eq:3.2.12}). Equation (\ref{eq:3.2.15}) indicates that the
amplitudes $F$ entering the conductance are random Gaussian
variables with zero average and variance (\ref{eq:3.2.12}). (This can
be proven by explicit consideration of the higher moments.) This
observation immediately allows one to compute the fluctuations of the
conductance in the valley,
\begin{equation}
\frac{\langle\delta G^2\rangle}
{\langle G \rangle^2}= \frac{2}{\beta}.
\label{eq:3.2.16}
\end{equation}
With the breaking of time reversal symmetry the fluctuations are
reduced by a factor of $2$ similar to the conductance fluctuations in
the bulk systems, $\langle G \rangle \gg e^2/(\pi\hbar)$. 
For the cases of $E_e \gg E_h$ or $E_e \ll E_h$
(vicinity to the Coulomb blockade peak), Eq.~(\ref{eq:3.2.16}) was
first obtained in \cite{AleinerGlazman96}. 

It is clear from Eq.~(\ref{eq:3.2.16}) that the fluctuations of the
conductance are of the order of the conductance itself, despite the
naive expectation that the large number of the contributing states 
produce the self-averaging. The reason for this, is that we have to
add amplitudes, not probabilities, to compute the 
conductance.\footnote{This statement is valid for the case when
the quantum dot is coupled to the electron reservoirs via tunneling
point contacts, which we consider here.
In case of a wide junction (i.e., a contact with many
contributing channels with small transparency each), the
conductance is a sum over all channels
in the contacts of the transmission probabilities for these
channels. In that case, the relative size of the fluctuations
is decreased with respect to the average.}

Because the fluctuations of the conductance are large, its
distribution function is not Gaussian. However, it can be found
analytically with the help of Eq.~(\ref{eq:3.2.6}) and the established
fact that $F$'s are Gaussian variables with the variances
(\ref{eq:3.2.12}). Straightforward calculation yields that the
distribution of $G$ normalized to its average is given by
\begin{equation}
W(\alpha) = \left\langle\delta\left(\alpha - \frac{G}{\langle
G\rangle}\right)\right\rangle
=\cases{\theta(\alpha)
\displaystyle{\frac{e^{-\alpha/2}}{\sqrt{2\pi\alpha}}}; &
$\beta=1$\cr 
\theta(\alpha){e^{-\alpha}} & $\beta = 2$ },
\label{eq:3.2.17}
\end{equation}
which coincides with Porter-Thomas distribution \cite{PorterThomas}.
This was to be expected from the central limit theorem, because 
the conductance is determined by a
large number of random amplitudes, which is exactly the assumption
behind the Porter-Thomas distribution.

Now, we turn to the study of the effect of the magnetic field on the
valley conductance. Our purpose is to find the correlation function
for the conductance fluctuations, similar to Eq.~(\ref{eq:3.1.15}),
and the distribution function on the crossover between orthogonal and
the unitary ensembles. Here we discuss only the situation in the
vicinity of the peaks, which corresponds to the condition $E_e \gg
E_h$ or $E_e \ll E_h$, so that only one contribution is dominant. The
results for the general case are presented in Appendix~\ref{ap:2}.
Because a large number of levels contribute to the transport of 
electrons through the quantum dot, we can
use Eq.~(\ref{eq:3.2.8}) and then apply Eq.~(\ref{eq:3.2.18}) to
average the product of the Green functions. We thus find that the
amplitudes $F$ are still Gaussian variables, while their
averages are modified in comparison with Eq.~(\ref{eq:3.2.12}):
\begin{eqnarray}
\langle F_eF_e^\ast \rangle =
\frac{\Delta}{E_e}
{\Lambda\left(\frac{N_h^D\Delta}{2\pi E_e}\right)}
, \quad
\langle F_eF_e\rangle=\frac{\Delta}{E_e}
{\Lambda\left(\frac{N_h^C\Delta}{2\pi E_e}\right)};
\label{eq:3.2.19}\\
\langle F_hF_h^\ast \rangle =
\frac{\Delta}{E_h}
{\Lambda\left(\frac{N_h^D\Delta}{2\pi E_h}\right)}
, \quad
\langle F_hF_h\rangle=\frac{\Delta}{E_h}
{\Lambda\left(\frac{N_h^C\Delta}{2\pi E_h}\right)}.
\end{eqnarray}
Here the dimensionless function $\Lambda(x)$ is
given by
\begin{eqnarray}
\Lambda (x) &=& \frac{1}{\pi x}
\left[\ln x\ln (1+x^2) +\pi
\arctan x +
\frac{1}{2} {\mathrm Li}_2(-x^2)
\right],
\label{lambda}
\label{eq:3.2.20}
\end{eqnarray}
with ${\mathrm Li}_2(x)$ being the second polylogarithm function
\cite{Ryzhik}.  The asymptotic behavior of function $\Lambda$ is 
$\Lambda(x) = 1+(x\ln x)/\pi$, for $x\ll 1$, and $\Lambda(x) =(\pi
x)^{-1} \ln^2x$, for $x\gg 1$. The parameters $N_h^D$ and $N_h^C$ are
defined in Section~\ref{sec:mf}. The limits of the pure orthogonal
(unitary) ensembles (\ref{eq:3.2.12}) are recovered by putting
$N_h^D=0$ and $N_h^C=0(\infty)$ in Eq.~(\ref{eq:3.2.19}).

The correlation function acquires a universal form
\cite{AleinerGlazman96} ({\em i.e.}, all the dependences for
different $E_{e(h)}$ can be collapsed to a single curve upon rescaling of the
magnetic field):
\begin{eqnarray}
C(B_1,B_2)&=&\frac{\langle\delta G(B_1)\delta G(B_2) \rangle}
{\langle G\rangle^2} \nonumber \\
&=&
\left\{
\Lambda\left[\left(\frac{\Phi_1 + \Phi_2}{\Phi_c}\right)^2\right]
\right\}^2
+
\left\{
\Lambda\left[\left(\frac{\Phi_1 - \Phi_2}{\Phi_c}\right)^2\right]
\right\}^2
,
 \label{eq:3.2.22}
\label{c}
\end{eqnarray}
where the scaling function $\Lambda(x)$ is defined in
Eq.~(\ref{eq:3.2.20}), $\Phi_i$ is the magnetic flux through the dot
due to the field $B_i$, and the correlation magnetic flux $\Phi_c$ is
controlled by the charging energy
\begin{equation}
\Phi_c = \frac{1}{\sqrt{\chi g}}
\left(\frac{2\pi
E}{\Delta}\right)^{1/2}\Phi_0,
 \quad 
E= {\mathrm min} \left( E_e,\ E_h
\right),
\label{Bc}
\label{eq:3.2.23}
\end{equation}
where $\Phi_0$ is the flux quantum, $g$ is the dimensionless
conductance of the closed dot, see Sec.~\ref{sec:2}, and the dimensionless
coefficient $\chi$ is discussed after Eq.~(\ref{eq:2.101}). 
The dependence of the
correlation flux on the charging energy is caused by the fact that the
transport is contributed to by a large number of states in the energy
strip of the order of $E$ (see the discussion in Sec.~\ref{sec:mf}),
which depends on the gate voltage.

We emphasize that the functional form of $C(\Delta B)$ is different
from the results for the peak heights fluctuations (\ref{eq:3.1.16}),
see Fig.~\ref{Fig11}.  It is worth noticing from
Eq.~(\ref{eq:3.2.13}) and Eq.~(\ref{Bc}) that in this regime, the
correlation magnetic flux $\Phi_c$ drops with approaching a charge
degeneracy point, whereas the quantity $\langle G \rangle \Phi_c^2$
remains invariant.

{\begin{figure}
{\epsfxsize=10cm\centerline{\epsfbox{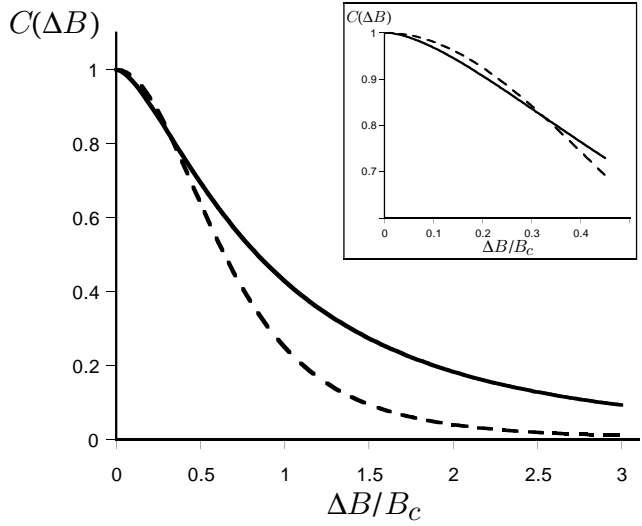}}}
\caption{The correlation function $C(\Delta B=B_1-B_2)$ for the conductance
fluctuations in the elastic co-tunneling regime (solid line) and for
the peak height fluctuations $C=[1+(\Delta B/B_c)^2]^{-2}$ (dashed
line) in the regime $\Phi_1,\Phi_2\gg \Phi_c$. For the elastic co-tunneling, 
$C(\Delta B)$ is non-analytic at $\Delta B \to 0$, see the inset, and
Eq.~(\protect\ref{c}).}
\label{Fig11}
\end{figure}
}

The distribution function of the conductance in the crossover regime 
can be found from Eq.~(\ref{eq:3.2.6}) and the fact that the
amplitudes $F$ are Gaussian with the correlators (\ref{eq:3.2.19}).
The calculation yields, instead of (\ref{eq:3.2.17}),
a distribution function characterized by a single crossover
parameter $\lambda$,
\begin{equation} W(\alpha) =
\frac{\theta(\alpha)}{\sqrt{1-\lambda^2}}\exp\left(-\frac{\alpha}
{1-\lambda^2}\right)
{\mathrm I}_0\left(\frac{\alpha\lambda}{1-\lambda^2}\right);
\label{Presult}
\label{eq:3.2.24}
\end{equation} 
here ${\mathrm I}_0(x)$ is the zeroth order modified Bessel
function of the first kind. In the limiting cases $\lambda = 1$ and
$\lambda =0$, the distribution function $P(g)$ coincides with the
Porter-Thomas distribution (\ref{eq:3.2.17}) for the
orthogonal and unitary ensembles respectively.

The dependence of the crossover parameter
$\lambda$ on the magnetic flux $\Phi$ in our case is quite nontrivial
 \cite{AleinerGlazman96}
\begin{equation}
\lambda = \Lambda \left(\frac{4\Phi^2 }{\Phi_c^2}\right),
\label{eq:3.2.26}
\end{equation}
where the correlation flux is given by Eq.~(\ref{Bc}), and function
$\Lambda(x)$ is defined in Eq.~(\ref{eq:3.2.20}). The crossover in the
mesoscopic fluctuations of conductance in the valleys far from the
peaks is discussed in Appendix~\ref{ap:2}.

\begin{figure}
{\epsfxsize=10cm\centerline{\epsfbox{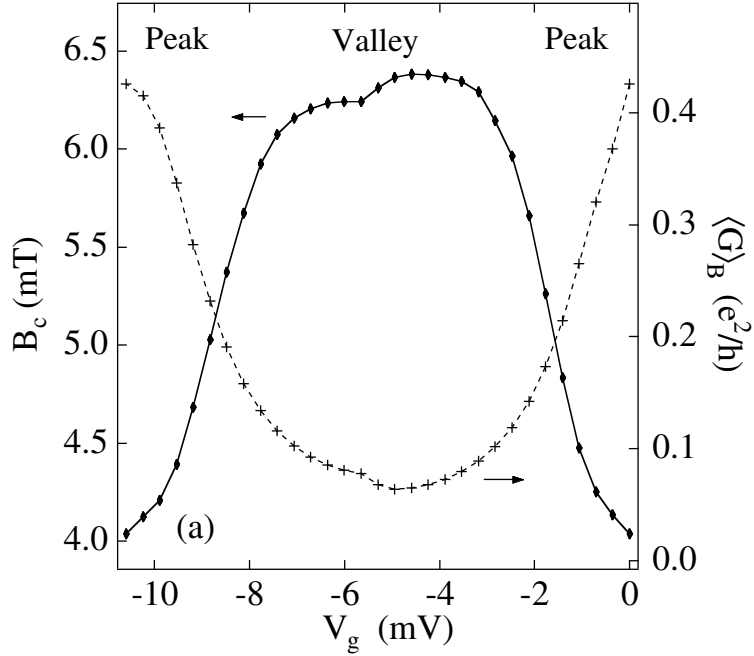}}}
\caption{\label{Fig:sara}
Ensemble-averaged characteristic correlation field (solid)
  and average conductance (dashed) across peak-valley-peak for $\sim
  14$ independent data sets.}
\end{figure}

The interference effects in the Coulomb blockade valleys were studied
experimentally in Ref.~\cite{Sara97}. In agreement with the theory of
this section it was found that the average conductance does not show
a weak localization correction (it remains the same for the unitary and
orthogonal ensembles).  The statistics of mesoscopic fluctuations of
the valley conductance also agrees reasonably well with the theory.
Extracted with the help of Eq.~(\ref{eq:3.2.22}), the dependence of the
correlation magnetic field on the gate voltage across a valley is
shown in Fig.~\ref{Fig:sara}.  One can see that the correlation
magnetic field increases with the deviation from a conductance peak,
in agreement with Eq.~(\ref{eq:3.2.23}). However, the value of the
ratio of the correlation fields for peaks, see Sec.~\ref{sec:3.1}, and
for the valleys was found somewhat smaller than the theoretical
prediction.

We recall that a relatively large correlation magnetic field in a
valley results from a wide energy band ($\sim \Ec/\Delta$) of virtual
discrete states in the dot. By the same token, the conductance in adjacent
valleys is also correlated, because all but one virtual states
participating in transport are the same for such valleys. In fact
the conductance remains correlated over a large number $\sim
\Ec/\Delta$ of valleys.
The corresponding corellation function was calculated in
Ref.~\cite{Baltin}, and the valley-valley correlation function for the
differential capacitances was found in Ref.~\cite{Kaminski}.

Considering electron transport through a blockaded dot, we have
used so far finite-order perturbation theory in the dot-lead tunneling
amplitudes. If the dot carries a non-zero spin, like in the case
of an odd number of electrons on the dot, such a treatment may miss
specific effects caused by the degeneracy of the spin state of an
isolated dot. Tunneling results in the exchange interaction between
the spins of the dot and leads. The exchange, in turn, leads to a
many-body phenomenon, the Kondo effect. This effect results in
an unexpected temperature dependence of the conductance across the
dot at temperatures $T\ll\Delta$.

\subsection{Kondo effect in a strongly blockaded dot}
\label{sec:Kondo}

The Kondo effect is one of the most studied and best understood
problems of many-body physics. Initially, the theory was developed to
explain the increase of resistivity of a bulk metal with magnetic
impurities at low temperatures~\cite{Kondo}. Soon it was realized that
Kondo's mechanism works not only for electron scattering, but also for
tunneling through barriers with magnetic
impurities~\cite{Appelbaum,Anderson66,Rowell}. A non-perturbative
theory of the Kondo effect has predicted that the cross-section of
scattering off a magnetic impurity in the bulk reaches the unitary
limit at zero temperature~\cite{Nozieres}. Similarly, the tunneling
cross-section should approach the unitary limit at low temperature and
bias~\cite{Ng,Raikh} in the Kondo regime.

The Kondo problem can be discussed in the framework of Anderson's
impurity model~\cite{Anderson}.  The three parameters defining this
model are: the on-site electron repulsion energy $U$, the one-electron
on-site energy $\varepsilon_0$, and the level width $\Gamma$ formed by
hybridization of the discrete level with the states in the bulk.  The
non-trivial behavior of the conductance occurs if the level is singly
occupied, $\Gamma < |\varepsilon_0| < U$, and the temperature $T$ is
below the Kondo temperature
\begin{equation}
T_K \simeq \sqrt{U\Gamma} \exp \left\{\frac{\pi
\varepsilon_0(\varepsilon_0+U)}{2\Gamma U}\right\}, 
\label{eq:5.4}
\end{equation}
where $\varepsilon_0<0$ is measured from the Fermi level~\cite{Haldane79}.

\begin{figure}
{\epsfxsize=12cm\centerline{\epsfbox{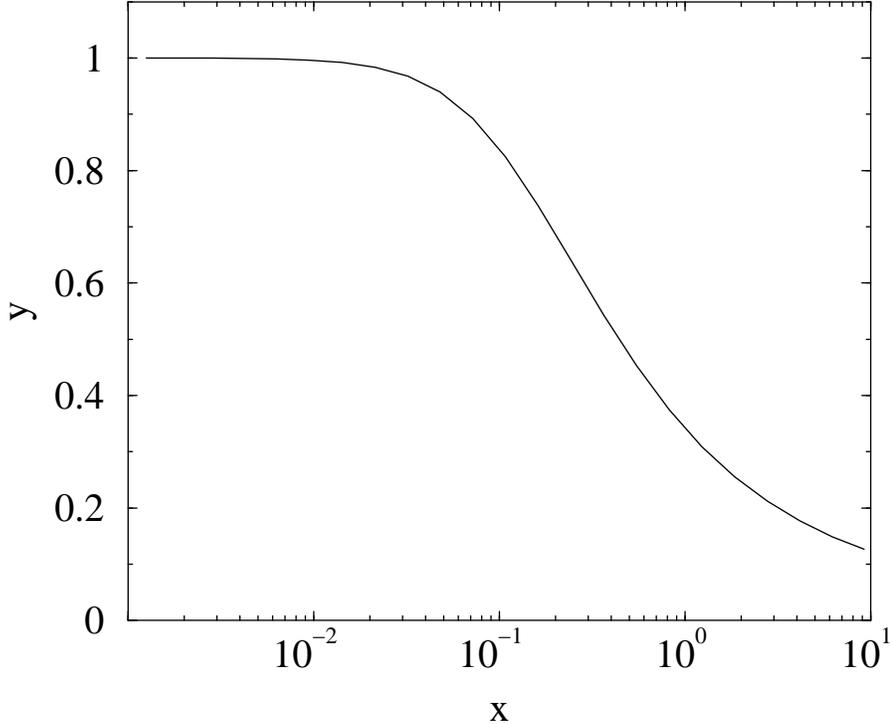}}}
\caption{Plot of the universal function $y=f(x)$ vs. $x=T/T_K$ from
  Ref.~(\cite{Costi}).}
\label{Fig20}
\end{figure}

Before considering the Kondo effect in a quantum dot, we first review
it for tunneling through a single localized level with on-site
repulsion $U$ using the Hamiltonian of the Anderson model, 
\begin{eqnarray}
  {\hat H}_A &=&\sum_{q,\sigma}\xi_q
(a_{1q\sigma}^\dagger a_{1q\sigma}
+a_{2q\sigma}^\dagger a_{2q\sigma})
+\sum_{\sigma}\varepsilon_0 a_{0\sigma}^\dagger a_{0\sigma}
+U{\hat n}_\uparrow{\hat n}_\downarrow
\\ && 
  \mbox{} +\sum_{q,\sigma}(t_1 a_{1q\sigma}^\dagger 
+ t_2a_{2q\sigma}^\dagger)a_{0\sigma}
+a_{0\sigma}^\dagger (t_1 a_{1q\sigma} + t_2a_{2q\sigma}),
\label{eq:5.1} \\ {\hat n}_\sigma &=& a_{0\sigma}^\dagger 
  a_{0\sigma}\nonumber.
\end{eqnarray}
Here $a_{1q\sigma}^\dagger$, $a_{2q\sigma}^\dagger$, and
$a_{0\sigma}^\dagger$ are the electron creation operators in the 
left and right leads ($1$ and $2$), and on the localized level, 
respectively;
$\xi_q$ are the corresponding energies in the electron continuum. For
brevity, the tunneling matrix elements $t_1$ and $t_2$ connecting the
localized state in the dot with the states in the leads are taken to
be $q$-independent. The widths $\Gamma_{i}$ are related to these
tunneling matrix elements: $\Gamma_i=2\pi\nu_{i} |t_i|^2$, where
$\nu_{i}$ is the density of states of lead $i=1,2$.  At first sight,
an Anderson impurity in a two-band model (\ref{eq:5.1}) may be
associated with a two-channel Kondo problem~\cite{Blandin}. However,
it is easy to show that in our case the parameters of this two-channel
problem are such that it can be reduced to the conventional
single-channel one.  Indeed, by a unitary transformation
\begin{equation}
  \label{transform}
  \left\{
\begin{array}{l}
\alpha_{q\sigma}\\
\beta_{q\sigma}
\end{array}
\right\}
=u a_{1q\sigma}\pm v a_{2q\sigma},\;\mbox{with}\quad
\left\{
\begin{array}{l}
u\\
v
\end{array}
\right\}
=
\frac{1}{\sqrt{|t_1|^2+|t_2|^2}}
\left\{
\begin{array}{l}
t_1\\
t_2
\end{array}
\right\},
\end{equation}
The Hamiltonian (\ref{eq:5.1}) can be converted~\cite{Raikh} to the
conventional one-band model with Anderson impurity~\cite{Anderson}.
The localized state and band described by the fermion operators
$\alpha_{q\sigma}$ form the usual Anderson impurity model, which is
characterized by three parameters: $U$, $\varepsilon_0$, and $\Gamma =
\Gamma_1+\Gamma_2$. The band described by the operators
$\beta_{q\sigma}$ is entirely decoupled from the impurity. The
scattering amplitude ${\cal A}(1,q\to 2,q')$ between two states in the
opposite leads is related to the scattering amplitudes ${\cal
A}(\alpha,q\to \alpha, q')$ and ${\cal A}(\beta,q\to \beta, q')$
within the bands $\alpha$ and $\beta$, respectively, by relation
\[{\cal A}(1,q\to 2,q')= uv[{\cal A}(\alpha,q\to \alpha,q')-{\cal
A}(\beta,q\to \beta, q')].\] The scattering amplitude in the
$\alpha$--band is directly related to the scattering phase $\delta_K$
for the conventional Kondo problem, ${\cal A}(\alpha,q\to
\alpha,q)=\exp (2i\delta_K)$; the scattering problem for the
$\beta$--band is trivial, ${\cal A}(\beta,q\to \beta, q)=1$. Using
Eq.~(\ref{transform}), we finally find
\[
|{\cal A}(1,q\to 2,q)|^2=4|uv|^2\sin^2\delta_K=
\frac{4g_1g_2}{(g_1+g_2)^2}\sin^2\delta_K.
\]
This way, the ``Kondo conductance'' $G_K$ associated with the
spin-degenerate localized level can be expressed in terms of the
problem of one-channel scattering off a single Kondo impurity in a
bulk material,\footnote{This
mapping is possible due to the complete decoupling of mode $\beta$
from the Anderson impurity spin. Mapping also works for quantum dots
which have spin $1/2$ and are described by the universal model
(\ref{eq:2.54a}). Corrections to this model, see
Eq.~(\ref{eq:2.54}), violate the mapping, but do not change
qualitatively the behavior of the Kondo conductance in the case of
$S=1/2$. If the spin $S>1/2$, then the temperature and magnetic field
dependence of $G_K$ may be quite different from the predictions of
Eq.~(\ref{eq:5.2}), even for a dot described by the universal model,
see Ref.~\cite{PG2001}.}
\begin{equation}
G_K=\frac{e^2}{\pi\hbar}\frac{4g_1 g_2}{(g_1+g_2)^2}
f\left(\frac{T}{T_K}\right).
\label{eq:5.2}
\end{equation}
Here $T_K$ is the characteristic Kondo temperature, and
$f(x)$ is a universal function. This function, found with the help of numerical
renormalization group in Ref.~\cite{Costi}, is plotted in Fig.~\ref{Fig20}.
A remarkable property of
scattering on a Kondo impurity, is that the corresponding
cross-section approaches the unitary limit at low energies, $f(0)=1$.
The low-temperature correction to the unitary limit is proportional to
$T^2$ and described by Nozi\`eres' Fermi-liquid theory~\cite{Nozieres},
\begin{equation}
f\left(\frac{T}{T_K}\right)=1-\frac{\pi^2T^2}{T_K^2},\quad T\ll T_K.
\label{eq:5.3}
\end{equation}

So far we considered the Kondo-enhanced tunneling through a single level.
The same situation can be realized in
tunneling through a quantum dot.
The number of electrons in the dot is controlled by the applied
gate voltage. If this number is an odd integer, $2n+1$, 
{\em i.e.}, the dimensionless gate voltage ${\cal N}=C_gV_g/e$ 
lies in the interval
\begin{equation}
|{\cal N}-(2n+1)|<\frac{1}{2},
\label{interval}
\end{equation}
the Kondo effect leads to a dramatic increase of the conductance at
low temperatures. The consequence is an even-odd alternation of the valley
conductance through the quantum dot: In ``even'' valleys, the low 
temperature conductance is due to elastic co-tunneling, while in ``odd'' 
valleys, the presence of an unpaired spin in the highest occupied
level leads to the Kondo effect, which causes a remarkable increase 
of the low temperature conductance compared to the even valleys. 

The advantage of using the quantum dots for the experimental studies
of the Kondo effect stems from an opportunity to effectively control
the parameters of the system, which is hardly possible
for a magnetic impurity embedded in a host material.

Quantitatively, the low-temperature limit for the conductance in the
odd valleys is given by Eqs.~(\ref{eq:5.2}) and (\ref{eq:5.3}). The 
Kondo temperature $T_K$, Eq.~(\ref{eq:5.4})) 
depends on the gate voltage through parameter
$\varepsilon_0$:
\begin{equation}
\varepsilon_0=2\Ec({\cal N}-{\cal N}^*) < 0, \quad U=2\Ec,
\label{eps0}
\end{equation}
where ${\cal N}^*=n+\frac{1}{2}$ is the degeneracy point at which two
different charge states of the dot have the same energy.

At a first glance, it appears that the Kondo temperature for a
quantum dot may be obtained by the substitution of the parameters
Eq.~(\ref{eps0}) into Eq.~(\ref{eq:5.4}). However, it is not exactly
the case, as we explain below.  Unlike the single-level Anderson
impurity model, the discrete energy spectrum of a dot is dense,
$\Delta\ll \Ec$. Still, if the junctions' conductances are small,
$G_{i}\ll e^2/\hbar$, (strong Coulomb blockade), those levels of the
dot which are doubly-filled or empty are important only at scales
larger than one-electron level spacing, and only the one single-occupied
level contributes to the dynamics of the system at energy scale
smaller than $\Delta$.  As a result, the model of the dot attached
to two leads can be truncated to the Anderson impurity model, however
the high energy cut-off is determined by $\Delta$ rather than by the
charging energy.  (See Sec.~\ref{sec:exchange} for more discussions).
This change in the energy cut-off replaces the first factor $U$ in
Eq.~(\ref{eq:5.4}) by $\Delta$. The level width $\Gamma$ can be related to
the sum $g_1+g_2$ for a given level in the quantum dot. Finally,
Eq. (\ref{eps0}) establishes the relation between $\varepsilon_0$ 
and $U$ on the one hand, and the
charging energy $\Ec$ and gate voltage ${\cal N}$ on the other hand. 
The resulting Kondo temperature (for the interval 
$0<{\cal N}-{\cal N}^*\simeq 1/2$) is found as
\begin{equation}
T_K\simeq \Delta\sqrt{(g_1+g_2)\frac{\Delta}{\Ec}}
\exp\left[-\frac{2\pi}{g_1+g_2}\frac{\Ec}{\Delta}
({\cal N}-{\cal N}^*)\right].
\label{eq:5.5}
\end{equation}

\begin{figure}
{\epsfxsize=10cm\centerline{\epsfbox{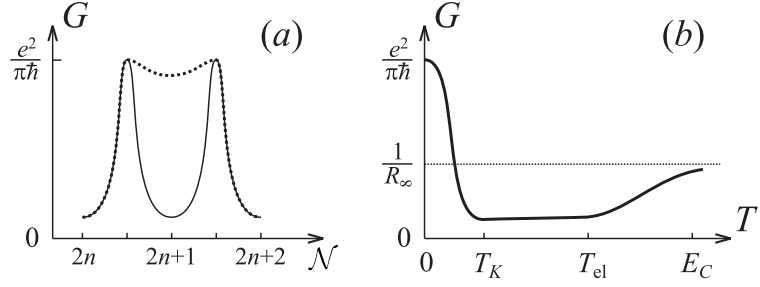}}}
\caption{({\it a}) Linear conductance at $T\gg T_K$ (solid line) and $T\lesssim
  T_K$ (dotted line). Due to the Kondo effect, the function $G({\cal
    N})$ develops plateaux in place of the ``odd'' valleys of the
  Coulomb blockade.  ({\it b}) Sketch of the temperature dependence of
  the linear conductance, $G(T)$, in the ``odd'' valley. The
  conductance decreases with the decrease of temperature from $T\sim
  \Ec$ down to $T\sim T_{\rm el}\simeq\sqrt{\Ec\Delta}$, see, {\it
    e.g.}, \protect\cite{AverinNazarov90,AleinerGlazman96}, and
  Eq.~(\protect\ref{eq:3.2.30}). At very low temperatures, $T\lesssim
  T_K$, the conductance grows again. If the singly-occupied level,
  which gives rise to the Kondo effect, has equal partial widths,
  $\Gamma_1=\Gamma_2$, the conductance reaches the unitary limit
  $G=e^2/\pi\hbar$ at low temperatures.}
\label{Fig15}
\end{figure}

Equations~(\ref{eq:5.2}) and (\ref{eq:5.3}) tell us that upon
sufficiently deep cooling, the gate voltage dependence of the
conductance through a quantum dot should exhibit a drastic change.
Instead of the ``odd'' valleys, which correspond to the intervals of
gate voltage (\ref{interval}), plateaux in the function $G({\cal N})$
develop, see Fig.~\ref{Fig15}a. In other words, the temperature
dependence of the conductance in the ``odd'' valleys should be very
different from the one in the ``even'' valleys. The conductance,
determined by the activation and inelastic co-tunneling mechanisms,
decreases monotonously with the decrease of temperature, and then
saturates at the value of $G_{\rm el}$ in the even valleys, see the
previous Section.  On the contrary, if the gate voltage is tuned to
one of the intervals (\ref{interval}), the $G(T)$ dependence is
non-monotonous, see Fig.~\ref{Fig15}b.  When temperature is lowered,
the conductance first drops due to the Coulomb blockade phenomenon. A
further decrease of temperature results in the increase of the
conductance at $T\lesssim T_K$.  Its $T=0$ saturation value depends on
the ratio of the partial level widths $\Gamma_1$ and $\Gamma_2$ of a
particular discrete level. The partial widths
are related to the tunneling matrix elements, $\Gamma_i=2\pi\nu
|t_i|^2\propto g_i$, where the matrix elements $t_i$, in turn, 
are proportional to
the values of the electron eigenfunctions within the dot,
$t_{i}\propto\psi_n(i)$, where $n$ labels the eigenstate, and
the argument $i$ of the wavefunction $\psi_n$ refers to the site
in the dot adjacent to junction $i$, see Eqs.~(\ref{eq:3.2})
and (\ref{gf}). In a disordered or chaotic dot, the electron wave
functions are random quantities, described by the Porter-Thomas
distribution~\cite{PorterThomas}. As we already discussed in the previous
section the randomness of the wave functions results in mesoscopic
fluctuations of the elastic co-tunneling
conductance~\cite{AleinerGlazman96}. The $T=0$ value of the
conductance in the Kondo regime should fluctuate as well, see
Eq.~(\ref{eq:5.2}). Experiments are performed, however, at a finite
temperature $T$, where fluctuations of conductance occur mostly due to
the fluctuation of the Kondo temperature (\ref{eq:5.5}), which has
an exponential dependence on the level width $\Gamma_1+\Gamma_2$.

\begin{figure}
{\epsfxsize=10cm\centerline{\epsfbox{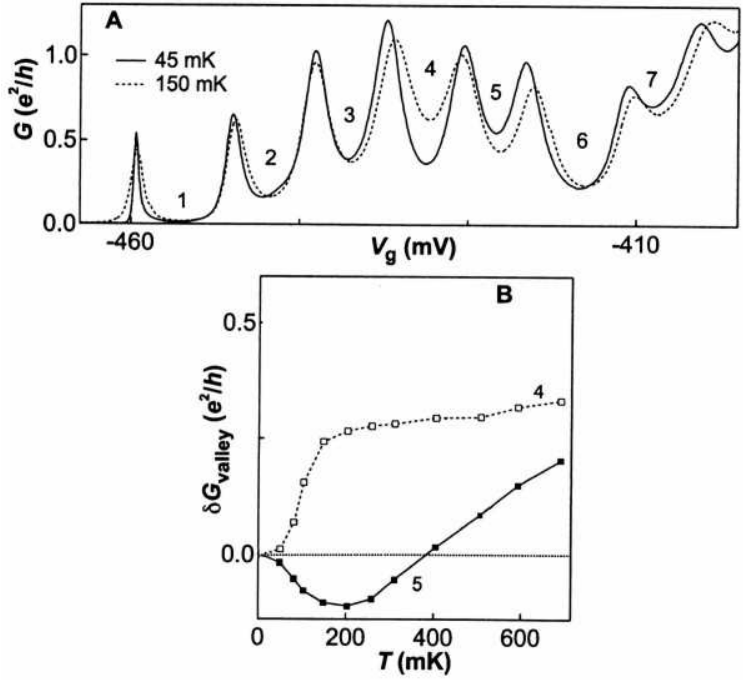}}}
\caption{The gate voltage dependence of the conductance at various
  temperatures ({\bf A}), and the temperature dependence of the conductance
  in the valleys labelled 4 and 5 ({\bf B}), see
  \protect\cite{SaraKondo}. Reprinted with permission from Science
  {\bf 281}, 540 (1998). Copyright 1998 American Association for the
  Advancement of Science.}
\label{Fig16}
\end{figure}

The experimental search for a tunable Kondo effect brought positive
results \cite{Goldhaber,SaraKondo,Schmid98} only recently. In a number of
experiments the difference in the low-temperature conductance behavior
in the ``odd'' and ``even'' valleys was clearly observed, see
Fig.~\ref{Fig16}. In retrospect it is clear, why such experiments were
hard to perform.  The negative exponent in Eq.~(\ref{eq:5.5}) contains
a product of two large parameters, $\Ec/\Delta$ and $e^2/h(G_1+G_2)$,
leading to a very small value of $T_K$.

To bring $T_K$ within the reach of a modern low-temperature
experiment, one may try smaller quantum dots in order to decrease
$\Ec/\Delta$; this route obviously has technological limitations. (In
the experiments~\cite{Goldhaber} this ratio was $\Ec/\Delta\gtrsim
7$.)  Another, complementary option is to increase the junction
conductances, so that $G_{1,2}$ come close to $e^2/\pi\hbar$, which is
the maximal conductance of a single-mode quantum point contact.  The
junctions in the experiments~\cite{Goldhaber} were tuned to $G\simeq
(0.3-0.5) e^2/\pi\hbar$. Under these conditions, one would expect
$T_K\gtrsim 100$mK only if ${\cal N}-{\cal N}^*\lesssim 0.07$, see
Eq.~(\ref{eq:5.5}); in the middle of the valley (${\cal N}-{\cal
  N}^*=0.5$) the Kondo temperature is unobservably small. Indeed, an
evidence for the Kondo effect was found~\cite{Goldhaber} in a
relatively narrow interval interval, ${\cal N}-{\cal N}^*\lesssim
0.15$. Only in this domain of gate voltages the anomalous increase of
conductance $G(T)$ with lowering the temperature $T$ was clearly
observed; the anomalous temperature dependence of the conductance was
not seen in the middle of the Coulomb blockade valley.

In conventional Kondo systems, a signature of the
effect is in the characteristic minimum of the resistivity. The
minimum comes from the competition of the two contributions to the
resistivity: the phonon contribution is decreasing with lowering the
temperature, while the Kondo contribution has the opposite temperature
dependence. A similar feature in tunneling through a quantum dot would
be a minimum in the conductance in an ``odd'' valley. However, such a
minimum is quite shallow at $G_1,G_2\ll e^2/\hbar$. Indeed, the
minimum comes from the competition of the temperature dependencies of
the inelastic co-tunneling and of the Kondo contribution. In the
domain $T\ll\Delta$ the inelastic contribution to the co-tunneling is
exponentially small, $G_{\rm in}\propto\exp (-\Delta/T)$. The proper
expansion of the function $f(T/T_K)$ in Eq.~(\ref{eq:5.2}) at $T\gg
T_K$ and the use of (\ref{eq:5.5}) yield the following temperature
dependence of the Kondo contribution to the valley conductance
\begin{equation}
\langle G(T)\rangle = 
\langle G_{\rm el}\rangle
\left[1+
\frac{\hbar (G_1+G_2)}{e^2}
\left(\frac{\Delta}{\Ec}\right)^2\ln\left(\frac{\Delta}{T}\right)
+\dots\right],
\label{ln}
\end{equation}
where $\langle G_{\rm el}\rangle = (\hbar\langle G_1\rangle\langle
G_2\rangle/{4\pi e^2})({\Delta}/{\Ec})$ is the average elastic
co-tunneling conductance, see Eq.~(\ref{eq:3.2.13}).  As one can see
from Eq.~(\ref{ln}), the Kondo correction to the conductance remains
particularly small everywhere in the temperature region 
$T\gtrsim T_K$.

In order to increase the Kondo temperature and to observe the anomaly of the
tempera\-ture-dependent conductance $G(T)$ under these unfavorable
conditions, one may try to make the junction conductances larger.
However, if $G_{1,2}$ come close to $e^2/\pi\hbar$, the discreteness
of the number of electrons on the dot is almost completely washed
out~\cite{Matveev95}.  Exercising this option, therefore, raises a
question about the nature of the Kondo effect in the absence of strong
charge quantization.  We address this question later on, in
Section~\ref{sec:exchange}.

\subsection{Overall temperature and gate voltage dependence of the
  conductance}
\label{sec:tdependence}

Closing this Section, we briefly discuss the gate voltage dependence
of the linear conductance through a strongly blockaded dot in
various temperature intervals.

As we have already mentioned in the very beginning of
Subsection~\ref{sec:3.2}, rate equations predict that the peaks in the
$G({\cal N})$ function should have width $\sim T/\Ec$; the conductance
in the valleys is exponentially small, with the activation energy set
by the charging effect. In the temperature interval $\Ec\gg
T\gg\Delta$, the rate equations yield~\cite{RSLG} for the peak shape
\begin{equation}
G({\cal N}, T)=\frac{e^2}{2\pi\hbar}
\bar{g}
\frac{\Ec({\cal N}-{\cal N}^*)/T}{\sinh[\Ec({\cal N}-{\cal N}^*)/T]},
\label{eq:3.2.27}
\end{equation}
where ${\cal N}^*$ is a half-integer number, 
and the parameter $\bar{g}$ is determined by the conductances of the
point contacts $\langle g_{1,2}\rangle$, 
the statistical properties of the wave-functions of
the dot, and by the rate of inelastic processes.

If the rate of inelastic processes $1/\tau_\varphi$ is small compared
to the inverse electron dwell time in the dot
$\left(g_{1}+g_{2}\right)\Delta$, then the parameter $\bar{g}$ is
given by
\begin{equation}
\bar{g}= \left\langle\frac{g_1 g_2}{g_1 + g_2}\right\rangle = 
\frac{\langle g_1\rangle\langle g_2 \rangle}
{\left(\langle g_1\rangle^{1/2}+\langle g_2 \rangle^{1/2}
\right)^2}
F\left[2- \beta  + \left(\beta -1\right) a \right].
\label{eq:3.2.270}
\end{equation}
Here, $\beta=1,2$ in the limits of no and strong magnetic fields
respectively; the parameter of asymmetry of the point contacts $a$ and 
the function $F$ are defined in Eqs.~(\ref{eq:3.1.11}) and
(\ref{eq:3.1.12}). The conductance (\ref{eq:3.2.27}) has
the same dependence on magnetic field as the averaged peak conductance
in the low-temperature ($T\ll\Delta$) regime, see
Eq.~(\ref{eq:3.1.12}).  Notice, that unlike the low-temperature peak
conductance, the mesoscopic fluctuations on top of
result~(\ref{eq:3.2.27}) are suppressed by a large factor of
$\Delta/T$ due to the large number of discrete levels contributing to
the transport; all the peaks have the same height, which equals the
half of the high-temperature ($T\gg \Ec$) value of conductance through
the dot. This is a very straightforward consequence of the Coulomb
blockade: out of the two elementary processes of charge transfer (one
is increasing, and another is decreasing the initial charge of the dot
by $e$), only one process is allowed.

The important difference between the case of sequential tunneling
through the quantum dot mentioned above and the process of
elastic co-tunneling considered in Section~\ref{sec:3.2}, is that 
for sequential tunneling each
level contributes to a real process, rather than serves as a virtual
state for tunneling as is the case for elastic cotunneling. 
Therefore, for sequential tunneling not
amplitudes, but the probabilities of tunneling through each level are
added.

In the case of rapid inelastic relaxation,
$1/\tau_\varphi\gg\left(g_{1}+g_{2}\right)\Delta $, the time elapsing
between the events of an electron tunneling into and out of the dot is
sufficiently long that an equilibrium electron distribution can be
established inside the dot. If, in addition, the temperature
$T\gg\Delta$, an electron entering the dot uses one of many ($\sim
\Delta/T$) discrete levels in it. Similarly, an equally large number
of levels participates when the electron exits the dot. Because the
energy level used for the tunneling from the dot is most probably
different from the one used for the tunneling into the dot, the
dimensionless conductances $g_1$ and $g_2$ defined in Eq.~(\ref{gf}),
has to be independently averaged over many levels,
yielding\footnote{Higher-order in $g_1$ and $g_2$ corrections to
Eq.~(\ref{eq:3.2.271}) were considered in
Refs.~\cite{Schoeller,Furusaki95} for the cases of single-channel and
multichannel junctions respectively. The latter case ($g_1,g_2\ll 1$
at $N_{rm ch}$) is typical for the metallic Coulomb blockade
systems. Summation of the leading series in $(g_1+g_2)\ln(\Ec/T)$,
performed in the limit $N_{\rm ch}\to\infty$ results~\cite{Schoeller}
in the replacement $\bar{g}\to 2\bar{g}/[2+(\langle g_1\rangle
+\langle g_2\rangle)\ln(\Ec/T)]$. It is noteworthy that the increase
of the contacts conductances result in the decrease of the peak
conductance $G$ through the device.}
\begin{equation}
\bar{g}=
\frac{\langle g_1\rangle \langle g_2 \rangle}
{\langle g_1\rangle+\langle g_2 \rangle}.
\label{eq:3.2.271}
\end{equation}

The two cases of rapid and slow inelastic relaxation can be 
distinguished through
their magnetoconductance, which, if normalized by the average
peak height, is temperature independent
as long as inelastic processes are not slow,
$1/\tau_{\varphi} \ll (g_1 + g_2) \Delta$, but quickly
disappears for $T \gtrsim \Delta$ when $1/\tau_{\varphi} \gg
(g_1 + g_2) \Delta$ \cite{SilvBeen}. Recent measurements of
the magnetoconductance of Coulomb blockade peak heights
show values between the two extremes of
rapid and slow inelastic relaxation \cite{wlcb}.

At lower temperatures $T\ll\Delta$, the peaks heights are proportional
to $1/T$, and exhibit strong mesoscopic fluctuations, see
Subsection~\ref{sec:3.1}. The peak height reaches
\begin{equation}
G_{\rm peak}\sim \frac{e^2}{\pi\hbar}\frac{g_1g_2}{(g_1+g_2)^2},
\label{eq:peak}
\end{equation}
when $T$ becomes of the order of the discrete level width
$(g_1+g_2)\Delta$. At the same time the dimensionless peak width
$\delta{\cal N}_{\rm peak}$ decreases down to the value\footnote{
Width of the peak at larger values of conductances is discussed in
Ref.~\cite{AKLG00}}
\begin{equation}
\delta{\cal N}_{\rm peak}\sim (g_1+g_2)\frac{\Delta}{\Ec}.
\label{eq:width}
\end{equation}
When the temperature is further lowered, the peak continues to grow,
and simultaneously broadens again and becomes
asymmetric. However,these effects become noticeable at much lower
temperature, $T\sim T_K$, when the Kondo effect starts playing role.

In the valleys, upon lowering the temperature,
the exponential decrease of the
conductance that was mentioned above is replaced by a weaker, power-law
($\propto T^2$) temperature dependence provided by the mechanism of
inelastic co-tunneling~\cite{AverinOdintsov,AverinNazarov90}. In the
middle of the valley this mechanism results in the conductance
\begin{equation}
G_{\rm in}=\frac{4}{3\pi}\frac{e^2}{\hbar}
\langle g_1\rangle\langle g_2\rangle
\left(\frac{T}{\Ec}\right)^2.
\label{eq:3.2.28}
\end{equation}
One can see that the crossover between the two regimes occurs at
temperatures
\begin{equation}
T\simeq T_{\rm in}\equiv \Ec/|\ln (\langle g_1\rangle+\langle
g_2\rangle)|.
\label{eq:3.2.29}
\end{equation}
A comparison of Eq.~(\ref{eq:3.2.29}) with the result (\ref{eq:3.2.13})
for elastic co-tunneling shows that the crossover to the
temperature-independent conductance occurs at
\begin{equation}
T\simeq T_{\rm el}\equiv\sqrt{\Ec\Delta}.
\label{eq:3.2.30}
\end{equation}
If the dot carries a finite spin, then at much lower temperatures,
$T\lesssim T_K$, the conductance increases again due to the Kondo effect,
see Subsection~\ref{sec:Kondo} and Fig~\ref{Fig15}.

In a typical experiment with a quantum dot formed in a semiconductor
heterostructure, the five relevant energy scales are related as:
\begin{equation}
T_K\lesssim\Delta \lesssim T_{\rm el}\lesssim T_{\rm in}\lesssim \Ec.
\label{eq:3.2.31}
\end{equation}
Temperatures $T\lesssim T_{\rm el}$ are easily accessible; at such
temperatures the conductance fluctuations are of the order of the
average conductance.  At higher temperatures, when the main conduction
mechanism switches to the inelastic co-tunneling, fluctuations are
still determined by the elastic mechanism as long as $T <
\left(\Ec^2\Delta\right)^{1/3}$. At even higher temperatures, the
inelastic contribution dominates also in the fluctuations.  Their
relative magnitude, however, becomes small, $\langle\delta
G_{\rm in}^2\rangle/\langle G_{\rm in}\rangle^2 \simeq \Delta/T$.
Simultaneously, the correlation magnetic field becomes independent of
the gate voltage, as in the inelastic co-tunneling regime it is
controlled by the temperature rather than by the charging energy.

\section{Weakly blockaded  dots}
\label{sec:5} \label{sec:4}

Charge quantization deteriorates gradually with the increase of the
conductance of the junctions connecting the quantum dot to the leads.
The characteristic resistance at which a substantial deterioration
occurs can be estimated from the following heuristic argument. The
energy of a state with one extra electron in the dot is $\Ec$.
Dispensing with the discreteness of the quasiparticle spectrum
($\Delta\to 0$), one can estimate the lifetime of a state with a given
charge as the $RC$-constant of a circuit consisting of a capacitor
with capacitance $C=e^2/2 \Ec$, imitating the dot, and a resistor with
resistance $R=1/G$, imitating
the dot-lead contact. According to the Heisenberg uncertainty
principle, this state is well-defined as long as $\Ec\gtrsim\hbar
G/C$. Therefore charge quantization requires a small lead-dot
conductance,
$G\lesssim e^2/\hbar$. We refer to dots with $G \lesssim e^2/h$,
for which the charge on the dot is quantized, as ``strongly blockaded''
dots. Dots with $G \gtrsim e^2/h$, for which charge is not quantized,
are referred to as ``weakly blockaded''.
While the constraint on $G$ involves only the
universal quantity $e^2/\hbar$, the crossover from ``strong''
to ``weak'' Coulomb blockade occurring at
$G\sim e^2/\hbar$ is not universal.  The detailed behavior of a
partially open dot depends on the number of modes propagating through
the junction and on their transparency, even in the limit 
$\Delta\to 0$. In the case of a tunnel
junction with large number of modes each of which is characterized by
a small transmission coefficient, charging effects vanish gradually at
$G\gtrsim e^2/\pi\hbar$. At large $G$, the remaining periodic
oscillations of the average
charge of the dot with the varying gate voltage were
shown to be exponentially small,
proportional to $\exp(-\pi\hbar G/e^2)$ \cite{PanyukovZaikin} 
(a controllable calculation
of the pre-exponential factor still remains an open problem). 

In the case of a single-mode junction, Coulomb blockade oscillations
of the average charge vanish exactly at $G=e^2/\pi\hbar$, see
Ref.~\cite{Flensberg,Matveev95}. Here we concentrate on the case of a
narrow channel\footnote{We use the word ``channel'' instead of
  ``contact'' to emphasize that the structure of the wavefunction in
  and near the contact is one-dimensional, see Sec.\ 
  \protect\ref{sec:lowen}.}  supporting a few propagating modes and
connected adiabatically~\cite{Glazman88} to the lead and quantum dot
at its ends. This is the conventional geometry of a quantum dot device
formed in a two-dimensional gas of a semiconductor
heterostructure~\cite{dots}. The theory for such a model is quite well
developed, but the calculations are very involved.\footnote{Coulomb
  blockade at $G\sim e^2/\pi\hbar$ in the case of large number of
  channels, $N_{\rm ch}\gg 1$, in our opinion, is less understood.
  General scaling ideas regarding the evolution of the Coulomb
  blockade with the increase of $G$ in this case are presented in
  Ref.~\cite{Falci}} In this review, our aim is twofold. On the one
hand, we want to present the detailed derivations that transform the
problem into a one that can be solved using well-developed formalisms.
On the other hand, we also want to give the general picture of the
thermodynamic and transport properties of a weakly blockaded dot, as
it emerges from the theory.  Therefore this Section is organized as
follows. We start with Subsection~\ref{sec:overall}, where an overview
of the final results is given.  For those readers who wish to enter
into the more detailed derivations, this Subsection may also serve as
a guide to the more detailed presentations, which can be found in
Sections~\ref{sec:rigorous}--\ref{sec:dos} and Appendices
\ref{Ap:3}--\ref{Ap:4}, and use a combination of the bosonization and
perturbation theory methods.  Before presenting a rigorous treatment
of the mesoscopic fluctuations, we describe a simplified version of
the theory in Sections~\ref{sec:finitesize}--\ref{sec:qualit}.  This
version is still sufficient for performing the estimates of the most
important observables. Going beyond the estimates, however, requires a
rigorous theory. It is necessary not only for fixing the numerical
values of the coefficients in formulas for various measurable
characteristics, but also for finding their dependence on control
parameters. One can view the result for the correlation function of
mesoscopic conductance fluctuations (\ref{eq:mesoosc}) as an example.
Formalism and theory presented in
Sections~\ref{sec:rigorous}--\ref{sec:dos} were needed to predict the
four-fold reduction of this function in the magnetic
field~\cite{AleinerGlazman98}; this prediction was later checked
experimentally~\cite{fryingpan}.

\subsection{Main results for the conductance and differential capacitance of 
an almost open quantum dot}
\label{sec:overall}

{\em Conductance through a quantum dot.} \hfill \\ 
We first discuss the case of a dot
connected to two leads by almost reflectionless single-mode
junctions. We assume that the reflection amplitudes
$r_1$ and $r_2$ in the two contacts connecting the dot to the
leads are of the same order,
\begin{equation}
r_1 \approx r_2 \equiv r,
\label{eq:sameorder}
\end{equation}
and obey the condition
\begin{equation}
\Ec |r|^4\lesssim \Delta.
\label{eq:veryweak}
\end{equation}
For such an open dot the conductance $G$ depends only weakly on
temperature and gate voltage as long as $T\gtrsim\Delta$. In the
regime $\Ec\gg T\gg\Delta$, the disorder averaged conductance 
$\langle G \rangle$ is
\begin{eqnarray}
\langle G\rangle &=& \frac{e^2}{2\pi\hbar}
\left\{1
-\frac{2\Gamma(3/4)}{\Gamma(1/4)}
\left( {\Ec e^{\bf C} \over \pi T} \right)^{1/2}
(|r_1|^2 + |r_2|^2)
\right. \nonumber \\ && \ \ \ \ \ \ \ \ \left. 
\vphantom{\left( {\Ec e^{\bf C} \over \pi T} \right)^{1/2}}
\mbox{}
+\frac{1}{3}\left(1-\frac{2}{\beta}\right)
\left[1+\frac{0.24\Delta}{T}\right]
\right\},
\label{Gveryweak}
\end{eqnarray}
where ${\bf C}\approx 0.577$ is the Euler constant, and $\Gamma (x)$
is the Gamma function.
This result summarizes Eqs.~(\ref{eq:70.9}) and (\ref{eq:70.12})
 below,
see also Ref.~\cite{Furusaki95}. 
The symmetry index $\beta=1$ 
($2,4$) for the orthogonal
(unitary, symplectic) ensemble. The two temperature-independent terms
in Eq.~(\ref{Gveryweak}) correspond to the classical resistance of
the quantum dot and
the interference (weak-localization) correction to it, and are known
in the context of mesoscopic systems without electronic
interactions. The temperature-dependent terms appear as a result
of electron-electron interactions. 
However, they are small compared to the temperature-independent
terms at temperatures
$T\gtrsim\Delta$. As a
matter of fact, the amplitude of ``usual'' mesoscopic fluctuations
characteristic of a non-interacting system exceeds the
interaction correction to the average conductance, and dominates the
variance and the conductance correlation function $\langle \delta G({\cal
N}_1)\delta G({\cal N}_2)\rangle$, as can be seen from
 Eqs.~(\ref{eq:70.12}) and (\ref{eq:UCF}) below [${\cal N}$ is the
dimensionless gate voltage, see Eq.\ (\ref{eq:2.54a})].

The temperature dependence of the interaction correction to the
conductance saturates at $T\sim\Delta$, when its magnitude is
comparable to the weak-localization correction. At these low
temperatures, interaction effects become important. Under the 
condition Eq.~(\ref{eq:veryweak}),
Coulomb blockade does not manifest itself in the form of
oscillations of the ensemble averaged conductance $\langle G({\cal
N})\rangle$ as a function of the dimensionless gate voltage 
${\cal N}$. Instead, Coulomb blockade
affects the mesoscopic
fluctuations of $G$, as described by the conductance correlation 
function $\langle \delta G({\cal N}_1)\delta G({\cal N}_2)\rangle$, 
which acquires an
oscillatory dependence on ${\cal N}_1-{\cal N}_2$. The amplitude of
oscillations is of the order of $e^2/\pi\hbar$. The
oscillations remain correlated over an interval $|{\cal N}_1-{\cal
N}_2|\sim \Ec/\Delta$.
Analytical formulae for this regime are not available yet.

In the case of stronger reflection,
\begin{equation}
\Ec|r|^4\gg\Delta,
\label{eq:weak}
\end{equation}
the above consideration and Eq.~(\ref{Gveryweak}) is applicable,
though in a narrower temperature interval,
\begin{equation}
\Ec\gg T\gtrsim\Ec|r|^4.
\label{eq:interval}
\end{equation}
At lower temperatures and for contacts with equal reflection
probability,
$|r_1|=|r_2|$, the conductance 
shows peaks with maxima of order $G\sim e^2/\pi\hbar$, see Ref.~
\cite{Furusaki95}. The maxima are well pronounced 
if Eq.~(\ref{eq:weak}) is satisfied: in the
valleys of Coulomb blockade the estimate for the conductance is
\begin{equation}
G_{\rm min}\sim \frac{e^2}{\pi\hbar}\frac{\Delta}{\Ec|r|^4}.
\label{eq:Gvalley}
\end{equation}
At even lower temperatures,
\begin{equation}
T\lesssim T_K,\, 
T_K\sim\Delta\sqrt{\frac{\Delta}{\Ec|r|^4}}
\exp\left[-\alpha({\cal N})\frac{\Ec|r|^4}{\Delta}\right],
\label{eq:evenodd}
\end{equation}
the Kondo effect develops, and the conductance in the ``odd'' valleys
approaches the unitary limit. The function $\alpha ({\cal N})$ is
positive with values $\sim 1$; the pre-exponential numerical factor is
$\sim 1$. While the estimates Eqs.~(\ref{eq:Gvalley}) and
(\ref{eq:evenodd}) are easily obtainable by the combination of methods
reviewed in Sections~\ref{sec:3.2} and \ref{sec:exchange}, the
explicit dependence of the conductance on gate voltage across a valley
and the explicit form of the functional dependence on ${\cal N}$ in
the exponential of Eq.~(\ref{eq:evenodd}) are not known.

If the reflections in the junctions are of different strengths, then
several temperature intervals appear, and the behavior of the
conductance is significantly different in each interval.  We will
review here only the case of a strongly asymmetric setup: in one
channel, backscattering is strong, $G_{1} \equiv 1- |r_{1}|^2 \ll 1$, 
while in the other channel, backscattering is 
weak,\footnote{The case of $|r_2|\ll |r_1|\ll 1$ is
considered in Ref.~\cite{GHL}.} but still not infinitesimally small:
\begin{equation}
\Delta/\Ec \ll |r|^2 \ll 1.
\label{eq:weakstrong}
\end{equation}
(Here we wrote $r$ for the backscattering amplidude $r_2$ in the
weakly backscattering channel.)
In this configuration, the first  contact can be treated within 
the tunneling Hamiltonian formalism, and all the results of
Sec.~\ref{sec:dos} are applicable.  We find from Eqs.~(\ref{eq:80.6})
and (\ref{eq:80.7}) of that section
\begin{eqnarray}
\langle G(T) \rangle \approx G_1
\left\{
\matrix{
1, & T \gg  \Ec; \cr \displaystyle{
\left({\pi^3 T \over 8\Ec e^{\bf C}}
\right)^{\vphantom{M^M_M}}_{\vphantom{M^M_M}}}, & 
\Ec \gg T \gg T_0({\cal N});
\cr \displaystyle{
\frac{8\pi^2}{3e^{\bf C}} \frac{T^2}{\Ec T_0({\cal N})}},
& T\ll T_0({\cal N}),
}
\right. 
\label{eq:ov1}
\end{eqnarray}
where $T_0({\cal N})$ is the energy scale at which crossover to the
Fermi liquid behavior, $G\propto T^2$, occurs [cf. Eq.~(\ref{T0ex})],
\begin{equation}
T_0({\cal N}) 
= {8 e^{\bf C} \over \pi} \Ec |r|^2\cos^2\pi{\cal N}.
\label{eq:ov2}
\end{equation}

The transport processes contributing to Eq. (\ref{eq:ov1}) are
inelastic.
At temperatures $T\ll T_0({\cal N})$ the inelastic mechanism
results in the Coulomb blockade oscillations. However,
this contribution to the conductance vanishes at $T \to 0$. At very
low
temperatures the elastic co-tunneling becomes dominant in the electron
transport.  This mechanism yields, see Eq.~(\ref{eq:elastic1}),
\begin{eqnarray}
\langle G_{\rm el} \rangle = 
G_1 \frac{\Delta e^{-\bf{C}}}{2 \Ec} 
\ln\left[\frac{\Ec}{T_0({\cal N})}\right],
\quad
T \ll \sqrt{\Delta T_0({\cal N})}.
\label{eq:ov3}
\end{eqnarray}
Here, in accordance with the left inequality of
Eq.~(\ref{eq:weakstrong}), we assumed $T_0({\cal N}) \gg \Delta$. In
this temperature regime, the mesoscopic fluctuations of the conductance
are of the order of conductance itself.

The result~(\ref{eq:ov3}) does not depend on the temperature and one
might think that this is the limiting low temperature behavior of the
conductance. This is indeed true if the average charge of the dot 
$\langle Q \rangle = e {\cal N}$, with $2j
- 1/2 < {\cal N} < 2j + 1/2$ (``even valley''), where $j$ is an
integer number.  In this case, the ground state of the closed dot is
non-degenerate and there is no new physics at low temperatures.

If however $2j + 1/2 < {\cal N} < 2j + 3/2$, (``odd valley''),
the conductance at low temperatures shows the re-entrance behavior due to the
Kondo effect, see Eq.~(\ref{GKlimit}),
\begin{equation}
G(T) = G_1 
\left[\alpha_1\frac{\Delta e^{-\bf{C}}}{2 \Ec} 
\ln\left(\frac{\Ec}{T_0({\cal N})}\right)+\alpha_2
\frac{32}{\pi}\frac{T_0({\cal N})}
{\Ec e^{\bf C}}f\left({T\over T_K(\cal{N})}\right)
\right].
\label{eq:ov4}
\end{equation}
Here, the first term represents elastic cotunneling, cf.\
Eq.\ (\ref{eq:ov3}), while the second term is due to the Kondo 
effect. Further, $f$ is the
universal scaling function ~\cite{Costi}, see Fig.~\ref{Fig20}, and
the Kondo temperature is given by
\begin{equation}
T_K\simeq\Delta 
\sqrt{\frac{\Delta}{T_0({\cal N})}}\exp\left\{- \frac{T_0({\cal
      N})}{\alpha_3\Delta}\right\} \ll \Delta.
\label{eq:ov5}
\end{equation}
The parameters $\alpha_1$, $\alpha_2$ and $\alpha_3$ are independent
random quantities\footnote{note, that in the case of asymmetric setup,
  the ${\cal N}$-dependence of the exponential in
  Eq.~(\protect\ref{eq:ov5}) is known, unlike the case
  $|r_1|^2=|r_2|^2$, cf. Eq.(\protect\ref{eq:evenodd}).} obeying
Porter-Thomas statistics~\cite{PorterThomas}, see
Eq.~(\ref{eq:3.2.17}). Therefore, the Kondo contribution also
experiences mesoscopic fluctuations that are of the same order as the
average conductance.  At $T \ll T_K$ the Kondo contribution exceeds
the elastic co-tunneling contribution to the conductance by the large
parameter $T_0({\cal N})/\Delta \gg 1$.

The above formulas for the co-tunneling contribution to the
conductance, see E's.~(\ref{eq:ov3}) and (\ref{eq:ov4}), are invalid in narrow
intervals around the conductance peaks, where $T_0({\cal N})$ becomes
smaller than $\Delta$. In these intervals, the factor
$T_0({\cal N})$ in the
argument of logarithm in Eqs.~(\ref{eq:ov3}) and (\ref{eq:ov4}) should
be replaced by $\Delta$. The schematic temperature dependence for the
conductance in even and odd valleys is similar to the one shown in
Fig.~\ref{Fig15}.

{\em Differential capacitance of an open dot.} \hfill \\ 
In addition to
complications stemming from quantum fluctuations of charge, the
transport phenomena discussed here
involve one more important part of
many-body physics: the Anderson orthogonality
catastrophe~\cite{Anderson67}. This element becomes especially
important in the analysis of conductance in a strongly asymmetric
setup, see the discussion following Eqs.~(\ref{eq:40.13}) and
(\ref{eq:80.4}). Thermodynamic quantities are free from that
complication. Therefore measurement of a differential capacitance
provides information additional to that found from the transport
measurements. Experimental investigation of the capacitance of a
quantum dot is quite a challenging task~\cite{Ashoori}, and by now
there were no systematic study of mesoscopic fluctuations of this
quantity. 

In presenting the theoretical results, we concentrate on the case of a
dot connected to a single lead. We also assume that the junction
allows for propagation of a single mode of electrons carrying both
directions of spin (small or no Zeeman splitting). Reflection of the
mode is characterized by the reflection amplitude $r$.

At a sufficiently high temperature, 
\begin{equation}
T\gg \Ec |r|^2,\Delta,
\label{eq:diff1}
\end{equation}
the result for the average capacitance can be found in Ref.\ 
\cite{Matveev95},
\begin{equation}
  {\langle C_{\rm diff}({\cal N}) \rangle \over C}
  \sim \ln(\Ec/T) |r|^2 \cos(2 \pi {\cal N}).
\label{diff2}
\end{equation}
The derivation of that result and its generalization to the case of
many modes (see also Ref.~\cite{Nazarov99}) is given in
Section~\ref{sec:cap}, see Eq.~(\ref{eq:deltaOmegaMultiN}) and the
discussion around it.

Upon lowering the temperature it becomes important, whether the
energy scale $\Ec |r|^2$ is larger or smaller than the mean level
spacing $\Delta$. If the condition
\begin{equation}
T_0({\cal N})\gg\Delta
\label{diff3}
\end{equation}
with $T_0({\cal N})$ from Eq.~(\ref{eq:ov2}) is satisfied, then
oscillations of the differential capacitance follow the
law~\cite{Matveev95} 
\begin{equation} 
{\langle C_{\rm diff}({\cal N}) \rangle \over C} 
\sim \ln\left(\frac{1}{|r|^2\cos^2\pi{\cal N}}\right) 
|r|^2 \cos(2 \pi {\cal N}).
\label{diff4}
\end{equation}
In the opposite case,
\begin{equation}
\Delta\gg\Ec |r|^2,
\label{diff5}
\end{equation}
mesoscopic fluctuations of the differential capacitance exceed
the amplitude of oscillations of the average 
$\langle C_{\rm diff}({\cal N})\rangle$.

To characterize the mesoscopic fluctuations of the differential
capacitance, we calculate the correlation function $K(s)=
\langle\delta C_{\rm diff}({\cal N}+s)\delta C_{\rm diff}({\cal
N})\rangle /C^2$, see Section~\ref{sec:cap}. For temperatures
$T\gg\Delta$, we find for the leading term in $K(s)$
\begin{equation}
K(s)=
\frac{16}{3\pi^2\beta^2}\cos (2\pi s)\left(\frac{\Delta}{\Ec}\right)^2
\ln^4\left(\frac{\Ec}{T}\right),
\label{eq:cdiffcorr}
\end{equation}
see Eq.~(\ref{eq:60.10}) in Section~\ref{sec:cap}; here $\beta=1(2)$
for a system with (without) time--reversal symmetry. The periodic
oscillations of $K(s)$ are due to the Coulomb blockade; they remain
correlated
over an interval $s\sim \Ec/\Delta$. Note, that for the case of one
single-mode junction we consider here, the periodic part of the
correlation function exceeds the non-periodic part [see
Eq.~(\ref{eq:varcdiff}) with $\Nch=2$] by a large factor $\sim
10 (T/\Delta)
\ln^4(\Ec/T)$.  If the temperature becomes lower than $\Delta$,
the periodic part of $K(s)$ can be estimated by
Eq.~(\ref{eq:cdiffcorr}) with $T$ replaced by $\Delta$ under the
logarithm, and for the estimate of the non-periodic part one may
replace $T$ by $\Delta$ in Eq.~(\ref{eq:varcdiff}). The periodic part
of the correlation function dominates at the lowest temperatures as
well. With the increase of the number of channels $\Nch$,
however, the relative magnitude of the periodic part drops rapidly (see
Section~\ref{sec:cap} for details).

\subsection{Finite-size open dot: Introduction to
  the bosonization technique and the relevant energy scales}
\label{sec:finitesize}
\label{sec:lowen}
\label{sec:reduction}

\begin{figure}
\vspace{0.2cm}
\hspace{0.3\hsize}
\epsfxsize=0.4\hsize
\epsfbox{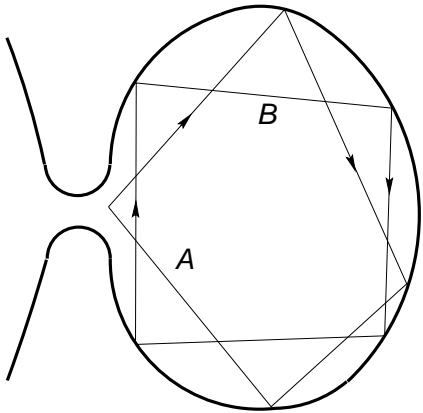}
\vspace{0.6cm}
\caption{Schematic view of a quantum dot with a single junction}
\label{Fig17}
\end{figure} 

We start with a single-junction system, see Fig~\ref{Fig17}. The
relevant electron states which facilitate the electron transfer
through the junction and thus determine the charge fluctuations, can
be labeled by the momenta measured in the narrow region at the
contact, referred to as channel. This way, the
problem of charge fluctuations in the system of interest is reduced to
an effective one-dimensional problem~\cite{Matveev95}.  The
corresponding Hamiltonian consists of two parts. The first part
describes non-interacting fermions moving in one spatial dimension.
Without backscattering in the channel, and in the limit $\Delta\to 0$,
this part takes form
\begin{equation}
\hat{H}_0=\sum_{\sigma}\int_{-\infty}^\infty
dx\left[\frac{1}{2m}\nabla\psi^\dagger_\sigma\nabla\psi_\sigma -
\mu\psi^\dagger_\sigma\psi_\sigma\right].
\label{eq:5.1.1}
\end{equation}
Here $\psi_\sigma^\dagger(x)$ and $\psi_\sigma(x)$ are the
creation and annihilation operators of the one-dimensional fermions in
spin state $\sigma$; the half-axes $(0;\infty]$ and $[-\infty; 0)$
correspond to the dot and the combined channel plus
lead system, respectively. We are
interested in the quantum dynamics of the system at energy scale $\sim
\Ec$, which is small compared to the Fermi energy. This allows us to
linearize the spectrum of the fermions. Writing
$\psi_\sigma(x)=e^{-ik_Fx}\psi_{L\sigma}(x)
+e^{ik_Fx}\psi_{R\sigma}(x)$, where $\psi_{L\sigma}$ and
$\psi_{R\sigma}$ are left- and right-moving fermions respectively,
we obtain from Eq.~(\ref{eq:5.1.1})
\begin{equation}
\hat{H}_0=iv_F
\sum_\sigma\int_{-\infty}^\infty dx
\left(\psi_{L\sigma}^\dagger\partial_x\psi_{L\sigma} -
  \psi_{R\sigma}^\dagger\partial_x\psi_{R\sigma} \right).
\label{eq:5.1.2}
\end{equation}
Here $v_F$ is the Fermi velocity of one-dimensional fermions. 
The possibility of
reflection in the channel, which results in a deviation of the channel
conductance from its perfect value $e^2/\pi\hbar$, can be taken into
account by addition of a backscattering term to the Hamiltonian 
(\ref{eq:5.1.2}),
\begin{equation}
\hat{H}_{\rm bs} = |r| v_F \sum_\sigma
\left(\psi_{L\sigma}^\dagger(0) \psi_{R\sigma}(0) + 
\psi_{R\sigma}^\dagger(0) \psi_{L\sigma}(0)\right),
\label{eq:5.1.3}
\end{equation}
where $|r|^2 \ll 1$ is the reflection coefficient.\footnote{The
operators $\psi(0)$ and $\psi^{\dagger}(0)$ in Eq.\ (\ref{eq:5.1.3})
are regularized as $\int dx f(x) \psi(x)$ and $\int dx f(x)
\psi^{\dagger}(x)$, respectively, where $f$ is a positive function, 
symmetric around $x=0$, with $\int dx f(x) = 1$, and with 
support close to the origin.
This same regularization is used throughout the remainder of
this section.}

The second part of the effective Hamiltonian represents the charging
energy. Until now, we were expressing it in terms of the number of
electrons inside the dot. Using the conservation of the total number
of electrons in the entire system, it is convenient here to
express this energy in terms of charge outside the dot,
\begin{equation}
{\hat H}_{\rm C}=\Ec\left[\int_{-\infty}^{0} dx\sum_\sigma\left(
:\psi_{L\sigma}^\dagger\psi_{L\sigma}
+\psi^\dagger_{R\sigma}\psi_{R\sigma}:\right)+{\cal N}\right]^2. 
\label{eq:5.1.4}
\end{equation}
(Here $:\ldots:$ denotes normal ordering of the fermion operators.)
We would like to treat the charging energy (\ref{eq:5.1.4})
non-perturbatively, while developing a perturbation theory in
$\hat{H}_{\rm bs}$. Refs.~\cite{Flensberg,Matveev95} suggest 
a scheme in which this can be done, using the bosonized 
representation for the
one-dimensional fermions in the channel. The 
boson representation makes the interaction
Hamiltonian (\ref{eq:5.1.4}) quadratic in the new variables, thus
removing the main obstacle in building the desired perturbation 
theory in the backscattering Hamiltonian (\ref{eq:5.1.3}).

In the boson representation, the two parts $\hat{H}_0$ and
$\hat{H}_{\rm bs}$ of the Hamiltonian take form:
\begin{eqnarray}
\hat{H}_0 &=&\frac{v_F}{2}\int_{-\infty}^{\infty} dx\sum_{\gamma=\rho,s}
\left[\frac{1}{2}(\nabla\phi_\gamma)^2+2(\nabla\theta_\gamma)^2\right],
\label{H0}\\
\hat{H}_{\rm bs}&=&-\frac{2}{\pi}|r|D\cos[2\sqrt{\pi}\theta_\rho(0)]
           \cos[2\sqrt{\pi}\theta_s(0)],
\label{Hr}
\end{eqnarray}
where $D$ is the energy bandwidth for the one-dimensional 
fermions, which are related to the
boson variables by the transformation~\cite{Haldane81}:
\begin{eqnarray*}
\psi^\dagger_{L\sigma(x)}=\hat{\eta}_\sigma\sqrt{\frac{D}{2\pi v_F}}
  \exp\left\{i\sqrt{\pi}\left[{1 \over 2} \phi_\rho(x)+
                              {1 \over 2} \sigma\phi_s(x)
+\theta_\rho(x)
+\sigma\theta_s(x)
\right]\right\},
\nonumber\\
\psi^\dagger_{R\sigma(x)}=\hat{\eta}_\sigma\sqrt{\frac{D}{2\pi v_F}}
  \exp\left\{i\sqrt{\pi}\left[{1 \over 2} \phi_\rho(x)+
                              {1 \over 2} \sigma\phi_s(x)
-\theta_\rho(x)
-\sigma\theta_s(x)
\right]\right\}
.
\label{psi}
\end{eqnarray*}
Here we introduced Majorana fermions $\hat{\eta}_{\pm 1}$ here to satisfy the
commutation relations for fermions with opposite spins,
$\{\hat{\eta}_{+1},\hat{\eta}_{-1}\}=0,\ \hat{\eta}_{\pm 1}^2=1
$. Anti-commutation of electrons of the
same spin ($\sigma=1\,\mbox{or}-1$), is ensured by the following
commutation relations between the canonically conjugated Bose fields:
\begin{eqnarray}
[\nabla\phi_{\gamma'}(x'),\theta_\gamma (x)]&=&
[\nabla\theta_{\gamma'}(x'),\phi_\gamma(x)]
\nonumber\\
&=&-i\delta_{\gamma\gamma'}\delta(x-x');\quad \gamma,\gamma'=\rho,s.
\label{algebra}
\end{eqnarray}
The interaction term (\ref{eq:5.1.4}) becomes also quadratic in the
boson representation:
\begin{equation}
\hat{H}_{\rm C}={\Ec}
\left[\frac{2\theta_\rho(0)}{\sqrt{\pi}}-{\cal N}\right]^2.
\label{HC}
\end{equation}
The operators $(2e/\sqrt{\pi})\nabla\theta_{\rho}(x)$ and
$(2/\sqrt{\pi})\nabla\theta_{s}(x)$ are the smooth parts of the
electron charge ($\rho$) and spin ($s$) densities, 
respectively.\footnote{Equation (\ref{HC}) contains no
contribution from the boson fields at $-\infty$, because of
the order of limits implied  in the interaction Hamiltonian
(\ref{eq:5.1.4}). First, the range of the interaction is taken
to infinity, then the length of the one-dimensional channel.}

The average charge of the dot can be determined from the 
${\cal N}$--dependence of the thermodynamic potential $\Omega$ of the full
Hamiltonian. Taking into account that only one of its parts, the charging
energy, depends on ${\cal N}$, we find:
\begin{equation}
\langle{\hat N}\rangle_q = {\cal N}-\frac{1}{2\Ec}
  \frac{\partial \Omega}{\partial {\cal N}},
\label{egr}
\end{equation}
(hereinafter $\langle \dots \rangle_q$ indicates the average over the
quantum state without disorder average).  In the absence of
backscattering, $r=0$, the Hamiltonian of the system
$\hat{H}_0+\hat{H}_{\rm C}$ is quadratic in the spin and charge densities.
The explicit ${\cal N}$-dependence of the Hamiltonian can be easily
removed by the transformation $\theta_{\rho}(x) \to \theta'_{\rho}(x)
+ (1/2) {\cal N} \pi^{1/2}$. Hence, in this case the thermodynamic
potential $\Omega$, and hence the ground state
energy, obviously have no ${\cal N}$-dependence,
the average charge is linear
in ${\cal N}$ and is not quantized, and 
\begin{equation}
  {2 \over \sqrt{\pi}} \langle\theta_\rho(0)\rangle_q
  = 
  \langle {\hat N}\rangle_q
  = {\cal N}.
  \label{eq:thetarhoavg}
\end{equation}

Note that scattering between the right- and left-moving states occurs
only at the point of a barrier in the channel $x=0$. The Hamiltonian
(\ref{eq:5.1.1})--(\ref{eq:5.1.3}) completely ignores the fact that
the dot has a finite size, and a particle that entered it, eventually
 ought to exit. However, these two events are separated by the dwell
time $\sim \hbar/\Delta$, which makes the model (\ref{H0})--(\ref{HC})
applicable at
$\Delta\to 0$. To illustrate the effect of a finite dwell time, we
replace the infinite interval $(0,\infty]$ by a finite segment
$(0,L]$, choosing the length of the effective one-dimensional system
in such a way that the time delay of a particle entering this segment
at $x=0$ and bouncing off the ``wall'' at $x=L$ equals
$\hbar/\Delta$. In other words, throughout the remainder of this
section we replace the upper limit of the integration in 
Eq.~(\ref{H0}) by $L$,
\begin{equation}
\hat{H}_0=\frac{v_F}{2}\int_{-\infty}^{L} dx\sum_{\gamma=\rho,s}
\left[\frac{1}{2}(\nabla\phi_\gamma)^2+2(\nabla\theta_\gamma)^2\right],
\label{eq:5.1.5}
\end{equation}
where $L\simeq\pi\hbar v_F/\Delta$, and set zero boundary conditions for
the displacement fields in the spin and charge mode:
\begin{equation}
\theta_\rho(L)=\theta_s(L)=0.
\label{boundary}
\end{equation}

When the reflection amplitude $r$ is finite, the Hamiltonian
of the system is 
\begin{equation}
\hat{H}=\hat{H}_0 + \hat{H}_{\rm C} + \hat{H}_{\rm bs}.
\label{eq:5.100}
\end{equation}
The largest energy scale appearing in the Hamiltonian (\ref{eq:5.100}) 
is the charging energy $\Ec$. At energies smaller
than $\Ec$ the charge is pinned to the value of gate voltage ${\cal
N}$, and does not exhibit quantum fluctuations. Alternatively, in
the language of Eqs.\ (\ref{H0})--(\ref{HC}),
the charge mode $\theta_{\rho}$ is pinned at $x=0$ by the charging
energy (\ref{HC}). The backscattering term (\ref{Hr}) induces 
only negligible perturbations in this mode, so that the dynamics
of the charge mode $\theta_{\rho}$ is described by the quadratic
Hamiltonian $\hat{H}_0 + \hat{H}_{\rm C}$, as in the absence of the
backscattering term. This observation allows us to reduce
the Hamiltonian (\ref{eq:5.100}) to an effective Hamiltonian
for the spin mode $\theta_s$, by averaging (\ref{Hr}) over the 
Gaussian fluctuations of $\theta_{\rho}$ around its average
(\ref{eq:thetarhoavg}).

This averaging yields \cite{Matveev95}
\begin{eqnarray*}
\langle\cos[2\sqrt{\pi}\theta_\rho(0)]\rangle_q &=&
\cos[2\sqrt{\pi}\langle\theta_\rho(0)\rangle_q]
\exp\left[-2{\pi} \left(\langle\theta_\rho(0)^2\rangle\right)\right]
\\ &=& 
\cos{\pi}{\cal N}\sqrt{\frac{2e^{\bf C} \Ec}{\pi D}},
\end{eqnarray*}
where ${\bf C}\approx 0.577$ is the Euler constant. As a result the
effective Hamiltonian for spin dynamics is given by
\begin{eqnarray}
\hat{H}^s_0&=&\frac{v_F}{2}\int_{-\infty}^{L} dx
\left[\frac{1}{2}(\nabla\phi_s)^2+2(\nabla\theta_s)^2\right] ,
\label{Hs0}
\label{eq:5.102}
\\
\hat{H}^s_{\rm bs}&=&
-\sqrt{D \Ec}|r|\cos(\pi{\cal N})
   \cos[2\sqrt{\pi}\theta_s(0)].
\label{eq:5.103}
\label{Hsr}
\end{eqnarray}
The Hamiltonian (\ref{Hs0}), (\ref{Hsr}) represents a
one-mode, $g=1/2$ Luttinger liquid with a barrier at $x=0$. 
We will now discuss the effect of backscattering at the barrier,
as described by the Hamiltonian $\hat{H}^s_{\rm bs}$.

First, at not too low temperatures (to be defined more precisely
below) the backscattering Hamiltonian (\ref{Hsr}) can be
considered perturbatively. In the lowest non-vanishing order one finds
for the backscattering-induced correction to the thermodynamic
potential:
\begin{equation}
\delta \Omega=
-\frac{1}{2}\int_0^{1/T} d\tau 
\frac{
{\mathrm Tr}
\left[
e^{\hat{H}^s_0(\tau -1/T)}\hat{H}^s_{\rm bs}
e^{-\hat{H}^s_0\tau}
\hat{H}^s_{\rm bs}
\right]
}
{
{\mathrm Tr}
\left[
e^{-\hat{H}^s_0/T}
\right]
}.
\label{eq:5.104}
\end{equation}

Calculation of the trace in the numerator of this formula reduces to
the evaluation of the correlation function:
\[
\langle\cos[2\sqrt{\pi}\theta_s(x=0,\tau)]
\cos[2\sqrt{\pi}\theta_s(x=0,0)]\rangle_q
\sim
\frac{\pi T}{D\sin\pi T\tau},
\]
which is readily performed with the help of quadratic Hamiltonian
(\ref{eq:5.102}). The result of this calculation is
\begin{equation}
\delta \Omega
({\cal N})
= -\frac{4e^{\bf C}}{\pi^3} |r|^2 \Ec \cos^2\pi{\cal N}
\ln \left(\frac{\Ec}{T}\right).
\label{eq:5.105}
\end{equation}

At low temperatures this correction diverges, which signals the
breakdown of the simple perturbation theory in the backscattering
Hamiltonian. In order to find the low
energy cut-off of the logarithm, we invoke the following
renormalization group arguments.\footnote{The model is exactly soluble
\cite{Matveev95}, however, the arguments we use seem to be more
physically transparent and suitable for further purposes.}  The
Hamiltonian Eqs.~(\ref{eq:5.102}), (\ref{eq:5.103}) represents spin
fluctuations in a wide energy band $\varepsilon \lesssim D$. We can
integrate out the high-energy part of this band, thus deriving an
effective Hamiltonian acting in a reduced band of width ${\tilde D}$.  
The backscattering at the barrier, described by the Hamiltonian
$\hat{H}^s_{\rm bs}$, is known to be a relevant perturbation~\cite{Kane}.
Indeed, averaging of the cosine term (\ref{Hsr}) over the ``fast''
degrees of freedom with energies
$D\gtrsim\varepsilon\gtrsim\tilde{D}$ yields
\begin{equation}
\left.\langle\cos[2\sqrt{\pi}\theta_s(0)]\rangle
\right|_{D\gtrsim\varepsilon\gtrsim\tilde{D}}\;\to\;
\exp\left[-\frac{1}{2}\int_{\tilde{D}}^{D}\frac {dk}{k}\right]
\cos[2\sqrt{\pi}\theta_s(0)]
\label{RG}
\end{equation}
(we use the same notation, $\theta_s$, for the fields defined in the
initial band and in the truncated band). Therefore, the amplitude of
the effective backscattering potential ${\cal U}_{\rm bs}$ scales with
the band width $\tilde{D}$ as
\begin{equation}
{\cal U}_{\rm bs}\propto\sqrt{\Ec\tilde{D}}\,|r|\cos(\pi{\cal N}),
\label{Ubs}
\end{equation}
and the effective scattering amplitude ${\cal U}_{\rm bs}/\tilde{D}$
grows proportionally to $\sqrt{\Ec/\tilde{D}}$ with the band width
reduction.  Eventually, at sufficiently small band widths, the
amplitude ${\cal U}_{\rm bs}$ exceeds $\tilde{D}$, signalling the
crossover from weak backscattering at higher energies, to rare
tunneling events at lower energies. The crossover energy $T_0({\cal
  N})$ can be found from the condition ${\cal U}_{\rm bs}\simeq
\tilde{D}$. It yields, up to a numerical factor, the following energy
scale:
\begin{equation}
T_0
\left(
{\cal N}
\right)
= (8e^{\bf C}/\pi) \Ec |r|^2\cos^2\pi{\cal N}.
\label{T0ex}
\end{equation} 
The numerical coefficient in Eq.\ (\ref{T0ex})
could not be found from the RG scheme
but results from the exact consideration. We will not describe the
calculation of this coefficient.
The renormalization procedure must be stopped when the band
width $\tilde{D}$ reaches $T_0({\cal N})$.

At lower energies the dynamics of the spin mode is described by the
renormalized Hamiltonian $\hat{H}^s=K\{\phi_s\}+U\{\theta_s\}$ with
the ``potential energy''
\begin{equation}
U\{\theta_s(x)\}=-\lambda_2T_0({\cal N})
\frac{\cos(\pi{\cal N})}{|\cos(\pi{\cal N})|}
\cos[2\sqrt{\pi}\theta_s(0)]+v_F\int_{-\infty}^L dx
(\nabla\theta_s)^2.
\label{utheta}
\end{equation}
The spin mode $\theta_s(0)$ is strongly pinned by the potential
energy. Physically, it means that the fluctuations of the spin of the
dot are rare events of hops between the local minima of the
potential~(\ref{utheta}) [the
numerical coefficient in Eq.~(\ref{utheta}) is $\lambda_2\sim 1$]. 

\begin{figure}
{\epsfxsize=12cm\centerline{\epsfbox{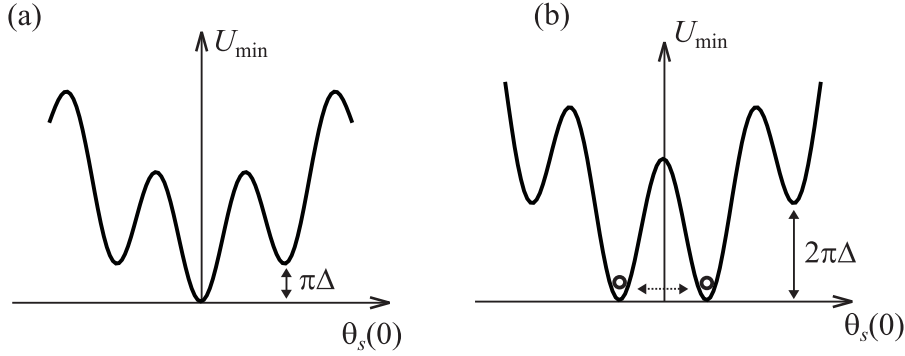}}}
\caption{The potential $U_{\rm min}[\theta_s(0)]$ at various gate
  voltages. ({\it a}) At $\cos(\pi{\cal N})>0$ the potential has one
  minimum (spin singlet state). ({\it b}) At $\cos(\pi{\cal N})<0$ the
  minimum is double-degenerate (spin doublet state).  Tunneling
  between the two minima corresponds to the hybridization of the spin
  states of the dot and lead. This hybridization and consequent
  lifting of the degeneracy essentially is the Kondo effect.}
\label{fig:520}
\end{figure}

The ``potential energy'' (\ref{utheta}) can be easily minimized with
respect to the possible spatial configurations of the field
$\theta_s(x)$, which satisfy the boundary condition (\ref{boundary}),
and have the value $\theta_s(0)$ at $x=0$:
\begin{equation}
U_{\rm min}[\theta_s(0)]=-\lambda_2T_0({\cal N})
\frac{\cos(\pi{\cal N})}{|\cos(\pi{\cal N})|}
\cos[2\sqrt{\pi}\theta_s(0)]+ \theta_s(0)^2 \Delta/\pi.
\label{potential}
\end{equation}
In the limit $L\to\infty$ ({\it i.e.}, $\Delta\to 0$) the second term
of Eq.~(\ref{potential}) does not contribute, and the minima of the
potential energy correspond to
\begin{eqnarray}
  \label{uuuu}
&  2\sqrt{\pi}\theta_s(0)=2\pi n\;,
&\quad \mbox{if}\; \cos(\pi{\cal N})>0\;,\\
& 2\sqrt{\pi}\theta_s(0)=(2n+1)\pi \;,
&\quad \mbox{if}\; \cos(\pi{\cal N})<0\;\nonumber
\end{eqnarray}
with integer $n$. In each of these two sets, the energy is the same
within a set for all $n$. At large but finite $L$, when the condition
\begin{equation}
T_0({\cal N})\gg\Delta
\label{deltato}
\end{equation}
is satisfied, the minima of energy (\ref{potential}) hardly shift, but
their degeneracy is reduced strongly. If the number of electrons on
the dot is close to an even number, $\cos(\pi{\cal N})>0$, then the
degeneracy is removed completely. If the charge is closer to an odd
number of electrons, $\cos(\pi{\cal N})<0$, the energy minimum
preserves double degeneracy ($n=-1$ and $n=0$), 
see Fig.~\ref{fig:520}.

The ``kinetic'' part $K\{\phi_s\}$ of the Hamiltonian $\hat{H}^s$ is
quadratic in $\nabla\phi_s$, see Eq.~(\ref{Hs0}); it does not commute
with the ``potential'' part, and causes tunneling between the
potential minima. The tunneling amplitude is
energy-dependent~\cite{LarkinLee}, and small at $E\ll T_0({\cal N})$.
To describe the low-energy dynamics of the spin mode, it is convenient
to project out all the states of the Luttinger liquid that are not
pinned to the minima of the potential (\ref{potential}). Transitions
within one of the sets (\ref{uuuu}) then can be described~\cite{Kane}
by a tunneling Hamiltonian
\begin{equation}
\hat{H}_{\pm}=
-\frac{\tilde{D}^2}{2\pi T_0({\cal N})}
\cos\left\{\sqrt{\pi}[\phi_s(+0)-\phi_s(-0)]\right\},
\label{Hpm}
\end{equation}
which operates in an energy band $\tilde{D}\ll T_0({\cal N})$.  Here a
discontinuity of the variable $\phi_s(x)$ at $x=0$ is allowed, and the
point $x=0$ is excluded from the region of integration in
Eq.~(\ref{Hs0}).  The Hamiltonian $\hat{H}_{xy}$, which is a sum of
two operators of finite shifts for the field $\theta_s(0)$, represents
hops $\theta_s(0)\to\theta_s(0)\pm\sqrt{\pi}$ between pinned states.
In other words, $\hat{H}_{\pm}$ is the finite shift operator with
respect to the variable $\theta$ as can be checked with the help of
the commutation relations (\ref{algebra}).

These hops correspond to a change by $1$ of the $z$--projection of
the dot's spin. At energy scales ${\tilde D}\gg \Delta$ the hops are
predominantly inelastic. The variation of the transition amplitude
with energy can be found by continuing the same RG procedure, until
the running cut-off ${\tilde D}$ reaches the value $\Delta$. At
energies below $\Delta$, fluctuations which involve the excitations of
spin mode in the dot ($0<x<L$) are suppressed. If the minimum of
energy (\ref{potential}) is non-degenerate (spin-singlet state of the
dot), then only one spin state in the dot remains available. However,
in the case $\cos(\pi{\cal N})<0$, there is an exact degeneracy
between two lowest-energy states, see Fig.~\ref{fig:520}; the ground
state is a spin-doublet. The Hamiltonian (\ref{Hpm}) hybridizes the spin
of the dot with the continuum of spin excitations in the lead. The
Kondo effect consists essentially of this hybridization, which lifts
the degeneracy and ultimately leads to the formation of a spin singlet
in the entire system. The energy scale at which the hybridization
occurs, is the Kondo temperature of the problem at hand.

Note that Eq.~(\ref{Hpm}) does not contain any unknown numerical
factor. Finding the multiplicative factor in (\ref{Hpm}) cannot be
done within the RG treatment we presented above. To find the numerical
constant, one has to use~\cite{GHL} the exact
result~\cite{Kane,Furusaki95} for the linear response of an
unconstrained ($L\to\infty$) Luttinger liquid with $g=1/2$. Matching
the low-temperature asymptote of the exact result with the result of
the calculation based on the tunneling Hamiltonian formalism, we were able
to find the constant in Eq.~(\ref{Hpm}). Also, we would like to notice
that the initial Hamiltonian (\ref{H0}), (\ref{Hr}), (\ref{HC})
commutes with the total spin of the system: scattering in the
junction, represented by the Hamiltonian (\ref{Hr}), is clearly
spin-conserving.  The spin Hamiltonian (\ref{Hs0})--(\ref{Hsr}),
obtained after the charge mode was integrated out, also preserves the
total spin.  However, the tunneling Hamiltonian (\ref{Hpm}) has the
symmetry of the operator $S_x(-0)S_x(+0)+S_y(-0)S_y(+0)$, and
represents the in-plane part of exchange between the spin densities of
the lead and dot [$S(+0)$ and $S(-0)$, respectively] at the point of
contact. To restore the $SU(2)$ symmetry of the Hamiltonian, we must
supplement $\hat{H}_\pm$ with another term, which represents the Ising
component of the exchange interaction:
\begin{equation}
\hat{H}_z=
\frac{v_F^2}{2 T_0({\cal N})}
\nabla\theta_s(-0)\nabla\theta_s(+0).
\label{Hz}
\end{equation}
This term is omitted usually in the DC tunneling problem for a
Luttinger liquid~\cite{Kane}. In agreement with the general idea about
$SU(2)$ symmetry of the $g=1/2$ Luttinger liquid~\cite{Tsvelik1}, both
operators (\ref{Hpm}) and (\ref{Hz}) have the same scaling dimension.
To find the numerical factor in Eq.~(\ref{Hz}), again a comparison
with the exact solution is necessary.

We have seen in this Subsection, that there is a hierarchy of
the energy scales characterizing an open dot. The charge of the dot
fluctuates only at energy scales exceeding $\Ec$; at lower energies
these fluctuations are suppressed. However, strong fluctuations of
the spin of the dot continues down to lower energies, $T_0({\cal N})$ see
Eq.~(\ref{T0ex}). If $T_0({\cal N})$ exceeds the level spacing
$\Delta$, then at energies $\Delta\lesssim\varepsilon\lesssim
T_0({\cal N})$ the spin of the dot is quantized, and may change only
in quanta of $\delta S=1$ due to the exchange interaction with the
electrons in the lead; at the same time, presence of many intra-dot
excitations (electron-hole pairs) is allowed. Finally, at energies
below $\Delta$ the intra-dot excitations are also suppressed. Under
these conditions, the spin of the dot is $S=0$ or $S=1/2$, depending
on the sign of $\cos\pi{\cal N}$. If this cosine is positive, then the
dot is in a singlet state, and there are no peculiarities in the
temperature dependence of observable quantities at $T\lesssim\Delta$.
However, if $\cos\pi{\cal N}<0$, then the dot carries a spin, and
the Kondo effect develops at sufficiently low temperatures. In the next
Subsection we will study the temperature domain $T\lesssim\Delta$,
find the Kondo temperature, and discuss the transport properties of
the dot in Kondo regime. Starting from Subsection~\ref{sec:open}, we return to
the intermediate temperature scale, $\Delta\lesssim T\lesssim \Ec$, in
order to discuss the dominant transport mechanisms there.

\subsection{The limit of low temperature: The effective exchange Hamiltonian 
and Kondo effect}
\label{sec:exchange}

We could continue using the same RG method applied to the bosonized
representation of the Hamiltonian, down to scale ${\tilde
D}\sim\Delta$; it would allow us to estimate the order of magnitude of
the effective exchange constant $J$ responsible for the
hybridization. It is possible, however, to develop a different method
applicable in the energy domain $T_0({\cal N})\gtrsim {\tilde
D}\gtrsim\Delta$, which allows us a more accurate determination of $J$
and of the Kondo temperature $T_K$.

At energies $E\ll T_0({\cal N})$, the spin field $\theta_s(0)$ is
pinned at the point contact. Recalling that $\theta_s(L)=0$, we see
that the spin of the dot indeed takes discrete values only, as was
mentioned above. 
At such energy scale, it is instructive to return to the fermionic
description of the problem. After the two parts of the Hamiltonian,
Eqs.~(\ref{Hpm}) and (\ref{Hz}), are found, one can explicitly see
that the initial $SU(2)$ symmetry of the problem is preserved.
Therefore, the effective Hamiltonian controlling the spin degrees of
freedom of the system dot$+$lead at low energies, corresponds to the
isotropic exchange interaction,
\begin{equation}
\hat{H}_{\rm ex}=J(\tilde{D}){\hat{\vec S}}{\hat{\vec S}}_d.
\label{exchange}
\end{equation}
Here ${\hat{\vec S}}_{d}=\half\hat{\chi}^\dagger_{\sigma_1}({\vec
  R}_0){\vec \sigma}_{\sigma_1\sigma_2}\hat{\chi}_{\sigma_2}({\vec
  R}_0)$ and ${\hat{\vec S}}=\half\hat{\psi}^\dagger_{\sigma_1}({\vec
  R}_0) {\vec \sigma}_{\sigma_1\sigma_2}\hat{\psi}_{\sigma_2}({\vec
  R}_0)$ are the operators of spin density in the dot and in the lead
  respectively, at the point ${\vec R}_0$ of their contact;
  ${\vec\sigma}_{\sigma_1\sigma_2}$ are the Pauli matrices. In what
  follows, we adopt this point to be the origin, ${\vec R}_0=0$. The
  dimension of the exhange constant $J(\tilde{D})$ depends on the way
  we normalize the electron wave functions; dimension-wise, it can be
  written as $J(\tilde{D})\propto\left[\nu_d\nu T_0\right]^{-1}$,
  where $\nu_d$ and $\nu$ are the densities of states in the dot and
  lead, respectively. 

In fact, the form of the Hamiltonian (\ref{exchange}) could be derived
without resorting to the explicit calculation performed above within
the bosonization technique. The only important notion needed for the
derivation, is the low-energy decoupling of the dot from the lead.
Pinning of both charge and spin modes is the manifestation of this
decoupling. The decoupling is complete, however, only in the absence of
excitations. To describe the behavior of the system at low but finite
temperatures, one needs to establish the least irrelevant operators
that violate the complete decoupling. The decoupled dot and lead are
described by independent Fermi liquids. The number of electrons in the
dot can not be changed, as such a variation of charge would result in a
large energy increase of order $\Ec$. That is why the leading irrelevant
operators preserving the initial $SU(2)$ symmetry are the interaction
in the charge and spin channels between the dot and the lead at the point
of contact,
\[
\hat{H}_{\rm irr} = A(\tilde{D}) \hat{\rho}\hat{\rho_d} + \hat{H}_{\rm ex},
\]
where $\hat{H}_{\rm ex}$ is defined in Eq.~(\ref{exchange}) and
${\hat{\rho}}_{d}=\hat{\chi}^\dagger_{\sigma}({\vec
R}_0)\hat{\chi}_{\sigma} ({\vec R}_0)$ and
${\hat{\rho}}=\hat{\psi}^\dagger_{\sigma}({\vec
R}_0)\hat{\psi}_{\sigma}({\vec R}_0)$ are the operators of the charge
density in the dot ($x>0$) and in the lead ($x<0$) respectively, at
the point ${\vec R}_0$ of their contact. The first term here
represents the interaction in the singlet channel (inelastic
scattering between an electron in the dot an electron in the lead).
Because the charge fluctuations are suppressed at energy $\Ec$ which
exceeds significantly the pinning energy of the spin mode $r^2\Ec$, one
finds $A/J \simeq r^2 \ll 1$. Therefore, we can neglect the
interaction in the singlet channel. Moreover, even if this mode is
taken into account, it does not lead to any non-trivial effect at low
energies.

The Hamiltonians (\ref{Hpm}), (\ref{Hz}) and (\ref{exchange})
describe, in the bosonic and fermionic representation respectively,
the low-energy properties of the very same system. Correlation
functions of observable quantities, calculated with the help of
Eqs.~(\ref{Hpm}), (\ref{Hz}) should coincide with the results for the
same functions obtained from the calculation which uses
Hamiltonian~(\ref{exchange}), at temperatures $\Delta\ll T\ll
T_0(\cal{N})$. This fact will enable us to find the exact value of the
coupling constant in the Hamiltonian (\ref{exchange}).
The advantage of the fermionic representation is that it  allows
for the description of the system at $T \simeq \Delta$ (ultimately
leading to the Kondo effect). We turn to this problem now.

We first rewrite the Hamiltonian (\ref{exchange}) in the basis of
electron eigenfunctions $\varphi_n({\bf R})$ and $\varphi_k({\bf R})$
in the dot and lead respectively. In this basis, the Hamiltonian takes
the form
\begin{equation}
\hat{H}_{\rm ex}=\frac{1}{4}\sum_{\{nk\sigma\}}{\tilde J}_{n_1n_2k_1k_2}
\left(
\vec\sigma_{\sigma_1\sigma_2}
\vec\sigma_{\sigma_3\sigma_4}\right)
c_{n_1\sigma_1}^\dagger c_{n_2\sigma_2}
c_{k_1\sigma_3}^\dagger c_{k_2\sigma_4}
\label{exchange1}
\end{equation}
with the matrix elements ${\tilde
  J}_{n_1n_2k_1k_2}=J(\tilde{D})\varphi^*_{n_1}(0)\varphi_{n_2}(0)
\varphi^*_{k_1}(0)\varphi_{k_2}(0)$. The detailed structure of the
one-electron functions in the continuum of states $\{k\}$ is not
important in the discussion of the Kondo effect. An examination of
the standard calculation~\cite{Kondo} shows that only the densities
$|\varphi_k(0)|^2$, averaged over the continuum of states $k$, matter.
Therefore we will suppress the indices $k_1$ and $k_2$ in the exchange
  constants hereinafter, assuming that the functions $\varphi_{k}$
  are normalized so that
\begin{equation}
|\varphi_k(0)|^2=\langle |\varphi_k(0)|^2\rangle =1.
\label{norm5}
\end{equation}

If the dot is in the spin-doublet state, the exchange constant gets
renormalized at low energies. Unlike the ``ordinary'' Kondo model with
only one localized orbital state involved, here the renormalization
occurs due to virtual transitions in both the continuum spectrum of
the lead and the discrete spectrum of the dot. To see the signature of
the Kondo effect, we calculate the second order correction to the
exchange constant $J_{00}$ where $0$ labels the singly occupied level
of the quantum dot.

\begin{figure}
{\epsfxsize=12cm\centerline{\epsfbox{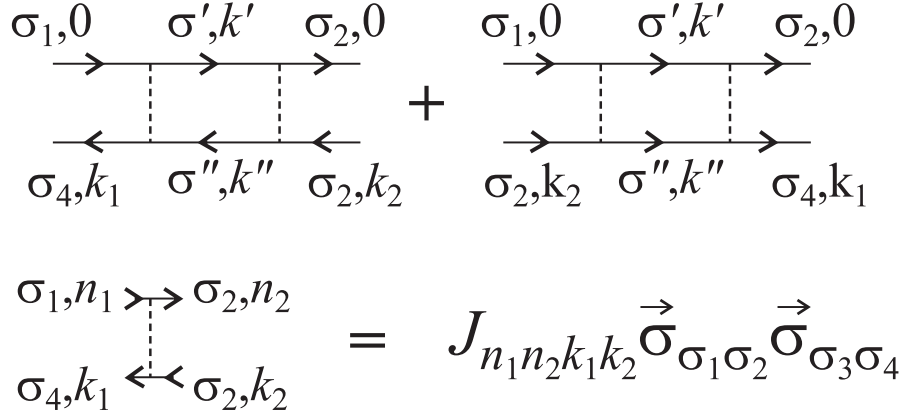}}}
\caption{The diagrammatic representation for the second order
correction to the exchange interaction constant. The electron states in the
intermediate lines include excitations within energy strip 
$\left[E, \tilde{D}\right]$.}
\label{fig:530}
\end{figure}

This second order correction, diagrammatically shown in
Fig.~\ref{fig:530}, has the form 
\begin{equation}
\delta J_{00}(E,{\tilde D})=J^2(\tilde{D})|\varphi_0(0)|^2
\int_{|E|}^{\tilde{D}}\nu d\xi\sum_{|\xi_d^{(l)}|<\tilde{D}}
\frac{|\varphi_l(0)|^2}{|\xi|+|\xi_d^{(l)}-\xi_d^{(0)}|},
\label{JE0}
\end{equation}
where the integral is taken over the continuum energy spectrum $\xi$
in the leads, $\nu$ is the corresponding density of electron states.
The sum over the discrete energy levels of electrons in the dot
includes the term involving the singly occupied level $l=0$.  This
term, which corresponds to a spin-flip within the same orbital level,
$\xi_d^{(0)}=0$, is responsible for the logarithmic singularity in the
``ordinary'' Kondo effect. The rest of the terms in the sum correspond
to the virtual transitions onto (from) the partially occupied level
from (to) doubly-occupied (empty) electron levels. 

At energies $E\gg\Delta$ the discreteness of the spectrum is not
important.  If one neglects this discreteness all together the
correction to the exchange constant is ultraviolet divergent $\propto
\tilde{D}$.  
\begin{equation}
\delta J=J^2
\int_{|E|}^{\tilde{D}}\nu d\xi\int_{-\tilde{D}}^{\tilde{D}}
\frac{\nu_dd\xi_d}{|\xi|+|\xi_d|},
\quad
E\gg\Delta.
\label{JD}
\end{equation}
Here $\nu_d=1/\Delta$ is the one particle density of states in the
dot. 

This divergent correction is similar to the irreducible vertex
function in Fermi liquid theory and can be incorporated into a new
constant $J$ which can be numerically different from the bare
constant. It is this ``new'' constant $J$ through which all
observables at low energies are expressed.  This condition allows us
to exclude a large portion of the quasi-continuous spectrum of states
in the dot at energies $E\lesssim \tilde D$, and to relate the ``new''
constant $J(E, {\tilde D})$ in the fermion Hamiltonian, which is valid
at energies $E\ll\tilde D$, to the constant in the boson Hamiltonian
(\ref{Hpm}), (\ref{Hz}), which does not have free parameters. To
establish such a relation, one needs to consider a quantity that does
not exhibit strong mesoscopic fluctuations, and therefore can be
expressed in terms, insensitive to the particular values of the wave
functions $\varphi_n(0)$.

The observable quantity to look at, is the spin magnetization of the
dot caused by a magnetic field $h_z$ applied only to the
lead. Application of such a field results in a spin polarization, due
to the Pauli susceptibility, of electrons in the lead. This
polarization, via the contact interaction, induces a finite spin
polarization of the dot. Such a calculation can be performed both in
the bosonic and fermionic variables. In the boson variables, the
Hamiltonian causing the perturbation of electrons in the lead has the
form $\int_{-\infty}^0dx\, h_z\nabla\theta_s(x)$. This form, together
with the Hamiltonian of the free spin mode, Eq.~(\ref{Hs0}), allows us
to find the induced non-zero value of
$\langle\nabla\theta_s(0)\rangle\sim h_z/v_F$. This polarization,
in turn, creates a perturbation $[v_Fh_z/T_0({\cal
N})]\nabla\theta_s(+0)$ for the spin mode in the dot, as seen from the
Hamiltonian (\ref{Hz}). Finally, the spin polarization of the dot
$S\equiv\int_0^Ldx\nabla\theta_s\sim h_z/T_0({\cal N})$ can be
evaluated by considering the just mentioned perturbation together with 
the unperturbed Hamiltonian (\ref{Hs0}). The final result is
$L$-independent and thus preserves its meaning in the limit $\Delta\to
0$. In a Fermion representation, one finds $S\sim h_zJ\nu_d\nu$ for
the spin of the dot by a straightforward first-order perturbation
theory in the exchange Hamiltonian (\ref{exchange1}). Comparing the
two results for $S$, we find
\begin{equation}
J(E,{\tilde D})=J\equiv\left[\nu_d\nu T_0({\cal
N})\right]^{-1},\quad
\tilde{D}\gg E\gg\Delta.
\label{J0}
\end{equation}
With the help of Eq.~(\ref{J0}) we can relate the
constant $J(\tilde{D})$ in the Hamiltonian (\ref{exchange}) with
$T_0({\cal N})$.

After we incorporated all ultraviolet contributions into the
definition of the exchange constant, we are ready to study the
Kondo effect per se. Taking the difference of Eqs.\ (\ref{JE0})
and (\ref{JD}) to find the second order correction to the ``new''
coupling constant (\ref{J0}), we find that the correction
now only involves a
contribution
from the difference  of the true (discrete) spectrum of the dot
and the approximation where the density of states is constant 
(and continuous):
\begin{eqnarray}
J_{00}=J |\varphi_0(0)|^2
&&\left[1 +
J
\int_{|E|}^{\tilde{D}}\nu d\xi
\left(
\sum_{|\xi_d^{(l)}|<\tilde{D}}
\frac{|\varphi_l(0)|^2}{|\xi|+|\xi_d^{(l)}-\xi_d^{(0)}|}
\right.\right.\label{JE}\\
&&\left.\left.-
\int_{-\tilde{D}}^{\tilde{D}}
\frac{\nu_dd\xi_d}{|\xi|+|\xi_d|}
\right)
\right],
\nonumber
\end{eqnarray}
where $\nu_d=1/\Delta$ is the one particle density of states in the
dot. The right hand side of Eq.~(\ref{JE}) is now perfectly convergent
in the high energy. On the other hand, the contribution of the term with
$l=0$ gives a logarithmic divergence at $E\to 0$. This is precisely
the term which describes the Kondo effect, since it corresponds to a
spin flip of an electron on an upper level without excitations of
electron-hole pairs in the dot.

We obtain from Eq.~(\ref{JE})
\begin{eqnarray}
&J_{00}(E,{\tilde D})=
J|\varphi_0(0)|^2+J^2\nu\left(|\varphi_0(0)|^4\ln\frac{\Delta}{E}
+\Lambda\right),
\label{Kondo}\\
&\Lambda=\lim_{\tilde{D}\to\infty}\lim_{E\to 0}
\displaystyle\int_{|E|}^{\tilde{D}}\nu d\xi \left[ 
\sum_{|\xi_d^{(l)}|<\tilde{D},l\neq 0}
\frac{|\varphi_l(0)|^2}{|\xi|+|\xi_d^{(l)}|}-
\int_{-\tilde{D}}^{\tilde{D}}\frac{\nu_dd\xi_d}{|\xi|+|\xi_d|}\right].
\label{Lambda1}
\end{eqnarray}
The limit $E=0$ exists in Eq.~(\ref{Lambda1}), and the constant
$\Lambda$ is of the order of unity and fluctuates from sample to
sample. We will not consider these fluctuations, because
fluctuations of the wave function $\varphi_0(0)$ controlling the logarithmic
Kondo correction to $J_{00}(E,{\tilde D})$ lead to much
stronger effects. Equation~(\ref{Kondo}) demonstrates that in the strong
tunneling regime the bandwidth for the effective Kondo problem at hand
is $\Delta$ rather than $\Ec$. Once Eq.~(\ref{Kondo}) is established,
one can obtain, following the lines of \cite{Haldane79}, the result
for Kondo temperature 
\begin{equation}
T_K\simeq\Delta 
\sqrt{\frac{\Delta}{T_0({\cal N})}}\exp\left\{- \frac{T_0({\cal
      N})}{\alpha\Delta}\right\} 
\label{TKsa}
\label{tksa}
\end{equation}
with $T_0({\cal N})$ defined in
(\ref{T0ex}) and 
$\alpha=|\varphi_0(0)|^2/\langle|\varphi_0(0)|^2\rangle$. 
The parameter $\alpha$  is a random
quantity obeying Porter-Thomas statistics~\cite{PorterThomas},
see Eq.~(\ref{eq:3.2.17}).

The above consideration of the Kondo effect in a partially open dot
shows us that the low-temperature
behavior of the blockaded dot has no qualitative changes with respect to
the case of a dot weakly coupled to the leads, as long as
\[
\Ec|r|^2\gg \Delta.
\]
The only difference developing at larger conductances, is that
the parameters defining the low-temperature behavior differ from the
bare conductances of the junction. Comparison of the results for Kondo
temperature (\ref{eq:5.5}) and (\ref{tksa}) illustrates this point. In
both cases, $T_K$ is controlled by the value of $J\nu$,
\[
T_K \simeq \Delta (J\nu)^{1/2}e^{-1/J\nu}.
\]
In the case of strong Coulomb blockade, this parameter is simply
related to the conductance of the junctions and to the ratio
$\Ec/\Delta$ by second-order perturbation theory in tunneling
amplitudes, see Eq.~(\ref{eq:5.5}). In the case of strong charge
fluctuations, we needed a more sophisticated procedure of
renormalization in order to relate $J\nu$ to the bare parameters of
the quantum dot, see Eq.~(\ref{J0}). The two expressions for the Kondo
temperature $T_K$ match each other at reflection coefficient $|r|^2
\simeq 1/2$.

The Kondo effect in a single-junction system results in a specific
behavior of the spin polarization. If the dot is in a singlet state,
the gap for its spin polarization is $\sim\Delta$. In the doublet
state, the contribution of the dot to the susceptibility at low
temperature and fields, $T,\mu_BH\ll \Delta$, is identical to that of
a Kondo impurity \cite{Nozieres} with $T_K$ given by Eq.~(\ref{TKsa});
here $\mu_B$ is the Bohr magneton for the electrons of the dot.  The
manifestation of the most interesting effect, the enhanced
low-temperature conductance, requires a two-junction dot geometry, and
we turn to the corresponding discussion now.

Our first goal is to write a fermionic Hamiltonian describing the
system at energies so small that, due to the pinning of all charge and
spin modes, the Fermi-liquid behavior is already restored. Similar
to our discussion after Eq.~(\ref{exchange}), we obtain such a
Hamiltonian using solely the symmetry arguments, and relegate the
explicit model calculation to Appendix~\ref{Ap:model}. 

Once again, the interaction of the charge degrees of freedom of the
dot and the lead are suppressed, so we can restrict ourselves to the
spin-spin interaction only. The difference from the one-contact
geometry, is that now processes which involve two contacts, as well as
the ones involving a single contact, are possible. The most general
form of the Hamiltonian describing all such processes is
\begin{equation}
\hat{H}_{\rm ex}=\frac{1}{4}
\sum_{i,j}J_{ij}{\vec \sigma}_{\sigma_1\sigma_2}{\vec\sigma}_{\sigma_3\sigma_4}
\hat{\psi}^\dagger_{\sigma_1,i}({\vec
R}_i)\hat{\chi}^\dagger_{\sigma_3}({\vec R}_j)\hat{\chi}_{\sigma_4}({\vec R}_i)
\hat{\psi}_{\sigma_2,j}({\vec R}_j),
\label{ex}
\end{equation}
where indices $i,j=1,2$ label the contacts, $J_{ij}=J_{ji}$ are
coupling constants, and ${\vec R}_i$ describes the location of the
point contact.

The exchange interaction constants $J_{ij}$ should be found from the
calculation of a set of low-energy correlation functions. We can do
the explicit evaluation of the constants for the strongly-assymetric
set-up, $|r_1|\to 1$, $|r_2|\ll 1$. In this case, the largest exchange
constant $J_{22}$ is determined by the single-junction result,
$J_{22}=J$, where $J$ is defined by Eq.~(\ref{J0}) with $r_2=r$. The constant
$J_{12}$, which characterizes the non-local exchange accompanied by the
charge transfer, is proportional to $\sqrt{g_1}$, and the exchange
constant at the weaker junction is $J_{11}\propto g_1$, where $g_1\sim
1-|r_1|^2$ is the dimensionless conductance of the weaker junction. As
a matter of fact, in the limit of strong asymmetry the numerical
coefficients can be found also, with the help of the exact
solution~\cite{Furusaki95} available for the case $\Delta=0$. To find
$J_{12}$, we can calculate the linear conductance $G(T)$ through the
dot using the Hamiltonian (\ref{ex}), and then compare it
with the exact result~\cite{Furusaki95}. In the lowest order of the
perturbation theory, transitions across the dot are induced by only
the terms $\propto J_{12}$ in the Hamiltonian (\ref{ex}). The conductance
can be calculated by applying the standard Fermi golden rule to the
problem. Because we neglect the level spacing at the moment, and
because of the four-fermion structure of the Hamiltonian (\ref{ex}),
the electron transition rate and the conductance $G$ at low
temperatures is proportional to $T^2$,
\[
G(T)=\frac{\pi^3e^2}{4\hbar}J_{12}^2\nu^2\nu_d^2T^2.
\]
The comparison with the exact result~\cite{Furusaki95}, see also
Eq.~(\ref{eq:80.7}), yields:
\begin{equation}
J_{12}^2=\frac{32}{3\pi^2e^{\bf C}}\frac{\pi\hbar}{e^2}G_1
\left[\Ec T_0({\cal N})\nu^2\nu_d^2\right]^{-1} .
\label{JLR}
\end{equation}
The smallest exchange constant in the problem, $J_{11}$, does not
affect the conductance in the strongly-assymetric set-up which we
consider here. For calculating the conductance in the absence of
strong asymmetry, one should know the value of $J_{11}$. For that
purpose, some other response function should be calculated with the
help of Hamiltonian (\ref{ex}) and then compared with an exact result;
we will not address this problem in this Review.  Instead, we utilize
Eq.~(\ref{JLR}) to determine the conductance through the dot in the
limit of low temperatures.  If the gate voltage is close to an odd
integer, $\cos\pi{\cal N}<0$, the lowest discrete energy level in the
dot is spin-degenerate. At $T\lesssim\Delta$, only this level remains
important. This way, the initial problem of the dot, which has a dense
spectrum of discrete levels, and is strongly coupled to the leads, is
reduced to the problem of a
single-level Kondo impurity considered in Section~\ref{sec:Kondo}. 
We can adapt the result (\ref{eq:5.2}) to
find the conductance. First, we express the temperature-independent
factor in terms of the exchange constants,\footnote{The following
  formula is valid only for the single-channel Kondo model, {\it
    i.e.}, if the condition $J_{12}^2=J_{11}J_{22}$ holds. Within the
  $\sqrt{g_1}$--accuracy of our calculation this relation trivially
  holds. Checking it to higher order in $g_1$ requires
  calculation of $J_{11}$.} rather than in terms of the conductances
$g_1$ and $g_2$:
\[
\frac{4g_1 g_2}{(g_1+g_2)^2}\to\frac{4J_{12}^2}{(J_{11}+J_{22})^2}.
\]
Next, we neglect the small term $J_{11}$ in the denominator here, and
find:
\[
G_K\left(\frac{T}{T_K}, {\cal N}\right)
=\frac{4e^2}{\pi\hbar}\left(\frac{J_{12}}{J_{22}}\right)^2
f\left(\frac{T}{T_K}\right).
\]
Now we are ready to find the Kondo conductance through the dot in the
regime of strong charge fluctuations which we consider in this
Section. For that, we substitute the exchange constants (\ref{J0}) and
(\ref{JLR}) into the above formula,
\begin{equation}
G_K\simeq
\left(\frac{16}{3}\right)
\left(\frac{4}{\pi}\right)^3
\frac{1024}{3\pi^3}G_1|r_2|^2(\cos\pi{\cal N})^2
f\left(\frac{T}{T_K({\cal N})}\right).
\label{GKlimit}
\end{equation}
Here, $T_K({\cal N})$ is given by Eq.~(\ref{tksa}), 
and $f(x)$ is the universal scaling function plotted in Fig.~\ref{Fig20}.
Note that the Kondo conductance (\ref{GKlimit}) in a strongly asymmetric
set-up is proportional to the product of the small conductance $G_1$
and small reflection coefficient $|r_2|^2$ and therefore is
significantly smaller than the conductance quantum $e^2/\pi\hbar$
even at $T=0$. 

Similar to the case of the closed quantum dot, the dominant mechanisms
switch from inelastic co-tunneling at high temperatures to the
elastic tempera\-ture-independent mechanism at lower temperatures, and
finally to the regime of the Kondo effect at the lowest temperatures. We
postpone a more detailed discussion of those crossovers until after we
have constructed a theory of elastic co-tunneling in open quantum
dots.  We turn to the outline of this theory in the following
subsections.

\subsection{Simplified model for mesoscopic fluctuations}
\label{sec:open}
\label{sec:qualit}

As was already mentioned in the beginning of this section, the main
difficulty for building a theory of interacting electrons in an open
dot stems from the need to treat both the Coulomb
interaction and the dot-lead conductance non-perturbatively. In the
limit $\Delta\to 0$, the
bosonization procedure allows one to treat exactly the interaction in
a dot connected to a lead by a perfect quantum channel. Then,
the effect of a finite
reflection amplitude $r$ in the channel can be treated by a systematic
perturbation theory \cite{Matveev95,Furusaki95}. An important
conceptual conclusion drawn from this
theory is that in the low-energy
sector (defined by $\varepsilon \lesssim \Ec |r|^2$),
the excitations inside and outside the
dot are independent from each other. This understanding allowed us to
consider the low-temperature effects ($T\lesssim\Delta$) in a
partially open dot with $\Ec|r|^2\gg\Delta$, because in this case
one is able to map the properties of a partially-open dot onto
the ``conventional'' case of strong Coulomb blockade (see the previous
subsection). Such a mapping allows one to consider the Kondo effect,
but is insufficient for evaluating the co-tunneling
contribution to the conductance, for which a broad range of energies
up to $\Ec$ is involved. (This was shown, e.g., in Sec.\
\ref{sec:3.2}, where we considered the valley conductance through
a strongly blockaded quantum dot.)
The aim of this and the following Sections is to build a
theory capable of describing simultaneously many-body effects and
mesoscopic fluctuations at the intermediate range of energies, 
$\Ec\gtrsim \varepsilon \gtrsim{\mathrm max}(\Ec|r|^2,\Delta)$.
This section deals with a qualitative picture of the role of the
mesoscopic fluctuations; a rigorous calculation to verify the
qualitative picture below is given in the following sections.

Let us first consider a completely opened channel ($r=0$). In the
limit $\Delta\to 0$, the electron charge of the dot varies with the
gate voltage as $\langle Q\rangle=e{\cal N}$, to ensure the minimum of
the electrostatic energy \cite{Matveev95}.  The interaction term in
Eq.~(\ref{eq:2.54a}) depends only on the number of electrons crossing
the dot-channel boundary. Since this information is represented as
well in the asymptotic behavior of the wave-functions far from the
entrance to the dot, such as the scattering phase shift, the
properties of the ground state can be characterized by an analysis of
those asymptotes.  For the sake of simplicity, we consider the case of
a dot connected to a lead by one channel, and consider spinless
fermions instead of real electrons with spin. The ground state
properties at low energy can then be understood from the following
argument. 

The entrance of an additional electron with energy $\varepsilon$ (all the
energies will be measured from the Fermi level) into the dot requires
the charging energy $\Ec$.  Therefore, the electron may spend a time
of the order of $\hbar/\Ec$ in the dot, after which the extra charge
accumulated on the dot has to relax. One can distinguish two processes
that lead to charge relaxation: (i) elastic process where the same
electron leaves the dot, and (ii) inelastic process where another
electron is emitted from the dot. At low energies the probability of
the inelastic process is small as $(\varepsilon/\Ec)^2$, by virtue of the
smallness of the phase volume. (The last statement assumes Fermi
liquid behavior at low energies and, as we will see later, is valid
only for the spinless one-channel case.) Therefore, for $\varepsilon \ll
\Ec$, we may conclude that an incoming electron spends a
time of the order $1/\Ec$ inside the dot, after which it bounces back 
into the channel without
creating excitations. The same consideration is applicable to an
electron attempting to leave the dot. The fact that within the energy
range $|\varepsilon|\lesssim \Ec$ from the Fermi level the electron
processes are predominantly elastic implies that the low energy
properties of the system can be mapped onto a dot effectively
decoupled from the channel, that there is a well-defined scattering
amplitude from the entrance of the dot, and that the phase of this
scattering amplitude is given by the Friedel sum rule
\begin{equation}
\delta =\pi\langle Q\rangle/e=\pi{\cal N}.
\label{phaseshift}
\end{equation}
Equation~(\ref{phaseshift}) can be applied to electrons incident from
inside the dot, as well as to electrons incident from the channel.
The description outlined here resembles closely Nozieres'
Fermi liquid description of
the unitary limit in the one-channel Kondo problem \cite{Nozieres}.

We now demonstrate that the above considerations are sufficient
to reproduce the result
\begin{equation}
E_g({\cal N})\simeq |r| \Ec \cos 2\pi {\cal N},
\label{Matveev}
\end{equation}
obtained by Matveev~\cite{Matveev95} for the ground state energy
$E_g({\cal N})$ of spinless fermions in the limit of zero level
spacing $\Delta$ in the dot and with a finite reflection amplitude $r$
in the contact, and then apply the same
scheme to find the corrections to
the ground state energy arising from a finite $\Delta$, but without
$r$. Those corrections will result in the mesoscopic fluctuations of
the ground state energy.

We start with the limit $\Delta=0$. First, we put
$r=0$ and calculate the density of electrons in the channel
$\rho(x)$. 
As we discussed, the Coulomb interaction leads to a perfect
reflection of electron at low energies; the
wavefunctions have the form
$\psi_k(x) = \cos \left(k|x| - \delta \right)$, with phase shift
$\delta$ given by Eq.~(\ref{phaseshift}). This form of the
wavefunctions leads to the Friedel
oscillation of the electron density 
$$
  \rho(x)=\sum_{v_F|k-k_F|\lesssim
\Ec}\left|\psi_k(x)\right|^2,
$$
where $v_F$ and $k_F$ are the Fermi velocity and Fermi wavevector 
respectively. 
(Here we omitted the irrelevant constant part of the electron
density; the restriction of wave vectors $k$ to the range
$\Ec/v_F$ around $k_F$ is because only for those wavevectors
elastic reflection from the contacts takes place.) We thus obtain
\begin{equation}
\rho(x) =\left\{
\begin{array}{ll}
\displaystyle{\frac{\Ec}{v_F} \cos (2k_F|x|- 2\delta)}, \quad & |x| <
v_F/\Ec, \\ \displaystyle{\frac{ \sin (2k_F|x|-2\delta)
^{\vphantom{M^M}}}{|x|}}, \quad
&|x| > v_F/\Ec.\\
\end{array}
\right.
\label{pr2}
\end{equation}
Next, we take into account the effect of a scattering potential 
$V(x)$ in the contact that generates a finite reflection
amplitude $r\neq 0$. In the first order of perturbation theory,
the presence of the potential $V$ gives a shift to the
ground state energy given by
\begin{equation}
E_g({\cal N}) = \int d x\rho(x)V(x).
\label{pr1}
\end{equation}
Substituting Eq.~(\ref{pr2}) into Eq.~(\ref{pr1}), assuming that the
effective range of the potential around $x=0$ is smaller than $v_F/\Ec$, and
using the standard expression $|r| = |\tilde{ V}(2k_F)|/v_F$, we
obtain Eq.\ (\ref{Matveev}). Here $\tilde{V}(k)$ is the Fourier
transform of the potential $V(x)$.

\begin{figure}
\epsfxsize=9.5cm
\centerline{\epsfbox{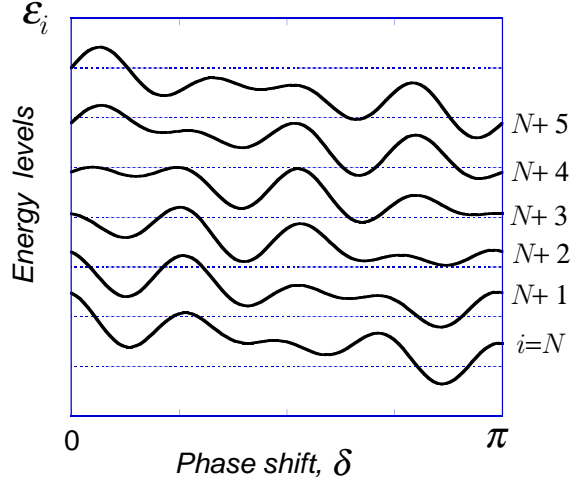}}
\vspace*{-4.4cm}
\caption{ Evolution of the energy levels of the quantum dot with the
scattering phase $\delta$. }
\label{Fig:2.1}
\label{fig:levels}
\end{figure}

Having verified the relation between scattering phase shifts and the
ground state energy for the case $\Delta=0$, we proceed with
evaluation of the ground state energy of a finite dot connected to a
reservoir by a perfect channel. According to the discussion preceding
Eq.\ (\ref{phaseshift}), the channel is effectively decoupled from the
dot due to the charging effect even though $r=0$. We now find the
ground state energy of the system by relating the ground state energy
of the closed dot to the scattering phase $\delta$ of
Eq.~(\ref{phaseshift}). For a chaotic dot, this problem is equivalent
to finding a variation of the eigenenergies by introduction of an
arbitrary scatterer~\cite{AleinerMatveev} with the same phase shift
$\delta$. The relevant contribution to the ground state energy is
given by
\begin{equation}
E_g = \sum_{-\Ec \lesssim \xi_i <0} \left[\xi_i(\delta) + \mu \right],
\label{E}
\end{equation}
where $\xi_i$ are the eigenenergies measured from the Fermi level
$\mu$.
(Again, we have restricted the summation to a window of size $\sim 
\Ec$ around the Fermi level, since the phase shift $\delta$ applies
only to particles in this energy range.)
As soon as the scattering phase changes by $\pi$, one more level
enters under the Fermi level.  Evolution of the energy levels with
changing $\delta$ is shown schematically in Fig.~\ref{Fig:2.1}. The
position of the level $\xi_i(\delta)$ satisfies the gluing condition
\begin{equation}
\xi_i(\delta +\pi) = \xi_{i+1}(\delta ).
\label{gluing}
\end{equation}
{}From Eqs.~(\ref{E}) and Eq.~(\ref{gluing}) we see that the ground
state energy depends almost periodically on $\delta$:
\begin{equation}
E_g(\delta) = E_g(\delta + \pi ) + {\cal O}(\Delta).
\label{periodic}
\end{equation}
As we will see below, the amplitude of the oscillation of the ground
state energy with the variation of $\delta$ is much larger than the
mean level spacing, so that we can neglect the last term in
Eq.~(\ref{periodic}).

In order to estimate the magnitude of the oscillations of the
ground state energy, we recall that
the correlation function of the level velocities is given
by \cite{Benreview}
\begin{equation}
\langle \partial _\delta \varepsilon_i \rangle =\frac{\Delta}{\pi}, \quad
\langle \partial _\delta \varepsilon_i \partial _\delta \varepsilon_j\rangle
=\delta_{ij}\frac{2}{\beta}\left(\frac{\Delta}{\pi}\right)^2,
\label{velocities}
\end{equation}
where $\beta = 1,2$ for the orthogonal and unitary ensembles
respectively, and $\langle\dots\rangle$ stands for the ensemble
averaging.  Formula (\ref{velocities}) can be easily understood from the
first order perturbation theory: At $\delta \ll 1$, we have
$\varepsilon_i (\delta) \approx \varepsilon_i(0) + (\delta/\pi \nu)
|\psi_i^2(0)| $. Using the fact that $\psi_i(0)$
is a Gaussian random
variable, see Eq.~(\ref{eq:3.2.10}), one obtains
Eq.~(\ref{velocities}).

Estimating the mesoscopic fluctuations of the ground state energy
(\ref{E}) with the help of Eq.~(\ref{velocities}), we obtain for
$\delta \ll 1$
\begin{equation}
\langle\left[E_g(\delta)-E_g(0)\right]^2\rangle =
\quad \Delta^2\left(\frac{\delta}{\pi}\right)^2  \sum_{-\Ec
\lesssim \xi_i,\xi_j <0}\frac{2}{\beta} \delta_{ij} \approx {\Delta
\Ec{\delta}^2 \over \beta}.  \label{Egcor} 
\end{equation}
As we have already explained, the energy (\ref{E}) is a periodic function
of $\delta$ with period $\pi $. On the other hand, for $\delta\lesssim
1$, Eq.~(\ref{Egcor}) is valid.  Therefore the characteristic
amplitude of the oscillations is of the order of $\sqrt{\Ec \Delta}$,
and it is plausible to assume that the correlation function of
energies at two different parameters ${\cal N}_1,\ {\cal N}_2$ takes
the form
\begin{equation}
\langle E_g\left({\cal N}_1\right) E_g\left({\cal N}_2\right)\rangle
\approx {\Delta \Ec \over \beta}
\cos 2\pi \left({\cal N}_1-{\cal N}_2\right),
\label{Egcor1}
\end{equation}
where we use Eq.~(\ref{phaseshift}). It is important to notice that
the variation of the energy of the ground state is much larger than
the mean level spacing $\Delta$.

A different, though equivalent way to arrive at Eq.~(\ref{Egcor1}) is
as follows. We have seen that because of the presence of the charging
energy $\Ec$, charge excitations in the dot relax elastically after a
time $\sim \hbar/\Ec$. If the dot size is infinitely big
(corresponding $\Delta = 0$), no backscattering from the dot's walls
occurs within this short time interval, and the only source of elastic
scattering is an eventual reflection amplitude $r$ in the contact.
However, if the dot has a finite size, there is a finite probability
that the incoming particle will scatter from its walls, or from impurities
inside the dot, and exit elastically before the time $\hbar/\Ec$. The
corresponding amplitude $r$ for reflection from the dot in a time
shorter than $\hbar/\Ec$ can be estimated as
\begin{equation}
  r \sim \int_0^{\hbar/\Ec} dt \Scat(t),
  \label{eq:rdot}
\end{equation}
where $\Scat(t)$ is the scattering matrix of the dot in time
representation, cf.\ Eq.~(\ref{eq:Stime}). Hence, in order
to include elastic scattering from the dot, we have to 
add the amplitude (\ref{eq:rdot}) to the reflection amplitude
$r_c$ from the contact in Eq.\ (\ref{Matveev}). The mesoscopic
fluctuations of $E_g$ can then be found from Eq.\ (\ref{eq:Savg})
for the mesoscopic fluctuations of $\Scat(t)$. One verifies that
Eq.\ (\ref{Egcor1}) is quickly recovered.
   
{}From the present derivation, it is clear that the charging energy
$\Ec$ enters into the final estimate (\ref{Egcor1}) as an energy
scale below which the dot is effectively closed. All higher lying energy
levels are not sensitive to the charging energy, and hence not
to the gate voltage ${\cal N}$. A similar picture
holds also for ``real'' electrons with spin. As we discussed in
Sec.~\ref{sec:finitesize}, see Eq.~(\ref{T0ex}), in
that case the dot effectively closes at the much lower
energy scale $\Ec |r|^2$. As a result, the amplitude of fluctuations for
spin-$1/2$ fermions acquires additional factor $r^2$, compared to
Eq.~(\ref{Egcor1}). Such an estimate is valid unless the
reflection coefficient $|r|^2$ becomes smaller than $\Delta/\Ec$. We
will see later that in the case of a perfectly transmitting channel
($r=0$), the energy scale $\Ec|r|^2$ should be replaced by $\simeq
\Delta \ln \left(\Ec/\Delta\right)$.

A magnetic field applied to the system affects the states in the
dot. Therefore, the ground state energy and, as we will see later,
conductance through the dot varies with the field. The corresponding
correlation magnetic field flux $\Phi_c$ is determined by the number
of levels that fit into the energy strip of one-particle levels
``blocked'' inside the dot. An estimate of $\Phi_c$ can be obtained
from Eq.~(\ref{eq:3.2.23}) with $E$ replaced by $\Ec$ or $\Ec|r|^2$
for spinless or spin-$1/2$ fermions, respectively.

\subsection{Rigorous theory of mesoscopic fluctuations: 
formalism for multichannel junctions}
\label{sec:rigorous}
\subsubsection{Derivation of the effective action}
\label{sec:action}
\addcontentsline{toc}{subsection}{
\protect\numberline{\ref{sec:action}}
{\ Derivation of the effective action}}

We now proceed with an exposition of the general formalism used to
calculate the capacitance, tunneling density of states, and
conductance of an interacting quantum dot, without neglecting the 
effect of mesoscopic fluctuations of the wavefunction in the dot. 
(That is,
without neglecting the possibility that an electron that has
entered the dot will return coherently to the point contact.) 
The first step is to reformulate the problem in terms of an 
effective action that accounts for these coherent returns.

Our formal starting point is the same as in Sec.\ \ref{sec:2.2}:
The Hamiltonian of the dot connected to the leads is written as
the sum of a Hamiltonian of the closed
dot, a Hamiltonian of the leads,
and a term $\hat{H}_{LD}$ coupling the lead and the dot,
\begin{equation}
\hat{H} = \hat{H}_D +  \hat{H}_{\rm C}+\hat{H}_L + \hat{H}_{LD}.
\label{eq:30.00}
\end{equation} 
Here, the Hamiltonian of the dot is, in turn, written as a sum of
the Hamiltonian $\hat{H}_D$
of the closed dot in the absence of electron-electron interactions
and the charging energy $\hat{H}_C$ 
[see also Eq.~(\ref{eq:2.54a}); here spin is included in the
indices $\alpha$ and $\gamma$]
\begin{equation}
\hat{H}_D= \sum_{\alpha, \gamma} {\cal H}_{\alpha\gamma}
\hat{\psi}^\dagger_{\alpha} \hat{\psi}_{\gamma},
\quad \hat{H}_{\rm C}=\Ec\left(\hat{n}-{\cal N}\right)^2.
\label{eq:HdHc}
\end{equation}
The Hamiltonian $\hat{H}_L$ of the leads is given by
Eq.~(\ref{eq:2.57}); the term
$\hat{H}_{LD}$ that couples the dot with the leads is given in
Eq.~(\ref{eq:2.58}).

The operator $\hat n$ in Eq.\ (\ref{eq:HdHc}) is for the total charge
$e \hat n$ on the dot. Conventionally, it is defined in terms of the
creation and annihilation operators $\hat{\psi}^{\dagger}_{\alpha}$
and $\hat{\psi}_{\alpha}$ of fermions in the dot, as in Eq.\
(\ref{eq:2.151}). For the purpose of the formal manipulations of this
section, it is more convenient, however, to change the definition
(\ref{eq:2.151}) of $\hat n$, and reformulate it in terms of the
operators $\hat{\psi}^{\dagger}_{j}(k)$ and $\hat{\psi}_{j}(k)$ for
the channel. This is possible, because the total number of particles in the
entire system (leads and dot) is an integer number which can be added
to the parameter ${\cal N}$ at no cost, so that we can write
\begin{equation}
  \hat{n} = - \sum_{j=1}^{\Nch} \int {dk \over 2 \pi} 
  \hat\psi^\dagger_{j}(k) \hat\psi_{j}(k).
\label{Q1}
\label{eq:30.40}
\end{equation}
Strictly speaking, the use of Eq.~(\ref{eq:30.40}) instead of
(\ref{eq:2.151}) is only permitted for a canonical description, 
in which the total number of particles in the system is held fixed.
Below, however, we use Eq.~(\ref{eq:30.40}) in a grand canonical
description. The difference between the two descriptions is only 
important if non-periodic fluctuations as a function of ${\cal N}$
are considered. The modifications of the theory in this
case are discussed in Appendix \ref{Ap:3}.

We now discuss the derivation of the effective action separately
for the three cases of interest.

{\em Free energy and differential capacitance.}
To calculate the ground state energy, we start with the thermodynamic
potential,
\begin{equation}
\Omega=-T\ln\left({\rm Tr}e^{-\hat{H}/T}\right).
\label{eq:30.5}
\label{Omega}
\end{equation}

We evaluate the trace in two steps, ${\rm Tr}\dots= {\rm Tr}_{L}{\rm
Tr}_D \dots$, where $L$ and $D$ indicate the fermionic operators
belonging to the leads and dot respectively.  
With the redefinition (\ref{eq:30.40}) of the charge $\hat n$, the
interaction Hamiltonian $H_C$ is now attributed to the leads. As a
result, the Hamiltonian of the system becomes
quadratic in the fermionic operators of the dot, so that this part 
of the system can be integrated out:
\begin{eqnarray}
  {\rm Tr}_D\, e^{-\hat{H}/T} &=& {\rm Tr}_D\,
e^{-\left(\hat{H}_L+\hat{H}_{\rm C} + \hat{H}_D + \hat{H}_{LD}\right)/T}
\label{eq:30.41}
\label{eq:Tr12}
\\
&=&\ \ 
e^{-\left(\hat{H}_L+\hat{H}_{\rm C} \right)/T}e^{-\Omega_D/T}
T_\tau e^{\frac{1}{2}\int_0^{1/T} d\tau_1d\tau_2 
\langle\hat{H}_{LD}(\tau_1)\hat{H}_{LD}(\tau_2) \rangle_D}.
\nonumber
\end{eqnarray}
Here $\hat{H}_{LD}(\tau)$ is the interaction representation of the
dot-lead coupling operator 
\[
\hat{H}_{LD}(\tau)= 
e^{\tau(\hat{H}_L+\hat{H}_{\rm C} + \hat{H}_D)}
\hat{H}_{LD}
e^{-\tau(\hat{H}_L+\hat{H}_{\rm C} + \hat{H}_D)},
\] 
$T_{\tau}$ denotes time ordering for the imaginary time $\tau$,
$\Omega_D$ is the thermodynamic potential of the closed dot,
\begin{equation}
\Omega_D =
-T\ln {\rm Tr} e^{- \hat{H}_D/T},
\label{eq:30.400}
\end{equation}
and the average $\langle \ldots \rangle_{D}$
over the Hamiltonian $H_{D}$ of the dot is defined
as 
\[
\langle \dots\rangle_D = e^{\Omega_D/T}{\rm Tr}_D\left(
e^{-\hat{H}_D/T}\dots\right). 
\]
Note that, as the interaction has been shifted to the leads,
the thermodynamic potential $\Omega_D$ of the closed dot
is formally independent of the gate voltage ${\cal N}$.

It remains to compute the average 
$  \langle\hat{H}_{LD}(\tau_1)\hat{H}_{LD}(\tau_2)\rangle_D$
in Eq.\ (\ref{eq:30.41}). 
Hereto we use the explicit form (\ref{eq:2.58}) of ${\hat H}_{LD}$
and the Green function ${\cal G}_{\alpha \beta}(\tau)$ of the
closed dot,
\begin{equation}
  {\cal G}_{\alpha \beta}(\tau) \equiv
  - \langle T_{\tau}\, \hat{\psi}_{\alpha}(\tau) \hat{\bar{\psi}}_{\beta}(0)
  \rangle_{D} = T \sum_{n=-\infty}^{\infty} e^{-i \omega_n \tau} \left(
  {1 \over i \omega_n - {\hat H}_D} \right)_{\alpha \beta},
\label{eq:dotGF}
\end{equation}
where $\omega_n = \pi T (2n + 1)$ is a fermionic Matsubara frequency
and $\hat{\bar{\psi}}(\tau) = \hat{\psi}^{\dagger}(-\tau)$. Hence we find
\begin{eqnarray}
  \langle\hat{H}_{LD}(\tau_1)\hat{H}_{LD}(\tau_2)\rangle_D &=& -2
  \sum_{j_1,j_2} \hat{\bar{\psi}}_{j_1} (\tau_1,0) \left[ W^{\dagger}
  {\cal G}(\tau_1-\tau_2) W \right]_{j_1j_2} \hat{\psi}_{j_2}(\tau_2,0),
  \nonumber \\ \label{eq:HH}
\end{eqnarray}
where the operators $\hat{\psi}_j(\tau,x)$ are the Fourier transform
of $\hat{\psi}_j(\tau,k)$,
$$
  \hat{\psi}_j(x) =
  \int {dk \over 2 \pi} e^{-i k x} \hat{\psi}_{j}(k).
$$

It is convenient to introduce a new matrix, $L_{j_1j_2}$, by the
following relation:
\begin{equation}
4 L(i \omega_n)_{j_1j_2}\equiv 
-{\mbox{sign}\, \omega_n \over i \pi\nu}\delta_{{j_1j_2}}
+\left[ W^{\dagger} {\cal G}(i\omega_n) W \right]_{j_1j_2}, 
\label{eq:Ldef}
\end{equation}
where $\nu = 1/2 \pi v_F$ is the density of states in the leads. 
With this definition, the dot Green function ${\cal G}$ and the
coupling matrix $W$ can be eliminated from Eq.\ (\ref{eq:HH}) in
favor of the matrix $L$, so that we obtain
\begin{eqnarray}
  \langle\hat{H}_{LD}(\tau_1)\hat{H}_{LD}(\tau_2)\rangle_D &=&
  {2 T \over \pi \nu \sin[\pi T (\tau_1 - \tau_2)]}
  \sum_{j}
  \hat{\bar{\psi}}_{j} (\tau_1,0) \hat{\psi}_{j}(\tau_2,0)
  \nonumber \\ && \mbox{}
  -{1 \over 2}  \sum_{j_1,j_2} \hat{\bar{\psi}}_{j_1} (\tau_1,0) 
  L_{j_1 j_2}(\tau_1 - \tau_2) \hat{\psi}_{j_2}(\tau_2,0).
  \label{eq:Ldef2}
\end{eqnarray}
The
first term in the r.h.s.\ comes from the diagonal part of the Green
function ${\cal G}$ ({\em i.e.}, the density of states in the dot); 
this term can be obtained by multiplying the
ensemble-averaged Green function by the product $W^{\dagger} W = \pi^2
\nu/M \Delta$ characterizing an ideal dot-lead
coupling. The second term, which contains the matrix
$L_{j_1,j_2}$, also has components off-diagonal in $j_1$ and $j_2$. It
accounts for a combined effect of electron trajectories backscattered
into the lead. Backscattering may be caused by electron reflection
off a barrier in the contact, impurities in the contact and the
dot, and the boundaries of the dot. This term is responsible for the
mesoscopic fluctuations of the observable quantities.

Except for the average
density of states, the first term in Eq.\ (\ref{eq:Ldef2}) 
does not contain information about
the dot. We could have obtained the very same term if the coupling to
the dot would have been replaced by an ideal coupling of each chiral
fermion field $\hat{\psi}_j(x)$ in the lead to a free chiral fermion field
$\hat{b}_j(x)$.  Writing these fields explicitly, we find that
Eq.~(\ref{eq:Tr12}) takes the form
\begin{equation}
  \mbox{Tr}_{D} e^{- \hat{H}/T} \propto
   \mbox{Tr}_{b}\,
  e^{- \Omega_{D}/T}e^{- \hat{H}_{\rm eff}/T } T_\tau e^{-\Action}
\label{eq:30.480}
\end{equation}
where $\mbox{Tr}_{b}$ denotes a trace over the chiral fermion fields
$\hat{b}_j$, $\hat{H}_{\rm eff}$ is an effective Hamiltonian that contains the 
fermion fields $\hat{\psi}_j(x)$ in the lead, the fictitious fermion fields 
$\hat{b}_j(x)$, and their coupling at $x=0$,
\begin{eqnarray}
  \hat{H}_{\rm eff} &=&
  i v_F \sum_{j} \int {dx} \,  
  \left[\hat{\psi}^\dagger_{j}(x) \partial_x
  \hat{\psi}_{j}(x) + \hat{b}^\dagger_{j}(x) \partial_x 
  \hat{b}_{j}(x)\right]
  \nonumber \\ && \mbox{}
  + \frac{1}{\pi\nu} \sum_j
  \left[ \hat{b}_j^{\dagger}(0) \psi_j^{\vphantom{X}}(0) +
  \psi_j^{\dagger}(0) \hat{b}_j^{\vphantom{X}}(0) \right]
  \nonumber \\ && \mbox{} 
  + \Ec 
  \left(\sum_{j} \int dx :\! 
  \hat{\psi}^\dagger_{j}(x) \hat{\psi}_{j(x)} \! : +{\cal
  N}\right)^2,
\label{eq:30.481}
\end{eqnarray}
and $\Action$ is an effective action that contains the
fields $\hat{\psi}_{j}(x)$ at $x=0$ only,
\begin{equation}
\Action=4 \sum_{j_1,j_2} \int_0^{1/T} d\tau_1 \int_0^{1/T} d\tau_2 
  \hat{\bar{\psi}}_{j_1}(\tau_1;0) L_{j_1j_2}(\tau_1-\tau_2)
  \hat{\psi}_{j_2}(\tau_2;0).
\label{eq:30.48}
\end{equation}
The equivalence of the representations (\ref{eq:30.41}) and
(\ref{eq:30.480}) -- (\ref{eq:30.48}) can be easily checked by tracing
out the fermions $\hat{b}_j$ in Eq.~(\ref{eq:30.480}), with the help of the
relation [cf.\ the first term on the r.h.s.\ of Eq.\ (\ref{eq:Ldef})]
$$
  \langle T_\tau 
  \hat{b}_{j_1}(\tau;0)\hat{\bar{b}}_{j_2}(0;0)\rangle
  =i\pi\nu T \delta_{j_1j_2}
  \sum_{\omega_n}e^{-i\omega_n\tau}{\rm sgn}\, \omega_n.
$$
The operators $\hat{\psi}_j(0)$ and $\hat{b}_{j}(0)$
appearing in the effective Hamiltonian (\ref{eq:30.481})
and in the effective action (\ref{eq:30.48}) are the 
fermion fields evaluated at
the origin $x=0$. {}To simplify notation, from now on 
we omit the coordinate in the 
argument of the one-dimensional operators $\psi_j$ (or, later, of 
the bosonic operators $\phi_j$), when they are taken at the origin,
\begin{equation}
{\psi}_{j}(\tau) \equiv {\psi}_{j}(\tau; x=0), \quad
{\phi}_{j}(\tau) \equiv {\phi}_{j}(\tau; x=0).
\label{eq:convention}
\end{equation}
It can be checked that the operator in the second line of
Eq.~(\ref{eq:30.481})
connecting the fermion fields $\hat{\psi}$ and $\hat{b}$ at $x=0$
corresponds to an ideal coupling between those fields.  This coupling
means that a $\psi$-particle is scattered to a $b$-particle with unit
probability. This motivates the following substitution of right and
left moving fermion fields,
\begin{eqnarray}
  \hat{\psi}_j(x) &=& \hat{\psi}_{L,j}(x)\theta(-x) + 
  \hat{\psi}_{R,j}(-x)\theta(x),
  \nonumber \\ 
  \hat{b}_j(x) &=& 
  i\left[ \hat{\psi}_{R,j}(-x)\theta(-x) - 
  \hat{\psi}_{L,j}(x)\theta(x)\right].
\end{eqnarray}
[Here $\theta(x) = 1$ if $x > 0$, $0$ if $x < 0$, and $\theta(0) = 1/2$.]
With this substitution we obtain from Eqs.~(\ref{eq:30.481}) -- (\ref{eq:30.48})
\begin{eqnarray}
  \hat{H}_{\rm eff} &=& 
  i v_F \sum_{j}\int_{-\infty}^{\infty}dx
  \left(\hat{\psi}^\dagger_{L,j}\partial_x\hat{\psi}_{L,j} -
  \hat{\psi}^\dagger_{R,j}\partial_x \hat{\psi}_{R,j}\right) 
  \nonumber\\ && 
  \mbox{} + {\Ec} \left(\sum_{j}\int_{-\infty}^{0} dx
  :\! \hat{\psi}^\dagger_{L,j}\hat{\psi}_{L,j} + 
  \hat{\psi}^\dagger_{R,j}\hat{\psi}_{R,j}\! : 
  + {\cal N} \right)^2,  \label{eq:30.482} \\
  \Action &=& 
  \sum_{j_1j_2}\int_0^{1/T} d\tau_1 \int_0^{1/T} d\tau_2\,
  L_{j_2j_2}\left(\tau_1-\tau_2\right) 
  \nonumber \\ && \mbox{} \times
  \left[\hat{\bar{\psi}}_{L,j_1}(\tau_1) + \hat{\bar{\psi}}_{R,j_1}(\tau_1)\right]
  \left[\vphantom{\hat{\bar{\psi}}_{L,j_1}}
  \hat{\psi}_{L,j_2}(\tau_2) + \hat{\psi}_{R,j_2}(\tau_2)\right],
  \label{eq:action}
\end{eqnarray}
where the time dependence of any operator $\hat{A}$ is defined
as
\begin{equation}
\hat{A}(\tau)= e^{\tau \hat{H}_{\rm eff}}
\hat{A}
e^{-\tau \hat{H}_{\rm eff}}.
\label{eq:intrep}
\end{equation}

In this formulation, all backscattering from the dot and from the
contact is characterized by the kernel $\Lkern$. 
To make contact with the theory of Sec.\ \ref{sec:2.2}
of a non-interacting quantum dot, it is advantageous to express $L$
in terms of the scattering matrix
$S$ of the non-interacting system. Comparing Eqs.~(\ref{eq:2.601})
with  Eq.~(\ref{eq:Ldef}), one immediately finds
\begin{equation}
 \Lkern(\omega_n) = \int_0^{1/T} e^{i \omega_n \tau} L(\tau) d\tau =
\left\{ \matrix{ \displaystyle
  -{1 \over 2 \pi i \nu} \frac{\Scat(i\omega_n)}{1
+\Scat(i\omega_n)_{\vphantom{M}}}, & \omega_n >0,\cr 
  \displaystyle {1 \over 2 \pi i \nu}
\frac{\Scat^\dagger(i\omega_n)^{\vphantom{M^M}}}{1 +\Scat^\dagger(i\omega_n)}, &
\omega_n < 0.  } \right.
\label{eq:LS}
\label{eq:30.49}
\end{equation}
This relation allows us to deduce the statistical distribution of the
kernel $L(\tau)$ from the better known statistical distribution of the
scattering matrix $S$, see Sec.\
\ref{sec:2.2}. For ideal contacts the ensemble average of $S^n$
(with integer $n$) is zero, cf.\ Eq.\ (\ref{eq:60.7}),
which implies that the ensemble average of
$L$ is zero as well.

Equations (\ref{eq:30.482}) and (\ref{eq:action}) provide an
effective description of the
combined system of the lead and the dots in terms of fermions
moving in one-dimensional wires, where the interaction only
exists on one end of the wire. The effective action $\Action$ provides
a coupling between the wires and between left and right moving
fermions.
The one dimensional description for the charging effect,
which
was already introduced in Sec.\ \ref{sec:lowen}, was first
proposed by Flensberg \cite{Flensberg} and Matveev \cite{Matveev95}. 
Their theory can be interpreted as an approximation of the 
effective action theory presented here, corresponding to the 
replacement of $S$ (and $L$) by its ensemble average,
\begin{equation}
\Scat_{j_1j_2} \to \langle \Scat_{j_1j_2} \rangle = -ir_{j_1}\delta_{j_1j_2},
\label{eq:30.490}
\end{equation}
where $r_j$ is the backscattering amplitude in the contact
for the $j$-th channel, see Sec.\ \ref{sec:2.2}. At
weak backscatterring, $r_i \ll 1$, Eq.~(\ref{eq:LS}) gives,
cf.\ Eq.\ (\ref{eq:Sdirect26}),
\begin{equation}
L_{ij}(\tau)=v_F\delta(\tau)\delta_{ij}r_j,
\label{eq:30.491}
\end{equation}
so that the action loses its time dependence and acquires the form 
of a backscattering Hamiltonian. The approximation (\ref{eq:30.490}) 
loses all the information about the trajectories where the electron
returns to the entrance of the dot,after having been reflected by 
the dot's walls or by impurities inside the
dot. Such returns are the cause of mesoscopic fluctuations of
the capacitance and other 
observable quantities. Mesoscopic fluctuations are present in
the full action (\ref{eq:action}), which is non-local in time.
This  action was first derived in
Ref.~\cite{AleinerGlazman98}.

{}From Eqs.~(\ref{Omega}) and (\ref{eq:30.480}), we find the expression
for the thermodynamic potential of the system
\begin{equation}
\Omega = \Omega_D - T\ln \mbox{Tr}
\left\{
e^{- \hat{H}_{\rm eff}/T } T_\tau e^{-\Action}
\right\}
\label{eq:30.50}
\end{equation}
where $\Omega_D$ is defined by Eq.~(\ref{eq:30.400}), and the
effective Hamiltonian $\hat{H}_{\rm eff}$ and the
action ${\cal S}_{\rm eff}$ are given by Eq.~(\ref{eq:30.482}).

Equation (\ref{eq:30.50}) allows the calculation of all
thermodynamic properties of the system within the framework of the
effective action. Such quantities involve, among others, 
the specific heat and
the differential capacitance,
\begin{equation}
C_{\mathrm diff}\left({\cal N}\right)= C\left(1 - 
\frac{1}{2 \Ec}\frac{\partial^2\Omega}{\partial{\cal N}^2}\right).
\label{Cdif}
\label{eq:30.51}
\end{equation}
We will perform the explicit calculation of this quantity in the next
subsubsection.\footnote{In Eq.~(\ref{eq:30.50}) the thermodynamic
potential of the dot $\Omega_D$ does not depend on gate voltage ${\cal
N}$  and, thus, does not contribute to the differential
capacitance. More careful analysis, see Appendix~\ref{Ap:3}, reveals
the dependence, which gives a contribution small as $\Delta/\Ec$ to
$C_{\mathrm diff}$. 
We relegate the corresponding discussion to the end of Sec.~\ref{sec:cap}
and Appendix~\ref{Ap:3}.} 

It is not difficult to extend the effective action approach to the
calculation of kinetic quantities, such as the two-terminal
conductance and the tunneling densities of states. 
We will sketch these extensions below.
The corresponding expressions at $\Ec = 0$ were discussed in
Sec.~\ref{sec:2.2}.

{\em Two-terminal conductance, the general formulation.} We will use
the usual Kubo formula to express the two-terminal conductance $G$ in
terms of the Matsubara correlation function
\begin{eqnarray}
 && G = \lim_{\omega \to 0}
\frac{
\Pi(i \Omega_n \to \omega + i0) - \Pi(i \Omega_n \to \omega -
i0)}
{2i\omega}
, \nonumber \\
&& \Pi(i\Omega_n) = 
\int_0^{1/T} d\tau 
\Pi(\tau)e^{i \Omega_n \tau}, \quad
\Pi(\tau)=
\langle T_{\tau} \hat{I}(\tau) \hat{I}(0) \rangle_q,
\label{eq:currentcorrelatorAG} 
\label{eq:currentcorrelator}
\label{eq:30.52}
\end{eqnarray}
where $\Omega_n = 2 \pi n T$ is the bosonic Matsubara frequency,
and 
\begin{equation} 
  \hat{I}(\tau) = \exp(\hat{H} \tau) \hat{I} \exp(- \hat{H} \tau)
\label{eq:30.53}
\end{equation}
is the  operator of the electric current in the Matsubara representation.
Because the total current through any cross-section of 
the system is conserved we can define the current operator in the
leads. For future technical convenience, we define the current $I$
through the dot as a weighed difference of the currents $I_1$ and
$I_2$ entering the dot through the two leads $1$ and $2$, 
\begin{equation}
  \hat{I} = {1 \over N_1+N_2}(N_2 \hat{I}_1 - N_1 \hat{I}_2),
\label{eq:30.54}
\end{equation}
where $\NL$ ($\NR$) is the number of channels in the corresponding
lead and
\begin{equation}
  I_{\alpha} =  e v_F \left. \sum_{j \in \alpha} \left(
        \psi_{L j}^{\dagger} (x)\psi_{L j}^{\vphantom{dagger}} (x)- 
        \psi_{R j}^{\dagger} (x)\psi_{R j}^{\vphantom{dagger}} (x)
        \right)
        \right|_{x \to 0}, \quad \alpha =1,2
\label{eq:30.55}
\end{equation}
is the operator of the current in lead $\alpha$ ($\alpha=1,2$).

The fact that the current operator
$\hat{I}$ is defined inside the leads allows us to copy the
derivation of the effective action from the 
thermodynamic potential: we shift the interaction from the 
quantum dot to the lead and
then take a partial trace over all dot states. 
As before, the result is
represented by means of the same effective action 
for the one-dimensional fermions $\hat{\psi}_L$, $\hat{\psi}_R$.
We thus find
\begin{equation}
  \Pi(\tau) = 
\frac{{\mathrm Tr}
\left[
e^{-\hat{H}_{\rm eff}/T} 
T_{\tau}  \hat{I}(\tau)
\hat{I}(0) e^{-\Action} \right]
}
{{\mathrm Tr}
\left[
e^{-\hat{H}_{\rm eff}/T} 
T_{\tau}
e^{-\Action}
\right]
}, 
\label{eq:Ieffective} 
\label{eq:30.56}
\end{equation}
where the effective Hamiltonian and action are defined in
Eq.~(\ref{eq:30.482}), and the time dependence of the operator 
$\hat{I}$ is given by Eq.~(\ref{eq:intrep}).

\begin{figure}
\epsfxsize=6.7cm
\hspace*{0.5cm}
\epsfbox{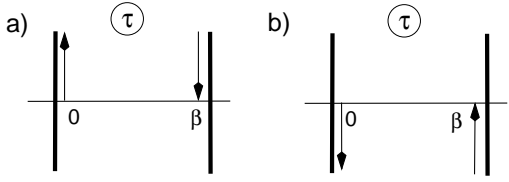}
\vspace{0.3 cm}
\caption{The integration contour used in the evaluation of
the conductance, see Eq.~(\protect\ref{eq:30.52}), for (a) $\Omega_n > 0$,
and (b) for $\Omega_n < 0$. Branch cuts of the analytic continuation
of $\Pi (\tau )$ are shown by  thick lines.}
\label{Fig:6.10}
\end{figure} 

Equations (\ref{eq:30.52}), (\ref{eq:30.56}) and (\ref{eq:30.482})
constitute the complete formulation of the two-terminal conductance 
problem within the effective action theory. The practical execution of the
analytic continuation in Eq.~(\ref{eq:30.52}) is more easily achieved
in the time domain. 
As we will see below, the function $\Pi(\tau)$ can be analytically continued
from the real axis to the complex plane, so that the result is analytic in a
strip $0 < {\rm Re }\,\tau < 1/T$, and has branch cuts along the lines
${\rm Re}\,\tau=0,1/T$. It allows one to deform the contour of
integration as shown in Fig.~\ref{Fig:6.10}, thus obtaining
\[
\Pi(i\Omega_n)=i
\int_{-\infty}^\infty \!dt e^{-\Omega_nt}
\left[\theta (\Omega_n) \theta(t)-
\theta (- \Omega_n) \theta(-t)  
\right]
\left[{\Pi (it+0) }
- {\Pi (it-0) }
\right].
\]

Now, the analytic continuation (\ref{eq:30.52}) can be performed, with
the result
\begin{equation}
G=\frac{i}{2}
\int_{-\infty}^\infty \!dt t\left[{\Pi (it+0) }
- {\Pi (it-0) }
\right].
\label{eq:30.57}
\end{equation}
Next, we  use the analyticity of $\Pi(\tau)$ in the strip $0<{\rm
Re}\,\tau <1/T$, and shift the integration variable $t \to t-i/2T$
in the first term in brackets in Eq.~(\ref{eq:30.57}), and $t \to
t+i/2T$ in the second term. Bearing in mind that
$\Pi(\tau)=\Pi(\tau +1/T )$, we find
\begin{equation}
G=\frac{1}{2T}
\int_{-\infty}^\infty dt{\Pi (it+1/2T) }.
\label{eq:30.58}
\end{equation}
This formula is the most convenient for practical calculations.

\begin{figure}
\vspace{0.2cm}
\hspace{0.3\hsize}
\epsfxsize=0.4\hsize
\epsfbox{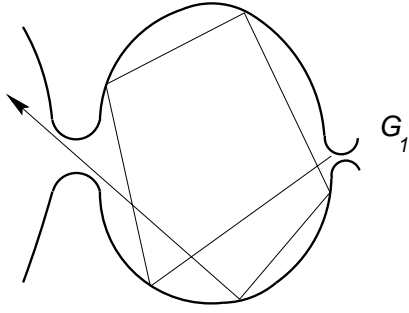}
\vspace{0.6cm}
\caption{Schematic view of the asymmetric two-terminal setup. The left
point contact has one channel almost open and the conductance of the
right point contact $G_1$ is much smaller than $e^2/2\pi\hbar$.
One of the electron trajectories contributing to elastic co-tunneling is 
also shown.
}
\label{Fig:6.1}
\end{figure}

{\em Tunneling density of states} -- In the case of a very asymmetric
setup, [$G_1\ll e^2/\pi\hbar$ and arbitrary $G_2$], the general
formula for two-terminal conductance (\ref{eq:30.58}) can be further
simplified. One can expand the scattering matrix in terms of small
transmission amplitude corresponding to $G_1$, like it was done for
the case of non-interacting electrons, see Section~\ref{sec:2.2},
Eqs.~(\ref{eq:2.65})--(\ref{eq:2.68}). This way the problem of
calculation of the two-terminal conductance is reduced to the one of
evaluation of the tunneling density of states for a dot strongly
coupled to one lead, see Figure~\ref{Fig:6.1}.

For the non-interacting system, the tunneling density of
states is defined in Eq.~(\ref{eq:2.66}). Because the transmission
coefficient $|t_1|^2 \ll 1$, we can consider the tunneling current $I$
as the function of applied voltage $V$ in second order 
perturbation theory in the tunneling Hamiltonian [first term in
Eq.~(\ref{eq:2.65})]. This gives us the standard result \cite{Mahan}
\begin{equation}
I(eV) =i\left[  
J\left(i\Omega_n\to eV+i0\right)-
J\left(i\Omega_n\to eV-i0\right) \right],
\label{eq:40.1} 
\label{eq:6.4}
\end{equation}
where $\Omega_n=2\pi Tn$ is the bosonic Matsubara frequency, and
the Matsubara current $J$ is defined as
\begin{equation}
J(i\Omega_n)={M\Delta}\frac{G_1}{2\pi e} 
\int_0^{1/T} d\tau e^{-i\Omega_n\tau} 
\Pi_{\rm tun} (\tau)\frac{\pi T}{\sin \pi T\tau},
\label{eq:40.2}
\label{eq:6.5}
\end{equation}
where the last factor corresponds to the one-particle Green function
of the electrons in the right lead (see Fig.\ \ref{Fig:6.1}; the
right lead is not affected by interactions or
mesoscopic fluctuations), and $\Pi_{\rm tun}$ corresponds to the
one-electron Green function of an electron in the dot at the point of
the tunnel junction, 
\begin{equation}
\Pi_{\rm tun}(\tau) = 
\langle T_\tau \hat{\bar{\psi}}_1\left(\tau\right)
\hat{\psi}_1(0)
\rangle_q. 
\label{eq:6.7}
\label{eq:40.3}
\end{equation}
Here we wrote $\hat{\bar{\psi}}_1$ and $\psi_1$ for the fermion creation
and annihilation operators in the dot at
the location of the tunnel junction, in keeping with the RMT
formulation of Sec.\ \ref{sec:2.2}, see Eq.~(\ref{eq:2.65}).

Analytic continuation in Eq.~(\ref{eq:40.1}) is performed 
similarly to the derivation of Eq.~(\ref{eq:30.58}), and one obtains
\begin{equation}
I=\left(T \sinh \frac{eV}{2T}\right) M\Delta 
G_1 \int_{-\infty}^\infty dt
e^{-ieVt}
\frac{\Pi_{\rm tun} \left(it+\frac{1}{2T}\right)}{\cosh \pi T t}.
\label{eq:40.4}
\label{eq:6.102}
\end{equation}
Here $\Pi_{\rm tun}(\tau)$ must be analytic in the strip $0<{\rm
Re}\,\tau <1/T$.
The linear conductance $G$ is therefore given by
\begin{equation}
G= G_1 M\Delta \int_{-\infty}^\infty dt
\frac{\Pi_{\rm tun} \left(it+\frac{1}{2T}\right)}{2\cosh \pi T t}.
\label{eq:6.103}
\label{eq:40.5}
\end{equation}
One can check that at $\Ec = 0$, Eqs.\ (\ref{eq:40.5}) and
(\ref{eq:40.3}) are equivalent to Eqs.~(\ref{eq:2.66}) and
(\ref{eq:2.68}). The Green function (\ref{eq:40.3}) is dramatically
affected by the interactions, and we would like to construct an
effective action theory to describe these effects.

Once again, we wish to get rid of the
fermionic degrees of freedom of the dot. Similar to Eq.~(\ref{Q1}),
it is convenient to rewrite the charge operator $e \hat n$
in terms of the variables
of the channel. However, here we have to keep in mind that the
tunneling events described by operators ${\psi}^\dagger_1$ and 
${\psi}_1$ change the charge in the dot by an amount
$+e$ and $-e$. This is not taken into account in the simple redefinition
(\ref{Q1}) of $\hat n$ that we used for the effective action theory
for the ground state energy and the two-terminal conductance.
Instead, to account for the fact that the total number of particles
in the system changes upon a tunneling event, one has to introduce
three additional operators \cite{Furusaki95}: A Hermitian operator  
$\hat{m}$ and two unitary operators $\hat{F}$ and $\hat{F}^\dagger$,
with the commutation relations
\begin{equation}
\left[\hat{m},\hat{F}^\dagger\right]=\hat{F}^\dagger.
\label{eq:6.11}
\label{eq:40.6}
\end{equation}
These operators serve to keep track of the total number of particles
in the system. They commute with all the fermionic degrees of freedom.
We now include $\hat m$ into the definition of the charge operator,
\begin{equation}
\hat{n} = - \sum_{j}
  \int {dk \over 2 \pi} \hat{\psi}^\dagger_{j} (k) \hat{\psi}_{j}(k)
+\hat{m},
\label{eq:6.12}
\label{eq:40.7}
\end{equation}
and rewrite Eq.~(\ref{eq:6.7}) as 
\begin{equation}
\Pi_{\rm tun}(\tau) = 
\langle T_\tau \hat{\bar{F}}(\tau ) \hat{\bar{\psi}}_1\left(\tau\right)
\hat{{F}}(0)\hat{\psi}_1\left(0\right)
\rangle. 
\label{eq:6.13}
\label{eq:40.8}
\end{equation}
It is easy to see from Eq.\ (\ref{eq:6.11}) that the operators 
$\hat{F}^\dagger, \hat{F}$ in Eq.~(\ref{eq:6.13}) change the charge,
as defined by Eq.\ (\ref{eq:6.12}), by $+e$ and $-e$ respectively, in
accordance with the initial definition (\ref{eq:2.151}) of charge
in terms of the fermionic operators of the dot.

After this manipulation, the Hamiltonian of the system becomes
quadratic in the fermionic operators of the dot, so that this part of
the system can be integrated out.  Manipulations analogous to the
derivation of Eq.~(\ref{eq:30.50}) give \cite{AleinerGlazman98}
\begin{eqnarray}
  \Pi_{\rm tun} (\tau) &=&  \Pi_{\rm in}(\tau) +  \Pi_{\rm el}(\tau),
\label{eq:6.18a}
\label{eq:40.9} \\
  \Pi_{\rm in}(\tau)&=&-\frac{{\cal G}_{11}(- \tau ) }
{ \langle T_\tau e^{-\hat\Action}\rangle_q}
\langle T_\tau e^{-\hat\Action}
\hat{\bar{F}}(\tau)\hat{F}(0)
\rangle_q,
\label{eq:40.10}
\label{eq:6.18b}\\
\Pi_{\rm el}(\tau) &=&
\sum_{\gamma,\gamma',ij}
\int_0^{1/T}\int_0^{1/T} d\tau_1d\tau_2 
  \nonumber \\ && \mbox{} \times
{\cal G}_{1\gamma}(\tau_1-\tau)W_{\gamma j}^{(2)} 
\Gamma_{ij}(\tau;\tau_1,\tau_2)
W^{(2)*}_{\gamma'i} 
{\cal G}_{\gamma'1}(-\tau_2)
\nonumber
\\
\Gamma_{ij}(\tau;\tau_1,\tau_2) &=&
{1 \over 4 \langle T_\tau e^{-\hat\Action}\rangle_q}
\left\langle \vphantom{\left[ \hat{\bar{\psi}}_{L,i} \right]}
  T_\tau e^{-\hat\Action}
\hat{\bar{F}}(\tau)\hat{F}(0)
  \right. \nonumber \\ && \left. \mbox{} \times
\left[\hat{\bar{\psi}}_{L,i}(\tau_1)+\hat{\bar{\psi}}_{R,i}(\tau_1)\right]
\left[\vphantom{\hat{\bar{\psi}}_{L,i}}
\hat{\psi}_{L,j} (\tau_2) + \hat{\psi}_{R,j} (\tau_2)\right]
\right\rangle_q,
\label{eq:6.18c}
\label{eq:40.11}
\end{eqnarray}
where the effective action is given by Eq.~(\ref{eq:30.482}),
the Green function of the dot, ${\cal G}_{\alpha\beta}(\tau)$,
is defined by Eq.~(\ref{eq:dotGF}),
and the coupling matrix $W_{\alpha j}^{(2)}$ corresponds to the
stronger (left) junction connecting the dot and a lead. Note also that
the scattering matrix $S$ entering in the expressions for the
effective action $S_{\rm eff}$, see Eqs.~(\ref{eq:action}) and
(\ref{eq:LS}), includes this stronger junction only.
The averages in Eqs.\ (\ref{eq:6.18a}) -- (\ref{eq:6.18c}) are
performed as 
$$
  \langle\dots\rangle_q = {\mathrm Tr} e^{-\hat{H}_{\rm eff}/T} \dots,
$$
with the effective Hamiltonian
\begin{eqnarray}
  \hat{H}_{\rm eff} &=& 
  i v_F \sum_{j}\int_{-\infty}^{\infty}dx
  \left(\hat{\psi}^\dagger_{L,j}\partial_x\hat{\psi}_{L,j} -
  \hat{\psi}^\dagger_{R,j}\partial_x \hat{\psi}_{R,j}\right) \nonumber\\ && 
  \mbox{} + {\Ec} \left(\sum_{j}\int_{-\infty}^{0} dx
  :\hat{\psi}^\dagger_{L,j}\hat{\psi}_{L,j} + \hat{\psi}^\dagger_{R,j}\hat{\psi}_{R,j}: 
  + {\cal N} - \hat{m}\right)^2.  \label{eq:40.13} 
\end{eqnarray}
The interaction representation of the operators is defined in 
Eq.~(\ref{eq:intrep}), with $\hat{H}_{\rm eff}$ from
Eq.~(\ref{eq:40.13}).
The difference
between Eqs.~(\ref{eq:40.13}) and (\ref{eq:30.482}) is
in the different definitions of the charge operators in
Eqs.~(\ref{Q1}) and (\ref{eq:6.12}). 

There is a good physical reason to distinguish the two contributions
to the tunneling density of states in Eq.\ (\ref{eq:40.9}). 
The first contribution in
(\ref{eq:6.18b}) is inelastic: this term does not allow the introduced
electron to leave the dot; the charge of the dot at the moment of
tunneling suddenly changes by $+e$ and all the other electrons have to
redistribute to accommodate this charge. We will see, that
the logarithmical divergence of the imaginary time action
corresponding to such evolution (orthogonality catastrophe) completely
suppresses this contribution at $T\to 0$. Conversely, the second
contribution, $\Pi_{\rm el}$, cf.\ Eq.~(\ref{eq:6.18c}), contains the
kernel $R(\tau)$ which promotes the tunneled electron through the dot
into the channel (left lead). Because the very same tunneling electron is
introduced to and then removed from the dot, there is no need in the
redistribution of other electrons, so no orthogonality catastrophe
occurs. As a result, the elastic contribution survives at $T\to 0$,
analogously to the elastic co-tunneling contribution to the conductance
of a strongly blockaded dot, see Sec.~\ref{sec:3.2}.

It can be shown, that if the charging energy vanishes, $\Ec=0$, all
physical results of the effective action theory contained in
Eqs.~(\ref{eq:30.482}) -- (\ref{eq:30.49}) and (\ref{eq:40.13}) are
equivalent to the non-interacting theory of 
Eqs.~(\ref{eq:2.63}), (\ref{eq:2.68}) and
(\ref{eq:2.80}). The advantage of the
effective action representation becomes clear when one has to treat
the effects of charging. The effective Hamiltonian (\ref{eq:30.482})
or (\ref{eq:40.13}) can then be 
diagonalized exactly and the effective action $\Action$
treated perturbatively. The technique for such treatment is described
in the following subsection.

\subsubsection{Bosonization of the effective Hamiltonian}
\label{sec:bosonization}
\addcontentsline{toc}{subsection}{
\protect\numberline{\ref{sec:bosonization}}
{\ Bosonization of the effective Hamiltonian}}

The interaction is treated by bosonization of the chiral Fermions. 
After bosonization, the Hamiltonian $\hat{H}_{\rm eff}$ is quadratic 
in the boson fields $\hat \phi$. The price we pay, however, is 
that the effective action $S_{\rm eff}$ is not quadratic in the
boson fields.

We
introduce boson fields $\hat{\varphi}_{Lj}$ and $\hat{\varphi}_{Rj}$
for left moving and right moving particles by\footnote{ These bosonic
  fields are related to the fields of Eq.~(\protect\ref{psi}) as
  $\theta_j(x) = (
  \hat{\varphi}_{Rj}-\hat{\varphi}_{Lj})/\sqrt{2\pi}$,
  $\hat{\varphi}_j(x) = (
  \hat{\varphi}_{Rj}+\hat{\varphi}_{Lj})/\sqrt{2\pi}$.  The linear
  scale $\lambda$ is related to the cut-off $D$ there by
  $\lambda=D/v_F$}
\begin{equation}
  \hat{\psi}_{Lj} = {\hat{\eta}_j \over \sqrt{2 \pi \lambda}} e^{-i \hat{\varphi}_{Lj}},\ \
  \hat{\psi}_{Rj} = {\hat{\eta}_j \over \sqrt{2 \pi \lambda}} e^{ i \hat{\varphi}_{Rj}},
\ \ j=1,\dots \Nch.
\label{eq:50.1}
\end{equation}
Here $\Nch$ is the total number of channels (including the spin
degeneracy), and $\hat{\eta}_j = \hat{\eta}_j^{\dagger}$ is a Majorana
fermion, $\ \left\{\hat{\eta}_j, \hat{\eta}_i\right\}=2 \delta_{ij}$. 
Since the Majorana fermions do not enter into the Hamiltonian, their
average is given by
\begin{equation}
\langle T_\tau \hat{\eta}_j(\tau_1), \hat{\eta}_i(\tau_2)\rangle_q =
\delta_{ij}{\mathrm sgn}\left(\tau_1-\tau_2\right).
\label{etas}
\end{equation}

The boson field $\hat{\varphi}_{Lj}$ and $\hat{\varphi}_{Rj}$ satisfy
the commutation relations
\begin{eqnarray}
  \left[ \hat{\varphi}_{Lj}(x), \hat{\varphi}_{Li}(y)\right] &=& -i \pi \delta_{ij}
  {\mathrm sgn}(x-y), \label{eq:50.2} \\
  \left[ \hat{\varphi}_{Rj}(x), \hat{\varphi}_{Ri}(y)\right] &=&  i \pi \delta_{ij}
  {\mathrm sgn}(x-y), \label{eq:50.3}\\
  \left[ \hat{\varphi}_{Rj}(x), \hat{\varphi}_{Li}(y)\right] &=& -i \pi \delta_{ij}.
  \label{eq:50.4}
\end{eqnarray}
The high energy cut-off $1/\lambda$ is of the order of $1/\lambda_F$ and 
is chosen consistently with the high-energy cut-off in the bosonic
correlation functions, see below Eq.~(\ref{eq:50.8}).

The products $\hat{\psi}_{Lj}^{\dagger}
\hat{\psi}_{Lj}^{\vphantom{\dagger}}$ and $\hat{\psi}_{Rj}^{\dagger}
\hat{\psi}_{Rj}^{\vphantom{\dagger}}$ read
\begin{equation}
  :\! \hat{\psi}_{Lj}^{\dagger}(x) \hat{\psi}_{Lj}^{\vphantom{\dagger}}(x)\! :\,
    = {1 \over 2 \pi} {\partial \over \partial x} \hat{\varphi}_{Lj}(x),\ \
  :\! \hat{\psi}_{Rj}^{\dagger}(x) \hat{\psi}_{Rj}^{\vphantom{\dagger}}(x)\! :\,
    = {1 \over 2 \pi} {\partial \over \partial x} \hat{\varphi}_{Rj}(x).
\label{eq:50.5}
\end{equation}
It is easy to check, using Eqs.~(\ref{eq:50.2}) -- (\ref{eq:50.4}),
that the fermionic fields (\ref{eq:50.1}) obey the standard
anti-commutation relations.  The commutation relation (\ref{eq:50.4})
guarantees the anticommutation relation $\left\{\hat{\psi}_{L,j},
  \hat{\psi}_{R,j}\right\}=0,\ \left\{\hat{\psi}_{L,j}^\dagger,
  \hat{\psi}_{R,j}\right\}=0 $ and the correct commutation relations between
the operator of the number of particles in the dot, see
Eqs.~(\ref{Q1}) and (\ref{eq:50.5}),\footnote{The absence of a term
  proportional to the bosonic fields at $x=-\infty$ in Eq.\ 
  (\ref{eq:nphi}) follows from the order of limits taken with respect
  the range of the interaction and the system size: First, the
  interaction is defined in terms of the charge of the entire length
  of the channel for $x< 0$, and only then the length of the channel
  is sent to infinity.}
\begin{eqnarray}
\hat{n} &=& \sum_{j=1}^{\Nch} \int_{-\infty}^{0}
:\! 
\hat{\psi}_{R,j}^\dagger \hat{\psi}_{R,j} + \hat{\psi}_{L,j}^\dagger \hat{\psi}_{L,j}
\! : dx \nonumber \\ &=&
  \frac{1}{2\pi}
  \sum_{j=1}^{\Nch} \left[\hat{\varphi}_{R,j}(x=0)+\hat{\varphi}_{L,j}(x=0)\right],
  \label{eq:nphi}
\end{eqnarray}
and the fermionic operators,
\[
\left[\hat{n},
\ \hat{\psi}^\dagger_{R,L}(x)
\right] = \hat{\psi}^\dagger_{R,L}(x) \theta(-x).
\]

In terms of the bosonic fields, the effective Hamiltonian
(\ref{eq:30.482}) acquires the form
\begin{eqnarray}
\hat{H}_0&=&\frac{v_F}{4\pi}
\sum_{j}\int_{-\infty}^{\infty}dx
\left[
\left(\frac{\partial \hat{\varphi}_{L,j}}{\partial x}\right)^2 +
\left(\frac{\partial \hat{\varphi}_{R,j}}{\partial x}\right)^2
\right] + \label{eq:50.6}\\
&& \frac{\Ec}{4\pi^2}
\left\{\sum_j\left[
\hat{\varphi}_{L,j}(0) +
\hat{\varphi}_{R,j}(0) 
\right]
+ 2\pi{\cal N}
\right\}^2.
\nonumber
\end{eqnarray}
The bosonized form of the Hamiltonian (\ref{eq:40.13}) is obtained by the
replacement ${\cal N} \to \left({\cal N} - \hat{m}\right)$.

As the Hamiltonian (\ref{eq:50.6}) is quadratic in the bosonic fields,
it can be easily diagonalized. One can immediately
notice \cite{Flensberg,Matveev95} that the eigenenergies of
(\ref{eq:50.6}) do not depend on the gate voltage ${\cal N}$
because the latter can be removed by the transformation
$\hat{\varphi}_{L(R),j}\to \hat{\varphi}_{L(R),j} - \pi {\cal N}/\Nch$. On the other
hand, the effective action (\ref{eq:30.482}) becomes a non-linear
functional of the bosonic operators
\begin{eqnarray*}
 \Action &=&
\frac{1}{2\pi\lambda}
  \sum_{ij}\int_0^{1/T} d\tau_1 \int_0^{1/T} d\tau_2\,
  L_{ij}\left(\tau_1-\tau_2\right) \hat{\eta}_i(\tau_1) \hat{\eta}_j(\tau_2)
  \\ && \mbox{} \times 
  \left[e^{-i \hat{\varphi}_{Li}(\tau_1)} + e^{i \hat{\varphi}_{Ri}(\tau_1)}\right]
 \left[e^{i \hat{\varphi}_{Lj}(\tau_2)} + e^{-i \hat{\varphi}_{Rj}(\tau_2)}\right]
\end{eqnarray*}
(hereinafter we use the convention (\ref{eq:convention}) for the
operators taken at $x=0$). Our general strategy will be to expand the
quantity of interest in powers of the effective action, and then to
average over the quadratic Hamiltonian (\ref{eq:50.6}). All the
relevant averages have the form $\exp(\hat{a})$, where $a$ is a linear
combination of the boson fields $\hat{\varphi}_{Lj}$ and
$\hat{\varphi}_{Rj}$, $j=1,\ldots,\Nch$. Such an average can be found
according the rule
\begin{equation}
  \langle \exp(\hat{a}) \rangle_q = 
\exp\left\{\langle\hat{a}\rangle_q + {1 \over 2} 
\left[\langle \hat {a}^2 \rangle_q - \langle \hat{a} \rangle_q^2
\right] \right\}.
\label{eq:50.7}
\end{equation}

The corresponding bosonic correlation functions can be found from the
Hamiltonian (\ref{eq:50.6}), see
e.g. Ref.~\cite{AleinerGlazman98}. For readers wishing to reproduce
the calculation, we notice only  that the
commutation relation (\ref{eq:50.4}) should be taken into account
to obtain the Green functions with correct analytic properties.
The result is
\begin{eqnarray}
\langle\hat{\varphi}_{L,j}\rangle_q & =& 
\langle\hat{\varphi}_{R,j}\rangle_q = 
-\frac {\pi {\cal N}}{\Nch}, \quad i,j =1,\dots,\Nch
\nonumber\\
\langle T_{\tau} \hat{\varphi}_{Li}(\tau_1) \hat{\varphi}_{Lj}(\tau_2) \rangle
  &=& \delta_{ij} \ln\left(\frac{\lambda}{v_F}\frac{\pi T}{|\sin \pi
T\left(\tau_1 -\tau_2 \right)|}\right) + A
  + \left( \delta_{ij} - {1 \over \Nch} \right) B
  \nonumber \\
  \langle T_{\tau} \hat{\varphi}_{Ri}(\tau_1) \hat{\varphi}_{Rj}(\tau_2) \rangle
  &=& \langle T_{\tau} \hat{\varphi}_{Li}(\tau_1) \hat{\varphi}_{Lj}(\tau_2) \rangle,
\nonumber\\
\langle T_{\tau} \hat{\varphi}_{Li}(\tau_1) \hat{\varphi}_{Rj}(\tau_2) \rangle
  &=&  \delta_{ij} {i \pi \over 2} {\mathrm sgn}\, (\tau_1-\tau_2)
    - {1 \over \Nch} \ln f(\tau_2-\tau_1) - A. 
 \label{eq:50.8}
 \label{eq:phiRphiR}
 \label{eq:phiLphiL}
\label{eq:phiLphiR}
\end{eqnarray}
Here all of the bosonic operators are taken at the origin $x=0$.
One should take the limit $A \to +\infty,$ $B \to +\infty$ at
the end of the calculation of each physical 
propagator.\footnote{The precise
meaning of these quantities is
$\Nch A/2 = \sum_{ij}\langle\hat{\varphi}_{L,i}(\tau)\hat{\varphi}_{L,j}(\tau)\rangle_q$,\\
$(\Nch-1)B + A/2 =\sum_{i}\langle\hat{\varphi}_{L,i}(\tau)\hat{\varphi}_{L,i}(\tau)\rangle_q$
} 

The function $\ln f(\tau)$ has the integral representation
\begin{equation}
 \ln f(\tau)  = 
   \ln \frac{\lambda \Ec \Nch e^{\bf C}}{ \pi  v_F }
    - \int_0^{\infty} dx e^{-x}
    \ln {\sinh[(i\tau +  \pi x/\Ec \Nch) \pi T] \over 
    \sinh ( \pi^2 T x/\Ec \Nch)},
\label{eq:50.9}
  \label{eq:orig} \label{eq:fexact}
\end{equation}
where ${\bf C} \approx 0.577$ is the Euler constant.
We note that $f(\tau)$ is analytic for $\mbox{Im}\, \tau < 0$. For $T
\ll \Ec$, to a good approximation, we set
\begin{equation} 
\label{eq:approx}
  f(\tau) \approx 
  {\lambda \pi T  \over i v_F \sin[(\tau - i t_0) \pi T]},
  \ \ t_0 = { \pi \over \Ec \Nch e^{\bf C}}. 
\label{eq:f}
\label{eq:50.10}
\end{equation}
This equation is valid both for $\tau=0$ and for $\tau \gg
1/\Nch \Ec$. It also provides the correct upper cut-off for the
arising logarithmic divergences. The approximation (\ref{eq:f})
fails when contributions from times $\tau \sim 1/\Nch \Ec$ are
important, as is the case for spinless fermions with $\Nch = 1$.

Using Eqs.~(\ref{eq:50.1}), (\ref{etas}), and (\ref{eq:50.8}) one can
find all the relevant correlation functions of the fermionic
operators. We will discuss some of them here in order to demonstrate the
analytic structure of the perturbation theory.

First of all, from Eq.\ (\ref{eq:50.8})
we immediately find that any Green's function which
involves {\em only} left (L) or {\em only} right (R) moving fermions
is not sensitive to the interaction, {\em i.e.},
\begin{equation}
(2\pi v_F)
\langle T_\tau \hat{\bar{\psi}}_{L,i}(\tau_1){\hat{\psi}}_{L,j}(\tau_2) \rangle_q
=\delta_{ij}\frac{\pi T}{\sin [\pi T(\tau_1-\tau_2)]},
\label{eq:50.11}
\end{equation}
and all the averages are given by the Wick theorem. All such Green
functions may have only pole type singularities in the plane of
the complex variable $\tau_i$.  The insensitivity of such type of the
Green functions to the interaction can be understood from the
chirality of the fermions and the causality principle.

\begin{figure}
\vspace{0.2cm}
\hspace{0.3\hsize}
\epsfxsize=0.4\hsize
\epsfbox{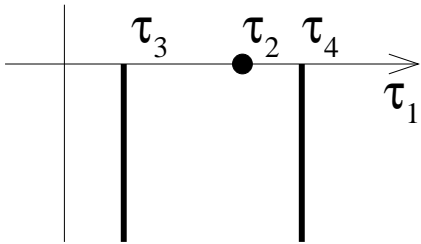}
\vspace{0.6cm}
\caption{Analytic structure of the correlator (\protect\ref{eq:50.12})
as a function of $\tau_1$. The dot indicates the pole singularity at
$\tau_1=\tau_2$; the thick lines the branch
cuts along the lines ${\mathrm Re}\, \tau_1 = \tau_3, \
{\mathrm Im} \tau_1 < 0$ and ${\mathrm Re}\, \tau_1 = \tau_4, \
{\mathrm Im} \tau_1 < 0$.
}
\label{fig:50.12}
\end{figure}

Let us now consider a more complicated average involving left movers
together with right movers,
\begin{eqnarray}
&&(2\pi v_F)^2\langle T_\tau
\hat{\bar{\psi}}_{L,i}(\tau_1){\hat{\psi}}_{L,i}(\tau_2)
\hat{\bar{\psi}}_{R,j}(\tau_3){\hat{\psi}}_{R,j}(\tau_4)
\rangle_q= \label{eq:50.12}\\
&&\quad \frac{\pi T}{\sin [\pi T(\tau_1-\tau_2)]}
\frac{\pi T}{\sin [\pi T(\tau_3-\tau_4)]}
\left(\frac{f(\tau_4-\tau_1)f(\tau_3-\tau_2)}
{f(\tau_3-\tau_1)f(\tau_4-\tau_2)}
\right)^{1/\Nch}.
\nonumber
\end{eqnarray}
We fix $\tau_2,\tau_3, \tau_4$ to be real numbers and consider
the correlator (\ref{eq:50.12}) as a function of the complex 
parameter $\tau_1$. It has a pole singularity at $\tau_1=\tau_2$ and
branch cuts along the lines ${\mathrm Re}\, \tau_1 = \tau_3, \
{\mathrm Im} \tau_1 < 0$, and ${\mathrm Re}\, \tau_1 = \tau_4, \
{\mathrm Im} \tau_1 < 0$, see Fig.~\ref{fig:50.12}. It is readily
seen, that the contribution to an integral over $\tau_1$ from the pole
does not depend on the interaction (it does no longer involve the
function $f$): the fractional-power factor in
Eq.~(\ref{eq:50.12}) turns into unity at the pole. Thus it is the branch
cut contribution that comes from the interaction. The same is true for
any higher-order correlation function. 

Imagine now that we expand an
observable quantity in power series in $L_{ij}$ and account in each
order of perturbation theory only for the pole contributions in the
integrals over the intermediate times $\tau_1, \tau_2,\dots$. The pole
contribution in each order does not depend on the interaction, and
constitutes just a Green function for non-interacting fermions. Such
contributions can be easily summed up, resulting in a geometric
series in $L$.\footnote{For the calculation of the ground state energy, the
geometric series appears only when one calculates the thermodynamic
potential (\protect\ref{eq:30.50}) of the entire system, including the
contribution $\Omega_D$ from the closed dot.} Since a geometric series
of the matrix $L$ is precisely the scattering matrix $S$, cf.\
Eq.~(\ref{eq:30.49}), we thus arrive at the following simple
result: in order to take into account the
pole contributions in {\em all} subsequent orders of perturbation
theory, one has to simply replace $L$ in the first non-vanishing
contribution by 
\begin{equation}
 \Lkern(\omega_n) \to
\left\{ \matrix{ \displaystyle 
  -{1 \over 2 \pi i \nu_{\vphantom{M}}} {\Scat(i\omega_n)}, & \omega_n >0,
\cr \displaystyle {1^{\vphantom{M}} \over 2 \pi i \nu}
{\Scat^\dagger(i\omega_n)}, &
\omega_n < 0.  } \right.
\label{eq:LSall}
\end{equation}
(Here $\omega_n$ is the fermionic Matsubara frequency.)  

Consider now
a correlation function which includes a branch cut in the leading
order ({\em e.g.}, as function of the times $\tau_1, \tau_2, \tau_3, 
\tau_4$ in second order in the effective action).  
Then, in each of the 
subsequent higher orders we select
the contributions including only one branch cut, all other
singularities in these terms being single poles. Since such a
contribution is factorised into a branch cut 
factor involving two times, and a product of 
Green functions for non-interacting
fermions for the remaining times, we can again sum up all such
contributions
as a geometric series. As before, the result
is that one has to take the first non-vanishing term in an expansion
in powers of $L$ that contains a single
branch cut and then make the
replacement (\ref{eq:LSall}) to account for all higher orders
in $L$ with the same branch cut singularity. In
such a way, we take into account all the orders of perturbation theory
in ${L}$ that contain only one branch cut singularity in each order.

The same analytic structure occurs in all higher orders of
perturbation theory: poles give a ``reducible'' average, whereas the
``irreducible'' part of the propagator has only branch cut
singularities. This leads to a very important statement: {\em the
perturbation theory in the action (\ref{eq:30.482}) can be constructed as
an expansion in powers of the scattering matrix $S$ rather than of the 
kernel $L$}. In this expansion all the reducible averages (poles) are
included automatically in all the orders of the perturbation theory,
whereas the branch cuts are considered perturbatively; each order of
of the perturbation theory in $S$ is characterized by the number of
branch cuts it includes.\footnote{
Regarding this statement, we point to the fact that for
noninteracting particles, 
the ground state energy, Eq.\ (\ref{eq:2.80}) and the
conductance, Eq.\ (\ref{eq:2.63}), are of second order in the 
scattering matrix $S$.
In the effective action theory, this result follows from 
the summation of the pole contributions to all orders in $L$, 
which produces a single term proportional to $S^2$.
The interaction corrections, however, can
be of higher order in $S$, depending on the number of branch
cut singularities that are included in the perturbation theory.}

The correlators (\ref{eq:50.11}) and Eq.~(\ref{eq:50.12}) do not
depend explicitly on the gate voltage ${\cal N}$. One can see from
Eq.~(\ref{eq:50.8}), that a correlator having such a
dependence must
include at least creation operators of left (right) moving particles
and annihilation operator for right (left)
moving particles in {\em all channels}. For a total number of $\Nch$
channels, the lowest order ${\cal N}$-dependent correlator is
\begin{eqnarray}
(2\pi v_F)^{\Nch} \left\langle T_\tau
\prod_{j=1}^{\Nch}
\hat{\bar{\psi}}_{L,j}(\tau_j^L){\hat{\psi}}_{R,j}(\tau_j^R)
\right\rangle_q &=&
\left(\frac{-i}{t_0}\right)^{\Nch} e^{-i 2\pi {\cal N}}
\label{eq:50.13}
\\ && \mbox{} \times
\left(\prod_{i,j=1}^{\Nch}\frac{
{f(\tau_i^R-\tau_j^L)}}{f(0)}
\right)^{1/\Nch}.
\nonumber
\end{eqnarray}
The expression for the correlator of an arbitrary number of the
fermionic operators as well as other necessary correlation functions
are given in Appendix~\ref{Ap:cor}.

To conclude this subsubsection, let us prove the assumption of
Sec.~\ref{sec:qualit} about the Fermi liquid behavior of the
one-channel system $\Nch=1$ at low energies. To this end we substitute
Eq.~(\ref{eq:50.10}) into Eqs.~(\ref{eq:50.12}) and (\ref{eq:50.13}),
and discover that the Green functions now have only the pole-type
singularities, which is the hallmark of the Fermi-liquid.  Moreover,
one finds from Eq.~(\ref{eq:50.13}) at $\tau \gg 1/\Ec$
\[
\langle T_\tau
\hat{\bar{\psi}}_{L}(0){\hat{\psi}}_{R}(\tau)
\rangle_q= - \frac{1}{2\pi v_F}
\frac{\pi T}{\sin\pi T\tau}
e^{-i 2\pi {\cal N}}.
\nonumber\\
\quad N=1,
\] 
which is exactly a Green function of one channel system with
unpenetrable barrier and the scattering phase $\delta = \pi \left(
{\cal N} +1/2 \right)$, see also Eq.~(\ref{phaseshift}).

\subsection{Results for the differential capacitance}

\label{sec:cap}
As we have already seen in the discussion of weak tunneling, the
existence of periodic
oscillations of measurable quantities with the gate voltage ${\cal
N}$ is a signature of Coulomb blockade. In this section
we consider such oscillations, which occur due to the incremental
charging of the dot, in the case of the differential capacitance
of a partially open dot. 

First we derive the relation between the ${\cal N}$-dependent
thermodynamic potential and the scattering matrix $S$ for an
interacting system, see Eqs.~(\ref{eq:60.1})--(\ref{eq:60.4}) 
and (\ref{eq:Omegalbee})--(\ref{eq:OmegaAperiodic}) below.
These formulae replace the known results for the non-interacting
electrons, see Eqs.~(\ref{eq:2.80})--(\ref{eq:2.71}) of
Section~\ref{sec:2.2}. 

Using the general result, we then find the ensemble-averaged
differential capacitance. The average capacitance oscillates
periodically with ${\cal N}$ as long as some reflection in the
channels connecting the dot with leads exists. The amplitude of
oscillations, however, decays with the number of channels as
$|r|^{\Nch}$, where $r$ is the reflection amplitude in a
channel, and the number of channels $\Nch$ accounts also for the
spin degree of freedom. The detailed formulae for the oscillations of
the average capacitance are given below in Eqs.~(\ref{eq:Savg}),
(\ref{eq:60.10}), and (\ref{eq:deltaOmegaMultiN}).

In the absence of reflection in the channels, the average differential
capacitance does not oscillate with ${\cal N}$. However, mesoscopic
fluctuations of the capacitance still bear a signature of charge
quantization. The general formulae
Eqs.~(\ref{eq:60.1})--(\ref{eq:60.4}) and
(\ref{eq:Omegalbee})--(\ref{eq:OmegaAperiodic})
allow us to examine the
two-point correlation function
\begin{equation}
K({\cal N}_1-{\cal N}_2)=\frac{\langle \delta C_{\rm diff}
({\cal N}_1)\delta C_{\rm diff}({\cal N}_2)\rangle}{C^2}
\label{eq:corrcc}
\end{equation}
and show that it consists of two additive parts,
\[
K(s)\equiv K_Q(s)+K_\mu(s).
\]
The first part, $K_Q(s)$, originates from the discreteness of charge, 
and can be represented as the product of an oscillatory function 
with unit period and
an envelope function $\Xi_{\Nch}(s\Delta/\Ec)$ which decays on
a scale $|s|\sim \Ec/\Delta$,
\begin{equation}
K_Q(s)\sim
  \frac{\Delta}{\Ec}
\cos(2\pi s)\, \Xi_{\Nch}\left(\frac{s\Delta}{\Ec}\right).
\label{eq:cospis}
\end{equation}
The explicit results for the special cases of a junction with one
propagating mode of spinless fermions and real spin $1/2$ electrons are
given below in Eqs.~(\ref{eq:60.8}) and (\ref{eq:60.10}), respectively.
The second part, $K_\mu (s)$, comes from the mesoscopic fluctuations
of the capacitance as a function of the  chemical potential
(which itself depends on ${\cal N}$). 
This part is identical with the one that
was found earlier in Ref.~\cite{BrouwerBuettiker96} within the RPA
approximation,
\begin{equation}
K_\mu(s)\sim\left(\frac{\Delta}{\Ec \Nch s}\right)^2,\quad s\gtrsim
1.
\label{eq:kamu}
\end{equation}
The derivation of $K_\mu (s)$ within the general formalism is also
outlined at the end of this Section.

With the increase of the number of channels $\Nch$ connecting
the dot and leads, the oscillatory part of the correlation function
quickly decays. It is experimentally detectable only for a small
number of channels.

{\it The oscillatory part $K_Q(s)$ of the capacitance autocorrelation
function}\\ 
As we have seen in the previous subsection, the ${\cal
N}$-dependent contribution to the thermodynamic potential appears in
$\Nch$th order perturbation theory ($\Nch$ is the total number of
channels):
\begin{equation}
\delta \Omega ({\cal N})= T \frac{(-1)^{\Nch+1}}{\Nch!}\langle
\left[
\Action \right]^{\Nch}\rangle_q.
\label{eq:60.1}
\end{equation} 
Substituting Eq.~(\ref{eq:30.482}) into Eq.~(\ref{eq:60.1}) and using
Eq.~(\ref{eq:50.13}), we obtain
\begin{eqnarray}
\delta \Omega ({\cal N}) &=&
-2\ (-1)^{\Nch}\, {\mathrm Re}\, 
e^{-i2\pi {\cal N}}
T
\nonumber \\ && \mbox{} \times
\int_0^{1/T}
{\mathrm det}[\Scat]
\left[\prod_{i,j=1}^{\Nch}
\left(
\frac{f(\tau_i^R-\tau_j^L)}{{2\pi t_0}f(0)}
\right)^{{1 \over \Nch}}
d\tau_i^R
d\tau_j^L
\right],
\label{eq:60.2}
\end{eqnarray}
where the function $f(\tau)$ is defined by Eq.~(\ref{eq:50.9}). The
determinant of the scattering matrix in the imaginary time domain is 
defined by
\begin{eqnarray}
{\mathrm det}[\Scat] &\equiv&
\left|
\matrix{
\Scat_{11}(\tau_1^L - \tau_1^R) & \Scat_{12}(\tau_1^L - \tau_2^R) & \dots &
\Scat_{1\Nch}(\tau_1^L - \tau_{\Nch}^R) \cr
\Scat_{21}(\tau_2^L - \tau_1^R) & \Scat_{22}(\tau_2^L - \tau_2^R) & \dots &
\Scat_{2\Nch}(\tau_2^L - \tau_{\Nch}^R) \cr
\dots &\dots &\dots & \dots\cr
\Scat_{\Nch1}(\tau_{\Nch}^L - \tau_1^R) & \Scat_{N2}(\tau_{\Nch}^L - \tau_2^R) & \dots &
\Scat_{\Nch\Nch}(\tau_{\Nch}^L - \tau_{\Nch}^R) 
}
\right|.
\nonumber \\
\label{eq:60.3}
\end{eqnarray}
The kernel $L$ was substituted with the
scattering matrix $S$ in Eq.~(\ref{eq:60.3}), 
in accordance with the previous subsubsection, and
\[
\Scat(\tau) = T\sum_{\omega_n=\pi T (2n+1)}
e^{-i\omega_n\tau}\left[\Scat(i\omega_n)\theta(\omega_n)
-\Scat^\dagger(i\omega_n)\theta(-\omega_n)
\right].
\]
The latter formula can be also written in a form of the Lehmann
representation
\begin{equation}
\Scat(\tau) ={1 \over 2 \pi i}
\int_{- \infty}^{\infty} dt\,  \left[
    \Scat(t) + \Scat^\dagger(t) \right] 
{\pi T \over \sin[(\tau - i t) \pi T]}.
\label{eq:60.4}
\end{equation}

Equations (\ref{eq:60.2}) and (\ref{eq:60.3}) express the gate
dependent interaction to the thermodynamic potential in a most general
form. It is valid for any scattering matrix of the dot. In what follows we
apply those equations for studying the different limiting cases.

{\em The oscillatory part of differential capacitance and $K_Q(s)$ for
the one-channel geometry (spinless fermions).} -- In this case the
scattering matrix is formed by its only element $\Scat(t)$.  Using the
fact that the function $f(\tau)$ is analytic in the lower complex
semiplane, see Eq.~(\ref{eq:fexact}), we find
\[
\delta \Omega ({\cal N})=\frac{\Ec e^{\bf C}}{\pi^2} 
{\mathrm Re}
\left[ 
e^{-i2\pi {\cal N}}
\int_{0}^\infty \Scat(t) \frac{f(-it)}{f(0)}dt
\right].
\]
With the help of Eq.~(\ref{eq:fexact}), one obtains for $T \ll \Ec$
\begin{eqnarray}
  \delta \Omega ({\cal N}) &=&
  \frac{\Ec e^{\bf C}}{\pi^2} 
  \nonumber \\ && \mbox{} \times
  {\mathrm Re}
  \left[ e^{-i2\pi {\cal N}} \int_{0}^\infty \Scat(t) \exp \left(
  -\int_0^\infty {dx}e^{-x}\ln\frac{x+t\Ec/\pi}{x} \right) dt \right].
  \label{eq:60.5}
\end{eqnarray}
One can use Eq.~(\ref{eq:60.5}) to analyze the statistics of the
differential capacitance (\ref{Cdif}). In the approximation
(\ref{eq:30.490}), one obtains \cite{Matveev95}
\begin{equation}
\frac{\langle\delta C_{\rm diff} ({\cal N})\rangle}{C} = 
2 e^{\bf C} |r| \cos 2\pi {\cal N},
\label{eq:5.70}
\label{eq:60.6}
\end{equation}
where $\langle \dots \rangle$ stands for the ensemble averaging
and $r$ is the reflection amplitude of the contact between the dot
and the lead.

The statistics of the mesoscopic fluctuations of the differential capacitance
for reflectionless contacts was analyzed in Ref.~\cite{AleinerGlazman98}
with the help of Eqs.\ (\ref{eq:60.7}), (\ref{eq:Savg}), and
(\ref{eq:60.5}).
The result is
\begin{equation}
K_Q(s)=\frac{5.59}{\beta}
\frac{\Delta}{\Ec}\cos 2\pi s.
\label{eq:5.7}
\label{eq:60.8}
\end{equation}
The correlation function for the capacitances at different magnetic
field can be found in Ref.~\cite{AleinerGlazman98}.
It is noteworthy that the characteristic amplitude of the fluctuations
is in agreement with the estimates of Sec.~\ref{sec:qualit}.

{\em The oscillatory part of differential capacitance and $K_Q(s)$ in
the one-channel geometry ( electrons with spin).} -- In the absence of
spin orbit interactions, the $2\times 2$ scattering matrix is given
by $\Scat(t)= {\mathrm diag}\{\ScatOrb(t),\ScatOrb(t)\}.$ In this
case, the main contribution comes from times larger than $\Ec^{-1}$,
so that one can use the approximation (\ref{eq:50.10}) for the
function $f(\tau)$ in Eq.~(\ref{eq:60.2}).  With the help of the
Lehmann representation (\ref{eq:60.4}), one finds
\begin{eqnarray}
&&\delta \Omega ({\cal N})=
-\frac{1}{2\pi^2}{\mathrm Re}
\left[
e^{-i2\pi {\cal N}}
\int_{0}^{\infty}{dt_1}{dt_2}
\Scat(t_1)\Scat(t_2)F(t_1,t_2),
\right]
\label{eq:60.9}\\
&&F(t_1,t_2)
=
\frac{\pi T}
{\left[\sinh (\pi T(t_1+t_0))\sinh (\pi T(t_2+t_0)\right]^{1/2}}
\nonumber\\
&&
\quad
\times
\int_0^{1/T} d\tau_3\frac{\pi T}{\left[\sinh\pi T 
(t_1+t_0-i\tau_3)\sinh\pi T
(t_2+t_0+ i\tau_3)\right]^{1/2}},
\nonumber\\
&&t_0 =  { \pi \over 2 \Ec  e^{\bf C}}.
\nonumber
\end{eqnarray}
Similarly to the model of the spinless fermions, Eq.~(\ref{eq:60.9})
can be applied to study of both the averaged differential capacitance
and the mesoscopic fluctuations. Using Eq.~(\ref{eq:30.490}) and
Eq.~(\ref{eq:60.7}), one obtains
\begin{eqnarray}
&&\frac{\langle\delta C_{\rm diff}({\cal N}\rangle)}{C} = \frac{4 e^{\bf
C}}{\pi}|r|^2 \cos 2\pi{\cal N} \ln \left(\frac{\Ec}{T}\right);
\label{eq:60.10}
\\
&&K_Q(s)=
\frac{16}{3\pi^2\beta}
\cos (2\pi s)
\frac{\Delta}{\Ec}
\ln^3\left(\frac{\Ec}{T}\right)
\left[
\frac{\Delta}{\beta \Ec}
\ln\left(\frac{\Ec}{T}\right)+
4e^{\bf C}|r|^2
\right].
\nonumber
\end{eqnarray}
The result for the averaged capacitance was first obtained in
Ref.~\cite{Matveev95}. The correlation function of the mesoscopic
fluctuations in magnetic field can be found in Ref.~\cite{AleinerGlazman98}. 
The origin of the logarithmic divergence in Eq.~(\ref{eq:60.10}) and
the cure for it is discussed in Sec.~\ref{sec:reduction}.

{\em Capacitance oscillations in the case of multichannel geometry
with reflection in the junction.} -- For the case of an ideal point
contact with many channels between the dot and the lead, the
periodic-in-${\cal N}$ capacitance fluctuations given by Eq.\
(\ref{eq:60.2}) will be dominated by non-periodic fluctuations, see
the following Subsection.
Periodic capacitance fluctuations, however,
remain important if the point contacts are non-ideal. In that case the
scattering matrix contains a non-fluctuating direct component, given
by the reflection matrix of the point contact, cf.\ Eq.\
(\ref{eq:Sdirect26}),
\begin{equation}
  \Scat_{ij}(t) = r_{ij} \delta(t) + \mbox{fluctuating
  part}. \label{eq:StdirectMulti}
\end{equation} 
Substitution of Eq.\ (\ref{eq:StdirectMulti}) into Eq.\
(\ref{eq:60.2}) gives a periodic contribution
to the thermodynamic potential,
\begin{equation}
  \delta \Omega({\cal N}) = {4 \over \pi^2 t_0} 
  c_{\Nch} \ln \left({\Ec \over T}\right)
  \mbox{Re}\, e^{-2 \pi i {\cal N}}
  \det r, \label{eq:deltaOmegaMultiN}
\end{equation}
where the numerical coefficient $c_{\Nch}$ depends weakly on the number
$\Nch$ of
channels; $c_2 = 1/4$, $c_3 = (9/8) \pi^{-3/2} 2^{1/3} \Gamma(1/3)
\Gamma(7/6)$.

Equation (\ref{eq:deltaOmegaMultiN}) demonstrates that the dependence
of the amplitude of the Coulomb blockade oscillations on the contact's
conductance $G$ is not universal; more detailed characteristics of the
contact are important. The determinant $\det r$ in
Eq.~(\ref{eq:deltaOmegaMultiN}) can be evaluated in terms of the
reflection eigenvalues $R_j$ of the contact, $|\det r| = \prod_{j}
R_{j}^{1/2}$ (here $R_j$ are defined as the eigenvalues of $r
r^{\dagger}$). The distribution of $R_j$ is known for a number of
relevant models, including the model of a short disordered
wire~\cite{Beenakker97}.  With the help of that distribution, one can
find from Eq.~(\ref{eq:deltaOmegaMultiN}) the amplitude of the
oscillations of $\Omega$ and of the differential capacitance, $\delta
C_{\rm diff}({\cal N})\sim \ln(\Ec/T)\exp(-2 \pi^3\hbar
G/e^2)\cos(2\pi{\cal N})$. This exponential dependence, and the
more general formula expressing the amplitude of oscillations in terms
of $\prod_{j} R_{j}^{1/2}$ (but both witout the logarithmically
divergent prefactor) were found in Ref.~\cite{Nazarov99}. There a
different method, based on an extension of the instanton
technique of Ref.~\cite{Korshunov} to the case of an
arbitrary (though energy independent) $S$- matrix of the contact
was used. The same results were reproduced within a non-linear
$\sigma$--model for interacting systems in Ref.~\cite{Kamenev2000}.

{\em Non-periodic interaction corrections to the
capacitance, $K_\mu (s)$}. --
In addition to the periodic-in-${\cal N}$ mesoscopic fluctuations
of the capacitance, resulting from charge discreteness, there 
are also non-periodic mesoscopic fluctuations of the capacitance.
These non-periodic fluctuations arise through mesoscopic
fluctuations of the thermodynamic potential $\Omega$ as a function
of the chemical potential $\mu$ in combination with the ${\cal
N}$-dependence of the chemical potential $\mu$. 
In the absence of direct reflection in the 
contacts and for temperatures $T \gg \Nch \Delta$, 
the non-periodic ${\cal N}$-dependent fluctuations
dominate over the periodic ones that occur in higher ($\Nch$)
order perturbation theory in the effective action.
[Higher orders in perturbation theory are small by the 
quantity $(\Nch \Delta/\pi^2 T)$.]

Mesoscopic fluctuations of the 
capacitance $C_{\rm diff}$ were studied in a series of papers 
 in a mean-field treatment of the interaction
\cite{BuettikerJCondensC,BTP,BPT,BrouwerBuettiker96,GMB},
with the result
\begin{equation}
  {C \over C_{\rm diff}} 
  = 1 + {1 \over 2 \Ec\, \nu(\mu;T)}.
  \label{eq:BuetC}
\end{equation}
Here $\nu(\varepsilon;T)$ is the non-interacting
density of states of the dot at temperature $T$,
cf.\ Eq.\ (\ref{eq:OmegaDopen}),
$$
  \nu(\varepsilon;T) =
  \int d\varepsilon \left( - {df_F \over d\varepsilon} \right) 
  {1 \over 2 \pi i} \mbox{Tr}\, \Scat^{\dagger}(\varepsilon)
  {\partial \Scat \over \partial \varepsilon},
$$
$f_F(\varepsilon) = 1/(1 + \exp(\varepsilon/T))$ being the
Fermi distribution function, and the ${\cal N}$-dependence of $\mu$ 
follows from the self-consistency condition
\begin{equation}
  {d \mu \over d {\cal N}} =
  - {C_{\rm diff} \over C \nu(\mu;T)}
  = \left[ \nu(\mu;T) + {1 \over 2 \Ec} \right]^{-1}.
  \label{eq:muNmf}
\end{equation}

The mesoscopic fluctuations of the capacitance (\ref{eq:BuetC}) are
caused by those of the (non-interacting) density of states 
$\nu(\varepsilon)$.
We now show that these capacitance fluctuations can be reproduced
within the effective action approach. The advantage of the effective
action theory, however, is that it allows us to calculate 
systematic interaction corrections to the capacitance that go beyond 
mean-field theory.

Formally, the ${\cal N}$-dependence of the chemical
potential $\mu$ in the effective action approach arises, because one
has to impose a canonical constraint on the ensemble, instead of a
grand canonical one, as was done in Sec.\ \ref{sec:action}. Details
of how this
can be done are given in Appendix \ref{Ap:3}. Here we only mention 
the result, which is that for temperatures $T \gg \Nch \Delta$ or 
for channel numbers $\Nch \gg 1$, where the fluctuations of
the density of states $\nu(\varepsilon;T)$ are small, one finds
\begin{equation}
  \mu = - {2 \Delta \Ec {\cal N} \over \Delta + 2 \Ec}. \label{eq:muN}
\end{equation}
Note that this is the same ${\cal N}$-dependence
as in the self-consistent theory, Eq.\ (\ref{eq:muNmf}).\footnote{The
difference between Eqs.\ (\ref{eq:muN}) and (\ref{eq:muNmf}) is
in the fluctuations of the density of states, which are small 
when $T \gg \Nch \Delta$ or $\Nch \gg 1$. Fluctuations of
$\nu(\varepsilon)$ can be included systematically in the effective
action formalism, see App. \ref{Ap:3}} In the limit $\Ec \gg \Delta$
that we consider here, Eq.\ (\ref{eq:muN}) simplifies to
$\mu \approx -\Delta {\cal N}$.

By Eq. (\ref{eq:30.50}), the thermodynamic potential $\Omega$ consists of two
parts, 
$$
  ~ \hspace*{3cm}
  \Omega = \Omega_D 
  -T \ln \mbox{Tr}\, e^{-\hat H_{\rm eff}/T} T_{\tau} e^{-\Action},
  \hspace*{3.7cm} (\protect\ref{eq:30.50})
$$
where $\Omega_D$ is the thermodynamic potential of the closed dot
without interactions and the second term accounts for the 
difference between the thermodynamic
potentials of an open dot with interactions and a closed dot
without. The thermodynamic potential $\Omega_D$ of the
closed dot without interactions can be written as an integral over 
the density of states 
$\nu_{\rm closed}(\varepsilon)$ of the closed dot,
\begin{equation}
  \Omega_D
  = -T  \int d\varepsilon \nu_{\rm closed}(\varepsilon)
  \ln \left(1 + e^{-\varepsilon/T} \right).
\end{equation}
The density of states $\nu_{\rm closed}(\varepsilon)$ 
is a sum of delta
functions, and thus differs from the density of states 
$\nu(\varepsilon)$
of the dot when it is coupled to the leads, see Eq.\
(\ref{eq:OmegaDopen}), which is a continuous function of 
energy. It was the latter density of states that
determined the thermodynamic potential (\ref{eq:2.80})
of the open quantum dot
in the absence of interactions, and that entered into the
mean-field description, see Eqs.\ (\ref{eq:BuetC})--(\ref{eq:muNmf})
above.
Using the equality
\begin{equation}
  \nu_{\rm closed}(\varepsilon) = 
  {\partial \over \partial \varepsilon}
  {1 \over 2 \pi i} \mbox{Tr}\, \ln \left[
  {1 + \Scat(\varepsilon - i 0^{+}) \over 1 + \Scat(\varepsilon + i 0^{+})}
  \right],
\end{equation}
together with unitarity of $S$,
we can write $\nu_{\rm closed}(\varepsilon)$ as
\begin{equation}
  \nu_{\rm closed}(\varepsilon)
  = \nu(\varepsilon) - {1 \over \pi}
   {\partial \over \partial \varepsilon} \mbox{Im}\,
   \ln[1 + S(\varepsilon + i 0^{+})].
\end{equation}
Hence, we can represent
$\Omega_D$, in turn, as
\begin{eqnarray}
  \Omega_D &=& \Omega_{{\rm lb}} +
  {1 \over \pi} \int_{-\infty}^{\infty} d\varepsilon
  f_F(\varepsilon) 
  \mbox{Im}\, \mbox{Tr}\, \ln[1 + \Scat(\varepsilon + i 0^{+})], 
  \label{eq:OmegaDopenclosed} 
\end{eqnarray}
where $\Omega_{{\rm lb}}$ is the thermodynamic potential of the 
dot coupled to the leads in the absence of interactions, see 
Eq.\ (\ref{eq:2.80}). (The subscript ``lb'' has been written
in analogy with the non-interacting contribution $G_{\rm lb}$
to the two-terminal conductance, see Subsection \ref{sec:g} below.)

The second term in Eq.\ (\ref{eq:30.50}) 
can be calculated by expansion in the effective action $\Action$, 
followed by an
average over the fermionic operators using Eqs.\
(\ref{eq:50.11}) and (\ref{eq:50.12}).
As explained below Eq.\ (\ref{eq:50.12}), one can distinguish 
two types of contributions: the
``pole contributions'', which are already present in the absence
of interactions, and the ``branch-cut contributions'', which 
represent the corrections to $\Omega$ due to the electron-electron
interactions in the dot. Since, in the absence of interactions,
the thermodynamic potential of the dot is given by $\Omega_{\rm lb}$, 
the pole contributions to the second term in
Eq.\ (\ref{eq:30.50}) and the second term of Eq.\ (\ref{eq:OmegaDopenclosed})
must cancel to all orders in the effective action $\Action$.
Hence, the thermodynamic potential is given by 
\begin{equation}
  \Omega = \Omega_{{\rm lb}} + 
\Omega_{{\rm ee}},
  \label{eq:Omegalbee}
\end{equation}
where the interaction correction $\Omega_{\rm ee}$ represents the 
``branch-cut contributions'' to 
$-T \ln \mbox{Tr}\, e^{-\hat H_{\rm eff}/T} T_{\tau} 
e^{-\Action}$. 
(The calculation of $\Omega_{\rm ee}$
proceeds along the same lines as the calculation
of the interaction correction to the two-terminal conductance; 
Details of the latter
calculation can be found in Sec.\ \ref{sec:g}
and Appendix \ref{Ap:4}.)
To second order in the action $\Action$ ({\em i.e.}, up to
one branch cut contribution), we find
\begin{eqnarray}
  \Omega_{\rm ee} &=&
  - {T^2 \over 2} \sin {\pi \over \Nch}\, 
    \int_0^{\infty} dt_1 dt_2
    \int_{t_0}^{\infty} ds
    {\mbox{Tr}\, \Scat(t_1) \Scat^{\dagger}(t_2) \over 
    \sinh[\pi T(t_1+s)] \sinh[\pi T(t_2+s)]}
  \nonumber \\ && \ \ \mbox{} 
    \times \left\{ 
    {\sinh[\pi T(s-t_0)] \sinh[\pi T(s+t_1+t_2+t_0)] \over
    \sinh[\pi T(t_1+t_0)] \sinh[\pi T(t_2 + t_0)]} 
  \right\}^{{1 \over \Nch}}.
  \label{eq:OmegaAperiodic}
\end{eqnarray}

In order to find the differential capacitance (\ref{Cdif}) from
Eqs.\ (\ref{eq:Omegalbee}), (\ref{eq:OmegaDopen}), and
(\ref{eq:OmegaAperiodic}), we need to include the dependence
on the chemical $\mu$, which is done via the relation
\begin{equation}
  \Scat(\mu,t) = \Scat(0,t) e^{-i \mu t}.
\end{equation}
Hence, with help of Eq.\ (\ref{eq:muN}) we find for $\Ec \gg \Delta$
\begin{eqnarray}
  C_{\rm diff} &=&  C_{\rm diff, lb} + C_{\rm diff, ee}, 
  \label{eq:Cdiff} \\
  {C_{\rm diff, lb} \over C} &=&
  1 - {\Delta \over 2 \Ec}
  + {\Delta^2 \over 8 \pi \Ec} 
  \int_0^{\infty} dt \int_0^{\infty} dt' \mbox{Tr}\, 
  [\Scat^{\dagger}(-t') \Scat(t) - \langle \Scat^{\dagger}(-t') \Scat(t)
    \rangle]
  \nonumber \\  && \mbox{} \times
  { \pi T (t^2-t'^2)  
    \over 
    \sinh[\pi T(t-t')]} 
  e^{i \Delta {\cal N} (t-t')},
  \label{eq:Cdifflb}
  \\ 
  {C_{\rm diff, ee} \over C} &=&
  - {\Delta^2 \over 4 \pi^2 \Ec} 
    \sin {\pi \over \Nch} 
    \int_0^{\infty} dt \int_0^{\infty} dt'
    \int_{t_0}^{\infty} ds\, \mbox{Tr}\, \Scat^{\dagger}(-t') \Scat(t)
  \nonumber \\  && \mbox{} \times
    { \pi^2 T^2 (t - t')^2 
     \over 
    \sinh[\pi T(t+s)] \sinh[\pi T(t'+s)]}
  e^{i \Delta {\cal N} (t-t')}
  \nonumber \\ && \mbox{} 
    \times \left\{ 
    {\sinh[\pi T(s-t_0)] \sinh[\pi T(s+t+t'+t_0)] \over
    \sinh[\pi T(t+t_0)] \sinh[\pi T(t' + t_0)]} 
  \right\}^{{1 \over \Nch}}.
\end{eqnarray}
In the first term, $C_{\rm diff, lb}$, we have separated the 
ensemble average
and the mesoscopic fluctuations. 
The interaction correction $C_{\rm diff, ee}$ goes beyond mean-field
theory, and reveals the onset of Coulomb blockade. Its
ensemble average is zero, because of the presence of the factor
$(t-t')^2$ in the integrand. 

For temperatures $T \gg \Nch \Delta$, the capacitance fluctuations
can be computed from Eq.\ (\ref{eq:Cdiff}). After averaging
over $S(t)$ with the help of Eq.\ (\ref{eq:Savg}), one finds
\begin{equation}
  { \mbox{var}\, (C_{\rm diff}/C)} =
  {\pi^2 \over 12 \beta}
  \left[{\Delta \over \Nch\Ec}\right]^2  
  \left( {\Nch \Delta \over 2\pi^2 T}\right)
  \left\{1- {c \over 4\Nch}
  \left( {\Nch \Delta \over \pi^2 T}\right)^3
    \right\},
\label{eq:varcdiff}
\end{equation}
where $c$ is a numerical factor that takes the value $c \approx 3.42$ 
for $\Nch \to \infty$. The first term between brackets
$\{ \ldots \}$ represents the fluctuations of $C_{\rm diff,lb}$
(calculated to leading order order in $\Nch \Delta/T$ only),
while the second term is the leading interaction correction.
We conclude that for $\Nch \gg 1$ and/or $T \gg \Nch \Delta$
the mesoscopic capacitance fluctuations 
are dominated by the fluctuations of $C_{\rm diff, lb}$, corresponding 
to the mean-field theory of Ref.\ \cite{BuettikerJCondensC,BTP,BPT},
the interaction corrections being a factor 
$\sim \Nch^{-1} (\Nch \Delta/ \pi^2 T)^3$
smaller. Note that at $\pi T \sim \Nch \Delta/\pi$, which is where 
the crossover to the low-temperature limit of the
theory is expected, the contribution of $C_{\rm ee}$ to the
capacitance fluctuations is still smaller by a factor $\Nch$.

We close this section with the correlators of the (mean-field)
capacitance at two different gate voltages ${\cal N}_1$ and ${\cal
N}_2 = {\cal N}_1+s$ for $\Nch \gg 1$. From Eq.\ (\ref{eq:Cdifflb})
and (\ref{eq:Savg}), one finds for $T \gg \Nch \Delta$,
\begin{eqnarray}
K_\mu(s)
 &=& {3\, \mbox{var}\, (C_{\rm diff} /C)
  \over
  \sinh^2(\Delta s/2 T)}
  \left[{\Delta \over 2 T}s
  \coth\left({\Delta \over 2 T} s\right)
  - 1\right].
  \nonumber 
\end{eqnarray}
The fluctuations of $C_{\rm diff, lb}$ for $\Nch \gg 1$
in the low temperature limit $\pi T \ll \Nch \Delta/\pi$ 
read \cite{BTP,BrouwerBuettiker96}
\begin{eqnarray}
K_\mu(s) \label{eq:Kmus}
 &=&
  \left( {\Delta^2 \over 2 \Ec} \right)^{2}
  \langle \delta \nu({\cal N} \Delta)
  \delta \nu(({\cal N}+s) \Delta) \rangle\label{eq:rpa}
   \\ &=& \nonumber
  {4 \over \beta} 
  \left( {2 \Delta \over \Nch \Ec} 
  \right)^2
  {1 - 4 \left(2s / \Nch\right)^2
  \over
  \Big[
  1 + 4 \left(2s/ \Nch
  \right)^2 \Big]^2}.
\end{eqnarray}
where the correlator of the density of states was taken from
Ref.\ \cite{FyodorovSommersJMP}.\footnote{To 
reproduce this result from Eq.\ (\ref{eq:Cdifflb})
it is not sufficient to use a Gaussian average with the 
correlator (\ref{eq:Savg}), as non-Gaussian contributions
are also important for times of order $t \sim 1/\Nch \Delta$.} 
(The argument of the density of states is ${\cal N} \Delta$,
since fluctuations in the self-consistency relation (\ref{eq:muNmf})
can be neglected for this correlator for $\Nch \gg 1$.)

\subsection{Conductance through a dot with two almost reflectionless 
junctions}
\label{sec:g}

This subsection is devoted to the effects of interaction on the
two-terminal conductance of a dot with junctions to the leads 
that are either
completely reflectionless, or have only a small reflection
coefficient. The statistics of the
two-terminal conductance in the absence of the interactions
is reviewed in Ref.\ \cite{Beenakker97}.  
The effect of interaction on this
quantity was analyzed in Ref.~\cite{BrouwerAleiner99} and we follow
this reference in this subsection.

The Kubo formula for the conductance $G$ is given by
Eqs.~(\ref{eq:30.58}), (\ref{eq:30.56}), and
(\ref{eq:30.54}). Similarly to the consideration of the capacitance,
we expand the current-current correlation function (\ref{eq:30.58}) in
powers of the effective action $\Action$ and calculate the
arising fermionic correlation functions with the help of
Eqs.\ (\ref{eq:50.11}) -- (\ref{eq:50.13}) and
(\ref{eq:II}) -- (\ref{eq:IIphi4}).

The lowest non-vanishing contribution to the current-current
correlation function $\Pi(\tau)$, and hence to the
two-terminal conductance, is of the second order in the action,
\begin{equation}
\Pi(\tau)=
\langle I(\tau) I(0) \rangle_q
+ {1 \over 2} \langle I(\tau) I(0)  \Action^2\rangle_q 
    -
    {1 \over 2} \langle I(\tau) I(0) \rangle_q
    \langle \Action^2 \rangle_q + \dots.
\label{eq:70.1}
\end{equation}
We substitute into Eq.~(\ref{eq:70.1}) the explicit form of the
current operators (\ref{eq:30.54}) and the effective action 
(\ref{eq:30.482}).
After calculation of the averages of the fermionic operators with 
the help
of Eqs.~(\ref{eq:50.11}) -- (\ref{eq:50.13}) and (\ref{eq:II}) --
(\ref{eq:IIphi4}),
and with the approximation (\ref{eq:50.10}),
we find
\begin{eqnarray}
  G &=&G_{\rm lb} + G_{\rm ee} + G_{\rm osc}
\label{eq:70.2},\\
  G_{\rm lb} &=&\frac{e^2}{2\pi\hbar}
\left[\frac{\NL\NR}{\Nch}
\label{eq:70.3}
  - \int_{0}^{\infty} d t_1
  \int_0^{\infty} d t_2  
  {(t_1 - t_2) \pi T\, 
  \mbox{Tr}\, \Scat^{\dagger}(-t_1) \Lambda \Scat(t_2) \Lambda
  \over
  \sinh[(t_1 - t_2) \pi T]}
\right],
\\
 G_{\rm ee} &=&-\frac{e^2}{2\pi\hbar}
\left(\frac{1}{\pi} \sin {\pi \over \Nch}\right)
\int_{0}^{\infty} d t_1
  \int_0^{\infty} d t_2 
\mbox{Tr}\, \Scat^{\dagger}(-t_1) \Lambda \Scat(t_2) \Lambda 
\nonumber\\
  &&\mbox{} \times
\int_{t_0}^{\infty} ds\,  
\frac{(2 s + t_2 + t_1)\pi^2 T^2} 
 {\sinh[(s+t_1)\pi T] \sinh[(s+t_2)\pi T]}
\nonumber
\\
&&
 \mbox{} \times
  \left( {
  \sinh[(t_1+t_2+s+t_0) \pi T]
  \sinh[(s - t_0) \pi T]
  \over
  \sinh[(  t_1 + t_0) \pi T]
  \sinh[(  t_2 + t_0) \pi T]
  } \right)^{1/\Nch}.
\label{eq:70.4}
\end{eqnarray}
The charging energy enters into the expressions
only through the time $t_0$, see Eq.~(\ref{eq:50.10}), which has the
meaning of the characteristic time of the classical recharging of the
dot ($RC$-time).  Notice, that the final results are written in terms
of the scattering matrix $S$
of the dot in the absence of the interactions, see
Sec.~\ref{sec:bosonization}. In this order of perturbation theory,
the term $G_{\rm osc}$ oscillating with gate voltage is present only
for the case $\NR=\NL=1$ (spinless fermions, and both contacts carry 
a single channel each):
\begin{eqnarray} 
G_{\rm osc} &=& 
   {e^2 \over
      4 \pi^2 \hbar} \int_{0}^{\infty} d t_1\, dt_2\, 
   \int_{t_0}^{\infty} ds\,
  { (2 s + t_2 + t_1) \pi^2 T^2 
  \over
  \sqrt{  \sinh[(  t_1 + t_0) \pi T]
  \sinh[(  t_2 + t_0) \pi T]}}
  \nonumber \\ && \mbox{} \times
  { 1 \over \sqrt{\sinh[(t_2 + t_1 + s + t_0) \pi T]
  \sinh[(s - t_0) \pi T] 
   }}
  \nonumber \\ && \mbox{} \times
  \mbox{Re}\, e^{ - 2 \pi i {\cal N}} 
  \left[ \Scat_{11}(t_1) \Scat_{22}(t_2) + \Scat_{12}(t_1) \Scat_{21}(t_2) \right].
\label{eq:70.5}
\end{eqnarray}
The derivation of Eqs.~(\ref{eq:70.2}) -- (\ref{eq:70.5})
is relegated to Appendix~\ref{Ap:4}.

Equation (\ref{eq:70.3}) coincides with the Landauer formula
for the
conductance of non-interacting electrons (\ref{eq:2.63a}).  Equation
(\ref{eq:70.4}) is the leading interaction correction at temperatures
much smaller than the charging energy. The power-law form-factor (the
last line in this formula) describes the non-Fermi liquid behavior of
the dot when $\Nch > 1$. Finally, the Coulomb blockade like
contribution (\ref{eq:70.4}) for $N_1=N_2=1$ is somewhat similar to
the oscillation of the capacitance, Eq.~(\ref{eq:60.2}).

To gain intuition about the structure of the interaction correction,
we start with the simplest example of instantaneous 
reflection from the contact, when $\Scat(t)$ is given by 
Eq.\ (\ref{eq:StdirectMulti}). 
Neglecting the fluctuating part of $S$ and substituting
Eq.~(\ref{eq:StdirectMulti}) into Eq.~(\ref{eq:70.3}), we find
\begin{equation}
G_{\rm lb} = \frac{e^2}{2\pi\hbar}
\left[\frac{\NL\NR}{\Nch}
-
 \mbox{Tr}\, r \Lambda r^{\dagger} \Lambda
\right],
\label{eq:70.7}
\end{equation}
The first term in Eq.~(\ref{eq:70.7}) is nothing but the ``classical''
conductance of the two reflectionless junctions connected in series,
and the second term comes from the correction to the transmission
amplitudes due to the finite backscattering in the contacts.

Substitution of Eq.~(\ref{eq:StdirectMulti}) into Eq.~(\ref{eq:70.4}) 
yields
\begin{equation}
G_{\rm ee} = - \frac{e^2}{2\pi\hbar}
 \mbox{Tr}\, r \Lambda r^{\dagger} \Lambda
\left[
\frac{c_{\Nch} \Nch\sin {\pi \over \Nch}}
{4\pi}
\left({\Nch \Ec e^{\bf C} \over \pi^2 T} \right)^{2/\Nch} 
-1
\right],
\label{eq:70.8}
\end{equation}
where $c_{\Nch}$ is a numerical coefficient that depends only weakly
on $\Nch$, ranging from $c_4 \approx 5.32$ to $c_{\infty} = 4$. 
Adding Eqs.\ (\ref{eq:70.7}) and (\ref{eq:70.8}), we obtain
\begin{equation}
G = \frac{e^2}{2\pi\hbar}
\left[\frac{\NL\NR}{\Nch}
-
\left({\Nch \Ec e^{\bf C} \over \pi^2 T} \right)^{2/\Nch}
\left(\frac{c_{\Nch} \Nch\sin {\pi \over \Nch}}{4\pi}\right)
\mbox{Tr}\, r \Lambda r^{\dagger} \Lambda
\right].
\label{eq:70.9}
\end{equation}
Comparing
Eq.~(\ref{eq:70.9}) with Eq.~(\ref{eq:70.7}), we see that the ``bare''
reflection amplitude entering in the Landauer formula is
renormalized by the interaction.
The renormalization is
temperature-dependent, manifesting the non-Fermi liquid behavior of
the open dot on the energy scale $\Ec\gtrsim T\gtrsim\Delta$.  For
$N_1=N_2=2$ (electrons with spin, and both contacts carry
one channel each) the result (\ref{eq:70.9}) has been obtained
previously in Ref.~\cite{Furusaki95}. 

The above example (\ref{eq:StdirectMulti}) of reflection at the moment
$t=0$ corresponds to case where the electron is backscattering
instantaneously, without entering the dot.
Now we are going to estimate the interaction-induced
renormalization of the amplitude for a backscattering event occurring
somewhere within the dot. In the language of the electron
trajectories, such an event corresponds to a trajectory returning
to the contact at some time $t_s > 0$,
\begin{equation}
  \Scat_{ij}(t) = r_{ij} \delta(t-t_s), \quad t_s>0. \label{eq:Stdelayed}
\label{eq:70.10}
\end{equation} 
It is easy to see that the noninteracting contribution 
(\ref{eq:70.7}) remains intact, whereas the interaction contribution 
depends significantly on $t_s$:
\begin{eqnarray}
  G_{\rm ee} &=& - \frac{e^2}{2\pi\hbar}
 \mbox{Tr}\, r \Lambda r^{\dagger} \Lambda \label{eq:70.11}
 \\
 &&\mbox{   }\times
\left\{ 
\matrix{ \displaystyle
\left( {\Nch \Ec  \over  T} \right)^{2/\Nch}_{\vphantom{M_M}}
-1, 
& t_s \ll \left(\Nch \Ec\right)^{-1} \cr
\displaystyle \left({ 1 \over t_s T} \right)^{2/\Nch} - 1, &
  (\Nch \Ec)^{-1} \ll 
t_s \ll T^{-1}
 \cr
\displaystyle e^{-2\pi Tt_s}, & t_s \gg T^{-1}.
}
\right.
\nonumber
\end{eqnarray}
(for brevity, we omitted here the numerical coefficients of the order
of unity). We see that for
short trajectories, with return times $t_s$ smaller than the $RC$-time
for the open dot, see Eq.~(\ref{eq:50.10}), the renormalization of the
scattering amplitude is complete, {\em i.e.}, the same as for the direct
reflection from the contact we studied before. 
On the other hand, the
effect of interaction on the scattering events occurring ``deep
inside'' the dot, $t_s\gg 1/T$ is exponentially weak. As long as
the temperature is sufficiently high, $T\gg \Delta \Nch$, and the 
scattering from the dot is ergodic, the majority
of the scattering events is not renormalized by the
interaction. Therefore, the Landauer part of the
conductance (\ref{eq:70.7}) remains intact at such temperatures. The
interaction correction affects only a small portion of scattering
events, and thus only slightly affects the statistics of the two
terminal conductance. Note that the characteristic
energy $\Nch\Delta$ that appears here
is nothing but the inverse escape time
for a quasiparticle from the dot into one of the leads.

To study the statistics of the two-terminal conductance, one should
use the statistical properties of the scattering matrix 
(\ref{eq:60.7}),
and Eqs.~(\ref{eq:70.2}) -- (\ref{eq:70.4}).
The result substantially depends on whether we are dealing with
spinless fermions, denoted by the spin-degeneracy factor $g_s =1$,
or electrons with spin, denoted by $g_s=2$. In the
latter case, the results are written in terms of the number of orbital
channels $\Nch^o = \Nch/g_s$ (and, similarly, $N_{1}^{o} = N_1/g_s$,
$N_{2}^{o} = N_2/g_s$). Below
we concentrate on the case of the reflectionless contacts; all
backscattering events occur within the dot. For the averaged
conductance, we find~\cite{BrouwerAleiner99}
\begin{eqnarray}
\langle G \rangle &=&
\frac{g_se^2}{2\pi \hbar}
\left\{\frac{\NL^o\NR^o}{\Nch^o}
\label{eq:70.12}
\right. \\ && \left. \mbox{} 
+ \left(1 - {2 \over \beta} \right) 
\frac{\NL^o\NR^o}{\Nch^o\left(\Nch^o+2/\beta-1\right)}
\left[1 + {c_{\Nch}  \over \Nch^o} 
\left(
{\Nch^o \Delta \over g_s \pi^2 T}
\right) \right]
\right\},
\nonumber
\end{eqnarray}
where $\beta=1$ ($2,4$) for the orthogonal (unitary, symplectic) 
ensemble, and
$c_{\Nch}$ is an $\Nch$-dependent numerical constant, ranging from 
$c_4 \approx 4.77$ to $c_{\infty} = \pi^2/6$. 
Equation (\ref{eq:70.12}) is valid for an arbitrary number of channels
except the case of $g_s=1$, $\NL=\NR=1$. For the latter case, one finds
\begin{equation}
\langle G \rangle
=
\frac{e^2}{2\pi \hbar}
\left\{\frac{1}{2}
- \delta_{\beta,1}
 \left[ \frac{1}{6}  +
 { \Delta \over 8 \pi T}
\ln\left( {4 \Ec e^{\bf C} \over c \pi^2 T} \right)
\right] \right\},
\label{eq:70.13}
\end{equation}
where $c \approx 2.81$.

Similar to Eq.~(\ref{eq:70.7}), here the first term in brackets is the
classical resistance of two point contacts connected in series.  The
second term is the familiar weak localization correction for
non-interacting electrons [this term replaces the correction to the
Kirchoff law coming from the backscattering in the junctions, see
Eq.~(\ref{eq:70.7})]. Finally the last term is the high-temperature
expansion of the interaction correction. This correction comes from
the renormalization of the amplitudes of the backscattering events
occurring in the dot within ``time horizon'' $t_s\sim 1/T$,
cf. Eq.~(\ref{eq:70.11}). At high temperature, the interaction
correction is less significant than the correction from the weak
localization.  Notice, however, that at $\pi T=\Delta \Nch/\pi$,
which is the lower limit of the applicability of the theory, the
interaction correction is smaller than the weak localization
correction only by parameter $1/\Nch$, and may be still important
for the temperature dependence of magnetoresistance.

One notices also that for the unitary ensemble, $\beta=2$, both the
weak localization and interaction corrections to the average
conductance vanish, in contrast with the usual situation in disordered
metals \cite{Altshuler85}. The reason is that the interaction enhances
any electron scattering, see Eq.~(\ref{eq:70.11}), no matter whether
it is scattering from one lead to the other ($1 \leftrightarrow 2$)
or backscattering into the same lead ($1\leftrightarrow 1$ and
$2\leftrightarrow 2$). For the unitary ensemble, the scattering
$1\leftrightarrow 1$ and $1\leftrightarrow 2$ occur with equal
probabilities and therefore neither reflection nor transmission are
enhanced in average. On the other hand for the orthogonal ensemble,
backscattering into the same lead 
is more probable due to the weak localization, and
the interaction further increases the backscattering probability.

In a similar manner, the mesoscopic fluctuations are also affected by
the electron-electron interaction. We find from Eqs.~(\ref{eq:70.2})
-- (\ref{eq:70.5}), and Eq.~(\ref{eq:60.7}), in the regime $T \gg
\Delta \Nch$,
\begin{eqnarray}
  \mbox{var}\, G &=& {1 \over \beta} 
  \left(\frac{g_s e^2}{2\pi\hbar}\right)^2
  \left( \NL^o \NR^o \over \Nch^{o2} \right)^2
  \left[{\Nch^o \Delta \over 6 T} + 
  {c'_{\Nch} \over \Nch^{o}} {\Nch^{o2} \Delta^2 \over g_s \pi^4 T^2}
\right],
  \label{eq:UCF}
\end{eqnarray}
with $c' \approx 6.49$ for $\Nch \gg 1$. The first term in
Eq.~(\ref{eq:UCF}) represents the high temperature asymptote of the
mesoscopic fluctuations of non-interacting electrons
\cite{Efetov95}. The RMT index $\beta=1$ ($2$) in the presence
(absence) of time-reversal symmetry; $\beta=4$ for the symplectic
ensemble, which corresponds to the presence of
both strong spin-orbit scattering and
time-reversal symmetry.  
For the case $\NL=\NR=1$, $g_s=1$, there is an explicit ${\cal
N}$-dependence of the conductance fluctuations,
\begin{eqnarray}
  \label{eq:GNNfluct}
  \langle G({\cal N}) G({\cal N}') \rangle &=& 
  \left( {e^2 \over 2 \pi \hbar} \right)^2 
  \\ && \mbox{} \times
  \left( 1 + 2 \delta_{\beta,1} \right)
  {\Delta^2 \over 64\, \pi^2T^2} \{1+\cos[2 \pi ({\cal N}-{\cal N}')]\}
  \ln^2 \left({\Ec \over T} \right), \nonumber
\end{eqnarray}
where $\Ec$
is the charging energy of the dot and $C\approx 0.577$ is the Euler
constant. 
In this order of perturbation theory, such an explicit
${\cal N}$-dependence, which reflects the discreteness of charge,
exists only for spinless fermions, but is absent for particles with spin.
For spin $1/2$ electrons, effects of the discreteness of charge appear
only in the $\Nch$-th order in perturbation theory in $S$,
similarly to Eq.~(\ref{eq:60.2}). For a large number of
reflectionless channels, the characteristic amplitude of mesoscopic
fluctuations goes down very rapidly as $(1/\Nch)^{\Nch}$.

The situation here is similar to that of the differential
capacitance, see Sec.\ \ref{sec:cap},
where we found that for $\Nch > 2$ the average and variance 
were dominated by the mesoscopic fluctuations of the non-interacting
density of states, while the interaction
corrections were smaller by a factor $\Nch^{-1} (\Nch \Delta/ \pi^2
T)^3$ for $T \gg \Nch \Delta$. The periodic-in-${\cal N}$ Coulomb 
blockade effects become
dominant for $\Nch \le 2$, thus limiting the range of
validity of the 
non-interacting (or mean-field) theories to larger 
values of $\Nch$. This limitation is more 
relevant for the
capacitance, where the case $\Nch=2$ can be realized experimentally
for electrons
with spin and a single-mode point contact, than for the conductance,
were the case $\Nch=2$ is mainly of academic interest, as it requires
spinless electrons.

\subsection{Conductance in the strongly asymmetric setup}
\label{sec:dos}

We consider here a dot with two leads, one of which is connected to
the dot by a tunneling contact with a small transmission coefficient,
while the other is coupled to the dot via an ideal (no reflection) or 
almost ideal (small $r$) point contact. The first lead serves as a 
tunneling probe of the ``open'' quantum dot. See the discussion in
Secs.\ \ref{sec:2.2} and \ref{sec:action}.

{\em Inelastic tunneling.} -- We begin with the analysis of the
inelastic contribution to the tunneling conductance.  According to
Sec.~\ref{sec:bosonization}, the order of perturbation theory is
characterized by the number of the branch cuts the corresponding
fermionic correlator includes.  In the leading order, we will allow
only one branch cut in the complex $\tau$ plane and collect all the pole
contributions in Eqs.~(\ref{eq:40.10}) and (\ref{eq:40.11}).
Physically, this corresponds to neglecting the process of escape of
the tunneled electron into another lead. With the help of
Eqs.~(\ref{eq:str1}), (\ref{eq:strall}), we find
\begin{eqnarray*}
&&\Pi_{\rm in}^{(0)}(\tau)=
-{\cal G}_{11}(-\tau)\left|\frac{f(\tau)}{f(0)}\right|^{2/\Nch},
\\
&&\Gamma_{ij}(\tau;\tau_1,\tau_2)
= T \left|\frac{f(\tau)}{f(0)}\right|^{2/\Nch}
\nonumber \\ && \hphantom{\Gamma_{ij}(\tau;\tau_1,\tau_2)=}
\mbox{} \times
\sum_{\omega_n}
e^{i\omega_n(\tau_1-\tau_2)}
\left[
i\pi\nu{\mathrm sgn}\omega_n 
+\pi^2\nu^2\Wmatrix^\dagger
\Greenmatrix_o(i\omega_n)\Wmatrix 
\right]_{ij},
\end{eqnarray*}
where $\omega_n$ is the fermionic Matsubara frequency,
and Green function for the open dot $\Greenmatrix_o$,
is defined in Eq.~(\ref{eq:Gopen}).\footnote{We remind that the
Matsubara Green function is expressed in terms of its  retarded and
advanced counterparts as $\Greenmatrix(i\omega_n) = 
\theta(\omega_n) \Greenmatrix^R(i\omega_n) +
\theta(-\omega_n) \Greenmatrix^A(i\omega_n)
$}
This gives with the help of Eq.~(\ref{eq:2.68})
\begin{equation}
\Pi_{\rm tunn}^{(0)}(\tau)=
-\nu_T(-\tau)\left|\frac{f(\tau)}{f(0)}\right|^{2/\Nch},
\label{eq:80.1}
\end{equation}
where the dimensionless function $f(\tau)$ is defined in
Eq.~(\ref{eq:50.9}), and the Matsubara counterpart of the tunneling
density of states $\nu_T(\tau)$ is related to the exact tunneling
density of state of noninteracting system $\nu_T(\varepsilon)$, see
Eq.~(\ref{eq:2.68}), by
\begin{equation}
\nu_T(\tau)= \frac{1}{2}\int d \varepsilon\, \nu_T(\varepsilon)
e^{-\varepsilon\tau}\left[{\rm sgn}\tau +\tanh\frac{\varepsilon}{2T}\right],
\label{eq:80.2}
\end{equation} 
at $-1/T < \tau < 1/T$. It is noteworthy, that the inelastic co-tunneling
contribution is expressed in terms of the Green functions of the open
dot, rather than that of the closed dot in accordance with the discussion
in the previous subsection. In the absence of the interaction, the prefactor
$f(\tau)/f(0)$ equals unity and Eqs.~(\ref{eq:80.1}), (\ref{eq:80.2}),
and Eq.~(\ref{eq:40.5}) reproduce the formula for non-interacting model
(\ref{eq:2.66}).

It follows from Eqs.~(\ref{eq:50.9}),
(\ref{eq:80.1}), and (\ref{eq:80.2}) that $\Pi(\tau)$ is an analytic
function of $\tau$ for $0 < {\rm Re}\, \tau < 1/T$, so that we can
use Eq.~(\ref{eq:40.5}) for the tunneling conductance. For $T \ll 
\Nch \Ec$ we find
\begin{eqnarray}
G_{\rm in} &=& G_1  \left({\pi^2 T \over \Ec \Nch e^{\bf C}}\right)^{2/\Nch}
\int_{-\infty}^\infty d\varepsilon 
\frac{ M\Delta \nu_T(\varepsilon)}{4 \cosh {\varepsilon \over 2T}} 
\int_{-\infty}^\infty dt e^{i \varepsilon t}
\left(\frac{1}{\cosh \pi t T}\right)^{2/\Nch + 1 }
, \nonumber \\
\label{eq:80.3}
\end{eqnarray}
where $C\approx 0.577$ is the Euler constant. 
Here we used the approximation (\ref{eq:50.10}), justified at $T \ll
\Nch \Ec$.

To find the averaged conductance, we notice that $\langle \nu_T \rangle
= 1/(M\Delta)$. Then, the integration in Eq.~(\ref{eq:80.3}) immediately
yields
\begin{equation}
\langle G_{\rm in}\rangle
= G_1 \left({\pi^2 T \over \Ec \Nch e^{\bf C}}\right)^{2/\Nch}
\left[
\frac{\sqrt{\pi} \Gamma \left(1+1/\Nch\right)}
{2  \Gamma \left(3/2 +1/\Nch\right)}
\right],
\label{eq:80.4}
\end{equation}
where $\Gamma(x)$ is the Gamma function.
Equation (\ref{eq:80.4}) is valid for reflectionless point contacts,
provided that $\pi^2 T \gg \Nch \Delta$.

Note that Eq.~(\ref{eq:80.4}) predicts a power-law temperature
dependence of the tunneling conductance. The power-law factor can be
understood in terms of the well-known orthogonality
catastrophe~\cite{Anderson67}. An addition of an electron without a
re-distribution of charge between the dot and lead would cost an
amount of energy $\sim \Ec$. It means that the
tunneling spectrum would have a ``hard'' gap
with the width of the order of charging energy, and the temperature
dependence of the conductance would display an activation
behavior. The power-law (\ref{eq:80.4}) indicates that a
redistribution of charge indeed occurs when an additional electron
tunnels onto the dot. In fact, each channel of the ideal dot-lead
junction carries away charge $e/\Nch$. According to the Friedel sum
rule, it corresponds to an additional phase shift $\delta=\pi/\Nch$
for the scattering in each of the channels. Tunneling of an electron
with energy $\varepsilon\ll \Ec$ creates an excited state of the
electron system consisting of the dot and lead. The overlap of this
state with the initial state of the system defines the probability of
tunneling. The smaller the energy $\varepsilon$, the smaller such
overlap is (orthogonality catastrophe~\cite{Anderson67}). This overlap
can be expressed in terms of the phase
shifts~\cite{Anderson67,Nozieres69}, and is proportional to
$(\varepsilon/\Ec)^\chi$, where $\chi=\sum (\delta/\pi)^2$, and the
sum is taken over all the modes. In our case, the charge is
``shifted'' through $\Nch$ channels inside the dot, and through the
same number of channels in the lead, yielding $2\Nch$ equal
contributions to the sum; $\chi=2/\Nch$. This exponent indeed
coincides with the one in the power law (\ref{eq:80.4}).

A particular case of Eq.~(\ref{eq:80.4}) for one-channel junctions
deserves additional discussion. For spinless fermions, $\Nch=1$, one
obtains
\begin{equation}
\langle G_{\rm in}\rangle
= \frac{2 G_1}{3}\left({\pi^2 T \over \Ec e^{\bf C}}\right)^{2}
, \quad \Nch =1
\label{eq:80.5}
\end{equation}
We see that in this case the temperature dependence is similar to
that for a strongly blockaded dot, see Eq.~(\ref{eq:3.2.28}).  The
$T^2$ dependence can be obtained from the usual phase space argument
for the quasiparticle lifetime, and is a manifestation of the Fermi
liquid behavior of a one-channel dot, see Sec.~\ref{sec:qualit} for a
qualitative discussion.  The temperature dependence of the tunneling
conductance for the electrons with spin, $\Nch=2$, is qualitatively
different \cite{Furusaki95}:
\begin{equation}
\langle G_{\rm in}\rangle
= {G_1}\left({\pi^3 T \over 8\Ec e^{\bf C}}\right), \quad \Nch =2.
\label{eq:80.6}
\end{equation}
The temperature exponent here is smaller than the one given by the
phase space argument. This indicates non-Fermi liquid behavior,
which we have already discussed in Sec.~\ref{sec:lowen}. If there is
no scattering in the junction ($r=0$), and, in addition, one neglects
backscattering from within the dot (corresponding to the limit
$\Delta/\Ec\to 0$), then
the result (\ref{eq:80.6}) is valid at arbitrarily low
temperature. The existence of a finite backscattering, however,
restores the Fermi-liquid behavior at sufficiently low temperature. We
have already discussed in Sec.~\ref{sec:lowen} the mechanism that
recovers the Fermi liquid behavior in the case of $r\neq
0$. Reflection is a relevant perturbation; at low energies it becomes
stronger, reaching $r_{\rm eff}\sim 1$ at $\varepsilon\sim T_0({\cal
N})$. The characteristic energy scale $T_0({\cal N})$ is given by
Eq.~(\ref{T0ex}). At smaller energy scales the dot effectively is
weakly coupled to both leads, and the strong Coulomb blockade behavior
[cf. Eq.~(\ref{eq:3.2.28})] is restored. The Fermi-liquid result
\begin{equation}
\langle G_{\rm in}\rangle
= \frac{8\pi^2}{3e^{\bf C}} G_1\frac{T^2}{\Ec T_0({\cal N})},
\quad T\ll T_0({\cal N})
\label{eq:80.7}
\end{equation}
matches Eq.~(\ref{eq:80.6}) at the upper limit of the allowed
temperature interval. We refer the reader to the original reference
\cite{Furusaki95} for the rigorous derivation of the numerical
coefficient in Eq.~(\ref{eq:80.7}).

We see that inelastic contribution vanishes at $T=0$ in a close
resemblance of the strong Coulomb blockade, see
Sec.~\ref{sec:3.2}. Pursuing the analogy with the strongly blockaded
regime further, one may expect, that there should be the another
contribution analogous to the elastic co-tunneling for the dots weakly
connected to the leads, and we turn to the analysis of this mechanism
now.

{\em Elastic tunneling.} -- To obtain the leading elastic contribution
we calculate correlator  Eq.~(\ref{eq:40.11})
taking into account all terms which include a branch cut in the 
complex $\tau$ plane and two additional branch cuts: the cuts
in the complex $\tau_1$ and $\tau_2$ planes in the
correlation function 
(\ref{eq:str2}), and similarly in higher order correlation functions,
see Eq.~(\ref{eq:strall}).
After analytic continuation (\ref{eq:40.5}), see also
Appendix~\ref{Ap:tun}, we find for $\Nch \geq 2$
\begin{eqnarray}
&&G_{\rm el}=\frac{ G_1 M\Delta}{4} \int_{-\infty}^\infty dt dt_1dt_2
 \frac{\pi T }{\cosh (\pi Tt) \cosh [\pi T (t+t_1-t_2)]}
\nonumber \\ &&\hphantom{G_{\rm el}=} \mbox{} \times
\left[
\left(\frac{f(-it_1)f(it_2)}{f(it_2-it-1/2T)f(-it_1+it+1/2T)
}\right)^{1/\Nch} -1 \right],
\nonumber \\ && \hphantom{G_{\rm el}=} \mbox{} \times
\left|\frac{f(it+1/2T)}{f(0)}\right|^{2/\Nch} 
{\rm Re}\, \left[{\cal G}_o^R(t_1)\nu WW^\dagger {\cal G}_o^A(t_2)
\right]_{11} 
\label{eq:80.9}
\end{eqnarray}
where ${\cal G}_o(t)^{R,A}$ are the exact Green functions for the open
dot without interactions, see Eq.~(\ref{eq:Gopen}). 

For the case of a one channel contact ($\Nch=1$; spinless fermions) an
additional contribution appears, which is reminiscent of the
Coulomb blockade oscillations
\begin{eqnarray}
&&G_{\rm el}=\frac{ G_1 M\Delta}{4} \int_{-\infty}^\infty dt dt_1dt_2
{\rm Re}\, e^{-i2\pi{\cal N} -i\pi/2}
\left[{\cal G}_o^R(t_1)\nu WW^\dagger {\cal G}_o^R(t_2)
\right]_{11}
 \nonumber\\
&&\mbox{}\times
\frac{f(it+it_1+it_2+1/2T)}{t_0f(0)\cosh \pi Tt}   
\nonumber\\
&&\mbox{}\times
\left|\frac{f(it+1/2T)}{f(0)}\right|^{2}
\frac{f(-it_1)f(-it_2)}{f(-it_2-it-1/2T)f(-it_1+it+1/2T)}.
\label{eq:80.90}
\end{eqnarray} 

Similar to the discussion of the case of two-terminal conductance,
we apply Eq.~(\ref{eq:80.9}) to the simplest model situation
first and only then consider the statistical properties of the
conductance. Imagine, there is a short semiclassical 
 trajectory connecting the tunneling contact and the lead:
\begin{equation}
\left[{\cal G}_o^R(t_1)\nu WW^\dagger {\cal G}_o^A(t_2)
\right]_{11} \simeq \frac{\gamma}{M}\delta(t_1-t_e)\delta(t_2+t_e),
\label{eq:80.10}
\end{equation}
where $t_e$ is the time for an electron to travel along this trajectory
and $\gamma$ is the constant characterizing the coupling of the
tunneling contact with this trajectory.
Substituting Eq.~(\ref{eq:80.10}) into Eq.~(\ref{eq:80.9}), 
we obtain for $T\ll \Ec$
\begin{equation}
G_{\rm el}\simeq\gamma G_1
\times\left\{
\matrix{1, & t_{e}\Nch \Ec <1,\cr
\left(t_{e}\Nch \Ec\right)^{-2/\Nch}, & t_0\Nch \Ec > 1,
}
\right.
\label{eq:80.11}
\end{equation}
where we omitted the non-essential numerical factors.

Equation (\ref{eq:80.11}) indicates that the contribution of the
processes where the time that the an electron spends in the dot 
time is smaller than the
recharging time $\hbar /(\Nch \Ec)$ is not affected by the interaction
at all. This was to be expected, because during such a short time 
the other
electrons do not manage to redistribute themselves. On the other hand,
the longer trajectories experience a power law suppression. The
physical reason for this is the orthogonality catastrophe with the
index $\chi=2/\Nch$ which we already faced in the discussion of the
inelastic contribution.

We are ready now to perform the statistical analysis for the tunneling
conductance of the system. We will limit ourselves to the case of
a single mode point contact and
electrons with spin ({\em i.e.}, $\Nch = 2$; results for spinless 
fermions can be found in
Ref.~\cite{AleinerGlazman98}). To find the average conductance we use
Eq.~(\ref{eq:GopenRA}), and the fact that for reflectionless contacts
$\nu {\rm Tr}WW^\dagger = \pi^2 \Nch \Delta M$.  Integration in
Eq.~(\ref{eq:80.9}) is then easily performed and gives for $\Delta \ll
T \ll \Ec$

\begin{equation}
\langle G_{\rm el} \rangle = G_1 \frac{ \Delta e^{- {\bf C}}}{2 \Ec}
\ln\left(\frac{\Ec}{T}\right),
\label{eq:elastic}
\end{equation}
where ${\bf C}\approx 0.577$ is the Euler constant. At very low
temperatures, $T$ should be substituted by $\Delta \ln(\Ec/\Delta)$.
The reason for the logarithmic divergence in Eq.\ (\ref{eq:elastic})
can be understood from Eq.~(\ref{eq:80.11}): the contribution of 
trajectories
with characteristic time $t_e$ scales as $1/t_e$, therefore the sum
over all the trajectories logarithmically diverges.

Similar to the case of the strong Coulomb blockade, the elastic co-tunneling
is a strongly fluctuating quantity,
\begin{equation}
\frac{\langle \delta G_{\rm el}^2 \rangle}{\langle G_{\rm el} \rangle^2}
=\frac{2}{\beta},
\label{eq:elasticmeso}
\end{equation}
where $\beta=1(2)$ for the orthogonal (unitary) ensemble.

If a finite backscattering in the channel is included, the energy scale
(\ref{T0ex}) appears. Below this scale, the Fermi-liquid behavior of
the system is restored, see also Eq.~(\ref{eq:80.7}), so that the low
temperature limit of the elastic co-tunneling becomes an oscillating
function of the gate voltage ${\cal N}$,
\begin{equation}
\langle G_{\rm el} \rangle = G_1 \frac{ \Delta e^{- {\bf C}}}{2 \Ec}
\ln\left(\frac{\Ec}{ T_0({\cal N})}\right).
\label{eq:elastic1}
\end{equation}
The relationships (\ref{eq:elasticmeso}) for mesoscopic fluctuations
still holds.

\begin{figure}
\epsfxsize=0.8\hsize
\centerline{\epsffile{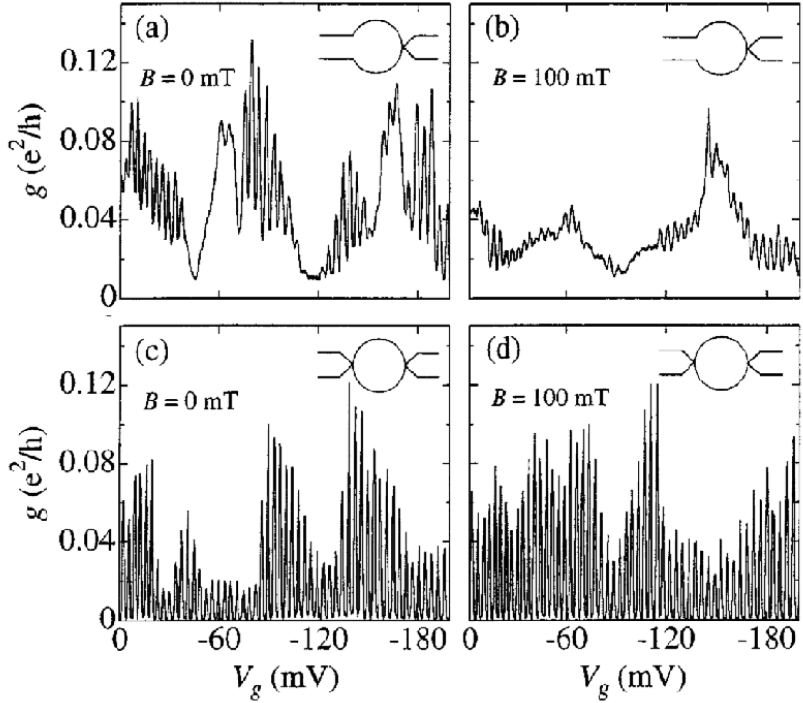}}
\caption{Conductance showing Coulomb blockade oscillations as a
  function of gate voltage $V_g$ in the open-channel regime [(a), (b)]
  and weak-tunneling regime [(c), (d)] at $B=0$mT, and $B=100$mT (the
  crossover between the Orthogonal and Unitary ensembles occurs at
  $B\approx 20$mT for this quantum dot), from
  Ref.~\protect\cite{fryingpan}. Note that the oscillations in the
  open-channel case are stronger at $B=0$mT compared to $B=100$mT,
  unlike the weak-tunneling Coulomb blockade.}
\label{Fig25}
\end{figure}

For the case of a reflectionless junction, the periodic-with-${\cal
N}$ oscillations
appear only in the next order perturbation theory in the
scattering matrix
$\Scat$. At large temperatures $T > \Delta \ln (\Ec/\Delta)$
they are smaller than the aperiodic part of the fluctuations. Calculation
of the periodic part of the fluctuations gives \cite{AleinerGlazman98}
\begin{equation}
\frac{\langle \delta G_{\rm el}({\cal N}_1) \delta G_{\rm el}({\cal N}_2)
\rangle}{\langle G_{\rm el} \rangle^2}
=\frac{0.33}{\beta^2}\frac{\Delta}{T}\ln\left(\frac{\Ec}{T}\right)
\cos 2\pi ({\cal N}_1-{\cal N}_2).
\label{eq:mesoosc}
\end{equation}
It is important to emphasize that this part is more sensitive to the
magnetic field than the aperiodic part: it reduces by a factor of four
rather than by a factor of two. (Noninteracting models for the 
quantum dot predict the aperiodic fluctuations only, and hence fail
to describe this anomalous magnetic field dependence.) This
prediction was checked experimentally in Ref.~\cite{fryingpan}. The
analysis of the Coulomb blockade oscillations at $B=0$ (Orthogonal
ensemble) and at $B=100$mT (Unitary ensemble), see Fig.~\ref{Fig25},
yielded a suppression factor $5.3$ for the fluctuations of the
conductance. This is somewhat larger than the predicted factor, for
reasons not yet understood. 

Also note that the average conductance in the case
of a symmetric weak-tunneling setup [panels (c) and (d) of
Fig.~\ref{Fig25}] grows with the application of the symmetry-breaking
magnetic field. This is in agreement with the low temperature
RMT predictions
discussed in Section~\ref{sec:3.1}, see
Eqs.~(\ref{eq:3.1.11})--(\ref{eq:3.1.13}). If the two junctions to the
dot are of the same average conductance, then the crossover from the
Orthogonal to the Unitary ensemble should result in the increase of
$\langle G_{\rm peak}\rangle$ by a factor of $4/3$ at low
temperatures.
More detailed measurements of the magnetic field dependence of
the average peak heights were reported recently in Ref.\
\cite{wlcb}.

\section{Conclusions}

In the beginning of this Review, see
Subsections~\ref{sec:2.1}-\ref{sec:hfinal}, we have derived the model
Hamiltonian of an isolated dot. One of the main assumptions in that
derivation was that the electron states are random. Randomness of the
one-electron states, a large dimensionless conductance inside the 
dot ($g\equiv
E_T/\Delta \gg 1$), and relative weakness of the electron-electron
interaction ($r_s\lesssim 1$) allowed for a relatively simple
description of an isolated dot.\footnote{This simple model is not
  applicable to dots with a small number of electrons where the
  dimensionless conductance is small. It is also not applicable for
  dots of integrable shape \cite{atom}, where the shell structure
  of the orbital states is dominant.} The main part of the Hamiltonian
of the dot is universal; it consists of a free-electron part
described by Random Matrix Theory and an interaction term, which
is not random, see Eq.~(\ref{eq:2.54}). Corrections to this
universal part of the Hamiltonian are small at $g\gg 1$. In the
discussion of the transport phenomena and of other effects which
result from the contact of the dot with external electron
reservoirs (leads), we have considered that universal part of the
Hamiltonian only.
Despite this extremely simple description of the physics inside the
quantum dot, a large variety of intricate physical effects is
uncovered when the dot is coupled to the outside world via point
contacts. These effects range from quantum
interference transport phenomena
in the regime strong Coulomb blockade (Subsections~\ref{sec:3.1} and
\ref{sec:3.2}) to the effects of quantum charge fluctuations and
non-Fermi liquid behavior of partially open quantum dots
(Subsection~\ref{sec:open}).  Remarkably, the low temperature behavior
of transport through a quantum dot with a finite spin is
always dominated by the Kondo effect, see Subsections~\ref{sec:Kondo}
and \ref{sec:exchange}.

In this concluding Section, we will mention three areas of research
on quantum effects in quantum dots that are not covered in this review.

The first area is related to the problem of electron relaxation in a
constrained geometry. It is well-known that in a bulk clean Fermi
liquid the relaxation rate of an electron with energy $\varepsilon$ is
$1/\tau_{\varepsilon}\sim\varepsilon^2/E_F$. In a confined system, an
electron occupying a discrete level with energy $\varepsilon$ can
decay with the excitation of an electron-hole pair. It was shown in
Ref.~\cite{SiImAr} that the relaxation rate
$1/\tau_{\varepsilon}\sim\Delta\varepsilon^2/E_T^2$ still scales as
$\propto\varepsilon^2$ with the electron energy.\footnote{This result
  can be obtained by considering the second term of
  Eq.~(\ref{eq:2.54}) in second order perturbation theory.} 
The quadratic dependence of the inelastic relaxation rate on the the
electron energy reflects the smallness of the phase volume of the
final state, and, therefore, is the same for the quantum dots and for
the clean Fermi liquid.  The proportionality coefficient is, however,
much larger than its  value for the clean system, which reflects the stronger
correlation of one-particle states of the disordered dot in comparison
with the plane waves. Nevertheless, $1/\tau_\varepsilon \ll \Delta$ for
$\varepsilon \ll E_T$, which justifies the one particle approximation as
the starting point for isolated quantum dot.

The treatment of Ref.~\cite{SiImAr} neglected the discreteness of the
spectrum of the final states. The question of the validity of such an
approach was first addressed in Ref.~\cite{KamenevLevitov} where it was shown
that the approximation of the one particle density of states by the
discrete levels broadened to $\hbar/\tau_\varepsilon$ is correct only at
energies larger than $E^*\simeq\sqrt{E_T\Delta}$. At such energies,
the three particle state obtained as a result of one-particle decay
will decay further into five-particle state, etc., thus resulting in
a true decay, without the possibility of a return to the initial
state. If the energies of the
initial state is smaller than $E^*$, the three particle state is not
strongly connected to the five particle state, so that the quantum
dot will return to the original state. In that case, the level is broadened
much less than it is found in Ref.~\cite{SiImAr}.\footnote {Clearly, in
a finite system each ``broadened'' one-electron level is in fact a
bunch of many discrete many-body levels enveloped around that
one-particle state. By the one-particle level width one means the
width of that envelope.}

Inelastic processes within the dot affect the conductance through the
dot. This is exemplified in the temperature dependence of the 
conductance of an open quantum dot, 
Eq.~(\ref{eq:70.12}). It follows from this formula, that
the weak localization correction (second term) does not depend on
temperature. This is, however, a consequence of neglecting the
inelastic process within the dot. Only due to the those processes, the
phase coherence inside the dot is destroyed.  Quantitatively, it is
described approximately \cite{BarangerMello2,AleinerLarkin,replacement}
by the replacement 
$\Nch^o+1 \to \Nch^o +
1 + 2\pi/(\tau_{\varphi}\Delta)$ with $\tau_\varphi$ 
equal to the inelastic relaxation time for an electron with energy
$T$. From the theory, one expects $\tau_\varphi \propto T^{-2}$.
Experiments with open quantum dots \cite{deph} give the result
$\tau_\varphi \propto T^{-1}$. The reasons for such discrepancy have
not been identified as of yet.

The second area of research omitted in this review is
related to the spin-spin
interactions. These effects become significant at $r_s\sim 1$, and may
result in a formation of a dot ground state with spin exceeding $1/2$.
Sufficiently strong electron-electron interaction (large $r_s$) should
result in a complete spin polarization (Stoner instability). However,
fluctuations of the density of orbital levels may result in an
anomalously large spin of a finite-size sample even at moderate values
of $r_s$.  This problem was studied in
Refs.~\cite{Oreg,Baranger,Kurlyand}. The simplest manifestation of the
anomalous spin polarization is the transition from a singlet to
triplet state in a dot when the number of electrons is even. Such a
transition can be achieved by a decrease of the spacing between the
doubly-occupied level and the next empty orbital state.
In very small quantum dots, the level separation can be controlled 
experimentally by application
of a magnetic field that affects the electron orbital
motion~\cite{Tarucha}. The singlet-triplet transition can be seen
experimentally in a measurement of the conductance through the
dot~\cite{singtrpexp} as it results in the increase of Kondo
temperature at the transition point~\cite{Eto,Pustilnik}. In larger
dots, it can
also be seen by the parametric dependence of Coulomb blockade peaks
on the magnetic field, which shows a kink at the single-triplet
transition \cite{Baranger}, or a change in the peak-to-peak 
correlations from neighboring peaks to next-neighboring peaks
\cite{Luescher}. Sofar, the effect
of the anomalous spin polarization on the conductance was
considered only in the regime of weak tunneling. The counterpart of this
effect for the strong tunneling regime has not been studied.

Finally, we have not discussed the effects of a time-dependent
change of the shape of the dot by external gates; this effect is
described by the first term in Eq.~(\ref{eq:2.54}). We mention here
only the suppression of the weak localization correction in quantum
dots and generation of a dc-current by a time depenedent perturbation
\cite{Vavilov,PumpMarcus,PumpZhou,PumpBrouwer,PumpShutenko}.

\section*{Acknowledgements}
We are grateful to Aspen Center for Physics, Max-Planck-Institute for
Complex Systems (Dresden) and Lorentz Centre at Leiden University
where a significant part of this review was written; to A. Chang,
L.P. Kouwenhoven, C.M. Marcus, and T. Costi for providing experimental
and numerical data graphs, and to A. Kaminski for the help with
preparation of the manuscript. We also acknowledge numerous
discussions of the subjects reflected in the review with our collaborators,
O. Agam, B.L. Altshuler, A.V. Andreev, H.U. Baranger,
C.W.J. Beenakker, M. B\"uttiker, B.I. Halperin, F.W.J. Hekking,
A. Kamenev, A.I. Larkin, K.A. Matveev, Yu. V. Nazarov, and M. Pustilnik.
This work was supported by A.P. Sloan and Packard fellowships at SUNY
at Stony Brook, NSF grants no.\ DMR-9416910, DMR-9630064, and
DMR-9714725 at Harvard University (where PWB has started working on
this review), and by NSF under grants no.\ DMR-9731756 and DMR-9812340
at the University of Minnesota.

\section*{Appendices}
\addcontentsline{toc}{section}{Appendices}

\appendix
\section{Correlation of the wave-functions in ballistic dots.}
\label{ap:0}

Consider a system confined by a potential $U(\vec{r})$ 
that is smooth
on the scale $a$ much larger than the 
Fermi wavelength. The Hamiltonian
function ${\cal H}(\vec{p},\vec{r})$, 
\begin{equation}
{\cal H}(\vec{p},\vec{r})=\frac{\vec{p}^2}{2m}
+
U(\vec{r}),
\label{ap0:1}
\end{equation}
serves as a
classical counterpart of the Hamiltonian of the system.
The classical diffusion
in space, Eq.~(\ref{eq:2.9}) is replaced by the classical evolution on
the energy shell ${\cal H}(\vec{p},\vec{r})=E_F$. 
This classical evolution 
is described by Perron-Frobenius equation
\begin{equation}
\left[
\frac{\partial {\cal H}}{\partial \vec{p}}
\frac{\partial}{\partial \vec{r}}
-\frac{\partial {\cal H}}{\partial \vec{r}}
\frac{\partial}{\partial \vec{p}}
-\frac{1}{\tau} \left(\vec{n}
\times \frac{\partial}{\partial\vec{n}}\right)^2
\right]f_m(\vec{n},\vec{r}) =
\gamma_m f_m(\vec{n},\vec{r}),
\label{ap0:2}
\end{equation}
where $\vec{n}$ is the unit vector in the momentum direction.
The last term on the left hand side of
Eq.~(\ref{ap0:1}) is necessary for the regularization of the
otherwise singular
eigenfunctions of the Perron-Frobenius operator. The calculation
of the quantum corrections ({\em e.g.,} weak localization) to 
the chaotic dynamics
requires a finite value of $1/\tau \simeq \hbar/(ma^2)$, 
\cite{AleinerLarkin,AleinerLarkin97}, however,
for the leading in $1/g$ approximation the limit ${\tau} \to \infty$
 should be taken  only in the end
of the calculation.

The normalization condition (\ref{eq:2.210}) is replaced with
\begin{equation}
\int  d\vec{r} d\vec{n}  f_m(\vec{n},\vec{r})  
f_{m^\prime}(-\vec{n},\vec{r}) = \Omega_d\delta_{mm^\prime},
\label{an}
\end{equation}
where $\Omega_d$ is the surface area of the unit sphere in 
$d$-dimensional space, $\Omega_2 =2\pi$, $\Omega_3 =4\pi$.

The product of two Green functions,
compare with Eq.~(\ref{eq:2.21}),
averaged over energy is given by,
see e.g. Ref.~\cite{AleinerLarkin},
\begin{eqnarray}
&&\langle
{\cal G}^R(\varepsilon +\omega; \vec{r}_1^+,\vec{r}_2^- )
{\cal G}^A(\varepsilon;
\vec{r}_1^-,\vec{r}_2^+ )
\rangle 
\label{ap0:3}\\
&&\quad =\frac{2\pi}{\Delta {\cal V}_d}
\sum_{\gamma_m}
\int \frac{d\vec{n}_1}{\Omega_d}\frac{d\vec{n}_2}{\Omega_d}
e^{ik_F\vec{n}_1\vec{r}_1}e^{ik_F\vec{n}_2\vec{r}_2}
\frac{ f_m(\vec{n}_1, \vec{R}_1)  f_m(\vec{n}_2, \vec{R}_2) }
{\left(-i\omega + \gamma_m\right)},
\nonumber\\
&& \quad
\vec{r}_{1,2}^\pm = \vec{R}_{1,2} \pm \frac{\vec{r}_{1,2}}{2}.
\nonumber
\end{eqnarray}
Repeating all the steps in the derivation of Eq.~(\ref{eq:2.27})
from Eq.~(\ref{eq:2.19}) we obtain instead of Eq.~(\ref{eq:2.27})
\begin{eqnarray}
&&\langle H_{\alpha\beta\gamma\delta}^{(1/g)}\rangle = c_1\lambda
\frac{\Delta}{g} \left(\delta_{\alpha\delta}\delta_{\beta\gamma}+
\delta_{\alpha\gamma}\delta_{\beta\delta}\right); \label{aeq:2.27}
\\
&&c_1= \lim_{\delta r \to +0 }
\frac{1}{\pi}\sum_{\gamma_m \neq 0} 
\left[
\frac{{\rm Re} \gamma_1}{\gamma_m} \int 
\frac{d\vec{n}_1}{\Omega_d}\frac{d\vec{n}_2}{\Omega_d}
\int d\vec{r}
f_m(\vec{n}_1, \vec{r}- \vec{n}_1 \delta r)f_m(\vec{n}_2, 
\vec{r})
\right],
\nonumber
\end{eqnarray}
where the dimensionless conductance of the dot $g$ is defined similarly
to Eq.~(\ref{g}),
\[
g=\frac{{\rm Re}\gamma_1}{\Delta}.
\] 
It is important to emphasize that, unlike in the diffusive case, 
the expression
for the matrix element involves also the eigenfunctions of the classical
evolution operator. This can be traced to the difference of the normaliztion
conditions  (\ref{eq:2.210}) and (\ref{an}).

Analogously, Eq.~(\ref{eq:2.28}) acquires the form,
\begin{eqnarray}
&&\langle\left[\delta H_{\alpha\beta\gamma\delta}^{(1/g)}\right]^2\rangle = 
c_2 \lambda^2 \left(\frac{\Delta}{g}\right)^2,
\label{aeq:2.28}\\
&&c_2 = \frac{2}{\pi^2}
\lim_{\delta r \to +0 }
\sum_{\gamma_m,\gamma_{m\prime}\neq 0}
\Bigg[
\left(\frac{{\rm Re}\gamma_1}{\gamma_m}\right)
\left(\frac{{\rm Re}\gamma_1}{\gamma_{m\prime}}\right)
\int
\frac{d\vec{n}_1}{\Omega_d}\frac{d\vec{n}_2}{\Omega_d}
\frac{d\vec{n}_3}{\Omega_d}\frac{d\vec{n}_4}{\Omega_d}
\nonumber\\
&& \hphantom{c_2=} \mbox{} \times
\int d\vec{r}_1 d\vec{r}_2
f_m(\vec{n}_1, \vec{r}_1- \vec{n}_1 \delta r)f_m(\vec{n}_2, 
\vec{r}_2)
f_{m\prime}(\vec{n}_3, \vec{r}_2- \vec{n}_3 \delta r)f_{m\prime}(\vec{n}_2, 
\vec{r}_1)
\Bigg]
.
\nonumber
\end{eqnarray}

\section{Effect of the Interaction in the Cooper channel}
\label{ap:1}

As we already emphasized in Sec.~\ref{sec:interaction}, the universal
 description of the interaction by Hamiltonian (\ref{eq:2.152}) is 
 valid provided
 that the coupling constants in this Hamiltonian are calculated with
 taking into account virtual transitions to the excited states which
 are beyond the energy strip in which the Hamiltonian (\ref{eq:2.152}) is
 defined. For a weak short range
 interaction, these transitions only insignificantly change the
 interaction constants $\Ec$ and $J_S$ for the charge and spin
 channels, respectively.
 The situation is different for the Cooper channel, where the
 renormalization is significant even for a weak interaction.

\begin{figure}
\epsfxsize=0.6\hsize
\centerline{\epsffile{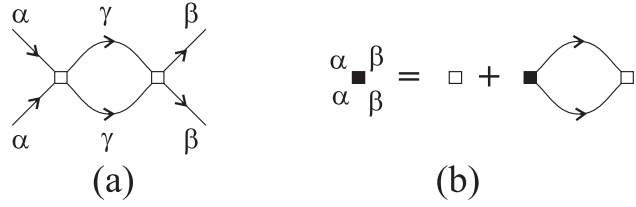}}
\caption{Diagrammatic expansion for the renormalization of the matrix
elements in the Cooper channel: (a) The first order correction;
(b) summation of the leading logarithm series for the renormalized coupling
constant. Intermediate states $\gamma$ are only those lying outside
the energy strip $E_T$.
}
\label{figap1}
\end{figure}

Consider the lowest order perturbation theory correction to the 
interaction matrix element
$\delta M_{\alpha\alpha\beta\beta}$ shown in Fig.~\ref{figap1}a.

The analytic expression for this correction is
\begin{equation}
\delta {\cal H}_{\alpha\alpha\beta\beta} \approx
- \sum_{|\varepsilon_{\gamma,\delta}| > E_T}
\frac{{\cal H}_{\alpha\alpha\gamma\delta}{\cal H}_{\delta\gamma\beta\beta}}
{|\varepsilon_\gamma + \varepsilon_\delta|}
\theta(\varepsilon_\gamma \varepsilon_\delta).
\label{eq:ap1}
\end{equation}
According to the structure of the matrix elements discussed in 
Sec.~\ref{shortrange}, the largest contribution comes from the terms
with $\gamma=\delta$, and one finds
\[
\delta {\cal H}_{\alpha\alpha\beta\beta} \approx -
 \Delta^{-1}\langle {\cal H}_{\alpha\alpha\beta\beta} \rangle^2 
\ln \left(\frac{\varepsilon^*}{E_T}\right),
\]
where $\varepsilon^*$ is a high energy cut-off above which the
approximation of the instantaneous interaction and equidistant
spectrum becomes non-applicable. Because of the large logarithmic 
factor here, the renormalization of the coupling constants in the 
Cooper channel
is significant,  even if the interaction is weak,
$\langle {\cal H}_{\alpha\alpha\beta\beta} \rangle \ll \Delta$.

To take into account this renormalization, one has to
sum the leading logarithmically divergent series shown in
Fig.~\ref{figap1} b. This results in the renormalization of the
interaction constant in comparison with its bare value,
\begin{equation}
J_c \to \tilde{J}_c= \frac{J_c}
{1+\frac{J_c}{\Delta} \ln\left(\frac{\varepsilon^*}{E_T}\right)}.
\label{eqap2}
\end{equation}

One can see, that for repulsive interaction, $J_c >0$, the
effective interaction in the Cooper channel is always weak, and for the
majority of effects it can be neglected form the very beginning. For
the attractive interaction, $J_c < 0$, the interaction constant
renormalizes to strong coupling limit and diverges at the energy scale
$\Delta_c = \varepsilon^* e^{-\Delta/|J_c|}$, which is nothing but the
superconducting gap in BCS theory. If the inequaltiy $\Delta_c < E_T$
holds, one can still use Eq.~(\ref{eq:2.152}) for the description of 
low-energy physics of a dot; in
particular, effects of finite level spacing on superconductivity can be 
accounted for in the formalism described here.
On the other hand, if $\Delta_c > E_T$, the random matrix 
description for the interaction effects is not applicable.

\section{Scattering states and derivation of 
Eq.\ (\protect\ref{eq:2.60})}
\label{Ap:5}

In this appendix we discuss the precise definitions of the
scattering states $\psi_j(k)$ in the lead Hamiltonian $H_L$ 
of Eq.\ (\ref{eq:2.57}) and the derivation of the relation 
(\ref{eq:2.60})
between the scattering matrix $S$ and the matrices $W$, $H$,
and $U$ parameterizing the Hamiltonian of the dot-lead system.
Our discussion follows that of Refs.\ 
\cite{LewenkopfWeidenmueller,MahauxWeidenmueller,IWZ,GMW,NishiokaWeidenmueller}.

\begin{figure}
\epsfxsize=0.5\hsize
\hspace{0.25\hsize}
\epsffile{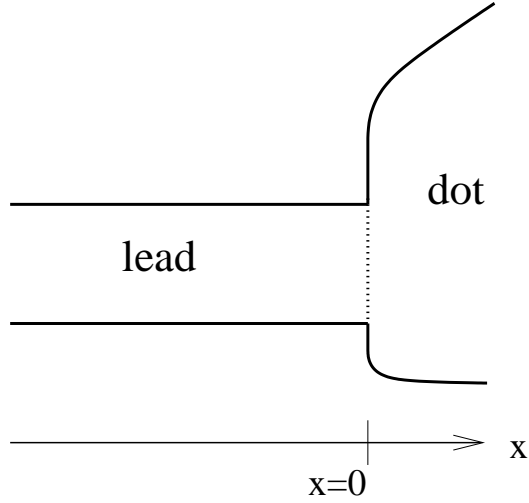}
\caption{Schematic picture of the dot with one of the leads.}
\label{Fig7}
\end{figure}

We illustrate a part of the Hamiltonian (\ref{eq:2.57})
corresponding to {\em one} of the leads in Fig~\ref{Fig7}. For each
particular lead, we take the direction towards the dot as positive
$x$, and locate the lead-dot interface near $x=0$. 
The electron operators then take the form
\begin{eqnarray}
  \hat{\psi}_e(\vec{r}) = \sum_{j=1}^{\Nch}  \Phi_j(\vec{r}_\perp) 
  \sum_{l=1}^{\Nch}
  \int \frac{dk}{2 \pi} 
  \left[ U_{jl}^{*}\, e^{i (k_F + k) x} +
  U_{jl}^{\vphantom{nothing}}\, e^{-i (k_F + k) x} \right] \hat \psi_{j}(k),
\label{eq:psielectron}
\end{eqnarray}
where $\Phi_j(\vec{r}_\perp)$ characterizes the structure of
the wave function in the transverse direction.
The unitary matrix $U_{jl}$ describes the scattering in the 
absence of any coupling to the quantum dot
\cite{LewenkopfWeidenmueller,NishiokaWeidenmueller}. 
If the contact between 
the lead and the dot is defined as in Fig.\ \ref{Fig7}, with
Dirichlet boundary conditions at the lead-dot interface, electrons
pick up a phase shift $\pi$ at the lead-dot interface at $x=0$, but
otherwise remain in the same transverse mode, hence $U_{jl} = i 
\delta_{jl}$ in that case.

In the absence of electron-electron interactions in the dot, 
the wave function in the leads at the energy
$\varepsilon = v_F k$ acquires the form 
\begin{eqnarray}
{\psi}_e (\vec{r}) = 
\sum_{j} \Phi_j(\vec{r}_\perp)
\left[a_j^{\rm in}
e^{i\left(k_F + \varepsilon/v_F \right)x} 
+ a_j^{\rm out} e^{-i\left(k_F + \varepsilon/v_F \right)x}
\right]
. 
\label{eq:2.59}
\end{eqnarray}
The amplitudes of the out-going waves $a_j^{\rm out}$ are related 
to the amplitudes of in-going waves $a_j^{\rm in}$ by the
$\Nch\times \Nch$ scattering matrix $\Scat(\varepsilon)$,
\begin{eqnarray}
a_i^{\rm out}= \sum_{j=1}^{\Nch}\Scat_{ij}(\varepsilon)a_j^{\rm in}.
  \label{eq:ainaoutb}
\end{eqnarray}
The matrix $S$ is unitary, as required by particle conservation.
In order to express the scattering 
matrix $S$ in terms of the Hamitonian of the closed dot, the
coupling matrix $W$ and the unitary matrix $U$,
we represent the Hamiltonian
(\ref{eq:2.56}) in terms of a Schr\"odinger equation for the
wavefunction $\psi_j(x)$, which corresponds to the Fourier
transform of the fermion fields $\hat \psi(k)$ over the 
entire real axis,
\begin{equation}
  \hat \psi_j(x) = \int {dk \over 2 \pi}
  \hat \psi_j(k) e^{-i k x},\ \ 
  j = 1,\ldots,\Nch.
\end{equation}
The operator $\hat \psi_j(x)$ is a mathematical construction
that is related, but not identical to the true electron operator
$\hat \psi_e(\vec r)$ defined in Eq.\ 
(\ref{eq:psielectron}). 
Then, denoting the wavefunction inside
the dot as $\phi_{\mu}$, we find from Eqs.\ (\ref{eq:2.54a}) and
(\ref{eq:2.56})--(\ref{eq:2.58}), that the Schr\"odinger equation
for the entire system reads
\begin{eqnarray}
  \varepsilon \psi_{j}(x) &=&
  i v_F {\partial \psi_j(x) \over \partial x} +
  \sum_{\nu=1}^{M} W_{\nu j}^{*} \delta(x) \phi_{\nu}, \nonumber \\
  \varepsilon \phi_{\mu} &=&
  \sum_{\nu=1}^{M} {\cal H}_{\mu\nu} \phi_{\nu} + 
  \sum_{l=1}^{\Nch} W_{\mu l} \psi_{l}(0).
  \label{eq:SchrodPsi}
\end{eqnarray}
For $x \neq 0$, Eq.\ (\ref{eq:SchrodPsi}) corresponds to a left
moving particle at velocity $v_F$. To find the scattering matrix 
$\Scat_{jl}$, we set, cf.\ Eq.\ (\ref{eq:2.59}), 
\begin{equation}
  \psi_{j}(x) = 
  \left\{ \begin{array}{ll}  \displaystyle
  e^{-i k x} \sum_{l=1}^{\Nch} U_{jl}^{\vphantom{nothing}} \,
  a^{\rm in}_{l}, & \ \ \ \ x > 0, \\ \displaystyle
  e^{-i k x} \sum_{l=1}^{\Nch} U_{jl}^{*} \,
  a^{\rm out}_{l}, & \ \ \ \ x < 0, \end{array} \right.
  \label{eq:psiansatz}
\end{equation}
where $k = \varepsilon/v_F$. At $x=0$ we use the standard 
regularization $\psi_j(0) = [\psi_j(-0) + \psi_j(+0)]/2$.
Substitution of Eq.\ (\ref{eq:psiansatz}) into the 
Schr\"odinger equation (\ref{eq:SchrodPsi}) yields
\begin{eqnarray*}
  \varepsilon \phi_{\mu} &=& \sum_{\nu} H_{\mu\nu} \phi_{\nu}
  + {\textstyle {1 \over 2}} \sum_{j,l=1}^{\Nch} W_{\mu j} 
  (U_{jl}^{\vphantom{nothing}} 
  a_{l}^{\rm in} + U_{jl}^{*} a_{l}^{\rm out}),  \\
  0 &=& i v_F (U_{jl}^{\vphantom{nothing}} 
  a_{l}^{\rm in} - U_{jl}^{*} a_{l}^{\rm out}) + \sum_{\nu} 
    W_{\nu j}^{*} \phi_{\nu}.
\end{eqnarray*}
The wavefunction $\phi_{\mu}$ of the dot can be eliminated from this
equation, resulting in the condition (\ref{eq:ainaoutb}) for the amplitudes
$a^{\rm in}_{j}$ and $a^{\rm out}_{j}$, 
with the $\Nch \times \Nch$ matrix $S$ given by
\begin{eqnarray*}
  \Scat(\varepsilon) = U \left[ 1 - 
  2\pi i \nu W^{\dagger}
  \left({\varepsilon -{\cal H}
  +i\pi\nu WW^\dagger}\right)^{-1} W \right] U^{\rm T}.
\end{eqnarray*}
This is Eq.\ (\ref{eq:2.60}) of the main text.

The coupling matrix $W$ can be represented in the form
\begin{eqnarray}
W = \sqrt{\frac{\Delta M}{\pi^2 \nu }} V O \tilde{W},
\label{eq:2.61}
\end{eqnarray} 
where $V$ is an orthogonal $M \times M$ matrix, $\tilde{W}$ is a real
$\Nch\times \Nch$ matrix, and $O$ is an $M\times \Nch$ projection
matrix, $O_{\alpha j}=\delta_{\alpha j}$, $\alpha=1,\ldots,M$ and
$j=1,\ldots,\Nch$. The first factor in
Eq.~(\ref{eq:2.61}) ensures a convenient normalization of the matrix
$\tilde{W}$. Because the distribution function of the random matrix
${\cal H}$ is invariant under rotations ${\cal H} \to V{\cal
H}V^\dagger$, the matrix $V$ in Eq.~(\ref{eq:2.61}) can be omitted.
The matrix $\tilde{W}$ can be related to the $(2\Nch)\times
(2\Nch)$ scattering matrix $\Scat_c$ of the point contacts between the
lead and the dot. 
Unlike the scattering matrix $\Scat(\varepsilon)$ which connects the
incoming and outgoing states for the entire device, the matrix $\Scat_c$
characterizes the properties of each point contact connecting a lead
to the dot separately. 
The corresponding transmission and reflection
amplitudes are defined in $\Nch\times \Nch$ matrices $r_c$,
$r'_c$, and $t_c$,
\begin{eqnarray}
\Scat_c = \left(\matrix{
r_c\quad & t^{\rm T}_c \cr
t_c\quad & r^\prime_c
}
\right),
\label{eq:2.62}
\end{eqnarray}
which can be expressed  \cite{Brouwer95} 
in terms of the coupling matrix $\tilde W$ and
the matrix $U$ of Eq.\ (\ref{eq:psielectron}),
\begin{eqnarray}
r_c &=& U \frac{1- \tilde{W}^{\dagger}\tilde{W}}{1+
\tilde{W}^{\dagger}\tilde{W}} U^{\rm T},
\quad r_c^\prime =-\tilde{W}\frac{1-
\tilde{W}^{\dagger}\tilde{W}} {1+ \tilde{W}^{\dagger}\tilde{W}}{\tilde
W}^{-1}, \quad t_c=
\tilde{W}\frac{2}{1+\tilde{W}^\dagger\tilde{W}} U. \nonumber \\
\label{eq:2.620}
\end{eqnarray}
The matrix $r_c$ expresses the reflection from a point contact to the
leads, $r_c^\prime$ corresponds to the reflection of an electron
entering the point contact from the dot, and the matrix $t_c$ describes the
transmission through the point contact from a lead to the dot. A
contact is called ideal if $r_c = r_c' = 0$, or, equivalently, $\tilde
W^{\dagger} \tilde{W}=1$.

\section{Mesoscopic fluctuations of elastic co-tunneling far from the peaks}
\label{ap:2}

The calculation of the magnetic field dependence of the mesoscopic
fluctuations of elastic co-tunneling is more complicated than in
the pure orthogonal and unitary ensembles, as
in the crossover between the orthogonal and unitary ensembles
the electron and hole amplitudes become correlated. This is because 
the wavefunctions for different
energies are not independent in the crossover regime.
The corresponding correlation functions are obtained from
 Eqs.~(\ref{eq:3.2.8}) and (\ref{eq:3.2.18}),
\begin{eqnarray}
\label{eq:a2.1}
\langle F_eF_h^\ast \rangle=
\frac{\Delta}{E_e+E_h}
\left[
\Lambda_1\left(\frac{N_h^D\Delta}{2\pi E_h}, 1+\frac{E_h}{E_e}\right)
+
\Lambda_1\left(\frac{N_h^D\Delta}{2\pi E_h}, 1+\frac{E_e}{E_h}\right)
\right],
\\
\langle F_eF_h \rangle=
\frac{\Delta}{E_e+E_h}
\left[
\Lambda_1\left(\frac{N_h^C\Delta}{2\pi E_h}, 1+\frac{E_h}{E_e}\right)
+
\Lambda_1\left(\frac{N_h^C\Delta}{2\pi E_h}, 1+\frac{E_e}{E_h}\right)
\right]
,
\nonumber
\end{eqnarray}
where the dimensionless functions $ \Lambda_1(x)$ is given by
\begin{eqnarray}
  \Lambda_1(x,y) &=&\frac{1}{\pi \left(1+x^2\right)}
\left\{
\frac{\pi^2x}{6} + \left(-\frac{\pi x}{2}+\ln xy\right)\arctan xy
- x {\mathrm Li}_2(1-y)
\right.\nonumber\\ 
&&
- \left.\frac{1}{2}\left(x\ln x + x +\frac{\pi}{2}\right)
\ln \left(1+x^2y^2\right) 
- {\mathrm Re}
\left[
(i+x) {\mathrm Li}_2(-ixy)
\right] 
\right\}, \nonumber \\
\label{eq:a2.2}
\end{eqnarray}
with ${\mathrm Li}_2(x)$ being the second polylogarithm function
\cite{Ryzhik}. The asymptotic behavior of functions 
$\Lambda_1$ is $\Lambda_1(x,y)=(xy/\pi)\ln (xy)$ for $x\ll 1$, and
$\Lambda_1(x,y)=\left[\frac{1}{2}\ln^2 (xy) -\ln^2
x\right]/(\pi x)$ for $x\gg 1$. The limits of the pure orthogonal
(unitary) ensembles (\ref{eq:3.2.12}) are recovered by putting
$N_h^D=0$ and $N_h^C=0(\infty)$ in Eq.~(\ref{eq:a2.1}).

The correlation function of the conductances at different values
of the magnetic field is somewhat involved,
\begin{eqnarray}
\frac{\langle\delta G(B_1)\delta G(B_2) \rangle}
{\langle G\rangle^2}
&=&\left\{
\frac{E_h}{E_e+E_h}
\Lambda\left(\frac{N_h^D\Delta}{2\pi E_e}\right)
\right. \label{eq:3.2.21}
\label{eq:a2.3}
\\
&& \mbox{} + 
\left.
\frac{E_hE_e}{\left(E_e+E_h\right)^2}
\Lambda_1\left(\frac{N_h^D\Delta}{2\pi E_h}, 1+\frac{E_h}{E_e}\right)
+ E_e \leftrightarrow E_h
\right\}^2 
\nonumber \\ && \mbox{} + N_h^D\leftrightarrow N_h^C .
\nonumber
\end{eqnarray}
In this general case, where the gate voltage can be anywhere in the
Coulomb blockade valley, it is impossible to point out the single scale
for the magnetic field. However, in the vicinity of the peaks, $E_e
\gg E_h$ or $E_e \ll E_h$, the situation simplifies significantly, and
one obtains the universal dependence (\ref{eq:3.2.22}).

The dependence of the crossover parameter $\lambda$ from
Eq.~(\ref{eq:3.2.24}) on the magnetic flux $\Phi$ in our case is
\begin{eqnarray}
\lambda&=&\left\{
\frac{E_h}{E_e+E_h}
\Lambda\left(\frac{N_h^C\Delta}{2\pi E_e}\right)
+ \frac{E_hE_e}{\left(E_e+E_h\right)^2}
\Lambda_1\left(\frac{N_h^C\Delta}{2\pi E_h}, 1+\frac{E_h}{E_e}\right)
\right. \nonumber \\ && \left. 
\vphantom{\left(\frac{N_h^C\Delta}{2\pi E_e}\right)}
  \mbox{} 
+  E_e \leftrightarrow E_h
\right\},
\label{eq:3.2.25}
\end{eqnarray}
where the functions $\Lambda$, $\Lambda_1$ are defined in
Eqs.~(\ref{eq:3.2.20}) and (\ref{eq:a2.2}), and $N_h^C$ is defined in
Eq.~(\ref{eq:2.101}) with $\Phi_1=\Phi_2 = \Phi$. As before, there
is no universal dependence on the magnetic field.  In the vicinity of
the peak, however, we recover the universal result (\ref{eq:3.2.26}).

\section{Derivation of the Hamiltonian (\protect\ref{ex})}
\label{Ap:model}

To include the second junction we should modify Hamiltonian given in
Eqs.~(\ref{Hr}), (\ref{HC}) and (\ref{eq:5.1.5}): in the charging energy,
Eq.~(\ref{HC}), the variable $\theta_\rho(0)$ should be replaced by
the difference $\theta_\rho(0)-\theta_\rho(L)$; we should also return
from Eq.~(\ref{eq:5.1.5}) to the generic form (\ref{H0}) of
the free-field Hamiltonian, and replace the boundary condition
Eq.~(\ref{boundary}) by a barrier described by the Hamiltonian
\begin{equation}
\hat{H}_{r1}=-\frac{2}{\pi}|r_1|D\cos[2\sqrt{\pi}\theta_\rho(L)]
           \cos[2\sqrt{\pi}\theta_s(L)],
\label{Hr1}\\
\end{equation}
which is similar to Eq.~(\ref{Hr}).  Similar to the one-junction case
[cf. Eq.~(\ref{eq:thetarhoavg})], in the energy range $E\ll\Ec$ the
fluctuations involving the total charge of the dot are pinned,
\begin{equation}
\frac{2e}{\sqrt{\pi}}\langle[\theta_\rho(0)-\theta_\rho(L)]\rangle_q
=e{\cal N},
\label{eq:thetarhoavg1}
\end{equation}
and fluctuations of the field $\theta_\rho$ have the same phase
at both junctions. If, on the other hand, $E\gg\Delta$, this is the
only constraint on the fluctuations of the charge and spin fields at
two different junctions. Therefore, in the energy domain $\Ec\gg
E\gg\Delta$, there are three independent modes left: the spin modes
$\theta_s$ and $\theta_{s1}$ at each of the junctions, and one charge
mode $\tilde{\theta}_\rho$ corresponding to passing of the charge
across the entire dot. It is more compact to describe the system in
terms of the effective action~\cite{Kane}, rather than a Hamiltonian,
\begin{eqnarray}
S&=&\sum_{i\omega_n}|\omega_n|\left[4|\tilde{\theta}_\rho(\omega_n)|^2+
2|\theta_s(\omega_n)|^2+2|\theta_{s1}(\omega_n)|^2\right]
\nonumber\\
&+&4\left[\frac{e^{-{\bf C}}}{\pi^3}\Ec D\right]^{\frac{1}{2}}
\int d\tau\left\{|r|\cos[\pi{\cal N}-2\sqrt{\pi}\tilde{\theta}_\rho(\tau)]
\cos[2\sqrt{\pi}\theta_s(\tau)]\right.
\nonumber\\
&+&\left.
|r_1|\cos[2\sqrt{\pi}\tilde{\theta}_\rho (\tau)]
\cos[2\sqrt{\pi}\theta_{s1}(\tau)]\right\},
\label{S}
\end{eqnarray}
here $\tau$ is the imaginary time, and $\omega_n$ is the Matsubara
frequency. The spin degrees of freedom at each junction are described
by $g=1/2$ Luttinger liquids, and the mode corresponding to the
remaining charge degree of freedom, is a Luttinger liquid with
$g=1/4$; its increased stiffness comes from the low-energy constraint
Eq.~(\ref{eq:thetarhoavg1}) of the charge neutrality of the dot. It is
easy to check that because of this increased stiffness [{\it i. e.},
because of ($g<1/2$)], the reflection at the point contacts is a
relevant perturbation.\footnote{The quadratic part of the effective
  action (\ref{S}) represents the fixed point of a four-channel Kondo
  problem\cite{Furusaki95}, and the reflection represents relevant
  perturbation near the fixed point.} If the bare reflection
amplitudes $|r|$ and $|r_1|$ are small, then with the reduction of the
energy scale $E$ they grow as $(\Ec/E)^{1/4}$, independently of each
other. We intend to consider the asymmetric geometry: the bare
reflection amplitudes are very different: $|r|\ll |r_1|$. Then, the
crossover to weak tunneling through the $x=L$ junction is either
occurs upon reaching the energy scale $\sim \Ec |r_1|^4$, or tunneling
is weak from the very beginning. At energies $E\lesssim \Ec |r_1|^4$,
the variables $\theta_{s1}$ and $\tilde{\theta}_\rho$ are fixed most
of the time, so that
$\cos(2\sqrt{\pi}\tilde{\theta}_\rho)\cos(2\sqrt{\pi}\theta_{s1})=1$,
to ensure the energy minimum of the junction. The system makes rare
hops between various configurations satisfying the energy minimum
condition. Upon further reduction of the band width, the effective
backscattering at $x=0$ follows the renormalization prescribed by
Eq.~(\ref{Ubs}), and reaches the limit of the weak tunneling at
energies $\sim \Ec |r_1r\cos\pi{\cal N}|^2$. This energy scale
coincides with Eq.~(\ref{T0ex}) if $|r_1|\sim 1$, and replaces it if
$|r_1|$ is small. Note that similar to the condition (\ref{deltato}),
the spatial quantization of the electron states in the dot does not
affect the crossover to the regime of weak tunneling, if
$\Delta\ll\Ec|r_1r\cos\pi{\cal N}|^2$. In that regime, the system
performs rare hops between the minima of the ``potential'' 
\begin{eqnarray}
U\{\theta_s,\theta_{s1},\tilde{\theta}_\rho\}=
&-&{\cal U}\cos(\pi{\cal N}-2\sqrt{\pi}\tilde{\theta}_\rho)
\cos(2\sqrt{\pi}\theta_s)
\nonumber\\
&-&
{\cal U}_1\cos(2\sqrt{\pi}\tilde{\theta}_\rho)
\cos(2\sqrt{\pi}\theta_{s1}),
\label{eq:potential}
\end{eqnarray}
represented by the last two lines in the action (\ref{S}). [Here the
renormalization-dependent energies ${\cal U}>0$ and ${\cal U}_1>0$
replace the corresponding bare values entering in Eq.~(\ref{S}).] In
the minima of the potential (\ref{eq:potential}), the spin of the dot
$(\sqrt{\pi}/2)(\theta_s-\theta_{s1})$ is integer or half-integer,
depending on the value of ${\cal N}$, cf.  Eq.~(\ref{uuuu}). There are
three types of hops which connect the minima of the potential
(\ref{eq:potential}). Two of the types,
$\theta_s(0)\to\theta_s(0)\pm\sqrt{\pi}$ and
$\theta_{s1}(0)\to\theta_{s1}(0)\pm\sqrt{\pi}$, are already familiar
to us from the single-junction case, see Equation~(\ref{Hpm}) and the
discussion which follows that Equation. These hops correspond to a
change by $1$ of the $z$-projection of the dot's spin via spin
exchange with one of the leads. The third type of hops involves a
simultaneous change of all three fields:
\[\theta_s(0)\to\theta_s(0)\pm\sqrt{\pi}/2,\,\,\,
\theta_{s1}(0)\to\theta_{s1}(0)\pm\sqrt{\pi}/2,\,\,\,
\tilde{\theta}_\rho(0)\to\tilde{\theta}_\rho(0)\pm\sqrt{\pi}/2.\] 
In such a hop, one electron (charge $e$ and spin $s_z=\pm 1/2$) is
transferred across the dot; the spin of this electron and the dot's spin
may flip in this process.  The corresponding tunneling Hamiltonian,
\begin{equation}
\hat{H}^{0L}_\pm\propto -
\cos\left(\!\frac{\sqrt{\pi}}{2}\tilde{\phi}_\rho^\pm\!\right)
\cos\left(\!\frac{\sqrt{\pi}}{2}\phi_{s1}^\pm\!\right)
\cos\left(\!\frac{\sqrt{\pi}}{2}\phi_{s}^\pm\!\right),
\label{Hpm2}
\end{equation}
supplements Eq.~(\ref{Hpm}) and a similar equation for the $x=L$
junction,
\begin{equation}
\hat{H}^{L}_{\pm}\propto
-\cos\left\{\sqrt{\pi}[\phi_{s1}(+0)-\phi_{s1}(-0)]\right\}.
\label{HpmL}
\end{equation}
Here $\phi_{s1}^\pm=\phi_{s1}(x=L+0)-\phi_{s1}(x=L-0)$ and
$\tilde{\phi}_\rho^\pm=\tilde{\phi}_\rho(x=+0)-\tilde{\phi}_{\rho}(x=-0)$
are the discontinuities of the respective fields at the junction (for
the field $\tilde{\phi}_\rho$ the discontinuity is the same at both
junctions). All three Hamiltonians, Eqs.~(\ref{Hpm2}), (\ref{HpmL})
and Eq.~(\ref{Hpm}), have the same scaling exponent, as can be easily
checked with the help of the action (\ref{S}). These three
Hamiltonians represent the easy-plane part of the exchange
interaction. The $SU(2)$ symmetry guarantees the existence of terms
like Eq.~(\ref{Hz}), which restore the isotropy of the exchange. The
exchange interaction changes the spin of the dot by an integer. If we
account for the finite level spacing, then the lowest energy states
would correspond to the smallest possible spin of the dot, like in the
one-junction geometry considered above. Kondo effect develops only if
the spin of the dot is $1/2$. This doublet state is realized in the
dot periodically with the gate voltage, when $\cos\pi{\cal N}<0$ (we
assume here $|r_1|\gg |r|$). In Subsection \ref{sec:exchange})
we concentrate on the doublet state
only. Returning to the fermionic variables at $E\ll
\Ec|r_1r\cos\pi{\cal N}|^2$, we find the exchange Hamiltonian
(\ref{ex}) which generalizes Eq.~(\ref{exchange}) to the case of two
junctions.

\section{Canonical versus grand canonical ensembles}
\label{Ap:3}

A crucial step in the effective action formalism is the redefinition
(\ref{eq:30.40}) of the charge operator $\hat n$ in terms of the
fermion operators in the leads. As was discussed below Eq.\
(\ref{eq:30.40}), strictly speaking, such a redefinition 
requires a canonical description of the entire system, {\em i.e.},
that the total number of particles in the
leads and the quantum dot is kept fixed in the ensemble.
A grand-canonical approach, as in Sec.\ \ref{sec:action},
can be used only when all observables are considered at a single 
value of the gate voltage ${\cal N}$, or when physical observables 
are a periodic function
of ${\cal N}$. For a non-periodic ${\cal N}$-dependence ({\em i.e.}, for
mesoscopic fluctuations of the differential capacitance), a canonical
description has to be used.
In this section, we discuss implementation of the
canonical constraint in the grand-canonical theory.
In particular, we show, 
within the effective action formalism, that for $T \gg \Nch \Delta$
or $\Nch \gg 1$, the canonical constraint
can be accounted for in the grand-canonical description by the 
dependence (\ref{eq:muN}) of the chemical 
potential $\mu$ on the gate voltage ${\cal N}$. 

To go between the canonical and grand-canonical 
descriptions, we use the inverse Laplace transform that
relates the thermodynamic potential $\Omega_{N_p}$ in a canonical
ensemble with $N_p$ particles to the thermodynamical potential
$\Omega(\mu)$ in a grand-canonical ensemble with chemical potential
$\mu$,
\begin{eqnarray}
e^{-\Omega_{N_p}/T} =
\mbox{Tr}_{N_p}\, e^{-\hat{H}/T} &=&
  \int_{-\pi T i}^{\pi T i}\frac{d\mu}{2\pi i T}\,
  \mbox{Tr}\, e^{-(\hat{H} - \mu \hat{N_p})/T} e^{\mu N_p/T}
\label{eq:muint}   
\end{eqnarray}
Here $\mbox{Tr}_{N_p}$ denotes a trace over all states with a fixed 
number of particles $N_p$, $\mbox{Tr}\,$ is a trace over all
states with all numbers of particles, and $\hat N_p$ is the operator
for the total number of particles.

Hence, what we need is the dependence of $\Omega$ on a complex
chemical
potential $\mu$ in a strip $\pi T < \mbox{Im}\, \mu < \pi T$. The
modification of the effective action theory of Sec.\ \ref{sec:action}
to include a complex chemical
potential is straightforward, because all intermediate steps are
analytic in $\mu$ for $\pi T < \mbox{Im}\, \mu < \pi T$. In the 
final result, the $\mu$-dependence appears through the $\mu$-dependence
of the scattering matrix, $\Scat(\mu,t) \to \Scat(0,t) e^{-i \mu t}$ and
through the $\mu$-dependence of the thermodynamic potential of
the lead. To find the latter, we include $\mu$ into the charging
energy, 
\begin{eqnarray}
  \hat{H}_{\rm eff} - \mu {\hat N}_{p,L} &=&  
  i v_F \sum_{j}\int_{-\infty}^{\infty}dx
  \left(\hat{\psi}^\dagger_{L,j}\partial_x\hat{\psi}_{L,j} -
  \hat{\psi}^\dagger_{R,j}\partial_x \hat{\psi}_{R,j}\right)
  \nonumber\\ && 
  \mbox{} + {\Ec} \left(\sum_{j}\int_{-\infty}^{0} dx
  :\hat{\psi}^\dagger_{L,j}\hat{\psi}_{L,j} + 
  \hat{\psi}^\dagger_{R,j}\hat{\psi}_{R,j}: 
  + {\cal N}' 
  \right)^2  
  \nonumber \\ && \mbox{} 
  - {\mu^2 \over 4 \Ec}
  - \mu {\cal N} - \mu \langle {\hat N}_{p,L} \rangle_q ,
  \label{eq:Heff2}
\end{eqnarray}
where ${\cal N}' = {\cal N} + \mu/(2 \Ec)$ and ${\hat N}_{p,L}$
is the number of particles in the leads. Hence, by the second line
of Eq.\ (\ref{eq:Heff2}), wherever we found an
explicit ${\cal N}$-dependence, ${\cal N}$ should be replaced by 
${\cal N} + \mu/(2 \Ec)$. The term $\mu \langle \hat{N}_{p,L}
\rangle_q$ denotes the quantum mechanical average of the number of
particles in the lead and arises due to the normal ordering of 
the $\hat{\psi}$-fields.

Let us now discuss the implication of this scheme for the calculation
of the mesoscopic fluctuations of the differential capacitance, 
that were the subject of the Subsection ``Non-periodic interaction 
corrections to the capacitance, $K_\mu(s)$''.
Using Eq.\ (\ref{eq:OmegaAperiodic}) and upon inclusion of the
$\mu$-dependent contributions
to $\Omega$ from Eq.\ (\ref{eq:Heff2}) and the $\mu$-dependence of 
the scattering matrix, we find that the $\mu$-dependence of the
thermodynamic potential of the entire system is given by 
\begin{eqnarray}
  \Omega(\mu) &=& \Omega(0) 
  - {\mu^2 \over 4 \Ec} - \mu {\cal N} -
  \mu \langle N_p \rangle
  \nonumber \\ && \mbox{} 
  - T 
  \int_0^{\infty} dt \int_0^{\infty} 
  dt' { (t+t') [1 - i \mu( t-t') - e^{-i \mu (t-t')}] 
    \mbox{Tr}\, \Scat^{\dagger}(t') \Scat(t)
    \over 
    4 (t-t')\sinh[\pi T(t-t')]}
  \nonumber \\ && \mbox{}
  + {1 \over 2} {T^2} \sin {\pi \over \Nch}
    \int_0^{\infty} dt \int_0^{\infty} dt'
    \int_{t_0}^{\infty} ds
    {\mbox{Tr}\, \Scat(t) \Scat^{\dagger}(t') [1-e^{-i \mu(t-t')}] \over 
    \sinh[\pi T(t+s)] \sinh[\pi T(t'+s)]}
  \nonumber \\ && \ \ \mbox{} 
    \times \left\{ 
    {\sinh[\pi T(s-t_0)] \sinh[\pi T(s+t+t'+t_0)] \over
    \sinh[\pi T(t+t_0)] \sinh[\pi T(t' + t_0)]} 
  \right\}^{{1 / \Nch}},
  \label{eq:OmegaAperiodicApp}
\end{eqnarray}
where $\langle \hat{N}_p \rangle$ is the expectation number of the total 
number of particles in the system (leads and dot)
at $\mu=0$. Mesoscopic fluctuations of $\Omega$ are dominated by
the fluctuations of the second line in Eq.\
(\ref{eq:OmegaAperiodicApp}), which are small as $N \Delta/T$
or $1/\Nch$ for $T \gg N \Delta$ or $\Nch \gg 1$, respectively.
Hence, to find the leading $\mu$-dependence of $\Omega(\mu)$,
one can replace $\Omega$ by its ensemble average, for which
one finds,\footnote{In this case, the approximate formula
(\protect\ref{eq:Savg}) for the average of the scattering matrix
gives a result that is wrong by a factor $1 + (2 - \beta)/
\beta \Nch^{o}$, because time scales $\sim \Nch \Delta$,
for which the approximation (\protect\ref{eq:Savg}) does not
hold, are important. Exact evaluation 
makes makes use of the relation
$\langle (2 \pi i)^{-1} \mbox{Tr}\, \Scat^{\dagger} (\partial S/\partial
\varepsilon)\rangle = \Delta^{-1}$.}
\begin{eqnarray}
  \langle \Omega(\mu) \rangle &=& \langle \Omega(0) \rangle 
  - {\mu^2 \over 4 \Ec} - \mu {\cal N} -
  \mu \langle N_p \rangle - {\mu^2 \over 2 \Delta}.
  \label{eq:Omegamu}
\end{eqnarray}
With this,
the $\mu$-integration in Eq.\ (\ref{eq:muint}) is straightforward
when we choose $N_p = \langle {\hat N}_p \rangle$; for $T \gg \Nch
\Delta$ the integration can be done by the saddle point method 
and amounts to the substitution (\ref{eq:muN}).
Interaction and fluctuation corrections to $\Omega$ can be treated
systematically by an expansion of $\exp(-\Omega/T)$ around the
average (\ref{eq:Omegamu}). The
differential capacitance (\ref{eq:Cdiff}) follows from
Eq.\ (\ref{eq:OmegaAperiodicApp}) by twofold 
differentiation to ${\cal N}$.

\section{Correlation functions}
\label{Ap:cor}

{\em Correlators of fermionic operators.} --
To complete the review of the properties of the fermionic correlation
function, we give the general formulas allowing one to compute any
non-vanishing average of the fermionic operators:

\begin{eqnarray}
&&\left\langle T_\tau
\prod_{j=1}^{\Nch}
\left[
\prod_{k=1}^{n_j^L+m\Nch}
\hat{\bar{\psi}}_{L,j}(\tau_{jk}^{\bar L})
\prod_{k=1}^{n_j^L}
{\hat{\psi}}_{L,j}(\tau_{jk}^L)
\prod_{k=1}^{n_j^R}
\hat{\bar{\psi}}_{R,j}(\tau_{jk}^{\bar R})
\prod_{k=1}^{n_j^R+m\Nch}
{\hat{\psi}}_{R,j}(\tau_{jk}^R)
\right]
\right\rangle_q=
\nonumber\\
&&\mbox{}
\frac{1}{(2\pi v_F)^{N_f}}
\left(\frac{1}{f(0)t_0}\right)^{|m|\Nch}
e^{-i2 \pi m \left( {\cal N } +\Nch/4\right)}
\prod_{j=1}^{\Nch}
\left[
\prod_{k>l}^{n_j^L+m\Nch}
\frac{\sin \pi T(\tau_{jk}^{\bar L}-\tau_{jl}^{\bar L} )}{\pi T}
\right.
\nonumber\\
&&\mbox{}
\times
\left.
\prod_{k>l}^{n_j^L}
\frac{\sin \pi T(\tau_{jk}^{L}-\tau_{jl}^{L} )}{\pi T}
\prod_{k>l}^{n_j^R}
\frac{\sin \pi T(\tau_{jk}^{\bar R}-\tau_{jl}^{\bar R} )}{\pi T}
\prod_{k>l}^{n_j^R +m\Nch}
\frac{\sin \pi T(\tau_{jk}^{R}-\tau_{jl}^{R} )}{\pi T}
\right.
\nonumber\\
&&\mbox{}
\times
\left.
\prod_{k=1}^{n_j^L+m\Nch}
\prod_{l=1}^{n_j^L}
\frac{\pi T}{\sin \pi T(\tau_{jk}^{\bar L}-\tau_{jl}^{L} )}
\prod_{k=1}^{n_j^R}
\prod_{l=1}^{n_j^R+m\Nch}
\frac{\pi T}{\sin \pi T(\tau_{jk}^{\bar R}-\tau_{jl}^{R} )}
\right]
\nonumber\\
&&\mbox{}
\times
\prod_{i,j=1}^{\Nch}
\left\{
\prod_{k=1}^{n_j^L+m\Nch}
\prod_{l=1}^{n_i^R+m\Nch}
\left[f(\tau_{il}^{R}- \tau_{jk}^{\bar L})
\right]^{1/\Nch}
\prod_{k=1}^{n_j^L}
\prod_{l=1}^{n_i^R}
\left[{f(\tau_{il}^{\bar R}- \tau_{jk}^{L})}
\right]^{1/\Nch}
\right. 
\nonumber\\
&&\mbox{}
\times
\left.
\prod_{k=1}^{n_j^L+m\Nch}
\prod_{l=1}^{n_i^R}
\left[\frac{1}{f(\tau_{il}^{\bar R}- \tau_{jk}^{\bar L})}
\right]^{1/\Nch}
\prod_{k=1}^{n_j^L}
\prod_{l=1}^{n_i^R+m\Nch}
\left[\frac{1}{f(\tau_{il}^{R}- \tau_{jk}^{L})}
\right]^{1/\Nch}
\right\},
\nonumber\\
\label{eq:50.130}
\end{eqnarray}
where $n_j^{L,R}$ are  arbitrary non-negative integers, $m$ is an
arbitrary integer, $N_f={\Nch m+\sum_j(n_j^L+n_j^R)}$,
and we use the convention $\prod_{j=1}^N a_j =1$, for $N=0$,
and $\prod_{j=1}^N a_j =0$, for $N<0$.
The function $f(\tau )$ and the time scale $t_0$ are defined by 
Eq.~(\ref{eq:50.10}).
One can easily check that Eqs.~(\ref{eq:50.11}), (\ref{eq:50.12})
(\ref{eq:50.13}) follow from Eqs.~(\ref{eq:50.130}).

{\em Correlators of fermionic operators and current operator.} --
For the calculation of the two-terminal conductance, we need
correlators that involve both the fermionic operators and the
current operator $I$, which is linear in the boson fields,
\begin{equation}
  I = {e v_F \over 2 \pi} {\partial \over \partial x}
        \left.
        \sum_{j} \Lambda_{jj}
        \left[ \hat{\varphi}_{Lj}(x) - \hat{\varphi}_{Rj}(x)\right]
        \right|_{x \uparrow 0}, \label{eq:Iphi}
\end{equation}
Here the diagonal matrix $\Lambda$ is defined in Eq.\ (\ref{P}).
The necessary bosonic correlators are obtained from those of the
free boson fields $\hat{\varphi}$ by substitution $\tau \to \tau \pm ix/v_F$, 
where the $+$ ($-$) sign is for left (right) movers. For the 
correlators of the current operators one thus finds
\begin{eqnarray}
  && \langle T_{\tau} I(\tau) I(0) \rangle_q =
\frac{1}{2\pi}
\left(
\frac{e^2}{2\pi}
\right)
\frac{N_1N_2}{\Nch}
  \sum_{\pm}
  { \pi^2T^2  \over  \sin^2[\pi T(\tau \pm i 0)]}
  .
  \label{eq:II}
\end{eqnarray}
The correlators of the current operators and two or four
fermionic operators are most conveniently expressed in the
correlators of the fermionic operators themselves,

\begin{eqnarray}
\lefteqn{\langle T_\tau I (\tau )
I (0)
\prod_{i=1}^{N_{\bar{L}}}
\hat{\bar{\psi}}_{L,n_i}(\tau_i^{\bar{L}})
\prod_{j=1}^{N_{{L}}}
\hat{{\psi}}_{L,n_j}(\tau_j^{{L}})
\prod_{k=1}^{N_{\bar{R}}}
\hat{\bar{\psi}}_{R,n_k}(\tau_k^{\bar{R}})
\prod_{l=1}^{N_{{R}}}
\hat{{\psi}}_{R,n_l}(\tau_l^{{R}})
\rangle_q=} \nonumber\\
&&\mbox{}
\langle T_\tau 
I (\tau )
I (0)
\rangle_q
\langle T_\tau
\prod_{i=1}^{N_{\bar{L}}}
\hat{\bar{\psi}}_{L,n_i}(\tau_i^{\bar{L}})
\prod_{j=1}^{N_{{L}}}
\hat{{\psi}}_{L,n_j}(\tau_j^{{L}})
\prod_{k=1}^{N_{\bar{R}}}
\hat{\bar{\psi}}_{R,n_k}(\tau_k^{\bar{R}})
\prod_{l=1}^{N_{{R}}}
\hat{{\psi}}_{R,n_l}(\tau_l^{{R}})
\rangle_q
\nonumber\\
&&\mbox{}
+
\frac{1}{2\pi}\left(\frac{e^2}{2\pi}\right)
\langle T_\tau
\prod_{i=1}^{N_{\bar{L}}}
\hat{\bar{\psi}}_{L,n_i}(\tau_i^{\bar{L}})
\prod_{j=1}^{N_{{L}}}
\hat{{\psi}}_{L,n_j}(\tau_j^{{L}})
\prod_{k=1}^{N_{\bar{R}}}
\hat{\bar{\psi}}_{R,n_k}(\tau_k^{\bar{R}})
\prod_{l=1}^{N_{{R}}}
\hat{{\psi}}_{R,n_l}(\tau_l^{{R}})
\rangle_q
\nonumber\\
&&\mbox{}
\times
\left[
\sum_{i=1}^{N_{\bar{L}}}
 g_{n_i}(\tau_i^{\bar{L}}-\tau)
-
\sum_{j=1}^{N_{{L}}}
g_{n_j}(\tau_j^{{L}}-\tau)
+
\sum_{k=1}^{N_{\bar{R}}}
g_{n_k}(\tau-\tau_k^{\bar{R}})
-
\sum_{l=1}^{N_{{R}}}
g_{n_l}(\tau-\tau_l^{{R}})
\right]
\nonumber\\
&&\mbox{}
\times
\left[
\sum_{i=1}^{N_{\bar{L}}}
 g_{n_i}(\tau_i^{\bar{L}})
-
\sum_{j=1}^{N_{{L}}}
g_{n_j}(\tau_j^{{L}})
+
\sum_{k=1}^{N_{\bar{R}}}
g_{n_k}(-\tau_k^{\bar{R}})
-
\sum_{l=1}^{N_{{R}}}
g_{n_l}(-\tau_l^{{R}})
\right];
\nonumber\\
&& g_n(\tau)\equiv \Lambda_{nn}{ \pi T}{ \cot[\pi T(\tau  + i 0) ]},
\label{eq:Iall}
\label{eq:IIphi4}
 \label{eq:IIphiR}
 \label{eq:IIphiL}
\end{eqnarray}
where $N_{\bar{L}},\ N_{\bar{L}}, \ {N_{\bar{R}}},\ \ N_{{R}} $
are the arbitrary non-negative integers, and $n_i =1,\dots, \Nch$
labels the channels. The expressions for the average of the
fermionic operators
were given earlier, see Eqs.~ (\ref{eq:50.11}) -- (\ref{eq:50.13}),
and (\ref{eq:50.130}).

{\em Correlators of fermionic operators and the charging operators
$\hat F$, $\hat F^{\dagger}$.} --
Calculation of the tunneling density of states Eqs.~(\ref{eq:40.9}) --
(\ref{eq:40.11}) involves not only the correlation function of the fermionic
operators but also the products of the operators $\hat{F}, \ \hat{F}^\dagger$
that increase or decrease the charge of the dot by one electron.
The corresponding calculation is facilitated by the observation that
\[
\hat{\bar{F}}(\tau) \hat{F}(0) =e^{\hat{H}_{\rm eff}\tau}
\hat{F}^\dagger e^{-\hat{H}_{\rm eff}\tau} \hat{F}=
T_\tau \exp\left\{\int_0^\tau d \tau_1
\left[\hat{H}_{\rm eff} - \hat{F}^\dagger \hat{H}_{\rm eff}(\tau_1) 
  \hat{F}\right]
\right\}, 
\]
where the one dimensional Hamiltonian is given by Eq.~(\ref{eq:40.13})
and the commutation relation for $\hat{F}, \ \hat{F}^\dagger$ are
given by Eq.~(\ref{eq:40.6}). In bosonized representation one finds
\begin{eqnarray}
\hat{H}_{\rm eff} - \hat{F}^\dagger \hat{H}_{\rm eff} \hat{F}
  &=& \frac{\Ec}{\pi}
\left\{\sum_j\left[
\hat{\varphi}_{L,j}(0) +
\hat{\varphi}_{R,j}(0) 
\right]
+ 2\pi\left({\cal N}-\hat{m} -1/2\right)
\right\}.
  \nonumber \\
\label{eq:string}
\end{eqnarray}
Because this term is linear in the bosonic fields, further calculation
proceeds without any difficulty with the help of Eq.~(\ref{eq:50.8}).
The operator $\hat{m}$ in Eq.~(\ref{eq:string}) does not have 
its own dynamics and can be fixed to be any integer, $e^{i2\pi\hat{m}}=1$.

We obtain
\begin{equation}
\langle T_\tau 
\hat{\bar{F}}(\tau )
\hat{{F}}(0)
\rangle_q = \left|\frac{f(\tau)}{f(0)}\right|^{2/\Nch},
\label{eq:str1}
\end{equation}
where the dimensionless function $f(\tau)$ is defined in
Eq.~(\ref{eq:50.9}).

For the averages involving fermionic operators as well as the charge changing
operators we find [compare with Eq.~(\ref{eq:50.11})],
\begin{eqnarray}
\lefteqn{(2\pi v_F)\langle T_\tau 
\hat{\bar{F}}(\tau )
\hat{{F}}(0)\hat{\bar{\psi}}_{L,i}(\tau_1){\hat{\psi}}_{L,j}(\tau_2)
\rangle_q = }
  \nonumber \\ \ \ && \delta_{ij}
\left|\frac{f(\tau)}{f(0)}\right|^{2/\Nch}
\frac{\pi T}{\sin [\pi T(\tau_1-\tau_2)]}
\left(\frac{f(\tau-\tau_1)f(-\tau_2)}
{f(-\tau_1)f(\tau-\tau_2)}
\right)^{1/\Nch},
\label{eq:str2}\\
\nonumber
\lefteqn{(2\pi v_F)\langle T_\tau 
\hat{\bar{F}}(\tau )
\hat{{F}}(0)\hat{\bar{\psi}}_{R,i}(\tau_1){\hat{\psi}}_{R,j}(\tau_2)
\rangle_q = } \nonumber \\ && \delta_{ij}
\left|\frac{f(\tau)}{f(0)}\right|^{2/\Nch}
\frac{\pi T}{\sin [\pi T(\tau_1-\tau_2)]}
\left(\frac{f(\tau_1-\tau)f(\tau_2)}{f(\tau_1)f(\tau_2-\tau)}
\right)^{1/\Nch}. \nonumber
\end{eqnarray}
Notice that the structure of this expression is equivalent to the form
factors in the problem of the orthogonality catastrophe~\cite{Nozieres69}.

Averages involving larger number of the fermionic operatos are found
in the same manner. We will give the formula expressing correlation
function involving the charging operator and the arbitrary number of
the fermionic operators in terms of the correlation function involving
the fermionic operators only:
\begin{eqnarray}
\lefteqn{\langle T_\tau \hat{\bar{F}}(\tau )
\hat{{F}}(0)
\prod_{i=1}^{N_{\bar{L}}}
\hat{\bar{\psi}}_{L,n_i}(\tau_i^{\bar{L}})
\prod_{j=1}^{N_{{L}}}
\hat{{\psi}}_{L,n_j}(\tau_j^{{L}})
\prod_{k=1}^{N_{\bar{R}}}
\hat{\bar{\psi}}_{R,n_k}(\tau_k^{\bar{R}})
\prod_{l=1}^{N_{{R}}}
\hat{{\psi}}_{R,n_l}(\tau_l^{{R}})
\rangle_q=} \nonumber\\
&&\mbox{}
\left|\frac{f(\tau)}{f(0)}\right|^{2/\Nch}
\langle T_\tau
\prod_{i=1}^{N_{\bar{L}}}
\hat{\bar{\psi}}_{L,n_i}(\tau_i^{\bar{L}})
\prod_{j=1}^{N_{{L}}}
\hat{{\psi}}_{L,n_j}(\tau_j^{{L}})
\prod_{k=1}^{N_{\bar{R}}}
\hat{\bar{\psi}}_{R,n_k}(\tau_k^{\bar{R}})
\prod_{l=1}^{N_{{R}}}
\hat{{\psi}}_{R,n_l}(\tau_l^{{R}})
\rangle_q \nonumber
\\
&&\mbox{}
\times
\left[
\prod_{i=1}^{N_{\bar{L}}}
\frac{f(\tau - \tau_i^{\bar{L}})}{f( - \tau_i^{\bar{L}})}
\prod_{j=1}^{N_{{L}}}
\frac{f( - \tau_j^{{L}})}{f(\tau - \tau_j^{{L}})}
\prod_{k=1}^{N_{\bar{R}}}
\frac{f(\tau_k^{\bar{R}}-\tau)}{f(  \tau_k^{\bar{R}})}
\prod_{l=1}^{N_{{R}}}
\frac{f(  \tau_j^{{R}})}{f( \tau_j^{{R}} -\tau)}
\right]^{1/\Nch}
\nonumber\\
\label{eq:strall}
\end{eqnarray}
where $N_{\bar{L}},\ N_{\bar{L}}, \ {N_{\bar{R}}},\ \ N_{{R}} $
are the arbitrary non-negative integers, and $n_i =1,\dots, \Nch$
labels the channels. The expressions for the correlation functions of
fermionic operators
were given earlier, see Eqs.~ (\ref{eq:50.11}) -- (\ref{eq:50.13}),
and (\ref{eq:50.130}).

\section{Derivation of Eqs.~(\protect\ref{eq:70.2}) --
(\protect\ref{eq:70.5})}
\label{Ap:4}

Here we present the details of the calculation of the current-current
correlator $\Pi(\tau)$ and the two-terminal conductance $G$ up to
second order in the effective action $\Action$.

To zeroth order in the interaction, $\Pi(\tau)$ is given by 
Eq.\ (\ref{eq:II}).
The correction $\delta \Pi(\tau)^{(1)}$ to first order in the 
action follows from Eqs.\ (\ref{eq:50.11}) and (\ref{eq:IIphiR}) as 
\begin{eqnarray}
  \delta \Pi(\tau)^{(1)} &=&
  \langle T_{\tau} I(\tau) I(0)\rangle_0 
  \langle \Action \rangle_0 -
  \langle T_{\tau} I(\tau) I(0) \Action \rangle_0
  \nonumber \\ &=&
  {e^2 T^3 \over 8 i} \sum_{\pm} 
  \int_0^{1/T} d\tau_1 \int_0^{1/T} d\tau_2
       {\mbox{Tr}\, S(\tau_1 - \tau_2) \Lambda^2 \over 
    \sin [\pi T (\tau_1 - \tau_2)]}
   \nonumber \\ && \mbox{} \times
    \left\{ \cot[\pi T (\tau_1 - \tau \pm i 0)]
           -\cot[\pi T (\tau_1 - \tau \pm i 0)] \right\}
  \nonumber \\ && \mbox{} \times
    \left\{ \cot[\pi T (     \tau_1 \pm i 0)]
           -\cot[\pi T (     \tau_2 \pm i 0)] \right\}.
\end{eqnarray}
We change to variables $\tau_1 = \sigma' + \sigma$ and $\tau_2 =
\sigma' - \sigma$ and integrate over $\sigma'$. Since the integrand
is analytic in the upper (lower) half of the complex plane for the
$+$ ($-$) signs in the denominators, one directly finds 
\begin{equation}
  \delta \Pi(\tau)^{(1)} = 0.
\end{equation}

\begin{figure}
\epsfxsize=0.8\hsize
\hspace{0.1\hsize}
\epsffile{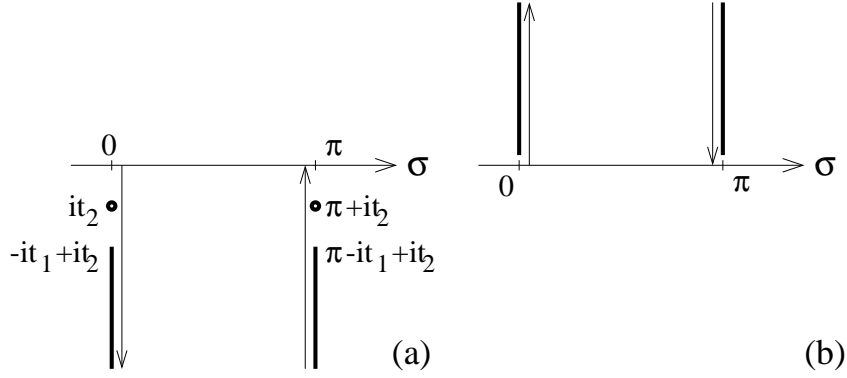}
\caption{\label{fig:sigmacur} Shift of the integration contours for
the variable $\sigma$ in Eq.\ (\protect\ref{eq:Lehmancur}). Branch
cuts of the integrand are indicated by thick lines, poles by dots.
The integration contour for first term between brackets $\{\ldots\}$
in Eq.\ (\protect\ref{eq:Lehmancur}) is shifted below the real axis
(a); the integration contour for the second term is shifted above the
real axis (b).}
\end{figure}

The second order correction to $\Pi(\tau)$ reads
\begin{eqnarray}
  \delta \Pi(\tau)^{(2)} &=&
    {1 \over 2} \langle I(\tau) I(0) \Action^2 \rangle_0 -
    {1 \over 2} \langle I(\tau) I(0) \rangle_0
    \langle \Action^2 \rangle_0.
\end{eqnarray}
Here we find for $\Nch > 2$
\begin{eqnarray}
  \delta \Pi(\tau)^{(2)} &=& 
  -{e^2 T^4 \over 16}  \sum_{\pm}
  \int_0^{1/T} d\tau_1\, d\tau_2\, d\tau_3\, d\tau_4
  \mbox{Tr}\, S(\tau_1 - \tau_2) \Lambda S(\tau_3 - \tau_4) \Lambda
  \nonumber \\ && \mbox{} \times
  {1 \over 
  \sin[(\tau_1 \pm \tau + i 0)\pi T]
  \sin[(\tau_4 \pm \tau + i 0)\pi T]}
  \nonumber \\ && \mbox{} \times
  {1 \over 
   \sin[(\tau_2 - i 0)\pi T]
   \sin[(\tau_3 - i 0)\pi T]} 
  \nonumber \\ && \mbox{} \times
    \left[
    {f(\tau_2 - \tau_1) f(\tau_3 - \tau_4) \over
     f(\tau_3 - \tau_1) f(\tau_2 - \tau_4)} \right]^{{1 / \Nch}}.
\end{eqnarray}
We left out terms that have all poles at the same
side of the real axis, because they
vanish after integration over the times
$\tau_1$, \ldots, $\tau_4$.    
We now parameterize $\tau_1$, 
$\tau_2$, $\tau_3$, and $\tau_4$ according to
\begin{eqnarray}
  \tau_1 &=& \sigma' + \sigma/2 + \sigma_1, \nonumber \\
  \tau_2 &=& \sigma' + \sigma/2 \nonumber, \\
  \tau_3 &=& \sigma' - \sigma/2 + \sigma_2, \label{eq:sigma} \\
  \tau_4 &=& \sigma' - \sigma/2 \nonumber,
\end{eqnarray}
perform the integration over $\sigma'$, and use the Lehmann
representation (\ref{eq:60.4}) for $S(\tau)$,
\begin{eqnarray}
  \delta \Pi^{(2)}(\tau)  &=& 
  {e^2 T^3 i \over 8} \sum_{\pm}
  \int_0^{\infty} d t_1
  \int_{-\infty}^{0} d t_2
  \int_0^{1/T} d\sigma 
  {\mbox{Tr}\, S(t_1) \Lambda S^{\dagger}(t_2) \Lambda
  \over
  \sin[(i t_2 - \sigma) \pi T]
  }
  \nonumber \\ && \mbox{} \times
  \left\{
  {1 \over \sin[(\pm \tau + i t_1) \pi T]
           \sin[(\pm \tau - \sigma + i 0) \pi T]}
  \right. \nonumber \\ && \left. \mbox{}
  - {1 \over \sin[(\pm \tau - i t_2) \pi T]
    \sin[(\pm \tau + \sigma + i t_1 - i t_2) \pi T]}
  \right\}
  \nonumber \\ && \mbox{} \times
  \left[ {
  f(i t_2 - i t_1 - \sigma)
  f(\sigma) 
  \over
  f(- i t_1)
  f(  i t_2) } \right]^{1/{\Nch}}
\label{eq:Lehmancur} 
\end{eqnarray}
Finally we shift the integration over $\sigma$ to the complex plane,
using the approximation (\ref{eq:approx}) for $f$. 
For the first term between brackets, we choose the integration path
below the real axis, while for the second term the integration path
is chosen above the real axis, see Fig.\ \ref{fig:sigmacur}.
The result is
\begin{eqnarray}
  \delta \Pi^{(2)}(\tau) &=&
  {e^2 T^2 \over 4} \sum_{\pm}
  \int_0^{\infty}  d t_1
  \int_{-\infty}^0 d t_2 \mbox{Tr}\, S(t_1) \Lambda S^{\dagger}(t_2) 
  \Lambda 
  \nonumber \\ && \mbox{} \times
  \left\{
  \vphantom{ \left( {\sinh \over \sinh } \right)^{1/2N}}
  {1 \over \sinh[(t_1 \pm i\tau) \pi T] 
           \sinh[(-t_2 \pm i\tau) \pi T]}
  \right. \nonumber \\ && \mbox{} + \left.
  T  \int_{t_0}^{\infty} d s\,
  { \sin(\pi/{\Nch}) \over \sinh[(s + t_1 - t_2 \pm i \tau) \pi T]}
  \right. \nonumber \\ && \ \ \  \left. \mbox{} \times
  \left[
    {1 \over \sinh[(t_1 \pm i\tau) \pi T]
             \sinh[(s + t_1) \pi T]}
  \right. \right. \nonumber \\ && \ \ \ \left. \left. \mbox{}
  + {1 \over \sinh[(- t_2 \pm i\tau) \pi T]
             \sinh[(s - t_2) \pi T]}
  \right]
  \right. \nonumber \\ && \ \ \ \left. \mbox{} \times
  \left( {
  \sinh[(t_1 - t_2 + s + t_0) \pi T]
  \sinh[(s - t_0) \pi T]
  \over
  \sinh[(  t_1 + t_0) \pi T]
  \sinh[(- t_2 + t_0) \pi T]
  } \right)^{1/{\Nch}}
  \right\}. \label{eq:dPiRLLRfinal}
\end{eqnarray}
The first term is the pole contribution, the second term contains
a contribution from the branch cut. Corrections to $\Pi(\tau)$ to
higher orders in perturbation theory can be classified in the same
way as terms with pole contributions, one branch cut, two
branch cuts, etc. As
discussed in Sec.\ \ref{sec:bosonization}, there are no 
corrections of higher order in the scattering matrix $S$ with zero 
or one branch cut --- these have been accounted for here;
all higher order corrections have more than one branch cut.
Now, using the conductance formula (\ref{eq:30.58}), 
one directly finds the interaction correction (\ref{eq:70.4}) to the 
conductance.

For $\Nch = 2$ there is an additional, periodic in ${\cal N}$,
contribution to the current-current correlator $\Pi(\tau)$ to 
second order in the action $\Action$. This origin of
this extra ${\cal N}$-dependent contribution is the fact that for
$\Nch = 2$ the product (\ref{eq:50.13}) of four different Fermion
operators has a nonzero average. We thus find
\begin{eqnarray}
  \delta \Pi(\tau)_{\rm osc} &=&
  -{e^2 T^2 e^{-2 \pi i {\cal N}} \over 64 \pi^2 t_0^2}  
  \int_0^{1/T} d\tau_1 d\tau_2 d\tau_3 d\tau_4
  \nonumber \\ && \mbox{} \times
  f(0)^{-2} \left[ {f(\tau_2 - \tau_1) f(\tau_4 - \tau_3) f(\tau_4 - \tau_1)
  f(\tau_2 - \tau_3)  }\right]^{1/2}
  \nonumber \\ && \mbox{} \times
  \left[S_{11}(\tau_1 - \tau_2) S_{22}(\tau_3 - \tau_4) 
  + S_{12}(\tau_1 - \tau_2) S_{21}(\tau_3 - \tau_4) \right]
  \nonumber \\ && \mbox{} \times
  \sum_{\pm} {\sin[(\tau_1 - \tau_3)\pi T] \over
  \sin[(\tau_1 \pm \tau + i 0)\pi T]
  \sin[(\tau_3 \pm \tau + i 0)\pi T]}
  \nonumber \\ && \mbox{} \times
  {\sin[(\tau_2 - \tau_4)\pi T] \over  \sin[(\tau_2 - i 0)\pi T]
  \sin[(\tau_4 - i 0)\pi T]}
  + {\rm c.c.}
\end{eqnarray}
[Note that the products $S_{11} S_{22}$ and $S_{12} S_{21}$ appear
with the same sign, because the latter product has an extra minus 
sign both in the fermionic correlator (\ref{eq:50.13}) and in the 
current-fermion correlators (\ref{eq:IIphi4}).]
Again, terms that vanish after integration have been left out.

We now use the parameterization (\ref{eq:sigma}) 
of the times $\tau_1$, \ldots, $\tau_4$,
integrate over $\sigma'$, and implement the
Lehmann representation (\ref{eq:60.4}), to find
 \begin{eqnarray}
  \delta \Pi(\tau)_{\rm osc} &=& 
  {e^2 T e^{-2 \pi i {\cal N}} \over 32 i \pi^2 t_0^2}
  \int_0^{\infty} dt_1 dt_2 \int_0^{1/T} d\sigma
  \left[S_{11}(t_1) S_{22}(t_2) 
  + S_{12}(t_1) S_{21}(t_2) \right]
  \nonumber \\ && \mbox{} \times
  f(0)^{-2} \left[ {f(-i t_1) f(-i t_2) f(-\sigma - i t_1)
  f(\sigma - i t_2)}\right]^{1/2}  
  \nonumber \\ && \mbox{} \times
  \sum_{\pm} \left\{ {  \sin[(\sigma + i t_1 - i t_2)\pi T]
  \over 
  \sin[(\sigma + i t_1 \pm \tau)\pi T]
  \sin[(i t_2 \pm \tau)\pi T]}
  \right. \nonumber \\ && \ \ \ \ \left. \mbox{}
  - {  \sin[(\sigma + i t_1 - i t_2)\pi T]
  \over \sin[(-\sigma + i t_2 \pm \tau)\pi T]
  \sin[(i t_1 \pm \tau)\pi T]} \right\}
  + \mbox{c.c.}
\end{eqnarray}
Next we integrate over $\sigma$ by shifting the integration contour
into the upper (lower) half of the complex plane for the first
(second) term between brackets $\{ \ldots \}$. The only
contribution is from the branch cut of the integrand, which reads,
with the approximation (\ref{eq:approx}) for $f$,
\begin{eqnarray}
  \delta \Pi(\tau)_{\rm osc} &=& 
  {e^2 T^3 e^{-2 \pi i {\cal N}} \over 16}
  \int_0^{\infty} dt_1 dt_2 \int_{t_0}^{\infty} ds
  \left[S_{11}(t_1) S_{22}(t_2) 
  + S_{12}(t_1) S_{21}(t_2) \right] 
  \nonumber \\ && \mbox{} \times
  {1 \over \sqrt{\sinh[(t_1 + t_0)\pi T] \sinh[(t_2 + t_0)\pi T]}}
  \nonumber \\ && \mbox{} \times    
  {1 \over \sqrt{\sinh[(t_1 + t_2 + t_0 + s)\pi T] 
  \sinh[(s - t_0)\pi T]} }
  \nonumber \\ && \mbox{} \times    
  \sum_{\pm} \left\{ {\sinh[(s + t_1 - t_2)\pi T] \over 
  \sin[(i s + i t_1 \pm \tau)\pi T]
  \sin[(i t_2 \pm \tau)\pi T]}
  \right. \nonumber \\ && \ \ \ \ \left. \mbox{}
  + {\sinh[(s + t_1 - t_2)\pi T]
  \over \sin[(i s + i t_2 \pm \tau)\pi T]
  \sin[(i t_1 \pm \tau)\pi T]} \right\}
  + \mbox{c.c.}
\end{eqnarray}  
Finally, using the conductance formula (\ref{eq:30.58}),
we find the oscillating contribution (\ref{eq:70.5})
to the conductance.

\section{Derivation of Eqs.~(\protect\ref{eq:80.9}) --
(\protect\ref{eq:80.90})
 }
\label{Ap:tun}

\begin{figure}
\epsfxsize=0.8\hsize
\hspace{0.1\hsize}
\epsffile{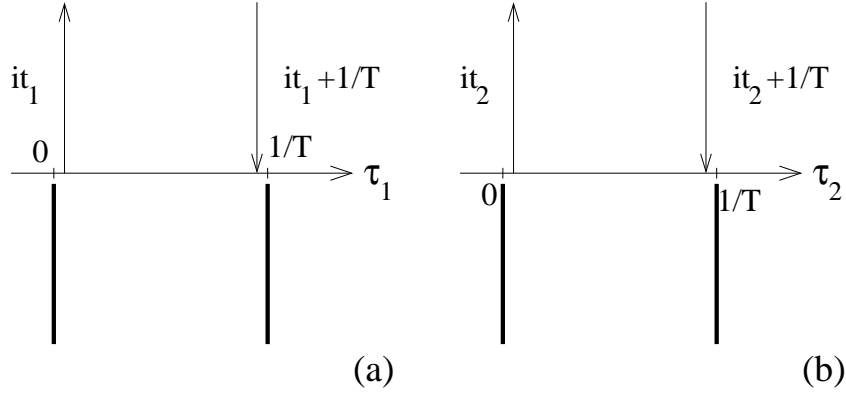}
\caption{\label{Figtun} Deformation of the integration contours for
the variables $\tau_1$ (a) and $\tau_2$ (b) in 
Eq.\ (\protect\ref{eq:80.8}). Branch
cuts coming from functions $f(\protect\tau)$ are shown by thick lines,
and the integration contour is represented by arrowed vertical thin lines.
}
\end{figure}

To obtain the leading elastic contribution
we calculate the correlator (\ref{eq:40.11}),
taking into account all the terms which include branch cut in 
the complex $\tau$
plane and the cuts
in the complex $\tau_1$ and $\tau_2$ planes in the correlation function 
(\ref{eq:str2}), and similarly in higher order correlation function,
see Eq.~(\ref{eq:strall}).
This gives \cite{AleinerGlazman98} for $\Nch \geq 2$
\begin{eqnarray}
\Pi_{\rm el}^{(1)}(\tau)&=&\frac{1}{4}
\left|\frac{f(\tau)}{f(0)}\right|^{2/\Nch}
\int_0^{1/T} d\tau_1d\tau_2 
\left[{\cal G}_o(\tau_1)\nu WW^\dagger 
{\cal G}_o(-\tau_2)
\right]_{11}
\nonumber \\ && \mbox{} \times
\frac{\pi T}{\sin \pi T (\tau+\tau_1-\tau_2)}
\left[
\left(
\frac{f(-\tau_1)f(-\tau_2)}{f(-\tau_1-\tau)f(\tau-\tau_2)}
\right)^{2/\Nch}
  \right. \nonumber \\ && \ \ \ \ \left. \mbox{}
+ \left(
\frac{f(\tau_1)f(\tau_2)}{f(\tau_1+\tau)f(\tau_2-\tau)}
\right)^{2/\Nch}
-2\right],
\label{eq:80.8}
\end{eqnarray}
where ${\cal G}_o(\tau)$ is the Matsubara Green function for the open
dot.  One can easily, see that the pole contribution at
$\tau+\tau_1-\tau_2=0$ is excluded from Eq.~(\ref{eq:80.8}), because
it is already taken into account in the inelastic contribution
(\ref{eq:80.1}).
One can now deform the contours of the integration over $\tau_1$
and $\tau_2$ 
in Eq.~(\ref{eq:80.8}) using analyticity of function $f(\tau)$ at 
${\rm Im} \tau <0$, see Fig.~\ref{Figtun}.

\begin{eqnarray}
\Pi_{\rm el}^{(1)}(\tau) &=& \frac{1}{2}
\left|\frac{f(\tau)}{f(0)}\right|^{2/\Nch}
\int_0^{\infty} dt_1dt_2 
\left[
\left(
\frac{f(-it_1)f(-it_2)}{f(-it_1-\tau)f(\tau-it_2)}
\right)^{2/\Nch} - 1 \right]
\nonumber\\
&&\mbox{}\times
{\rm Re} \left\{ \left[{\cal G}_o^R(t_1)\nu WW^\dagger 
{\cal G}_o^A(-t_2)
\right]_{11}\frac{\pi T}{\sin \pi T (\tau+it_1-it_2)}
\right\},
\label{eq:80.80}
\end{eqnarray}
where ${\cal G}_o(t)^{R,A}$ are the retarded and advanced
 Green function for the open
dot. Deriving Eq.~(\ref{eq:80.80}), we use the fact that $f(\tau+1/T )=
f(\tau)$, ${\cal G}(\tau+1/T) = - {\cal G}(\tau)$, and
${\cal G}(it+0)- {\cal G}(it-0)=i [{\cal G}^R(t)- {\cal G}^A(t)]$.

The resulting function (\ref{eq:80.80}) is analytic for $0 < \tau <
1/T$, and we can use the analytic continuation Eq.~(\ref{eq:40.5}) to
find the tunneling conductance Eq.~(\ref{eq:80.9}).

For the case of a one channel contact $\Nch=1$ (spinless fermions) the
additional contribution appears due to the anomalous average
(\ref{eq:50.13}).
\begin{eqnarray}
  G_{\rm el}^{\rm osc}(\tau)
&=&\frac{ G_1 M\Delta}{2}\left|\frac{f(\tau)}{f(0)}\right|^{2}
 \int_0^{1/T} d\tau_1d\tau_2
\frac{f(\tau+\tau_1-\tau_2)}{t_0f(0)}   
\frac{f(-\tau_1)f(\tau_2)}{f(\tau_2 -\tau)f(-\tau-\tau_1)}
\nonumber\\
&&\mbox{}\times
{\rm Re} \left\{ e^{-i2\pi{\cal N} -i\pi/2}
\left[{\cal G}_o(\tau_1)\nu WW^\dagger {\cal G}_o(-\tau_2)
\right]_{11} \right\}.
\label{eq:80.901}
\end{eqnarray} 
After deformations of the integration contours 
analogous to derivation of Eq.~(\ref{eq:80.80})
and analytic continuation  we arrive to Eq.~(\ref{eq:80.9}).

\end{document}